\documentclass[a4paper, 11pt, oneside]{Thesis}  
\graphicspath{{Figures/}}  

\usepackage[square, numbers, comma, sort&compress]{natbib}  
\usepackage{verbatim}  
\usepackage{vector}  
\usepackage{mathrsfs}
\def\met{{\slash\!\!\!\!\!\:E}_T}

\def\D0{\slash\!\!\!\!\!\!\!\!\!\:D0}
\hypersetup{urlcolor=blue, colorlinks=true}  

\begin{document}
\frontmatter	  

\title  {Phenomenology of the minimal $B-L$ Model: the Higgs sector at
  the Large Hadron Collider and future Linear Colliders}
\authors  {\texorpdfstring
            {\href{quarkonio@tiscali.it}{Giovanni Marco Pruna}}
            {Giovanni Marco Pruna}
            }
\addresses  {\groupname\\\deptname\\\univname}  
\date       {\today}
\subject    {}

\maketitle

\setstretch{1.3}  

\fancyhead{}  
\rhead{\thepage}  
\lhead{}  

\pagestyle{fancy}  

\addtotoc{Abstract}  
\abstract{
\addtocontents{toc}{\vspace{1em}}  

This Thesis is devoted to the study of the phenomenology of the Higgs
sector of the minimal $B-L$ extension of the Standard Model at present
and future colliders. Firstly, the motivations that call for the
minimal $B-L$ extension are summarised. In addition, the
model is analysed in its salient parts. Moreover, a detailed review of
the phenomenological allowed Higgs sector parameter space is
given. Finally, a complete survey of the distinctive Higgs boson
signatures of the model at both the Large Hadron Collider and the
future linear colliders is presented.

}

\clearpage  


\pagestyle{fancy}  

\lhead{\emph{Contents}}  
\tableofcontents  

\lhead{\emph{List of Figures}}  
\listoffigures  

\lhead{\emph{List of Tables}}  
\listoftables  
\Declaration{

\addtocontents{toc}{\vspace{1em}}  

I, Giovanni Marco Pruna, declare that this thesis titled,
``Phenomenology of the minimal $B-L$ Model: the Higgs sector at the
Large Hadron Collider and future Linear Colliders'' and the work
presented in it are my own. I confirm that:

\begin{itemize} 
\item[\tiny{$\blacksquare$}] This work was done wholly or mainly while
  in candidature for a research degree at this University. 
 
\item[\tiny{$\blacksquare$}] Where any part of this thesis has
  previously been submitted for a degree or any other qualification at
  this University or any other institution, this has been clearly
  stated. 
 
\item[\tiny{$\blacksquare$}] Where I have consulted the published work
  of others, this is always clearly attributed. 
 
\item[\tiny{$\blacksquare$}] Where I have quoted from the work of
  others, the source is always given. With the exception of such
  quotations, this thesis is entirely my own work. 
 
\item[\tiny{$\blacksquare$}] I have acknowledged all main sources of
  help. 
 
\item[\tiny{$\blacksquare$}] Where the thesis is based on work done by
  myself jointly with others, I have made clear exactly what was done
  by others and what I have contributed myself. 
\\
\end{itemize}

Signed:\\
\rule[1em]{25em}{0.5pt}  
 
Date:\\
\rule[1em]{25em}{0.5pt}  
}
\clearpage  

\pagestyle{empty}  

\null\vfill
\textit{``Se non \`e vero, \`e molto ben trovato.''} \\
(\textit{``If it is not true, it is very well invented.''})

\begin{flushright}
Giordano Bruno (1548-1600),\\
burnt at the stake as a heretic by the Catholic Church.
\end{flushright}

\vfill\vfill\vfill\vfill\vfill\vfill\null
\clearpage  

\setstretch{1.3}  

\acknowledgements{
\addtocontents{toc}{\vspace{1em}}  

With the oversight of my supervisor Professor Stefano Moretti,
editorial advice has been sought. No changes of intellectual content
were made as a result of this advice. For this and all the scientific
and human support that I have received in the last years, I owe my
deepest gratitude to him.

Then, I would like to thank Dr. Lorenzo Basso for sharing with me
these years of (both domestic and scientific!) difficulties and
challenges. He provided his support in many occasions, and this Thesis
would not have been possible without his help.

Moreover, I will be always grateful to Dr. Alexander S. Belyaev,
Professor Ulrich Ellwanger, Professor Per Osland, Professor Jonathan
Flynn.

This PhD has been partially supported by the NExT Institute.

Questo PhD \`e stato finanziato dal progetto ``Master and Back'' della
Regione Autonoma della Sardegna, pertanto un ringraziamento speciale
va a Renato Soru senza il quale tale risorsa sarebbe senza dubbio
mancata.

Luciano, Silvana, Francesca, e poi Daniele, Sara, Jacopo, Gabriella,
Tonino, Carlo.

A voi.

Mi traboccano dal cuore alcune parole, che si attaccano ad una carta
che forse nessuno legger\`a, come le vostre vite.

Tu, cara nonna, che vestita di nero sin da giovane hai cresciuto tre
bambine, e la prima e la pi\`u bella era mia madre. 

Tu, che ti sei consumata fino a che, quel giorno, mi chiedesti di
aiutarti a sollevare quel letto enorme, grandissimo, molto pi\`u grande
di me che avevo solo sette anni. 

Non riuscivi nemmeno a stare in piedi senza tremare, dalla vita di
merda che hai fatto. 

Tu, caro nonno, che sei dovuto scappare di casa, e andare a scavare
nelle orribili miniere iglesienti sin da quando eri bambino, per
sposarti e dare alla luce cinque bambini, ed il terzo e pi\`u bello era
mio padre. 

Tu che hai sputato i polmoni silicotici, ti sei consumato una spalla,
al punto che quando avevi gi\`a ottant'anni suonati piangevi per non
riuscire nemmeno a sollevare la mano destra per farti la barba.
 
Quando l'Alzheimer ti ha portato via, ho provato a convincermi che il
motivo del tuo sorriso era il Segreto pi\`u grande, lo stesso che mi
ruba le notti. 

Voi, che non l'avete fatto per arricchire i signorotti del Nord, con
quei ridicoli e falsi accenti e le teste vuote come cocomeri rotti al
sole, che non hanno mai imparato nulla da Noi e men che mai sono stati
in grado di insegnarci alcunch\'e (``gente di nulla''). 

Voi, che l'avete fatto per dare un futuro ai vostri figli, che a loro
volta l'hanno fatto affinch\'e quel futuro lo ereditassi io, e per fare
in modo che lo possa lasciare in eredit\`a ai miei figli. 

A voi, che mi avete insegnato come nessun titolo di studio possa mai
garantire ``quel tanto'' che rende l'uomo degno di appartenere alla sua
specie, \`e dedicata questa tesi.

Vi penso sempre.


\newpage

}
\clearpage  

\addtocontents{toc}{\vspace{2em}}  

\mainmatter	  
\pagestyle{fancy}  



\chapter{Introduction} 
\label{chap:1}
\lhead{Chapter 1. \emph{Introduction}} 

\section{The Standard Model}\label{sect:1-1}

Currently, the Standard Model ($SM$) of the electroweak ($EW$) and
strong interactions of elementary particles represents one of the
highest achievement of human kind in understanding the fundamental
laws of Nature.

The $EW$ theory is based on the gauge symmetry group $SU(2)_L \times
U(1)_Y$ of left–handed isospin and hypercharge. The Quantum
Chromo-Dynamics ($QCD$)\footnote{The theory of the strong interactions
  between the colored quarks.}, instead, is based on the symmetry
group $SU(3)_C$. These two theories provide a beautifully consistent
picture of the particle physics phenomena observed up to now.

The gauge symmetry groups of the $SM$ give raise to a quantum field
theory that is perturbative at sufficiently high energies and
renormalisable. For a complete review of the model in its deepest
theoretical and phenomenological implications, see
\cite{Peskin:1995ev,Djouadi:2005gi}.

Starting from this framework, it is widely accepted that each $SM$
particle (in particular, each massive gauge boson) obtains mass by
mean of the so-called Higgs mechanism, i.e., the mechanism of
spontaneous $EW$ symmetry breaking ($EWSB$) realised by adding a
theoretically consistent Higgs field: the minimal\footnote{It
  respects the requirement of renormalisability.} choice is
represented by the introduction of a $SU(2)$ doublet of complex scalar
fields (see \cite{Higgs:1964pj}).

After the $EWSB$, the $SU(2)_L \times U(1)_Y$ symmetry is
spontaneously broken to the electromagnetic $U(1)_{Q}$ symmetry (where
$Q$ is the electric charge quantum number). Three of the four degrees
of freedom of the doublet scalar field turn out to be absorbed in the
longitudinal polarisation component of each of the three weak gauge
bosons, $W^\pm$ and $Z$, whilst the fourth one is the physical $SM$
Higgs state $h$. At this stage, the fermion masses are generated
through the Yukawa interaction with the same scalar field and its
conjugate one.

Although the phenomenological power of this theory is remarkable (it
has been deeply tested in various experimental scenarios), there is
still no (either direct or indirect) evidence of the scalar sector
impact in the probed phenomenology, in other words the Higgs boson has
not been observed yet.

The urgency of discovering the Higgs boson (and consequently
establishing the nature of $EWSB$) has driven the Physics
community to devote major efforts to the realisation of more and more
powerful accelerators.

For this, since decades a large number of machines (as LEP, SLC,
Tevatron, etc.) has tried to investigate the $SM$ to
the high-precision measurement level (per mille accuracy), and apart
from the astonishing agreement with the theoretical prediction
concerning the fermion and vector boson sectors of the theory, the
Higgs boson (the most important ingredient of the theoretical picture)
is currently missing.

At the same time, the recent observation of the pattern of neutrino
oscillations (for a complete review, see
\cite{Bilenky:1987ty,Altarelli:2004za,Strumia:2006db}) produced the 
evidence of the inadequateness of the $SM$ (in its minimal version) in
describing the mass properties of neutrinos.

Besides, dark matter ($DM$) evidence and related cosmological
observations (see \cite{Bertone:2004pz}) gave an hard blow to the
minimal $SM$ that, as well as for the massive neutrino case, can not
be considered as a satisfactory theoretical framework for these
phenomena.

Other than by experimental facts, the $SM$ is also affected by several
theoretical problems: firstly, quantum gravity is manifestely not
included; secondly, the theory is affected by the so-called
``Hierarchy problem'' (see \cite{Gildener:1976ih,Weinberg:1978ym}).

However, while finding a solution to the former could be ``postponed''
to energy scales that are close to the Plank-scale
$\mathcal{O}(10^{19}) GeV$, the latter is an indication of consistency
problems already arising at the TeV energy scale. 

All these aforementioned theoretical and experimental reasons call for
some extension of the $SM$.

In fact, one could procede by extending the minimal $SM$ by either
top-down or bottom-up approaches. The former consist in the
formulation of a theory that solves all the $SM$ issues and can be
broken down to a symmetry pattern that finally includes the $SM$ as
effective low-energy theory: a typical example of this approach is
represented by supersymmetric scenarios as the
(Next to) Minimal Supersymmetryc Standard Model ($N$)$MSSM$, see
\cite{Martin:1997ns}(\cite{Ellwanger:2009dp}).

The latter approach consists in piecing together a system of minimal
extensions to give
rise to the grandest possible system. In the context of this
work, we choose to follow this approach in realising the so-called
minimal $B-L$ extension of the $SM$.

\section{A bottom-up approach: the minimal $B-L$ model}\label{sect:1-2}

In order to realise a consistent extension of the $SM$, we exploit the
fact that both the baryon ($B$) and the lepton ($L$) number are
conserved quantities of the theory, as well as the $B-L$
one. Furthermore, it is important to notice that the $B-L$
number could be gauged in a $U(1)$ symmetry group in combination with
an augmented neutrino sector, and this creates a model
that is free from anomalies.

Hence, the minimal $B-L$ extension of the $SM$ consists of
augmenting the $SM$ gauge groups $SU(3)_C\times SU(2)_L \times U(1)_Y$
by the aforementioned $U(1)_{B-L}$ symmetry (see
\cite{Jenkins:1987ue,Buchmuller:1991ce} and more recently
\cite{Khalil:2006yi,Emam:2007dy,Basso:2008iv,Emam:2008zz,Huitu:2008gf,Basso:2009hf,Basso:2010pe,Basso:2010jt,Basso:2010jm,Basso:2010hk,Basso:2010yz,Basso:2010si}).

This choice is minimal on three respects:
\begin{itemize}
\item it is minimal in the gauge sector, enlarging the gauge group by
  adding one spontaneously broken $U(1)$ factor, which provides one
  new (neutral) gauge boson;
\item it is minimal in the fermion sector, adding one new heavy
  neutrino per generation;
\item it is minimal in the scalar sector, adding one new complex
  Higgs field, singlet under the $SM$ gauge group.
\end{itemize}

This extension gives rise to a model:
\begin{itemize}
\item that is anomaly-free and gauge-invariant (see
  Section~\ref{sect:2-1} for details);
\item that provides an ``elegant'' way to generate the light neutrino
  masses by means of the so-called ``see-saw'' mechanism (see
  \cite{Minkowski:1977sc,VanNieuwenhuizen:1979hm,Yanagida:1979as,GellMann:1980vs,S.L.Glashow,Mohapatra:1979ia}
  and Section~\ref{sect:2-4} for details).
\end{itemize}

This rather simple approach fulfils the phenomenological requirement
of having a renormalisable theory that provides a mechanism for giving
mass to light neutrinos as well as a good candidate for $DM$
(see \cite{Okada:2010wd,Khalil:2011tb}).

In addiction, it is important to notice that $B-L$ symmetry
breaking takes place at the TeV energy scale, hence leaving open the
possibility of being part of a Grand Unified Theory ($GUT$) and giving
rise to new and interesting TeV scale phenomenology.

In this work, we focus on the phenomenology arising from the main
ingredients of $EW$ plus $B-L$
spontaneous symmetry breaking, i.e., the Higgs sector phenomenology of
the minimal $B-L$ model at present and future colliders.

\section{Experimental frameworks: present and future
  colliders}\label{sect:1-3}

In this Section we introduce the present and future machines that
could possibly allow us to test the minimal $B-L$ model.

For this, two experimental frameworks have been taken into account:
\begin{itemize}
\item the Large Hadron Collider (LHC). It is a hadron-hadron
  collider. It is the only high-energy accelerator that is currently
  working. Up to now, it is the world's largest and highest-energy
  particle accelerator ever made. 
\item The International Linear Collider (ILC) and the Compact Linear
  Collider (CLIC). They are planned to be $e^+e^-$
  colliders. They are two proposed accelerators and the projects have
  not yet been approved. If realised, they will be the largest and
  highest-energy linear particle accelerators ever made.
\end{itemize}

\subsection{The LHC}\label{subs:1-3-1}

The LHC was built by the European Organization for Nuclear Research
(CERN) with the intention of testing various predictions of
high-energy physics, with emphasis on addressing the nature of the
$EWSB$.

It is a circular collider and it is designed to investigate processes
in which the initial state is characterised by either protons or heavy
ions.

It consists of several detectors/experiments:
\begin{itemize}
\item ATLAS (A Toroidal LHC ApparatuS): together with CMS is one of
  the two general purpose detectors;
\item CMS (Compact Muon Solenoid): together with ATLAS is one of the
  two general purpose detectors;
\item ALICE (A Large Ion Collider Experiment): it is designed to
  investigate the so-called ``quark-gluon plasma'' and the
  (de)confinement scenarios;
\item LHCb (Large Hadron Collider $b$): it is mainly designed to
  investigate the nature of $CP$-violation in interactions of
  $b$-hadrons.
\end{itemize}

In this work we focus on discovery physics, i.e., we are interested
on proton-proton collisions and possible minimal $B-L$ signatures at
both ATLAS and CMS.

\subsection{Future linear collider prototypes}\label{subs:1-3-2}

Although the LHC has finally entered its operational stage, a
considerable part of the
international physics community is focusing efforts in order to plan
what the future of high-energy particle phenomenology and accelerators
could be.

Nowadays, the community is working on the proposal of two linear
accelerator prototypes: the ILC and the CLIC. Both of them represent
the new generation of electron-positron colliders.

While the LHC and its multi-purpose detectors have been built with the
aim of discovering new particles, the main goal that the new
generation of linear colliders (LCs) is supposed to achieve is the
subsequent profiling of the new physics whose evidence could arise
in the next years at the LHC.

Specifically, there are many phenomenological aspects on which the LHC
is unsufficient in with respect to either ILC or CLIC:
\begin{itemize}
\item if existing, direct measuring of mass, spin and coupling of the
  Higgs and new gauge bosons;
\item if existing, a complete profiling of any TeV scale
  extra-dimension;
\item if existing, a better understanding of any supersymmetric
  scenario and related dark matter aspects.
\end{itemize}

In particular, in this work we are only interested in the minimal $B-L$
model Higgs sector. Therefore we will consider the two linear
accelerator configurations only for the scope of probing the existence
of a Higgs boson.

However, it is important to mention the fact that, if realised, the two
accelerators will represent a unique way to precisely profile the
Higgs sector of the minimal $B-L$ model.

\section{Organisation of the work}\label{sect:1-4}

This Thesis is organised as follows:
\begin{itemize}
\item in Chapter~\ref{chap:2} we describe the minimal $B-L$ model in
  all its salient parts, with emphasis on the extended sectors, the
  $B-L$ symmetry breaking mechanism, some formal aspects related to
  the gauge-invariance and ghost fields, and the ``see-saw'' mechanism
  for generating the light neutrino masses;
\item in Chapter~\ref{chap:3} we perform a complete analysis of the
  experimental and theoretical constraints on the Higgs sector of the
  minimal $B-L$ model, focusing on the bounds on the Higgs masses
  coming from direct searches at LEP, unitarity and triviality
  arguments; then, we also present the experimental and theoretical
  constraints on the other parameters of the model and discuss the
  ``fine-tuning'' in the minimal $B-L$ model;
\item in Chapter~\ref{chap:4} we present our phenomenological results
  of the investigation of the minimal $B-L$ Higgs sector at colliders;
  we start with the study of Higgs branching ratios ($BR$), then we
  discuss some peculiar signatures at both the LHC and future LCs;
\item in Chapter~\ref{chap:5} we summarise our result and we also
  conclude our work discussing some open issues of our research;
\item in Appendix~\ref{appe:a} we list the potential of the minimal
  $B-L$ scalar effective-Lagrangian; this part is
  useful in analysing formal aspects of theory, as it will be clear in
  Section~\ref{sect:3-2};
\item in Appendix~\ref{appe:b} we list the Feynman rules of the
  minimal $B-L$ model adopting the Feynman-gauge.
\end{itemize}


\chapter{The minimal $B-L$ model} 
\label{chap:2}
\lhead{Chapter 2. \emph{The minimal $B-L$ model}} 

In this Chapter we introduce the minimal $B-L$ model by a detailed
description of the terms of its Lagrangian.

As we have already intimated, this model is an extension of the $SM$,
following the symmetry pattern:
\begin{equation}
SU(3)_C \times SU(2)_L \times
U(1)_Y \rightarrow SU(3)_C \times SU(2)_L \times U(1)_Y \times
U(1)_{B-L}.
\end{equation}

As we shall see, the charges of the two U(1) factors of the extended
gauge group are associated with the $SM$ weak hypercharge $Y$, and
with the total (Baryon − Lepton) number $B-L$\footnote{In the general
  $B-L$ model, this identification is not generally possible since the
  presence of ``mixing'' makes it basis-dependent, therefore the
  covariant derivative is generally non-diagonal. Indeed, it is always
  possible to diagonalize the covariant derivative by defining an
  ``effective charge'' $Y^P$ (with the corresponding coupling $g^P$),
  which is a linear combination of $Y$ and $B-L$, but in this case a
  more careful treatment should be done and it lie outside of our
  purposes.}.

The quantum consistency (anomaly cancellation) of the theory,
together with the phenomenological requirement of generating neutrino
masses, are fully satisfied by extending also the fermion content with
a right-handed neutrino $\nu_R$ (singlet under the $SM$ gauge group)
for each family.

Finally, the requirement of giving mass to the extra neutral gauge
boson $Z'$ is fulfilled by a minimal extension of the scalar sector
as well, with the introduction of a scalar field $\chi$, singlet under
the $SM$ gauge group but not under the $U(1)_{B-L}$ (since it carries
$B-L$ charge).

In Section~\ref{sect:2-1} we begin by identifying the model Lagrangian
in all its parts. In Section~\ref{sect:2-2} we then proceed with a
discussion of the extended Higgs mechanism, breaking $SU(2)_L \times
U(1)_Y \times U(1)_{B-L}$ down to $U(1)_{Q}$, and of the ensuing
spectrum of gauge bosons, fermions and scalars. In
Section~\ref{sect:2-3} we present a brief analysis of the minimal
$B-L$
gauge-fixing. In Section~\ref{sect:2-4} we introduce the principles of
the so-called ``see-saw'' mechanism. Finally, in
Section~\ref{sect:2-5} we conclude this Chapter with a summary of the
introduced parameters in comparison with the $SM$ case.

It is important to mention the fact that the content of this Chapter
will allow one to get all the necessary elements to implement the
minimal $B-L$ model in any Feynman-rules generator (as $LanHEP$
\cite{Semenov:1996es}, $FeynRules$ \cite{Christensen:2008py},
etc.). In Appendix~\ref{appe:b} we collect the Feynman rules of the
model that we have obtained by mean of the $LanHEP$ package.

\section{The Lagrangian and the parameterisation}\label{sect:2-1}

The minimal $U(1)_{B-L}$ extension of the $SM$ is realised by
augmenting the symmetry structure by a $U(1)$ gauge group, that is
related to the $B-L$ symmetry.

Without loss of generality, the classical gauge-invariant Lagrangian
for the class of models under consideration (obeying the $SU(3)_C
\times SU(2)_L \times U(1)_Y \times U(1)_{B-L}$) can be decomposed as:
\begin{eqnarray}\label{lag}
\mathscr{L} = \mathscr{L}_s + \mathscr{L}_{YM} + \mathscr{L}_f +
\mathscr{L}_Y ,
\end{eqnarray}
where the various terms represents the scalar, Yang-Mills, fermion and
Yukawa part, respectively.

The general structure is similar to the $SM$ Lagrangian, although (due
to the different symmetry structure) each term takes into account the
differences in the gauge, fermion and scalar sectors.

In the following, sector by sector, we will highlight analogies and
differences with respect to the $SM$ case.

\subsection{The Yang-Mills sector}\label{subs:2-1-1}

As in the $SM$, the vector fields are uniquely determined by the
choice of the gauge group, and by the transformation in their adjoint
representation. Hence, in the $\mathscr{L}_{YM}$, the non-Abelian
field strengths therein are the same (as in the $SM$) and the only
difference is contained in the Abelian terms.

Explicitely, it is formalised as follows:
\begin{equation}\label{lag:YM}
\mathscr{L}_{YM} = -\frac{1}{4} G_{\mu \nu}^{\phantom{\mu
    \nu} \alpha} G^{\mu \nu \alpha} -\frac{1}{4} W_{\mu \nu}^{\phantom{\mu
    \nu} a} W^{\mu \nu a} -\frac{1}{4} F^{\mu\nu} F_{\mu\nu} -
\frac{1}{4} F^{\prime\mu\nu} F^\prime _{\mu\nu}\, ,
\end{equation}
where
\begin{eqnarray}
\label{lag:fsB}
F_{\mu\nu} &=& \partial _{\mu} B_{\nu} - \partial _{\nu} B_{\mu},
\\
\label{lag:fsBp}
F^\prime_{\mu\nu}	&=& \partial _{\mu} B^\prime_{\nu} - \partial
_{\nu} B^\prime_{\mu} ;
\end{eqnarray}
here, $B$ and $B^\prime $ are the gauge fields associated with
$U(1)_Y$ and $U(1)_{B-L}$, respectively.

\subsection{The fermion sector}\label{subs:2-1-2}

The fermion content of the model is the same of the minimal $SM$,
except for the addition of a right-handed neutrino $\nu_R$ (singlet
under the $SM$ gauge group) for each of the three lepton families.

As we already mentioned, this addition is essential both for anomaly
cancellation and preserving gauge invariance.

Following the field-basis notation introduced in
Subsection~\ref{subs:2-1-1}, the covariant derivatives of the $B-L$
model are
defined as the usual $SM$ non-Abelian part plus an Abelian part:
\begin{equation}\label{cov_der_nm}
D_{\mu}\equiv \partial _{\mu} + i g_S \mathcal{T}^{\alpha}
G_{\mu}^{\phantom{o}\alpha} + i g T^a W_{\mu}^{\phantom{o}a} + i g_1 Y
B_{\mu} + i (\widetilde{g} Y + g_1' Y_{B-L}) B'_{\mu}.
\end{equation}

In the minimal version of the $B-L$, $\widetilde{g}=0$ is assumed
(i.e., no mixing between the two $U(1)$ factors)\footnote{It is
  important to highlight that this condition holds at the $EW$ scale
  only: if we assume the running of $\widetilde{g}$, then it will
  monotonically grow spoiling the ``minimality'' of the model (see
  Subsection~\ref{subs:3-2-2} for details).} and the covariant
derivative becomes:
\begin{equation}\label{cov_der}
D_{\mu}\equiv \partial _{\mu} + i g_S \mathcal{T}^{\alpha}
G_{\mu}^{\phantom{o}\alpha} + i g T^a W_{\mu}^{\phantom{o}a} + i g_1 Y
B_{\mu} + i g_1' Y_{B-L} B'_{\mu}.
\end{equation}

Then, the fermionic Lagrangian is given by:
\begin{eqnarray}
\label{lag:f} \nonumber
\mathscr{L}_f &=& \sum _{k=1}^3 \Big( i \overline {q_{kL}} \gamma
_{\mu} D^{\mu} q_{kL} + i\overline {u_{kR}} \gamma _{\mu} D^{\mu}
u_{kR} + i \overline {d_{kR}} \gamma _{\mu} D^{\mu} d_{kR} + \\
&\ & + i \overline {l_{kL}} \gamma _{\mu} D^{\mu} l_{kL} + i \overline {e_{kR}}
\gamma _{\mu} D^{\mu} e_{kR} + i\overline {\nu _{kR}} \gamma _{\mu}
D^{\mu} \nu _{kR} \Big)  \,,
\end{eqnarray}
where the charges of the fields are the usual $SM$ ones, plus the
$B-L$ ones:
\begin{itemize}
\item $Y_{B-L} = 1/3$ for quarks,
\item $Y_{B-L}=-1$ for leptons.
\end{itemize}

Assuming that the conjecture of ``universality'' is true, no
distinction between generations has been made.

\subsection{The scalar sector}\label{subs:2-1-3}

The model under study has an extended gauge sector, with an additional
neutral gauge boson $Z'$ with respect to the $SM$. In order to realise
a consistent Higgs mechanism (giving mass not only to the $SM$ weak
bosons but also to the $Z'$) it is necessary to enlarge the $SM$ Higgs
sector by means of a further complex Higgs singlet $\chi$.

The $B-L$ charges of the two scalar fields are set to be:
\begin{itemize}
\item $Y_{B-L}^{H}=0$,
\item $Y_{B-L}^{\chi}=+2$;
\end{itemize}
this choice is essential to preserve the gauge invariance of the
model.

The most general gauge-invariant and renormalisable scalar Lagrangian
is:
\begin{equation}\label{lag:s}
\mathscr{L}_s = \left( D^{\mu} H \right) ^{\dagger} D_{\mu} H + \left(
D^{\mu} \chi \right) ^{\dagger} D_{\mu} \chi - V(H,\chi ) \, ,
\end{equation}
with the scalar potential given by
\begin{eqnarray}\label{potential}\nonumber
V(H,\chi ) &=& m^2 H^{\dagger} H + \mu ^2 \mid \chi \mid ^2 +
\left(
\begin{array}{cc}
H^{\dagger} H & \mid \chi \mid ^2
\end{array}
\right)
\left(
\begin{array}{cc}
\lambda_1 & \frac{\lambda_3}{2} \\
\frac{\lambda_3}{2} & \lambda _2 \\
\end{array}
\right)
\left(
\begin{array}{c}
H^{\dagger} H \\ \mid \chi \mid^2 \\
\end{array}
\right) \\
\nonumber \\ 
&=& m^2 H^{\dagger} H + \mu ^2 \mid \chi \mid^2 + \lambda_1
(H^{\dagger} H)^2 + \lambda_2 \mid \chi \mid^4 + \lambda_3 H^{\dagger}
H \mid \chi \mid^2.
\end{eqnarray}

\subsection{The Yukawa term}\label{subs:2-1-4}

To complete the description of the Lagrangian in equation~(\ref{lag}),
it is necessary to perform a treatment of the Yukawa couplings.

In addition to the Yukawa couplings of the minimal $SM$, we have two
new types of Yukawa interactions involving right-handed neutrinos:
\begin{eqnarray}\label{lag:Y} \nonumber
\mathscr{L}_Y &=& -y^d_{jk} \overline {q_{Lj}} d_{Rk} H - y^u_{jk}
\overline {q_{Lj}} u_{Rk} \widetilde H - y^e_{jk} \overline {l_{Lj}}
e_{Rk} H \\
&\ & - y^{\nu}_{jk} \overline {l_{Lj}} \nu _{Rk} \widetilde H -
y^M_{jk} \overline {(\nu_R)^c_j} \nu _{Rk} \chi + {\rm h.c.},
\end{eqnarray}
where $\widetilde{H} = i \sigma^2 H^*$ and $i$, $j$, $k$ run from $1$
to $3$.

It is important to notice that the Yukawa interactions can generate
both Dirac mass terms and Majorana mass
terms for right-handed neutrinos (both of them in the secon line of
equation~(\ref{lag:Y})). As we will see
in Section~\ref{sect:2-4} these are the essential ingredients of the
``see-saw'' mechanism for giving mass to light and heavy neutrinos.

As aforementioned, it is now clear, from the terms involving the
$\chi$ scalar field, that $Y^\chi_{B-L}=+2$ is needed in order to
ensure the gauge invariance.

\section{Spontaneous $SU(2)_L \times U(1)_Y \times U(1)_{B-L}$
  breaking}\label{sect:2-2}

We generalise the $SM$ discussion of spontaneous $EWSB$ to the more
complicated case represented by the potential of
equation~(\ref{potential}).

To determine the condition for $V(H,\chi )$ to be bounded from below,
it is sufficient to study its behaviour for large field values,
controlled by the matrix in the first line of the aforementioned
equation. Requiring such a matrix to be positive-definite gives the
conditions:
\begin{eqnarray}\label{bound_pot}
4 \lambda_1 \lambda_2 - \lambda_3^2 &>& 0 , \\
\label{pos_pot}
\lambda_1, \lambda_2 &>& 0.
\end{eqnarray}

If the above conditions are satisfied, the choice of parameters is
consistent with a well-defined potential, hence we can proceed to the
minimisation of $V$ as a function of constant Vacuum Expectation
Values ($VEVs$) for the two Higgs fields.

Since the minimisation procedure is not affected by the choice of the
gauge, it is not restrictive to define the two $VEVs$ in the following
way:
\begin{equation}\label{VEVs}
\left< H \right> \equiv
\left(
\begin{gathered}
0 \\
\frac{v}{\sqrt{2}}
\end{gathered} \right), 
\hspace{2cm}
\left< \chi \right> \equiv \frac{x}{\sqrt{2}},
\end{equation} 
with $v$ and $x$ real and non-negative.

Then, the search for extrema of $V$ is made by mean of the following
differential set of equations:
\begin{equation}\label{minimisation}
\left\{
\begin{aligned}
\frac{\partial V}{\partial v}(v,x) &=& v \cdot \left( m^2 \lambda_1 v^2 +
\frac{\lambda_3^2}{2}x^2 \right)=0 \\
 \frac{\partial V}{\partial x}(v,x) &=& x \cdot \left( \mu^2 \lambda_2 x^2 +
\frac{\lambda_3^2}{2}v^2 \right)=0
\end{aligned}
\right.
\end{equation}

The physically interesting solutions are the ones obtained for $v$,
$x>0$:
\begin{eqnarray}\label{min_sol1}
v^2 &=& \frac{-\lambda_2 m^2 + \frac{\lambda_3}{2} \mu ^2}{\lambda_1
  \lambda_2 - \frac{\lambda_3^{\phantom{o}2}}{4}}, \\
\nonumber  \\ 
\label{min_sol2}
x^2 &=& \frac{-\lambda_1 \mu^2 + \frac{\lambda_3}{2} m ^2}{\lambda_1
  \lambda_2 - \frac{\lambda_3^{\phantom{o}2}}{4}}.
\end{eqnarray}

Since the denominator in equations~(\ref{min_sol1})-(\ref{min_sol2})
is always positive (assuming that the potential is well-defined), it
follows that the numerators are forced to be positive in order to
guarantee a positive-definite non-vanishing solution for $v$ and $x$.

In order to identify the extrema, we need to evaluate the Hessian
matrix:
\begin{equation}\label{hessian}
\mathcal{H}(v,x)\equiv \left(
\begin{aligned}
\frac{\partial^2 V}{\partial v^2} &\ & \frac{\partial^2 V}{\partial v
  \partial x} \\
\frac{\partial^2 V}{\partial v \partial x} &\ & \frac{\partial^2
  V}{\partial x^2}
\end{aligned}
\right) = \left(
\begin{aligned}
2 \lambda_1 v^2 &\ & \lambda_3 v x \\
\lambda_3 v x &\ & 2 \lambda_2 x^2
\end{aligned}
\right) .
\end{equation}

From this equation, it is straightforward to verify that the solutions
are minima if and only if equations~(\ref{bound_pot})-(\ref{pos_pot}) are
satisfied.

\subsection{The scalar mass spectrum}\label{subs:2-2-1}

To compute the scalar masses, one must expand the potential in
equation~(\ref{potential}) around the minima found in
equations~(\ref{min_sol1})-(\ref{min_sol2}).

Since the physical mass eigenvalues are gauge invariant, we define the
Higgs fields following the unitary-gauge prescription:
\begin{equation}\label{unit_higgs}
H \equiv
\left(
\begin{gathered}
0 \\
\frac{h+v}{\sqrt{2}}
\end{gathered} \right), 
\hspace{2cm}
\chi \equiv \frac{h'+x}{\sqrt{2}}.
\end{equation} 

After standard manipulations, the explicit expressions for the scalar
mass eigenvalues are:
\begin{eqnarray}\label{mh1}
M^2_{h_1} &=& \lambda_1 v^2 + \lambda_2 x^2 - \sqrt{(\lambda_1 v^2 -
  \lambda_2 x^2)^2 + (\lambda_3 x v)^2}, \\
\label{mh2}
M^2_{h_2} &=& \lambda_1 v^2 + \lambda_2 x^2 + \sqrt{(\lambda_1 v^2 -
  \lambda_2 x^2)^2 + (\lambda_3 x v)^2},
\end{eqnarray}
where $h_1$ and $h_2$ are the scalar fields of definite masses
$M_{h_1}$ and $M_{h_2}$ respectively, and we conventionally choose
$M^2_{h_1} < M^2_{h_2}$.

These eigenvalues are related to the following eigenvectors:
\begin{equation}\label{eigenstates}
\left(
\begin{array}{c}
h_1 \\
h_2
\end{array}
\right) = \left(
\begin{array}{cc}
\cos{\alpha} & -\sin{\alpha} \\
\sin{\alpha} & \cos{\alpha}
\end{array}
\right) \left(
\begin{array}{c}
h \\
h'
\end{array}
\right),
\end{equation}
where $-\frac{\pi}{2} \leq \alpha \leq \frac{\pi}{2}$
fulfils\footnote{In all generality, the whole interval $0 \leq \alpha
  < 2\pi$ is halved because an orthogonal transformation is invariant
  under $\alpha \rightarrow \alpha + \pi$.}:
\begin{eqnarray}\label{sin2a}
\sin{2\alpha} &=& \frac{\lambda_3 x v}{\sqrt{(\lambda_1 v^2 -
    \lambda_2 x^2)^2 + (\lambda_3 x v)^2}}, \\
\label{cos2a}
\cos{2\alpha} &=& \frac{\lambda_2 x^2 - \lambda_1
  v^2}{\sqrt{(\lambda_1 v^2 - \lambda_2 x^2)^2 + (\lambda_3 x v)^2}}.
\end{eqnarray}

For completeness, it is useful to write the isomorphic transformation
between the two $\lambda_1$-$\lambda_2$-$\lambda_3$ and
$M_{h_1}$-$M_{h_2}$-$\alpha$ spaces.

From equations~(\ref{mh1})-(\ref{mh2})-(\ref{sin2a}), it is straightforward
to have:
\begin{eqnarray}\label{isomorphism}
\lambda_1&=&\frac{M_{h_1}^2}{2 v^2} + \frac{ \left(
M_{h_2}^2 - M_{h_1}^2 \right)}{2 v^2}\sin^2{\alpha} =
\frac{ M_{h_1}^2}{2v^2}\cos^2{\alpha} +  \frac{M_{h_2}^2}{2
v^2}\sin^2{\alpha}\nonumber \\
\lambda_2&=&\frac{M_{h_1}^2}{2 x^2} + \frac{ \left(
M_{h_2}^2 - M_{h_1}^2 \right)}{2 x^2}\cos^2{\alpha} =
\frac{ M_{h_1}^2}{2x^2}\sin^2{\alpha} +  \frac{M_{h_2}^2}{2
x^2}\cos^2{\alpha} \nonumber \\
\lambda_3&=&\frac{ \left(
M_{h_2}^2 - M_{h_1}^2 \right)}{ 2vx}\sin{(2\alpha)}.
\end{eqnarray}

\subsection{The gauge mass spectrum}\label{subs:2-2-2}

To determine the gauge boson spectrum, we have to expand the scalar
kinetic terms as in the $SM$ case. From consistency arguments, we
expect that there exists a massless gauge boson (the photon) and the
other gauge bosons acquiring mass.

Moreover, we expect that the $SM$ charged vector boson spectrum is not
affected by the extension (since it involves the non-Abelian part
only). As for the Abelian vector bosons, using the unitary-gauge
parameterisation, we write the kinetic terms in equation~(\ref{lag:s})
as:
\begin{eqnarray}\nonumber
&\ &\left( D^{\mu} H \right) ^\dagger D_{\mu} H = \nonumber \\
&=& \frac{1}{2} \partial
^{\mu} h \partial_{\mu} h + \frac{1}{8} (h+v)^2 \big(0\; 1 \big) \Big[
  g W_a^{\phantom{o} \mu} \sigma_a + g_1 B^{\mu} \Big]^2
\left(
\begin{array}{c}
0 \\
1
\end{array}
\right) \nonumber \\
\label{boson_spec1}
&=& \frac{1}{2} \partial^{\mu} h \partial_{\mu} h + \frac{1}{8}
(h+v)^2 \left[ g^2 \left| W_1 ^{\phantom{o} \mu} - i W_2^{\phantom{o}
    \mu} \right|^2 + \left( g W_3^{\phantom{o} \mu} - g_1 B^{\mu}
  \right) ^2 \right],
\end{eqnarray}
and
\begin{eqnarray}\label{boson_spec2}
\left( D^{\mu} \chi \right) ^\dagger D_{\mu} \chi &=& \frac{1}{2}
\partial^{\mu} h' \partial_{\mu} h' + \frac{1}{2} (h'+x)^2 (g_1' 2
B'^{\mu})^2.
\end{eqnarray}

In equation~(\ref{boson_spec1}) we can easily identify the $SM$ charged
gauge bosons $W^\pm$, with $\displaystyle M_W=gv/2$.

As for the other fields, the vanishing mixing between the two $U(1)$
factors (i.e., the assumption $\widetilde{g}=0$) allow us to
immediately identify both the $SM$-like pieces and the new vector
boson $Z'\equiv B'$.

Once we set the $B^\mu$, $W_3^\mu$ and $B'^\mu$ as field basis, the
explicit expression for the squared mass matrix is:
\begin{eqnarray}
\mathcal{M}^2&=&\frac{v^2}{4}
\left(
\begin{matrix}
g^2_1 & -g g_1 & 0 \\
-g g_1 & g^2 & 0 \\
0 & 0 & 16\frac{x^2}{v^2}(g'_1)^2
\end{matrix}
\right) \nonumber \\
\label{boson_mass}
&=& \frac{v^2}{4}(g^2+g^2_1)
\left(
\begin{matrix}
\sin^2{\vartheta_w} & -\cos{\vartheta_w}\sin{\vartheta_w} & 0 \\
-\cos{\vartheta_w}\sin{\vartheta_w} & \cos^2{\vartheta_w} & 0 \\
0 & 0 & \frac{16 x^2 (g'_1)^2}{v^2 (g^2+g^2_1)}
\end{matrix}
\right),
\end{eqnarray}
where we have made use of the well-known relations:
\begin{equation}
\cos{\vartheta_w}=\frac{g}{\sqrt{g^2+g_1^2}}, \qquad
\sin{\vartheta_w}=\frac{g_1}{\sqrt{g^2+g_1^2}}.
\end{equation}

As in the $SM$ case, if we want to diagonalise the $2\times 2$
sub-matrix in equation~(\ref{boson_mass}), we need to apply a rotation
along the $B'$($Z'$) field, defined by:
\begin{equation}\label{rotation}
R_{EW}(\vartheta_w)=
\left(
\begin{matrix}
\cos{\vartheta_w} & -\sin{\vartheta_w} & 0 \\
\sin{\vartheta_w} & \cos{\vartheta_w} & 0 \\
0 & 0 & 1
\end{matrix}
\right),
\end{equation}
and this allow us to isolate each mass eigenvalue:
\begin{equation}\label{after_rot}
R_{EW}(\vartheta_w) \mathcal{M}^2 [R_{EW}(\vartheta_w)]^{-1} =
\left(
\begin{matrix}
0 & 0 & 0 \\
0 & \frac{v^2}{4}(g^2+g_1^2) & 0 \\
0 & 0 & 4 x^2 (g_1')^2
\end{matrix}
\right).
\end{equation}

Finally, we can associate the mass eigenvalues with the corresponding
physical vector boson eigenstates:
\begin{eqnarray}\nonumber
M_A &=& 0,\\
\label{Mzz'}
M_{Z} &=& \frac{v}{2}\sqrt{g^2 + g_1^2}, \\
M_{Z'} &=& 2 x g_1' . \nonumber
\end{eqnarray}

\section{The Feynman-gauge and the ghost Lagrangian}\label{sect:2-3}

Once the Lagrangian is established and the gauge symmetries are
spontaneously broken, the gauge bosons acquire their masses and we
have constructed a consistent theory.

However, the way we treated the Higgs mechanism so far is too
simplicist: assuming the unitary-gauge framework, we have not depicted
any exhaustive description of the computational Feynman rules of the
minimal $B-L$ model, hence we are not ready for any phenomenological
analysis yet.

Moreover, as we shall see in Subsection~\ref{subs:3-2-1}, we are
missing a correct treatment of formal unitarity aspects of the
theory.

Finally, it is well known that if the definition of each model is
delivered in the Feynman-gauge the public softwares dedicated to the
computation of amplitudes/cross-sections (as $CalcHEP$
\cite{calchep_man,Pukhov:2004ca}, $MadGraph$ \cite{Alwall:2007st},
$FeynArts$-$FormCalc$ \cite{Hahn:2000jm}, etc.) increase their
computational power, hence with an explicit parametrisation of the
Goldstone contribution within the Higgs potential.

For this, in the Subsection~\ref{subs:2-3-1} we will introduce a
standard parametrisation for the Higgs fields in the Feynman-gauge.

As for the aforementioned unitarity of the theory, it will be
apparently spoiled by the fact that the Feynman gauge will introduce
the effects of unphysical particles in the computational formalism. In
order to cancel this effects, in the Subsection~\ref{subs:2-3-2} we
will define a gauge-fixing Lagrangian, using the Fadeev-Popov method,
in order to restore the computational consistency of the the minimal
$B-L$ model.

\subsection{The Feynman gauge}\label{subs:2-3-1}

We focus again on the scalar sector of the Lagrangian described in
Subsection~\ref{subs:2-1-3}.

Following the prescription of the Feynman-gauge parametrisation of the
Higgs doublet and singlet of the minimal $B-L$ model, we consider not only
the Higgs fields and their $VEVs$, but also the so-called Goldstone
bosons.

Hence, $H$ and $\chi$ are now defined in the following way:
\begin{eqnarray}\label{feyn:par}
H=\frac{1}{\sqrt{2}}
\left(
\begin{gathered}
-i(w^1-iw^2) \\
v+(h+iz)
\end{gathered}
\right), \qquad
\chi =
\frac{1}{\sqrt{2}}
(x+(h'+iz')),
\end{eqnarray}
where $w^{\pm}=w^1\mp iw^2$, $z$ and $z'$ are the Goldstone
bosons of $W^{\pm}$, $Z$ and $Z'$, respectively.

From this definition and the set of relations described by
equations~(\ref{mh1})-(\ref{isomorphism}), we can calculate the
explicit rules of the interactions (i.e., the Feynman rules) in terms
of mass eigenstates and couplings.

As we will show in Subsection~\ref{subs:3-2-1}, a particular
interesting aspect of the structure of the scalar Lagrangian in the
Feynman gauge is that it fully describes the interaction of
longitudinally polarised vector bosons and Higgs bosons in the high
energy limit (``equivalence theorem'', see the Chapter~21 of
\cite{Peskin:1995ev} and \cite{Chanowitz:1985hj}).

Moreover, if we assume that the couplings of the theory are
perturbative and small then we can also apply the prescription
$D_\mu\simeq \partial_\mu$ in order to obtain the would-be-Goldstone
scalar sector: to all intents it works as an effective high energy
theory of the longitudinally polarised vector bosons and Higgs bosons
interactions.

For this, the would-be-Goldstone effective theory is a particulary
useful tool for analysing the perturbative unitarity stability of
longitudinally polarised vector boson scatterings in the high energy
limit, and we list the complete set of the functions appearing in the
would-be-Goldstone Lagrangian, in Appendix~\ref{appe:a}.

\subsection{The Fadeev-Popov Lagrangian}\label{subs:2-3-2}

Since there is no mixing in the neutral boson sector, the gauge-fixing
in the minimal $B-L$ model is fulfilled by a rather easy procedure.

Firstly, the Higgs doublet gauge-fixing Lagrangian is not affected by
the $B-L$ extension. Hence, following the notation of
\cite{Peskin:1995ev}, the ghost Lagrangian in the Feynman gauge is the
usual $SM$-like one:
\begin{equation}\label{lag:SMghost}
\mathscr{L}_{FP_{H}}=\bar{c}^a
\left[
-(\partial_\mu D^\mu)^{ab}-g^2(T^a \left< H \right> )\cdot T^b H
\right]c^b,
\end{equation}
where the $c$'s are the gauge-fixing fields corresponding to the $W$'s
and $Z$ gauge bosons.

Due to the fact that the $\chi$ field belongs to an Abelian gauge
symmetry structure, we have to extract the gauge-fixing for the
Abelian case in the Feynman gauge.

To begin, we focus on the spontaneously broken term of the scalar
Lagrangian in equation~(\ref{lag:s}):
\begin{equation}
\mathscr{L}_{s_{Ab}}=\left( D^{\mu} \chi \right) ^{\dagger} D_{\mu}
\chi - V(H,\chi),
\end{equation}
with $D_\mu=\partial_\mu+2ig_1'Z_\mu'$. From the second equation
of~(\ref{feyn:par}), we obtain:
\begin{equation}
\mathscr{L}_{s_{Ab}}=\frac{1}{2} (\partial_\mu h' - 2 g_1' Z_\mu'
z')^2 + \frac{1}{2} (\partial_\mu z' + 2 g_1' Z_\mu'(x + h'))^2 -
V(H,\chi).
\end{equation}

This Lagrangian is invariant under an exact local symmetry:
\begin{equation}\label{brst}
\delta h'= - \alpha(x)z', \qquad
\delta z'= \alpha(x)(x+h'), \qquad
\delta Z_\mu' = -\frac{1}{2g_1'}\partial_\mu \alpha(x).
\end{equation}

Due to the existence of this local symmetry, in order to define the
functional integral over the variables $h'$, $z'$, $Z'$, we must
introduce the Fadeev-Popov gauge-fixing.

Following the standard techniques (see the Chapter~9
of~\cite{Peskin:1995ev}), we define the functional integral:
\begin{equation}
Z=\int \mathcal{D}(Z') \mathcal{D}(h') \mathcal{D}(z')\ \exp \left[ i \int
  \mathscr{L}_{s_{Ab}} \right].
\end{equation}

Introducing a gauge-fixing constraint we find:
\begin{equation}
Z=C\int \mathcal{D}(Z') \mathcal{D}(h') \mathcal{D}(z')\ \exp \left[ i \int
  \mathscr{L}_{s_{Ab}} \right]\delta(G(Z',h',z')) \det \left(
\frac{\delta G}{\delta \alpha} \right),
\end{equation}
where $C$ is a constant proportional to the ``volume'' of the gauge
group and $G$ is a gauge-fixing condition. At this point, we can
introduce the gauge-fixing constraint as $\delta (G(x)-\omega(x))$ and
integrate over $\omega(x)$ with a Gaussian weight, to obtain:
\begin{equation}
Z=C'\int \mathcal{D}(Z') \mathcal{D}(h') \mathcal{D}(z')\ \exp \left[
  i \int d^4x
  (\mathscr{L}_{s_{Ab}} - \frac{1}{2}G^2) \right] \det \left(
\frac{\delta G}{\delta \alpha} \right).
\end{equation}

The gauge-fixing function $G$ is arbitrary, and a common choice in
the Feynman-gauge is:
\begin{equation}\label{gauge-fixing}
G=\partial^\mu Z_\mu'- 2 g_1' x z'.
\end{equation}

From the gauge-fixing condition, it is straightforward to obtain the
Lagrangian of ghost for the Abelian terms of the scalar
Lagrangian. Firstly, we evaluate the gauge variation of $G$:
\begin{equation}
\frac{\delta G}{\delta \alpha} = - \frac{1}{2 g_1'} \partial^2 - 2
g_1' x (x + h') = \frac{1}{2 g_1'} 
\left(
-\partial^2 - (2 g_1' x)^2 \left( 1 + \frac{h'}{x} \right)
\right).
\end{equation}

The determinant of this operator is related to the ghost Lagrangian by
the equation:
\begin{equation}
\det \left( \frac{\delta G}{\delta \alpha} \right) =
\int \mathcal{D}c \mathcal{D}\bar{c}\exp
\left[
-i \int d^4x \mathscr{L}_{FP_{\chi}}
\right],
\end{equation}
where
\begin{equation}
\mathscr{L}_{FP_{\chi}}= \bar{c}^{Z'} \left( 2 g_1' \frac{\delta G}{\delta
  \alpha} \right) c^{Z'} = \bar{c}^{Z'} \left(-\partial^2 - M_{Z'}^2 \left(
1 + \frac{h'}{x} \right) \right) c^{Z'}.
\end{equation}

In the last equation, $M_{Z'}=2g_1'x$ and $c^{Z'}$'s are the ghost
fields related to the $Z'$ boson.

It is interesting to notice that since this belongs to an Abelian
gauge structure, the ghost fields does not couple directly to the
gauge field, but only to the physical Higgs field.

\section{See-saw mechanism and neutrino masses}\label{sect:2-4}

In the $SM$ framework, there is not a straightforward way to generate
the experimentally observed neutrino masses and oscillations: any
isolated extension that solve this open issue (effective Majorana mass
terms, sterile right-handed neutrinos, etc.) is affected by
consistency problems (they spoil either renormalisability or gauge
invariance, see \cite{Bilenky:1987ty} for details).

The minimal $B-L$ model provides an elegant ``natural'' solution: the
presence of right-handed neutrinos in the Yukawa Lagrangian
(equation~(\ref{lag:Y})) gives raise to the so-called ``see-saw''
mechanism.

In details, after the spontaneous breaking of the gauge symmetry,
the Dirac neutrinos combine to six Majorana mass eigenstates, and the
mass matrix is:
\begin{equation}\label{neu_mass}
\mathscr{M}=\left(
\begin{matrix}
0 & m_D \\
m_D^T & M
\end{matrix} \right),
\end{equation}
where $m_D$ and $M$ are respectively the Dirac and Majorana mass
matrices, defined as:
\begin{equation}\label{dir_maj}
m_D=\frac{(y^\nu)^*}{\sqrt{2}}v, \qquad M=\sqrt{2} y^M x.
\end{equation}

Once assumed that the hierarchy $\Lambda_D \ll \Lambda_M$ (where
$\Lambda$ indicates the energy scale) is true, the diagonalization of
the mass matrix realises the ``see-saw'' mechanism.

After this procedure, we are left with three light Majorana neutrinos
$\nu_l$ and three heavy Majorana neutrinos $\nu_h$, whose $3\times 3$
mass matrices, denoted by $M_l$ and $M_h$ respectively, are given by:
\begin{eqnarray}\label{neu_mat}
M_l&\simeq &m_D M^{-1} m_D^T = \frac{1}{2\sqrt{2}} y^\nu (y^M)^{-1}
  (y^\nu)^T \frac{v^2}{x}, \nonumber \\
M_h&\simeq &M = \sqrt{2} y^M x.
\end{eqnarray}

In equations~(\ref{neu_mat}) we can appreciate the ``see-saw'' effect:
the greater is $M$, the smaller is $M_l$.

As explicit example we consider the case in which there is no mixing
between the three generations, hence the matrices $m_D$ and $M$ are
two diagonal matrices: in this case the $6 \times 6$ matrix splits in
three $2\times 2$ matrices. 

Thereafter, the diagonalisation is realised by the transformation
\begin{equation}
\left(
\begin{matrix}
\cos{\alpha_i} & -\sin{\alpha_i} \\
\sin{\alpha_i} & \cos{\alpha_i}
\end{matrix}
\right)
\left(
\begin{matrix}
0 & m_D^i \\
m_D^i & M^i
\end{matrix}
\right)
\left(
\begin{matrix}
\cos{\alpha_i} & \sin{\alpha_i} \\
-\sin{\alpha_i} & \cos{\alpha_i}
\end{matrix}
\right)
\simeq
{\rm diag}
\left( 
-\frac{(m_D^i)^2}{M^i},
M^i
\right),
\end{equation}
where $i=1,2,3$ denotes one of the three generations and
$\alpha^i=\arcsin{\left( m^i_D/M^i \right) }$.

We can roughly estimate the mass-scale $\Lambda_M$ needed to obtain
neutrino masses in agreement with phenomenological constraints
\cite{Fogli:2005cq,Fogli:2006yq}.

By taking $\Lambda_l < 1$ eV and $\Lambda_D \sim EW$ scale we get:
\begin{equation}
\Lambda_l \simeq \frac{\Lambda_D^2}{\Lambda_M} < 1\  {\rm eV} 
\Rightarrow \Lambda_M > 10^{13}\ {\rm GeV}.
\end{equation}

We should keep in mind, however, that $\Lambda_D$ could well be
several orders of magnitudes smaller than the weak scale (for example,
the electron mass): in such a case, much smaller scales for
$\Lambda_M$ are allowed.

Anyway, starting from equations~(\ref{dir_maj}), a generalised
condition could be:
\begin{equation}
v|y^\nu|\ll x |y^M|.
\end{equation}

It is important to recall that in the general case, in order to
extract the individual mass eigenvalues and eigenstates of the mixed
neutrinos eigensystem, we must separately diagonalise the two mass
matrices $M_l$ and $M_h$. In the case of $M_l$, the diagonalisation
gives rise to the well-known mixing $U_{PMNS}$ (see
\cite{Pontecorvo:1957cp,Pontecorvo:1957qd,Maki:1962mu,Pontecorvo:1967fh})
matrix with $6$ real
independent parameters, $3$ angles and $3$ phases.

\section{Counting of parameters}\label{sect:2-5}

The $SM$ contains 18 real parameters (3 in the gauge sector, 2 in the
Higgs sector, 13 in the fermion sector).

Once we add neutrino masses to the $SM$ via the see-saw mechanism, and
assume no mixing in the right-handed neutrino sector\footnote{This is
  realised by neglecting the additional complex parameters of their
  Yukawa couplings. Let's remind that in this work we are mainly
  interested in colliders and discovery physics, hence allowing any
  mixing either in the heavy or light neutrino sector does not affect
  our conclusions.}, 12 more real parameters are introduced (6
neutrino masses and the 6 parameters of $U_{PMNS}$), for a total of 30.

The Yang-Mills sector contains 1 more parameter than in the $SM$
model: the gauge coupling constant $g_1'$.

Finally, The scalar potential depends on 7 real parameters, the two
scalar masses ($m^2$, $\mu^2$), the three couplings ($\lambda_1$,
$\lambda_2$, $\lambda_3$) and the two $VEVs$ ($v$, $x$): three of them
are equivalent to the corresponding $SM$ parameters, thus we have here
4 additional parameters with respect to the $SM$.

In conclusion, the total number of parameters is 35.

\chapter{The Higgs sector parameter space} 
\label{chap:3}
\lhead{Chapter 3. \emph{The Higgs sector parameter space}} 

As we have already seen in Section~\ref{sect:2-5}, the $B-L$ extension
of the $SM$ introduces a new set of parameters.

Apart from the Higgs boson mass, all the parameters
involved in the definition of the $SM$ Lagrangian have been set by
experiments; when we call for the minimal $B-L$ extension, it opens
a new parameter space defined by 12 parameters (from the neutrino
sector\footnote{The 6 neutrino masses and the 6 parameters of
  $U_{PMSN}$.}), 1 parameter (from the $B-L$ gauge boson
sector\footnote{The $g_1'$ coupling.}) and 4 parameters (from the
Higgs bosons sector\footnote{The 3 $\lambda$'s and the $x$.}).

In principle, this could allow to explore potentially infinite new
phenomenological implications, nevertheless one has to take into
account the experimental and theoretical constraints that affects each
sector.

As for the Higgs sector, in the past four decades an enormous effort
has been made to improve both the theoretical and experimental
techniques that allow us to understand what is the forbidden
parameter space.

We can split these procedures in two sets: ``experimental'' and
``theoretical''.

The ones belonging to the former give constraints:
\begin{itemize}
\item by $EW$ precision tests (Subsection~\ref{subs:3-1-1}),
\item by direct searches (Subsection~\ref{subs:3-1-2});
\end{itemize} 
the ones belonging to the latter give constraints:
\begin{itemize}
\item by perturbative unitarity arguments
  (Subsection~\ref{subs:3-2-1}-\ref{subs:3-2-3}),
\item by triviality and vacuum stability
  (Subsection~\ref{subs:3-2-2}-\ref{subs:3-2-3}), 
\item by ``naturalness'' arguments (Subsection~\ref{subs:3-2-4}).
\end{itemize}

In this Chapter we present the main results that we have obtained by
the application of these techniques to the minimal $B-L$ model and
their major implications.

\section{Experimental constraints on the Higgs boson sector}\label{sect:3-1}

Despite the fact that the Higgs boson(s) has(have) not been discovered
yet, we know by consistency that the Higgs sector should participate
to the known phenomenology.

Firstly, the Higgs bosons (as well as the other new particles of this
model: heavy neutrinos and $Z'$) must contribute to the quantum
corrections to the high-precision $EW$ observables\footnote{In the
  last twenty years an enormous amount of $EW$ precision data have
  been collected by the $e^+e^-$ colliders LEP and SLC.}, and this
could impose boundaries on the free parameters.

Moreover, there are constraints coming from direct searches of Higgs
bosons at colliders, in particular from LEP.

In this Section we will summarise the main results of these
indirect and direct (respectively) constraints.

\subsection{High-precision data and constraints on the Higgs
  sector}\label{subs:3-1-1}

Even if the $B-L$ model is a minimal extension of the $SM$, it shows a
phenomenological richness that imposes an accurate approach to the $EW$
precision data analysis.

In principle, if one assumes that the symmetry group of new physics
($NP$) is still $SU(3)_C\times SU(2)_L \times U(1)_Y$, it is possible
to parametrise the radiative corrections in such a way that the
contributions from $NP$ could be easily implemented and
confronted with $EW$ precision data (a popular example is the well-known
Peskin-Takeuchi parametrisation in \cite{Peskin:1990zt,Peskin:1991sw}). 

However, it is important to mention the fact that the parametrisation
is based on the fact that the Higgs bosons must be the only source of
quantum corrections.

Obviously, this is not the case of the minimal $B-L$ model, because of
its extra $U(1)$ extension.

Indeed, in this case we have not only a new gauge boson $Z'$ that
affects the choice of the precision parameters parametrisation, but
also the presence of heavy neutrinos that are not weakly coupled with
the boson sector: this spoils the first assumption.

In the absence of new gauge bosons or heavy neutrinos, many successful
attempts have been made to profile the constraints on the Higgs
sector parameter space (in the multiple Higgs scenario, some
interesting study has been performed in \cite{Dawson:2009yx}), but
unfortunately the $B-L$ model falls outside these conclusions.

Hence, in this work we have not considered any boundary condition
coming from the $EW$ precision analysis. However, we borrow a
qualitative conclusion from the aforementioned multiple Higgs scenario
paper: we assume that the Higgs bosons inversion limit
$\alpha\rightarrow \pi/2$ is not allowed.

\subsection{Direct searches and constraints on the Higgs
  sector}\label{subs:3-1-2} 

The minimal $SM$ Higgs boson has been searched in the LEP
experiments, firstly at energies close to the $Z$ boson peak (LEP1),
then with center of mass energies up to $\sqrt{s} \sim 200$ GeV
(LEP2).

The main production modes that have been explored at LEP1 are the
so-called Bjorken process ($Z \rightarrow HZ\rightarrow Hf\bar{f}$)
and the $Z\rightarrow H\gamma$ (through triangular loops of $t$'s and
$W$'s).

In order to avoid the large hadronic background, Higgs boson have been
unsuccessfully searched in the two aforementioned channels (see
\cite{Buskulic:1996hz,Abreu:1994rp,Acciarri:1996um,Alexander:1996ai}),
leading to a $95\%$ Confidence Level ($CL$) limit of $M_H>65$ GeV on
the $SM$ Higgs mass.

In the LEP2 energy regime, the Higgs boson have been searched in its
dominant production process: the Higgs-strahlung, $Z \rightarrow HZ$.

LEP collaborations have explored several topologies, combining
their results in the analysis of the $Z$ boson recoil energy that led
to the well known exclusion limit of the $SM$ Higgs boson mass (see
\cite{Barate:2003sz}):
\begin{eqnarray}
M_H>114.4\ \rm{GeV}
\end{eqnarray}
at the $95\%$ $CL$ from the non-observation.

In the $B-L$ model (as well as in other extensions of the $SM$), this
limit is shifted back in mass by the fact that Higgs bosons have
non-standard coupling to the $Z$ (it is reduced by $\cos{\alpha}$):
\begin{eqnarray}
\frac{\sigma_{SM}(e^+e^-\rightarrow
  HZ)}{\sigma_{B-L}(e^+e^-\rightarrow H_1Z)} =\cos^2{\alpha}.
\end{eqnarray}

\begin{figure}[!t]
  \centering
  \includegraphics[angle=0,width=0.74\textwidth ]{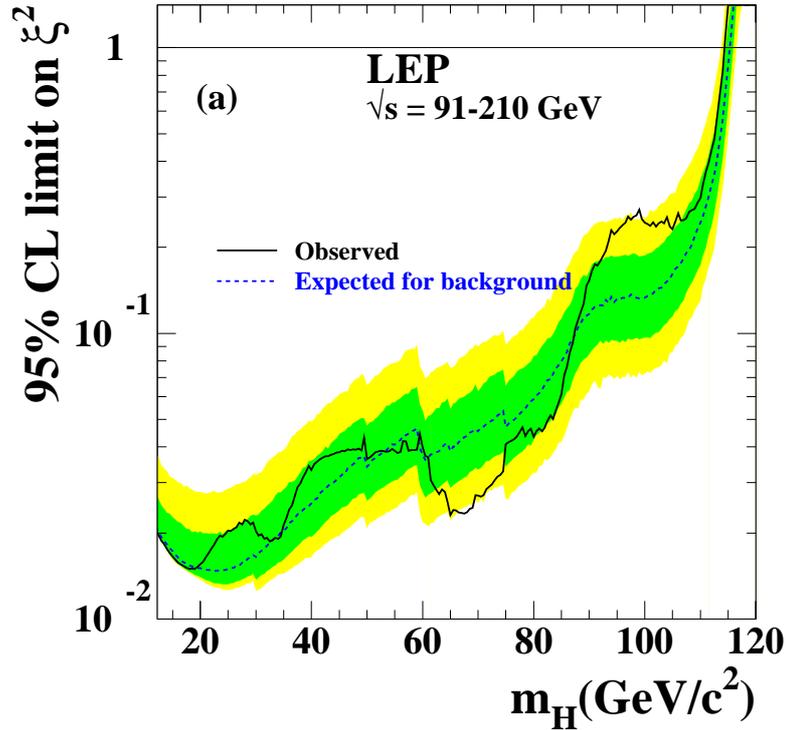}
  \caption[Experimental constraints on the Higgs boson sector - Direct
    searches and constraints on the Higgs sector]{The $95\%$
    $CL$ upper bound on
    the reduced coupling $\xi=g_{HZZ}/g^{SM}_{HZZ}$ set by the LEP
    collaborations; in the minimal $B-L$ model 
    $\xi=\cos{\alpha}(\sin{\alpha})$ for $H=h_1(h_2)$.}
  \label{LEP_lim}
\end{figure} 

Indeed, the LEP collaborations have also released
a detailed analysis of the $95\%$ $CL$ exclusion limits on the
(lightest) Higgs boson mass as function of the reduced coupling
$\xi^2=\cos^2{\alpha}(\sin^2{\alpha})$ for the $h_1(h_2)$, and
it is shown in figure~\ref{LEP_lim}.

In this work, we will consider the latter as the only experimental
constraint on the $B-L$ Higgs sector parameter space.

\subsection{Other experimental constraints}\label{subs:3-1-3}

In view of an extensive numerical analysis of the phenomenological
implications at colliders, one has to have a complete understanding of
how other experimental bounds could affect the $B-L$ parameter space.

For this, we briefly present the experimental constraints on the gauge
boson and neutrino sectors that we are going to consider in the rest
of this work.

\begin{itemize}
\item $M_{Z'}$, the new gauge boson mass. An indirect constraint on
  $M_{Z'}$ comes from analyses at LEP of precision $EW$ data (see
  \cite{Cacciapaglia:2006pk}, based on the analysis of experimental
  data published in 
  \cite{Anthony:2003ub,ew:2003ih,Azzi:2004rc,Woods:2004zr,Z-Pole})\footnote{A
    less conservative approach, based on Fermi-type effective
    four-fermions interactions, gives the weaker constraint
    $\frac{M_{Z'}}{g'_1} \geq 6\; \rm{TeV}$ \cite{Carena:2004xs}.}:
  \begin{equation}\label{LEP_bound}
    \frac{M_{Z'}}{g'_1} \geq 7\; \rm{TeV}\, .
  \end{equation}

Further limits have been obtained at Tevatron
\cite{Aaltonen:2008vx,Aaltonen:2008ah,Basso:2010pe}. Both have been
taken into account in this work.
\item $m_{\nu _l}$, the $SM$ (or light) neutrino masses. We use the
  cosmological upper bound $\sum_l m_{\nu _l}<1$ eV
  \cite{Fogli:2008ig}. Ultimately, they have been taken to be
  $m_{\nu_l}=10^{-2}$ eV. 
  Since we are mainly interested in the discovery and collider physics
  phenomenological aspects, we
  remind that in the rest of our work, for our illustrative purposes,
  we take all neutrino masses (both light and heavy) to be degenerate.
\end{itemize}

\section{Theoretical constraints on the Higgs boson
  sector}\label{sect:3-2} 

Several theoretical methods have been developed to establish what are
the constraints on the Higgs parameter space.

In this Section we apply them to the minimal $B-L$ model.

They tipically follow from general principles, like the assumption
that the perturbative unitarity and the couplings of a theory are
preserved to some energy scale (Subsection~\ref{subs:3-2-1}), as well
as the stability of the symmetry breaking vacuum expectation value
(Subsection~\ref{subs:3-2-2}).

These arguments could also be extended to different sectors of the
theory in order to constrain other free parameters, as $g'_1$, as we
will show in the Subsection~\ref{subs:3-2-3}.

Finally, one could also propose some conjecture about the
``naturalness'' of the model, finding reasonable to indicate some
``natural'' range of value for the allowed parameter space of the
theory in order to minimise the so-called ``fine-tuning'' problem
(Subsection~\ref{subs:3-2-4}).

\subsection{Perturbative unitarity}\label{subs:3-2-1}

Even if there is no evidence of the $(B)SM$ Higgs boson(s), the need of
such kind of particle in order to guarantee the perturbative unitarity
of a theory is of fundamental concern.

In fact, the potential problem that affects the $SM$ and its
extensions is the following: in the absence of the Higgs boson
contributions, the longitudinally polarised vector boson interactions
violate unitarity at some energy scale.

To see this, we must recall some general properties of the scattering
amplitudes; firstly, the amplitude $A$ of a $2\rightarrow 2$ process could
be decomposed into partial waves $a_l$ of orbital angular momentum $l$:
\begin{equation}
A=16\pi\sum_{l=0}^\infty (2l+1)P_l(\cos{\theta})a_l,
\end{equation}
where $P_l$ are the Legendre polynomials and $\theta$ is the diffusion
angle; this leads to the following expression for the cross-section:
\begin{equation}
\sigma=\frac{16\pi}{s}\sum_{l=0}^\infty (2l+1)|a_l|^2.
\end{equation}

From the optical theorem (see Chapter~7 of \cite{Peskin:1995ev}) we
also know that the cross-section is proportional to the imaginary part
of the amplitude in the forward direction:
\begin{equation}
\sigma = \frac{1}{s} {\rm Im}[A(\theta=0)] = \frac{16\pi}{s}
\sum_{l=0}^\infty (2l+1) |a_l|^2.
\end{equation}

From this, we have the relation $|a_l|^2={\rm Im}(a_l)$, that leads to
the well-known condition (see \cite{Luscher:1988gk})
\begin{eqnarray}\label{condition}
|\textrm{Re}(a_l(s))|\leq \frac{1}{2}.
\end{eqnarray}

In the $SM$ context, the pioneeristic work of
\cite{Lee:1977eg} showed that, when $M_H$ is greater than a
critical value $\simeq 1$ TeV (known as unitarity bound), the elastic
spherical wave describing the scattering of
the longitudinally polarised vector bosons at very high energy
($\sqrt{s} \rightarrow \infty$) violates the condition in
equation~\ref{condition}, and the perturbative stability of the theory
breaks down.


In the past, several efforts have been devoted to apply these
methodology to a variety of models, in order to extract any possible
information on their allowed parameter space. In particular, it has
been already applied to scenarios with extended scalar sectors yet
with same gauge structure as the $SM$, like those with additional
singlets (for example, see \cite{Cynolter:2004cq}), doublets (for
example, see \cite{Maalampi:1991fb} and \cite{Huffel:1980sk} for
non-Supersymmetric scenarios
and \cite{Casalbuoni:1987cz} for Supersymmetric ones),
triplets (for example, see \cite{Aoki:2007ah}). It has also
been shown that this approach is successful with respect to $U(1)$
gauge group extensions of the $SM$ (for example, for the case of $E_6$
superstring-inspired minimal $U(1)$ extensions, see
\cite{Robinett:1986nw}).

Then, our aim is to show how this methods could also be
successfully applied to the minimal $B-L$ case, taking into account
the presence of two Higgs fields and four massive vector bosons.

In fact, by mean of the so-called Equivalence Theorem we are allowed to
compute the amplitude of any process with external longitudinal vector
bosons $V_L$ ($V = W^\pm,Z,Z' $), in the high energy limit $M^2_V\ll s$,
by substituting each one of them with the related Goldstone bosons $v
= w^\pm,z,z'$ and its general validity is proven
(see \cite{Chanowitz:1985hj}); schematically, if we consider a
process with four longitudinal vector bosons: $A(V_L V_L \rightarrow
V_L V_L) = A(v v \rightarrow v v)+ O(m_V^2/s)$.

The intermediate vector boson exchange does not play a fundamental
role in the Higgs boson(s) limits\footnote{This is not
  generally true in gauge group extensions, nevertheless for this
  particular purpose we assume that $g_1'$ is perturbative and small,
  in such a way that any $t$-channel represents a subleading
  contribution to the scattering amplitude.}, hence, as we intimated
in Subsection~\ref{subs:2-3-1}, we simplify our
approach by employing a theory of interacting would-be Goldstone
bosons $v = w^\pm,z,z'$ described by the scalar Lagrangian in
Appendix~\ref{appe:a}.

In order to study the unitarity constraints in the $B-L$ model, we
calculate the tree-level amplitudes for all two-to-two processes 
involving the full set of possible (pseudo)scalar fields (the most
relevant subset is given by table~\ref{tab:channels}).

Given a tree-level scattering amplitude between two spin-$0$ particles
$A(s,\theta)$, where $\theta$ is the scattering (polar) angle, 
we know that the partial wave amplitude with angular
momentum $l$ is given by
\begin{eqnarray}\label{integral}
a_l = \frac{1}{32\pi} \int_{-1}^{1} d(\cos{\theta}) P_l(\cos{\theta})
A(s,\theta).
\end{eqnarray}

It turns out that only $l=0$ (corresponding to the spherical partial
wave contribution) leads to some bound, so we will not discuss the
higher partial waves any further. 

It is well known (and we have verified) that, in the high energy
limit, only the four-point
vertices (related to the four-point functions of the interacting
potential, equations~(\ref{4-goldstone})-(\ref{4-higgs}) of
Appendix~\ref{appe:a})
contribute to the $J=0$ partial wave amplitudes, and this
is consistent with many other aforementioned works that exploit the
same methodology. 

Hence, we present the main results of our study focusing only on
the relevant subset of all
spherical partial wave amplitudes that is shown in
table~\ref{tab:channels}. Here, we should notice that,
as one can conclude from direct computation, in the high energy limit
the contributions in table~\ref{tab:channels} ticked
with $\sim$ are just a double counting of the channels ticked with
$\surd$ or combinations of them.

\begin{table}[!htbp]
\begin{center}
\begin{tabular}{|c|c|c|c|c|c|c|}
\hline
\  & $zz$ & $w^+w^-$ & $z'z'$ & $h_1h_1$ & $h_1h_2$ & $h_2h_2$ \\
\hline
$zz$ & $\surd$ & $\surd$ & $\surd$ & $\surd$ & $\surd$ &  $\surd$ \\
\hline
$w^+w^-$ & $\sim$ & $\surd$ & $\sim$ & $\sim$ & $\sim$ & $\sim$ \\
\hline
$z'z'$ & $\sim$ & $\sim$ & $\surd$ & $\surd$ & $\surd$ & $\surd$ \\
\hline
$h_1h_1$ & $\sim$ & $\sim$ & $\sim$ & $\surd$ & $\surd$ & $\surd$ \\
\hline
$h_1h_2$ & $\sim$ & $\sim$ & $\sim$ & $\sim$ & $\surd$ & $\surd$ \\
\hline
$h_2h_2$ & $\sim$ & $\sim$ & $\sim$ & $\sim$ & $\sim$ & $\surd$ \\
\hline
\end{tabular}
\end{center}
\caption[Theoretical constraints on the Higgs boson sector -
  Perturbative unitarity (1)]{The most relevant subset of two-to-two
  scattering processes
  in the minimal $B-L$ model in the Higgs and would-be Goldstone boson
  sectors. The rows(columns) refer to the initial(final) state (or
  vice versa). The symbol $\sim$ refers to processes that can be
  computed by appropriate rearrangements of those symbolised by
  $\surd$.}
\label{tab:channels}
\end{table}

Moreover, the main contributions come 
from the so-called scattering eigenchannels, i.e., the diagonal
elements of the ``matrix'' in table~\ref{tab:channels}. In particular,
for our choice of method,
only $zz \to zz$ and $z'z' \to z'z'$,
and to a somewhat lesser extent also $h_1h_1\to h_1h_1$ and 
$h_2h_2\to h_2h_2$,
play a relevant role. For
completeness, we list here all the $a_0$'s, 
eigenchannel by eigenchannel\footnote{Actually, in the high energy
limit, $a_0(w^+w^-\rightarrow w^+w^-)$ differs from
equation~(\ref{a0ww}) by a quantity $\simeq \alpha_W$
due to photon and $Z$-boson exchange in
the $t$-channel, but since we are applying the condition in
equation~(\ref{condition}) and $\alpha_W \ll \frac{1}{2}$, this correction
does not change the picture of our Higgs boson mass limit search.}:
\begin{eqnarray}\label{a0zz}
a_0(zz\rightarrow zz)
&=& \frac{3 \alpha_W}{32
M_W^2}  \left[ M_{h_1}^2 + M_{h_2}^2
+ \left( M_{h_1}^2 -
M_{h_2}^2 \right) \cos{(2\alpha)} \right],
\end{eqnarray}
\begin{eqnarray}\label{a0ww}
 a_0(w^+w^-\rightarrow w^+w^-) &=&
\frac{\alpha_W}{16 M_W^2} \left[ M_{h_1}^2 + M_{h_2}^2 + \left( M_{h_1}^2
- M_{h_2}^2 \right) \cos{(2\alpha)} \right],
\end{eqnarray}
\begin{eqnarray}\label{a0zpzp}
a_0(z'z'\rightarrow z'z') &=&
\frac{3}{32 \pi  x^2} \left[ M_{h_1}^2 + M_{h_2}^2 - \left( M_{h_1}^2
- M_{h_2}^2 \right) \cos{(2\alpha)} \right],
\end{eqnarray}
\begin{eqnarray}\label{a0h1h1}
a_0(h_1h_1\rightarrow h_1h_1) &=&
\frac{3\alpha_W}{32 M_W^2}
\left[ M_{h_1}^2 + M_{h_2}^2 + \left( M_{h_1}^2 -
M_{h_2}^2 \right) \cos{(2\alpha)} \right] \cos^4{\alpha}  \nonumber \\
&-&
\frac{3\sqrt{\alpha_W}}{64 M_W \sqrt{\pi} x}
\left( M_{h_1}^2 -
M_{h_2}^2 \right) \sin^3{(2\alpha)}  \nonumber \\
&+&
\frac{3}{16  \pi  x^2}\left[
M_{h_1}^2  -
\left( M_{h_1}^2 -
M_{h_2}^2 \right) \cos^2{\alpha} \right] \sin^4{\alpha},
\end{eqnarray}
\begin{eqnarray}\label{a0h1h2}
a_0(h_1h_2\rightarrow h_1h_2) &=&
\frac{ \sqrt{\alpha_W }}{256 M_W \sqrt{\pi} x}
\left( M_{h_1}^2 - M_{h_2}^2 \right) ( \sin{(2\alpha)} -
3 \sin{(6\alpha)})  \nonumber \\
&+&
\frac{3}{64 \pi x^2}
\left[ M_{h_1}^2 - \left( M_{h_1}^2 -
M_{h_2}^2 \right) \cos^2{\alpha} \right] \sin^2{(2\alpha)} \nonumber \\
&+&
\frac{3 \alpha_W}{64 M_W^2}
\left[ M_{h_1}^2 - \left( M_{h_1}^2 -
M_{h_2}^2 \right) \sin^2{\alpha} \right]
\sin^2{(2\alpha)},
\end{eqnarray}
\begin{eqnarray}\label{a0h2h2}
a_0(h_2h_2\rightarrow h_2h_2)&=&
\frac{3}{16 \pi x^2}
 \left[ M_{h_1}^2 - \left( M_{h_1}^2 -
 M_{h_2}^2 \right) \cos^2{\alpha} \right] \cos^4{\alpha} \nonumber \\
&-&
\frac{3  \sqrt{\alpha_W} }{64 M_W \sqrt{\pi } x} \left( M_{h_1}^2
- M_{h_2}^2 \right) \sin^3{(2\alpha)} \nonumber \\
&+&
\frac{3 \alpha_W}{16 M_W^2} \left[ M_{h_1}^2 - \left( M_{h_1}^2 -
M_{h_2}^2 \right) \sin^2{\alpha} \right] \sin^4{\alpha},
\end{eqnarray}
where
\begin{eqnarray}
\alpha_W=\frac{M_W^2}{\pi v^2}.
\end{eqnarray}

We remark upon the fact that in the high energy limit, $\sqrt
s\rightarrow \infty$, only the $a_0$ partial wave amplitude (i.e., the
four-point function as one can conclude by direct comparison between
equations~(\ref{a0zz})-(\ref{a0h2h2})
and equations~(\ref{4-goldstone})-(\ref{4-higgs}) in
Appendix~\ref{appe:a}) does not 
vanish, instead it approaches a value depending only on $M_{h_1}$,
$M_{h_2}$ and $\alpha$. Therefore, by applying the condition in
equation~(\ref{condition}), we can obtain several different (correlated)
constraints on the Higgs masses and mixing angle, i.e., we can find the
$M_{h_1}$-$M_{h_2}$-$\alpha$ subspace in which the perturbative
unitarity of the theory is valid up to any energy scale.

The most relevant scattering channels for the unitarity analysis are
pure-$z$ and pure 
$z'$-bosons scatterings. As one can see from
equations~(\ref{a0zz})-(\ref{a0zpzp}), the limit coming from these two 
channels is unaffected by the transformation $\alpha \rightarrow
-\alpha$, hence it is not restrictive to consider the half domain
$\alpha \in [0,\frac{\pi}{2}]$ only.
Furthermore, we remind the reader that we are still not allowing the
inversion of the Higgs mass eigenvalues, i.e., we still
require $M_{h_1}<M_{h_2}$.


Afterwards, we analyse the space of the parameters $\alpha$,
$M_{h_1}$ and $M_{h_2}$, once it has been specified by the unitarity
condition applied to the spherical partial wave scattering amplitudes
listed in the previous Section in the very high energy limit.

We want to start our analysis in the $M_{h_1}$-$M_{h_2}$
subspace, hence we ``slice'' the $3$-dimensional parameter space we
are dealing with by keeping the Higgs mixing angle fixed.

\begin{figure}[!t]
  \subfigure[]{
  \label{a001}
  \includegraphics[angle=0,width=0.49\textwidth]{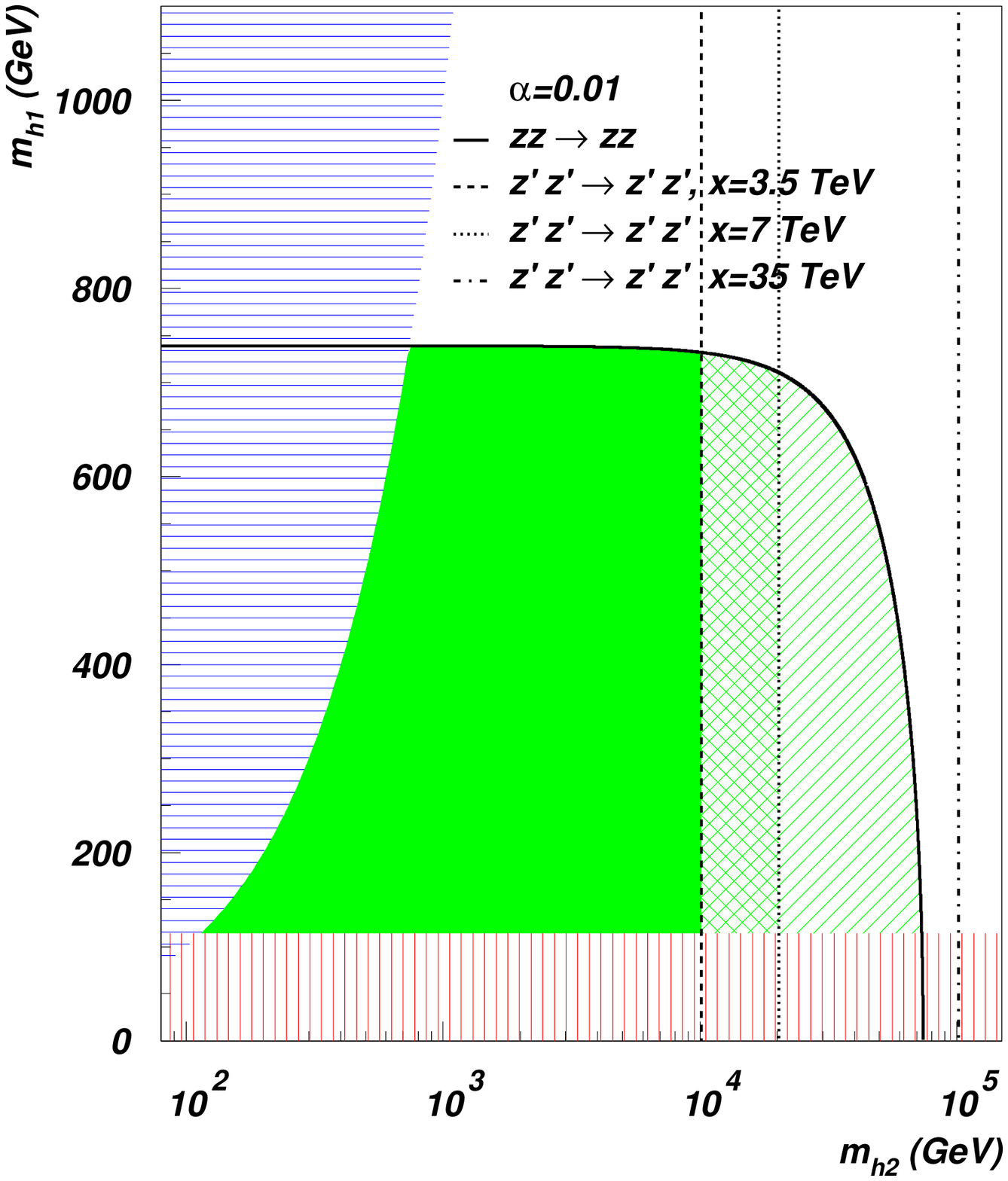}}
  \subfigure[]{
  \label{a01}
  \includegraphics[angle=0,width=0.49\textwidth ]{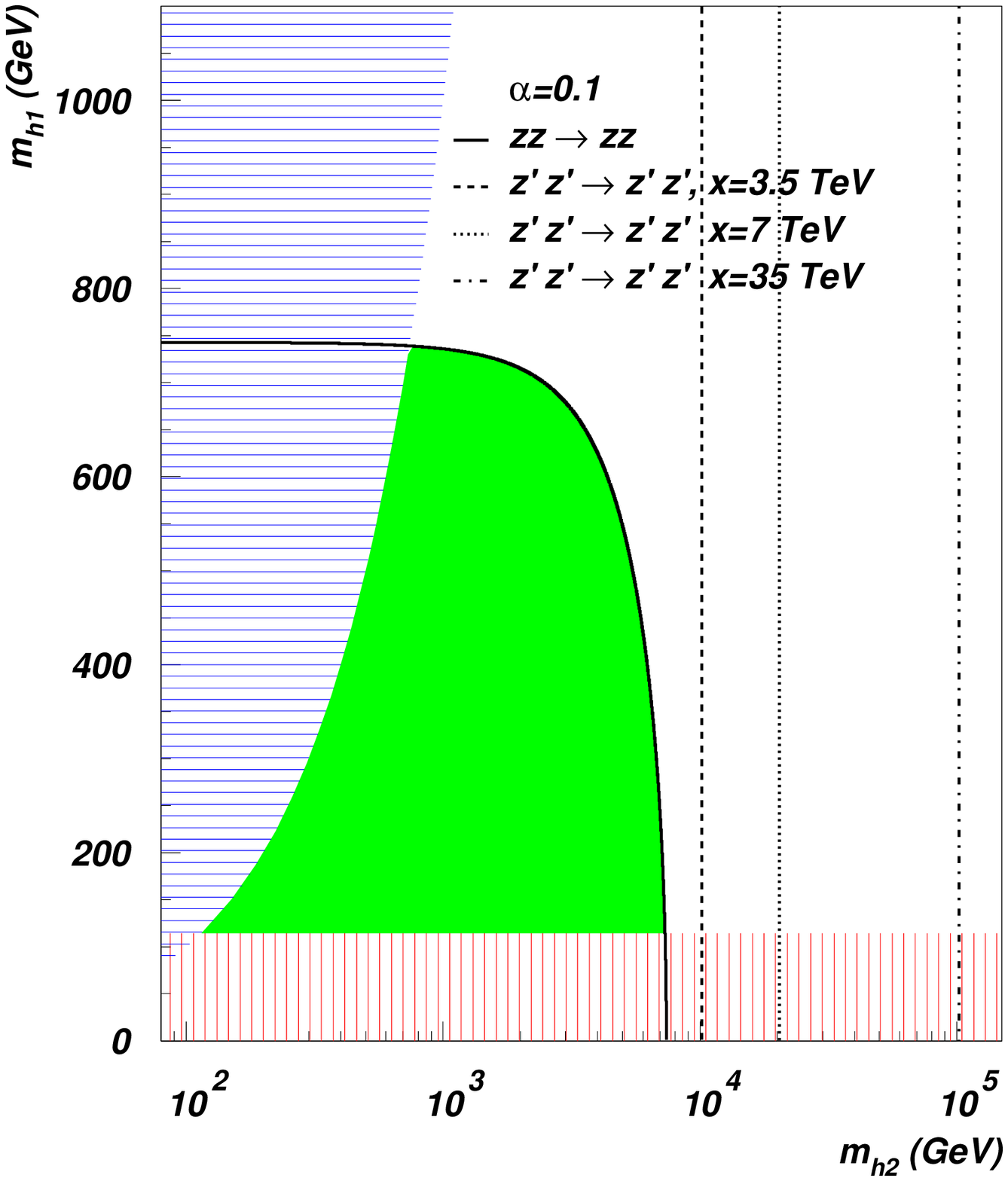}}
\\
  \subfigure[]{
  \label{api4}
  \includegraphics[angle=0,width=0.49\textwidth]{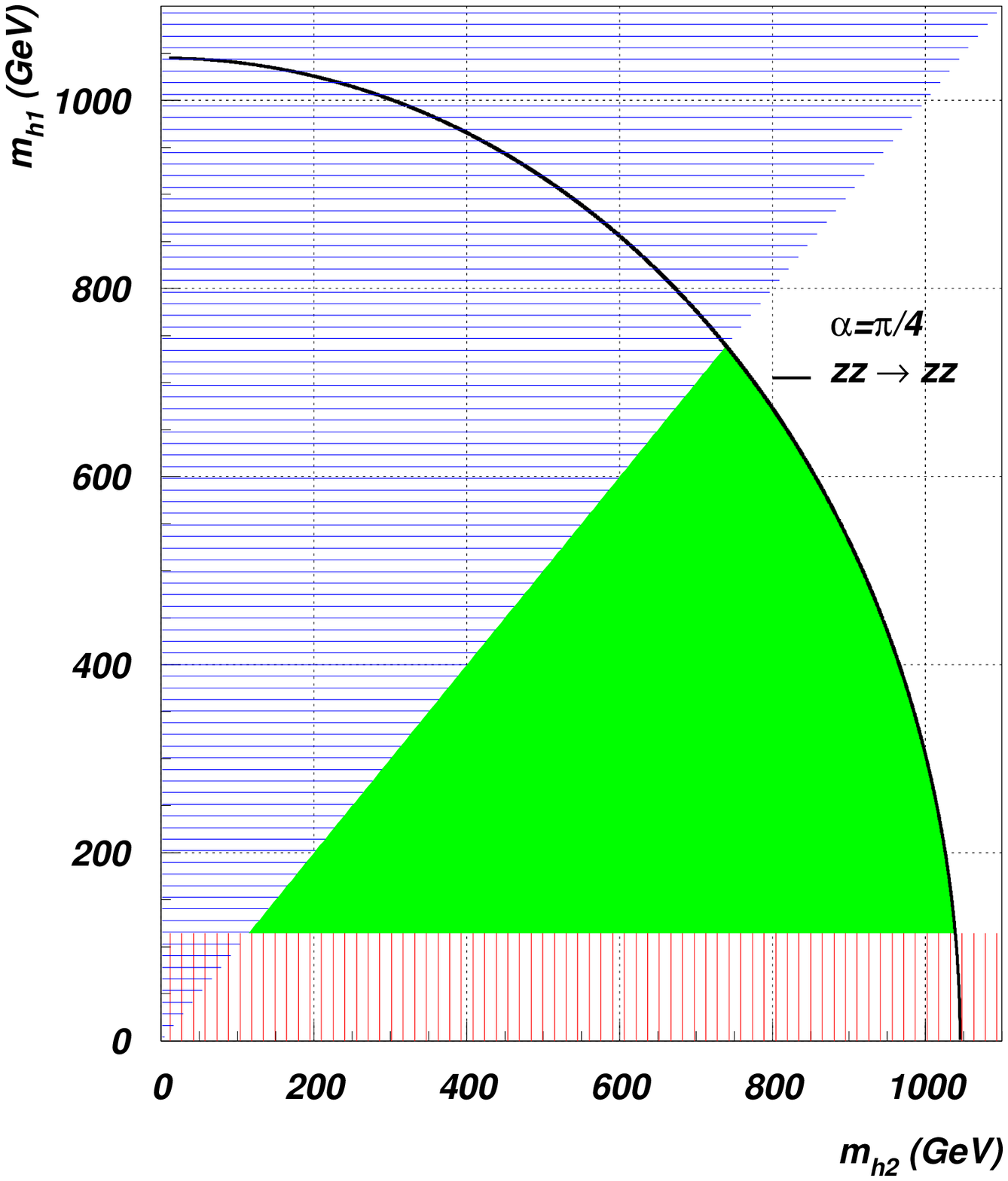}}
  \subfigure[]{
  \label{api2}
  \includegraphics[angle=0,width=0.49\textwidth]{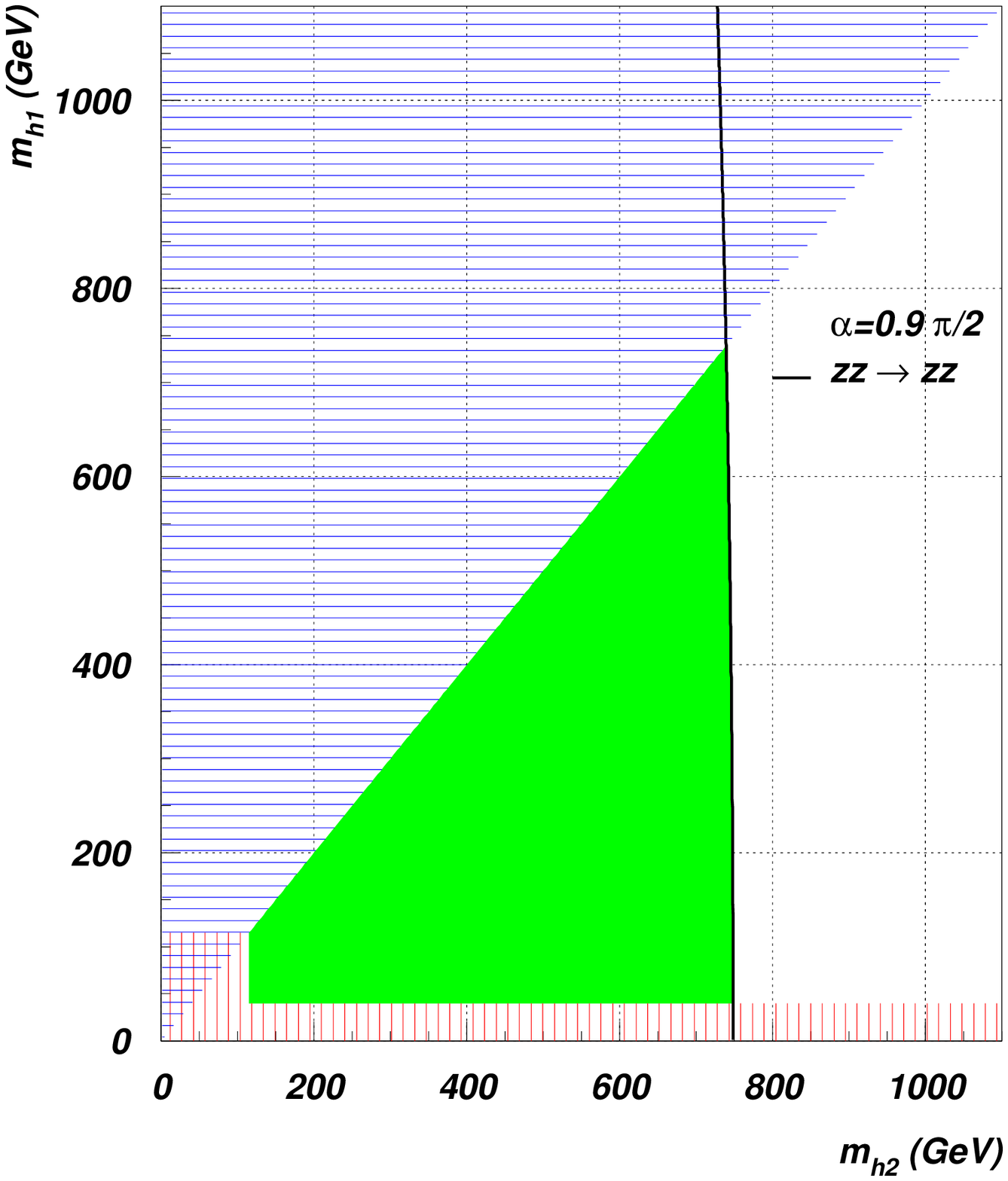}}
  \caption[Theoretical constraints on the Higgs boson sector -
  Perturbative unitarity (1)]{Higgs bosons mass limits in the $B-L$
    model coming from
  the unitarity condition $|\textrm{Re}(a_0)|\le \frac{1}{2}$ applied
  to the $zz \rightarrow zz$ and $z'z' \rightarrow z'z'$ scatterings for
  several values of $x$, for $\alpha=0.01$ (\ref{a001}), $\alpha=0.1$
  (\ref{a01}), $\alpha=\pi/4$ (\ref{api4}) and $\alpha=0.9\ \pi/2$
  (\ref{api2}). The (blue) horizontal shadowed region corresponds to the
  unphysical configuration $M_{h_1}>M_{h_2}$. The (red) vertical
  shadowed region is excluded by the LEP experiments.}
  \label{mh1vsmh2}
\end{figure}

By applying the unitarity constraint to the spherical partial waves
listed in the previous Section, one discovers that for a mixing angle
$\alpha$ such that
\begin{eqnarray}
{\rm{arctan}}\left( \frac{ M_W}{x\sqrt{\pi \alpha_W }} \right)
\le \alpha \le \frac{\pi}{2},
\end{eqnarray}
the allowed parameter space is completely defined by the
$zz\rightarrow zz$ eigenchannel.

We will call ``high-mixing domain'' the parameter space defined by a
choice of the mixing angle in this range, while the ``low-mixing
domain'' is the complementary one. For example, since $x \geq 3.5$ TeV
following the LEP analyses (Subsection~\ref{subs:3-1-3}), if we choose
$x$ to be exactly $3.5$ TeV, then we say that the high-mixing
domain, in this particular case, is the one for $ 0.073 \leq \alpha
\leq \frac{\pi}{2} $ (and, conversely, the low-mixing one is in the
interval $ 0 \leq \alpha < 0.073 $).

We can appreciate how the size of the Higgs mixing affects the limits
on the Higgs masses by looking at figure~\ref{mh1vsmh2}, in which we
plot the allowed space for the latter, limitedly to the two
eigenchannels $zz \rightarrow zz$ and $z'z' \rightarrow z'z'$, for
four different values of $\alpha$ and three of $x$ (the latter affects
only the limit coming from the $z'z' \to z'z'$ scattering).

We see that in both cases, as expected, the light Higgs mass upper
bound does not exceed the $SM$ one (which is $\simeq 700$ GeV, according
to~\cite{Luscher:1988gk}),
and it runs to the experimental lower limit from LEP
(figure~\ref{LEP_lim}) as the 
heavy Higgs mass increases. This is because the
two Higgses `cooperate' in the unitarisation of the eigenchannels so
that, if one Higgs mass tends to grow, the other one must become
lighter and lighter in order to keep the scattering matrix elements
unitarised.

While we are in the high-mixing domain, as in
figure~\ref{a01}-\ref{api4}-\ref{api2} (where
$\alpha=0.1$, $\alpha=\frac{\pi}{4}$,
$\alpha=0.9\ \frac{\pi}{2}$, respectively\footnote{For the last of
these values of the mixing angle, the lower limit from LEP experiments
on the light Higgs boson is $M_{h_1}>40$ GeV, while for the first it
is almost equal to the $SM$ lower limit ($M_{h_1}>115$ GeV),
as illustrated in figure~\ref{mh1vsmh2}.}), the
allowed region coming from the $zz\to zz$
scattering is completely included in the $z'z'\to z'z'$ allowed area,
and the highest value allowed for the heavy Higgs mass only depends
on the mixing angle via
\begin{eqnarray}\label{high-maxmh2}
{\rm{Max}}(M_{h_2})=2\sqrt{\frac{2}{3}}
\ \frac{M_W}{\sqrt{\alpha_W}\sin{\alpha}}.
\end{eqnarray}

When we move to figure~\ref{a001} (where $\alpha=0.01$, low-mixing
domain) we are able to appreciate some interplay between the two
scattering processes. In fact, in this case, while the $zz\to zz$
scattering eigenchannel allows the existence of a heavy Higgs of more
than $10$ TeV, 
the $z'z'\to z'z'$ scattering channel strongly limits the allowed mass
region, with a ``cut-off'' on the heavy Higgs mass almost insensible
to the light Higgs mass (and the value of the mixing angle, since we
are in the low-mixing domain), that is
\begin{eqnarray}\label{low-maxmh2}
{\rm{Max}}(M_{h_2})\simeq 2\sqrt{\frac{2\pi}{3}}x,
\end{eqnarray} 
which is in agreement (under different theoretical assumptions, though)
with the result in \cite{Robinett:1986nw}; from a
graphical point of view, in figure~\ref{a001} the (green) hollow area
represents the allowed configuration space at $x=3.5$ TeV, while at
$x=10$ TeV the allowed portion of the $M_{h_1}$-$M_{h_2}$ subspace
increases until the (green) double-lines shadowed region, finally the
constraint relaxes to the (green) single line shadowed region when
$x=35$ TeV.

\begin{figure}[!ht]
  \subfigure[]{ 
  \label{x35_mh1150}
  \includegraphics[angle=0,width=0.49\textwidth ]{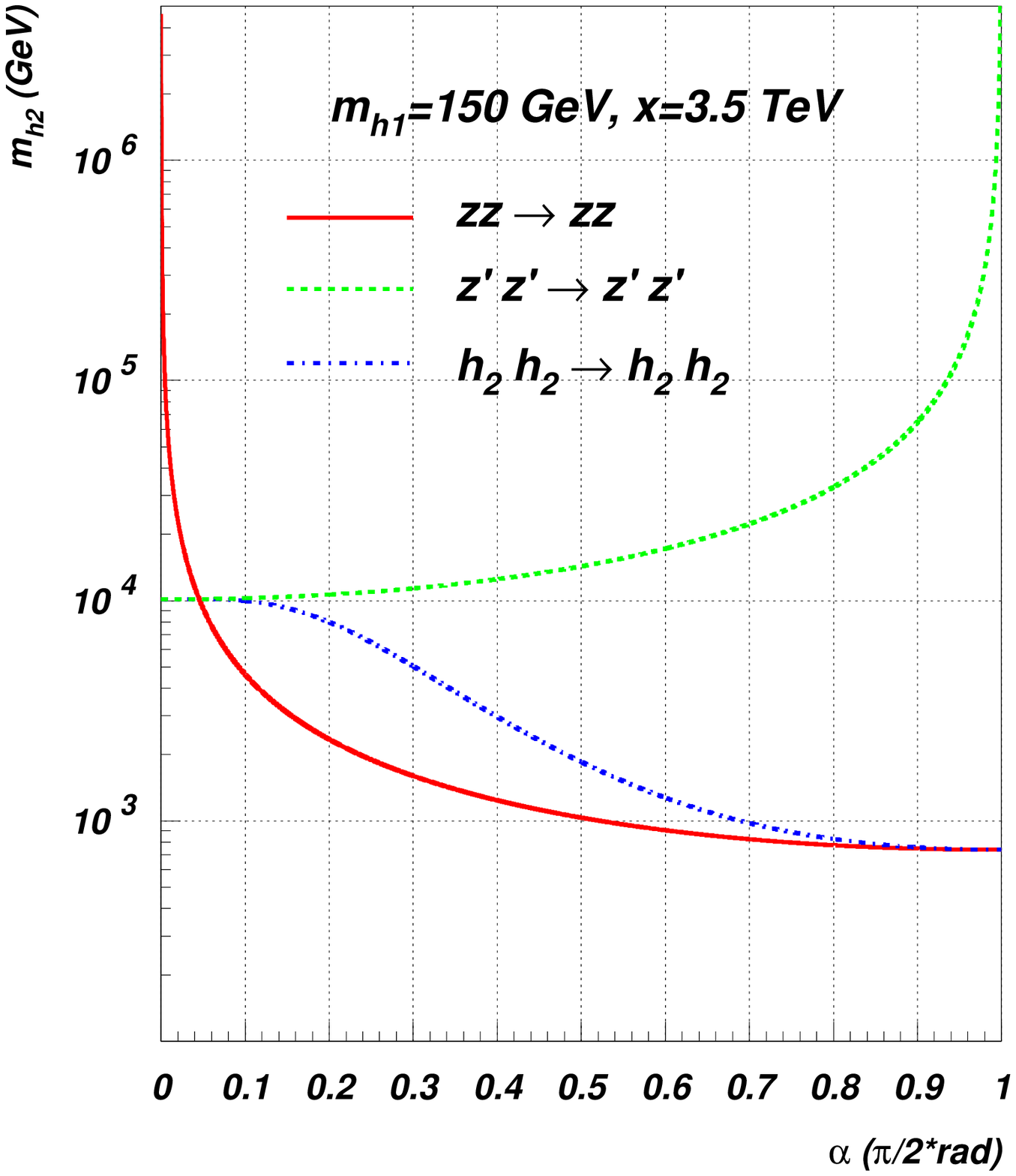}}
  \subfigure[]{
  \label{x35_mh1700}
  \includegraphics[angle=0,width=0.49\textwidth ]{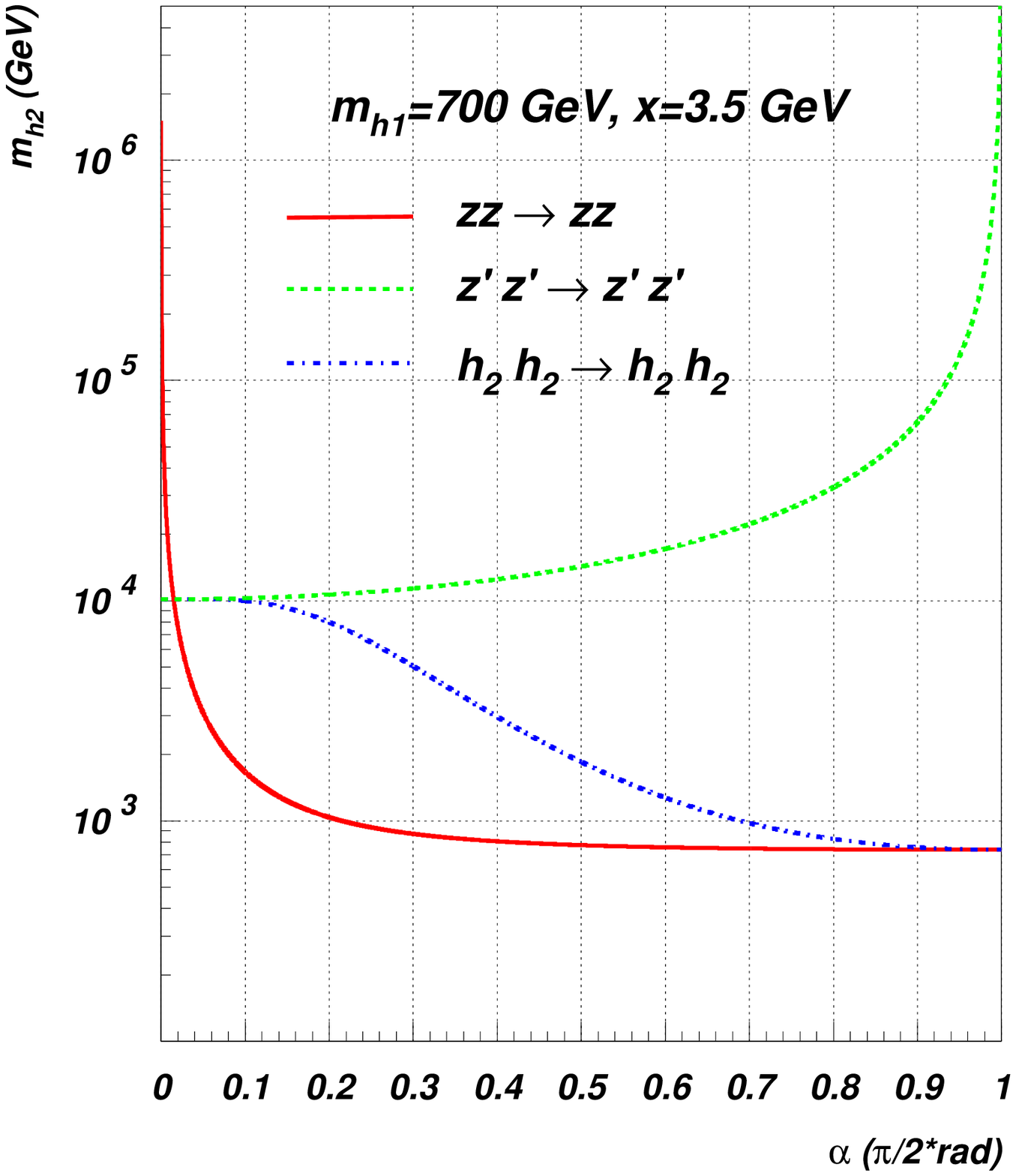}}
  \\
  \subfigure[]{ 
  \label{x350_mh1150}
  \includegraphics[angle=0,width=0.49\textwidth ]{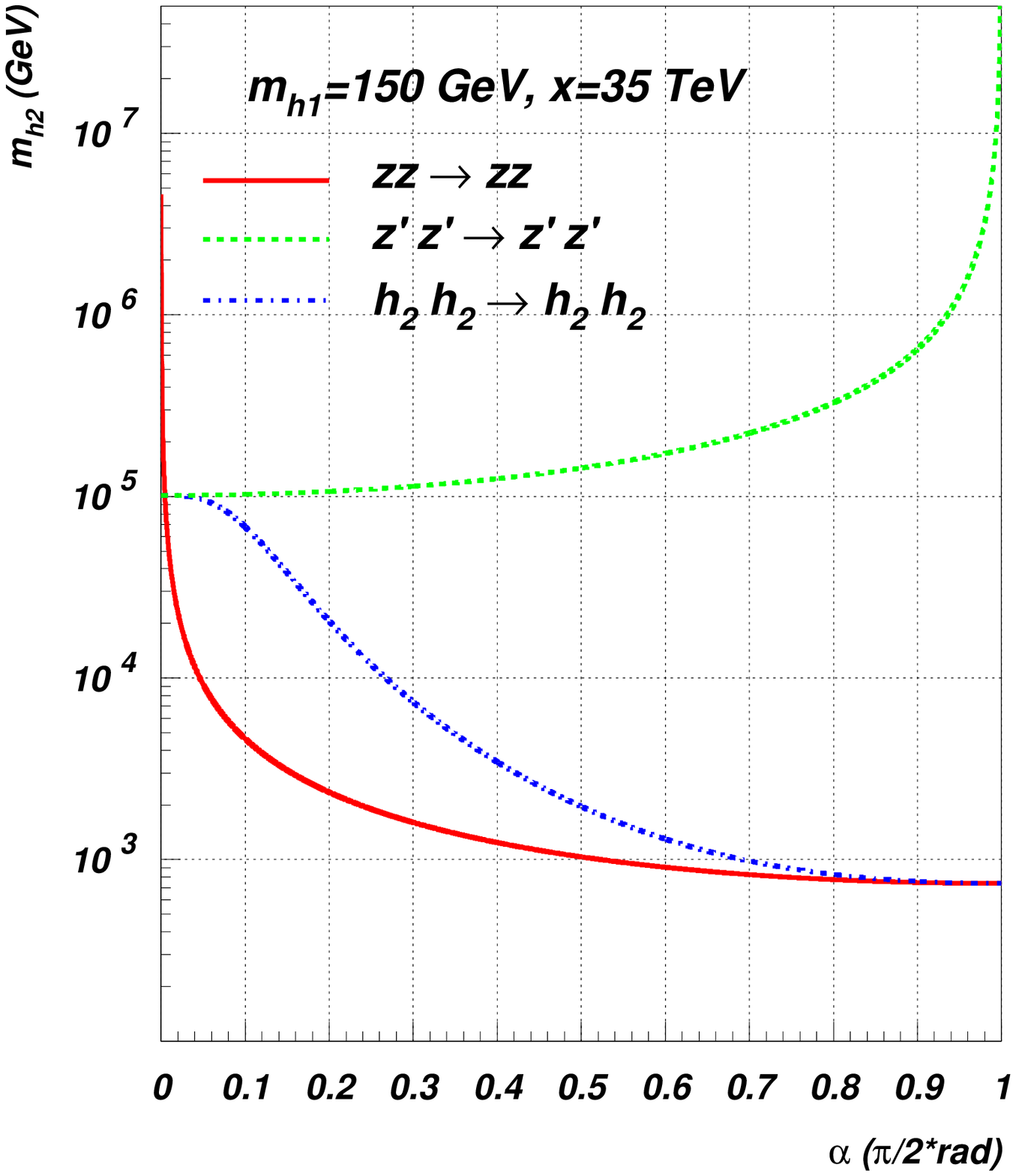}}
  \subfigure[]{
  \label{x350_mh1700}
  \includegraphics[angle=0,width=0.49\textwidth ]{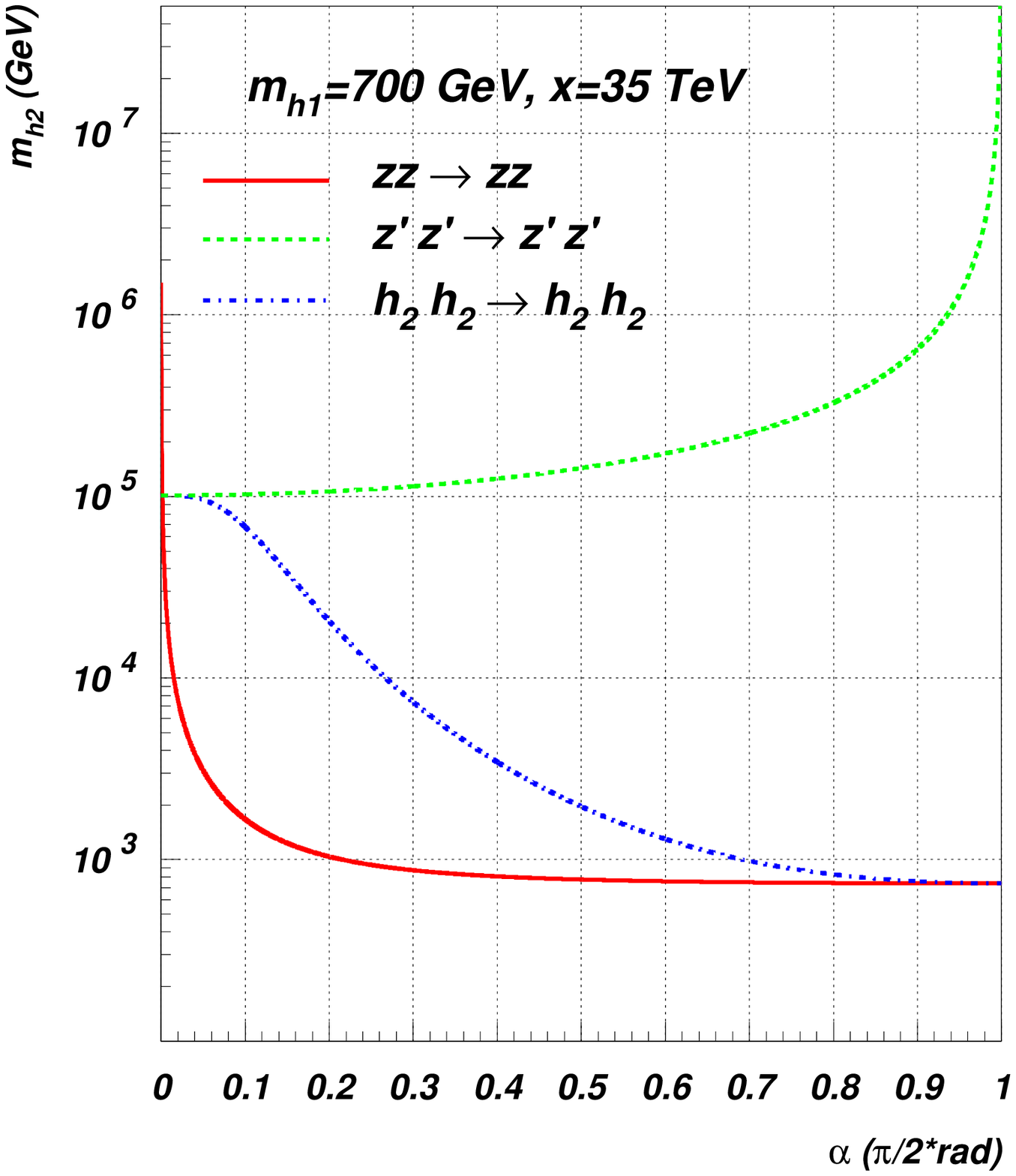}}
  \caption[Theoretical constraints on the Higgs boson sector -
  Perturbative unitarity (2)]{Heavy Higgs boson mass limits plotted against
    the mixing
  angle $\alpha$ in the minimal $B-L$ model. We have applied the
  unitarity condition $|\textrm{Re}(a_0)|\le \frac{1}{2}$ on $zz
  \rightarrow zz$ (red straight line), $z'z' \rightarrow z'z'$
  (green dashed line) and $h_2h_2 \rightarrow h_2h_2$ (blue
  dashed-dotted line) scatterings. This has been plotted for two fixed
  values of the light Higgs boson mass ($M_{h_1}=150$ GeV
  (\ref{x35_mh1150}, \ref{x350_mh1150}) and $M_{h_1}=700$ GeV
  (\ref{x35_mh1700}, \ref{x350_mh1700})) and of the singlet Higgs $VEV$
   ($x=M_{Z'}/(2g'_1)=3.5$ TeV (\ref{x35_mh1150}, \ref{x35_mh1700}) and
   $x=M_{Z'}/(2g'_1)=35$ TeV (\ref{x350_mh1150}, \ref{x350_mh1700})).}
  \label{mh2vsa}
\end{figure}

This interplay effect arising (somewhat unintuitively) for Higgs
low-mixing is due to the fact that the consequent decoupling between
the two Higgs states requires the light(heavy) Higgs state to
independently keep the scattering matrix elements of the
$z^{(')}z^{(')}\rightarrow z^{(')}z^{(')}$ process unitary, thus realising
two separate constraints: the first on the light ($SM$-like) Higgs mass 
due to the $zz\rightarrow zz$ unitarisation and the second on the
heavy Higgs mass due to the $z'z'\rightarrow z'z'$ unitarisation.

To summarise, given a value of the singlet Higgs $VEV$ $x$ (compatible
with experiment), the upper bound on the light Higgs boson mass varies
between the $SM$ limit and the experimental lower limit from LEP as long
as the upper bound for the heavy
Higgs mass increases. Moreover, when $\alpha$ assumes a value included
in the high-mixing domain, the strongest bound comes from the
unitarisation of the $z$-boson scattering, whilst in the low-mixing
domain the bound on the heavy Higgs mass coming from that channel
relaxes and the unitarisation induced by the $z'$-boson scattering
becomes so important to also impose a cut-off (which depends linearly
on $x$) on the heavy Higgs mass.

This is a very important result, because it allows us to conclude
that, whichever the Higgs mixing angle, both Higgs boson masses of the
$B-L$ model are bounded from above. As examples of typical values for
the heavy Higgs mass, in table~\ref{tab:mh2vsvev}, we show some upper
bounds that universally apply (i.e., no matter what the mixing angle
is) once the singlet Higgs $VEV$ is given.

\begin{table}[!htbp]
\begin{center}
\begin{tabular}{|c|c|}
\hline
$x$ (TeV) & Max$(M_{h_2})$ (TeV) \\
\hline
$3.5$ & $\simeq 10$ \\
\hline
$7$ & $\simeq 20$ \\
\hline
$10$ & $\simeq 30$ \\
\hline
$20$ & $\simeq 60$ \\
\hline
$35$ & $\simeq 100$ \\
\hline
\end{tabular}
\end{center}
\caption[Theoretical constraints on the Higgs boson sector -
  Perturbative unitarity (2)]{Universal upper
  bound on the heavy Higgs mass, $M_{h_2}$, in
  the $B-L$ model as a function of the singlet Higgs $VEV$, $x$.}
\label{tab:mh2vsvev}
\end{table}

Before we move on, it is also worth re-emphasising that,
if the Higgs mixing angle is such that we are in the high-mixing case,
the upper bound on the heavy Higgs boson mass coming from $z$-boson
scattering is more stringent
than the one coming from $z'$-boson scattering and it is totally
independent from the chosen singlet Higgs $VEV$.

Nowadays, it is important to refer in our analysis to the possibility of
a Higgs boson discovery at the Large Hadron Collider
(LHC). Thus, if we suppose that a light or heavy Higgs mass $M_{h_1}$
has been already measured by an experiment it is interesting to study
the $\alpha$-$M_{h_2}$ parameter space, to see whether an hitherto
unassigned Higgs state can be consistent with a minimal $B-L$
scenario.
%
%
%

To this end, in figure~\ref{mh2vsa} we fix $M_{h_1}$ and $x$ at two
extreme configurations: we take $M_{h_1}=150$ GeV as minimum value
(conservatively, taking a figure that is allowed by the experimental
lower bound established by LEP for a $SM$ Higgs boson) and
$M_{h_1}=700$ GeV as maximum value (close to the maximum allowed by
unitarity constraints, as we saw before). Then, we take $x=3.5$ TeV as
minimum value (that is, the lower limit established by LEP data
for the existence of a $Z'$ of ${B-L}$ origin) and $M_{h_1}=35$ TeV as
maximum value (that is, one order of magnitude bigger than the smallest
$VEV$ allowed by experiment).

Even in this case we can separate the $2$-dimensional subspace
in a
low-mixing region and a high-mixing region, as before. We can identify
the first(second) as the one in which the upper bound is established
by unitarisation through the $z'$($z$)-boson scattering. The value of
the mixing angle that separates the two regions in this case
is given by
\begin{eqnarray}
\alpha={\rm arccos}\sqrt{\frac{\left( 3 M_{h_1}^2 - 8 \pi x^2 \right)
      \alpha_W }{6 M_{h_1}^2 \alpha_W - 8 \pi x^2 \alpha_W - 8 M_{W}^2}}.
\end{eqnarray}

Once the light Higgs boson mass is fixed, we can see how the
heavy Higgs boson mass is bounded from above through the value defined 
by equation~(\ref{low-maxmh2}) through the $z'z'\rightarrow z'z'$
channel, and this  occurs in the low-mixing region.
In particular, we can notice how the $z'$-constraining function
reaches a plateau and overlaps with the $h_2 h_2 \rightarrow h_2h_2$
eigenchannel bound. Moreover, if we pay attention to the high-mixing
region, we see that, if $M_{h_1}$ is fixed to some low value, then
the bound on the heavy Higgs mass relaxes much more as the mixing gets
smaller and smaller with respect to the the situation in which
$M_{h_1}$ is large, where the unitarisation is shared almost
equally by $M_{h_2}$ and $M_{h_1}$.

\subsection{Triviality and vacuum stability bounds}\label{subs:3-2-2}

It is very well known that because of quantum corrections, the
parameters which appear in the $B-L$ Lagrangian (as well as in the
$SM$ Lagrangian) are energy-scale dependent (see \cite{Basso:2010jm}
for details).

This is also true for the quartic Higgs couplings as well as the gauge
couplings which will be monotonically increasing with the energy
scale, and this growth leads to constraints on the parameter
space.

It is important to highlight the fact that we have previously defined
the minimal $B-L$ by setting $\widetilde{g}=0$ in
equation~(\ref{cov_der_nm}). However, $\widetilde{g}$ is scale-dependent,
and it increases with the energy-scale, so it is more appropriate to
say that we define the minimal $B-L$ by setting $\widetilde{g}=0$ at
the $EW$ energy-scale.

From this, it is clear that a correct estimate of the evolution of the
couplings must include the $\widetilde{g}$ contribution.

Therefore, the variation of the parameters is described by the
Renormalisation Group Equation ($RGE$), and from
\cite{BL_master_thesis,Basso:2010jm} we know that the running of
$\lambda_i$'s is described by:
\begin{eqnarray}\label{RGE_lamda1}
\frac{d (\lambda_1)}{d(\log{\Lambda})} &\simeq &  \frac{1}{16\pi^2}
\left( 24\lambda_1^2 +\lambda_3^2 + 12 \lambda_1 y_t^2 - 9 \lambda_1
g^2 - 3 \lambda_1 g_1^2 - 3 \lambda_1 \widetilde{g}^2 \right)
,\nonumber  \\
\label{RGE_lambda2}
\frac{d (\lambda_2)}{d(\log{\Lambda})} &\simeq & \frac{1}{8 \pi^2}
\left( 10 \lambda_2^2 + \lambda_3^2 + 4 \lambda_2 {\rm Tr} \left[ (y^M)^2 \right]
-24 \lambda_2 (g_1')^{2} \right), \\ \nonumber \label{RGE_lamda3}
\frac{d (\lambda_3)}{d(\log{\Lambda})} &\simeq & \frac{\lambda_3}{8\pi
  ^2} \left( 6
\lambda_1 + 4\lambda_2 + 2\lambda_3  \right),
\end{eqnarray}
and the running of the gauge couplings is described by:
\begin{eqnarray}\label{RGE_g1}
\frac{d(g_1)}{d(\log{\Lambda})} &=& \frac{1}{16\pi
^2}\left[\frac{41}{6}g_1^3 \right], \nonumber \\ \label{RGE_g2} 
\frac{d(g_1')}{d(\log{\Lambda})} &=& \frac{1}{16\pi
^2}\left[ 12 g_1'^3 +
2 \frac{16}{3} g_1'^2\widetilde g+\frac{41}{6}g_1'\widetilde
g^2 \right], \\ \label{RGE_g_tilde} 
\frac{d(\widetilde g)}{d(\log{\Lambda})}&=& \frac{1}{16\pi
^2}\left[\frac{41}{6}\widetilde{g}\,(\widetilde
g^2+2g_1^2) + 2 \frac{16}{3} g_1' (\widetilde{g}^2 + g_1^2)
+ 12 g_1'^2\widetilde
g \right]. \nonumber 
\end{eqnarray}

Once we know the $RGEs$ in
equations~(\ref{RGE_lambda2})-(\ref{RGE_g2}), we must fix the
boundary conditions for the evolution of the parameters. We choose the
$EW$ gauge coupling to be set by experiment ($g_1(EW)\simeq 0,36$) as
well as the ``top''-Yukawa ($y_t\simeq 1$).

Regarding the neutrinos, for simplicity we
consider them degenerate and we fix their masses to $M^{1,2,3}_{\nu h}
= 200$ GeV (hence, $y^M\simeq m_{\nu
  h}/x\sqrt{2}$)\footnote{For 
  semplicity we assume that $y_t$ and $y^M$ do not evolve,
  nevertheless their evolution could in principle affect
  quantitatively (not qualitatively) our conclusions (see
  \cite{Basso:2010jm} for further details).}, since this value has
been proven to be related to interesting phenomenological scenarios
(see \cite{Basso:2008iv}).

Moreover, another boundary condition is set by the definition of
the model: $\widetilde{g}(EW)=0$.

Focusing on the scalar sector, the free parameters in our study are
then $M_{h_1}$, $M_{h_2}$, $\alpha$ and $x$. The
general philosophy is to fix in turn some of the free 
parameters and scan over the other ones, individuating the allowed
regions fulfilling the aforementioned set of conditions (as we
previously did in Subsection~\ref{subs:3-2-1}). 

We firstly define a parameter to be ``perturbative'' for values less
than unity. This is a conservative definition, as we could relax it by
an order of magnitude and still get values of the parameters for which
the perturbative series will converge\footnote{Notice that the
  parameters upon which the perturbative expansion is
  performed are usually of the form $\sqrt{\alpha} = g/ \sqrt{4\pi}$,
  rather then being $g$ itself.}. 

Then, $RGE$ evolution can constrain the parameter space of the scalar
sector in two complementary ways: from one side, the couplings must be
perturbative,
\begin{equation}\label{cond_1}
0 < \lambda _{1,2,3}(Q') < 1 \qquad \forall \; Q' \leq Q\, ,
\end{equation}
and it is usually referred to as the ``triviality'' condition; on the
other side, the vacuum of the theory must be well-defined at any
scale, that is, to guarantee the validity of
eqs.~(\ref{bound_pot}) and (\ref{pos_pot}) at any scale $Q'\leq
Q$: 
\begin{equation}\label{cond_2}
0 < \lambda _{1,2}(Q') \qquad \mbox{and} \qquad
4\lambda _1(Q')\lambda _2(Q')-\lambda _{3}^2(Q') > 0 \qquad \forall \;
Q' \leq Q\, ,
\end{equation}
and it is usually referred to as the ``vacuum stability'' condition.

Given the simplicity of the scalar sector in the $SM$, the triviality
and vacuum stability conditions can be studied independently and they
both constrain the Higgs boson masses, providing an upper bound and a
lower bound, respectively. In more complicated models than the one
considered here, it might be more convenient to study the overall
effect of equations~(\ref{cond_1})-(\ref{cond_2}), since there are
regions of the parameter space in which the constraints are evaded
simultaneously. This is the strategy we decided to follow.

\begin{figure}[!th]
  \subfigure[]{ 
  \label{mh1_mh2_a0}
  \includegraphics[angle=0,width=0.48\textwidth
  ]{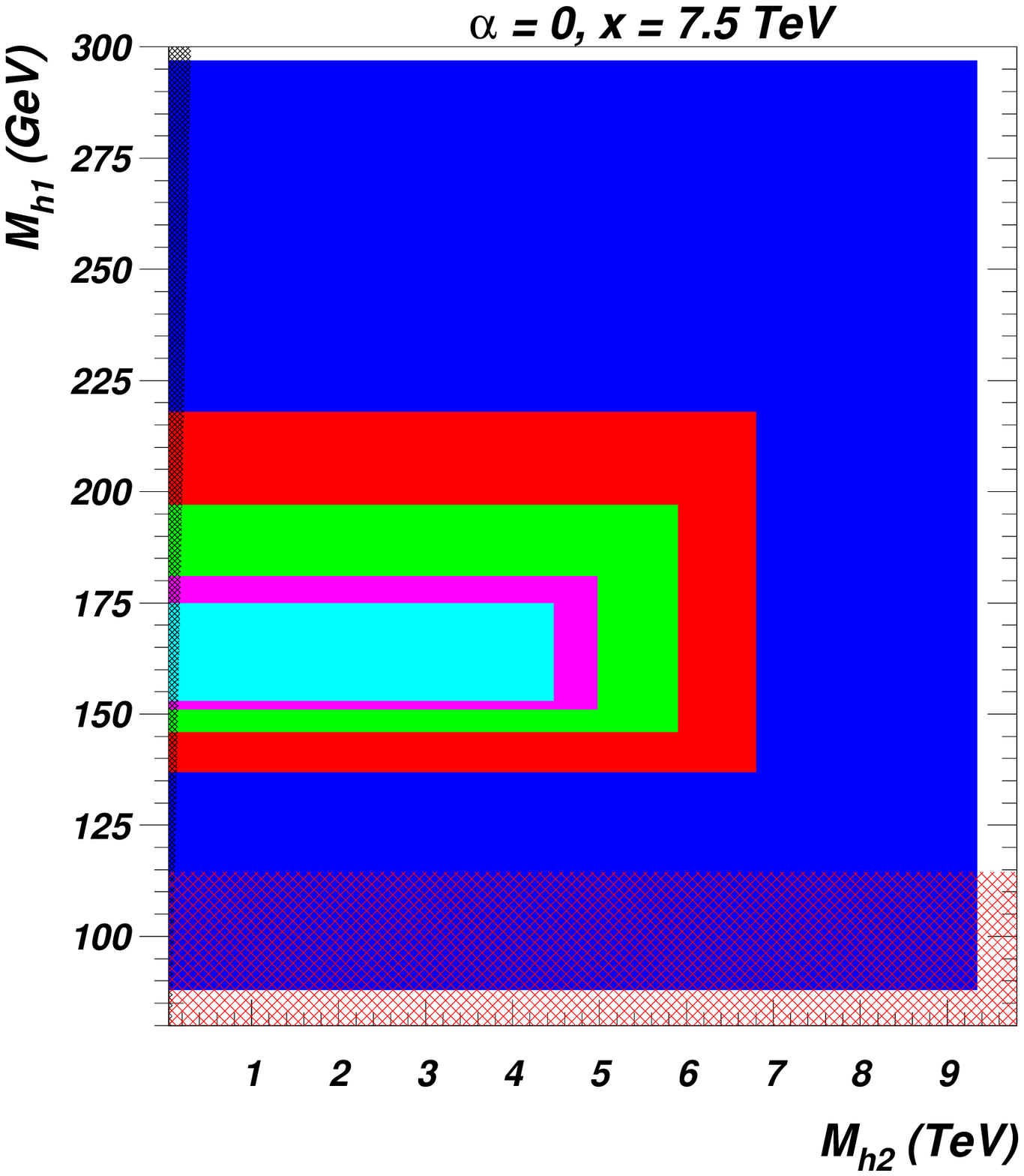}} 
  \subfigure[]{
  \label{mh1_mh2_a01}
  \includegraphics[angle=0,width=0.48\textwidth
  ]{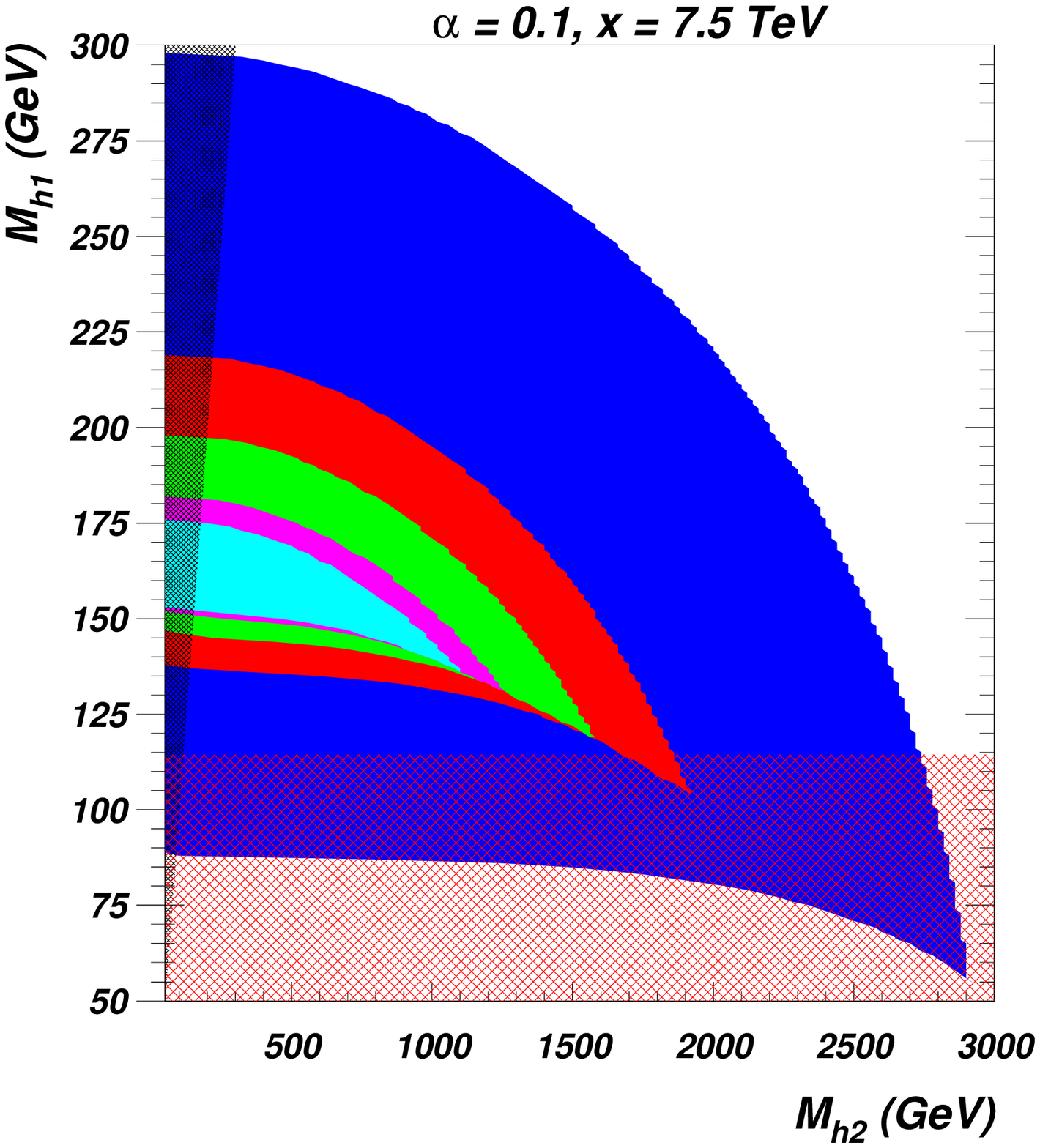}} 
  \\
  \subfigure[]{
  \label{mh1_mh2_api4}
  \includegraphics[angle=0,width=0.48\textwidth
  ]{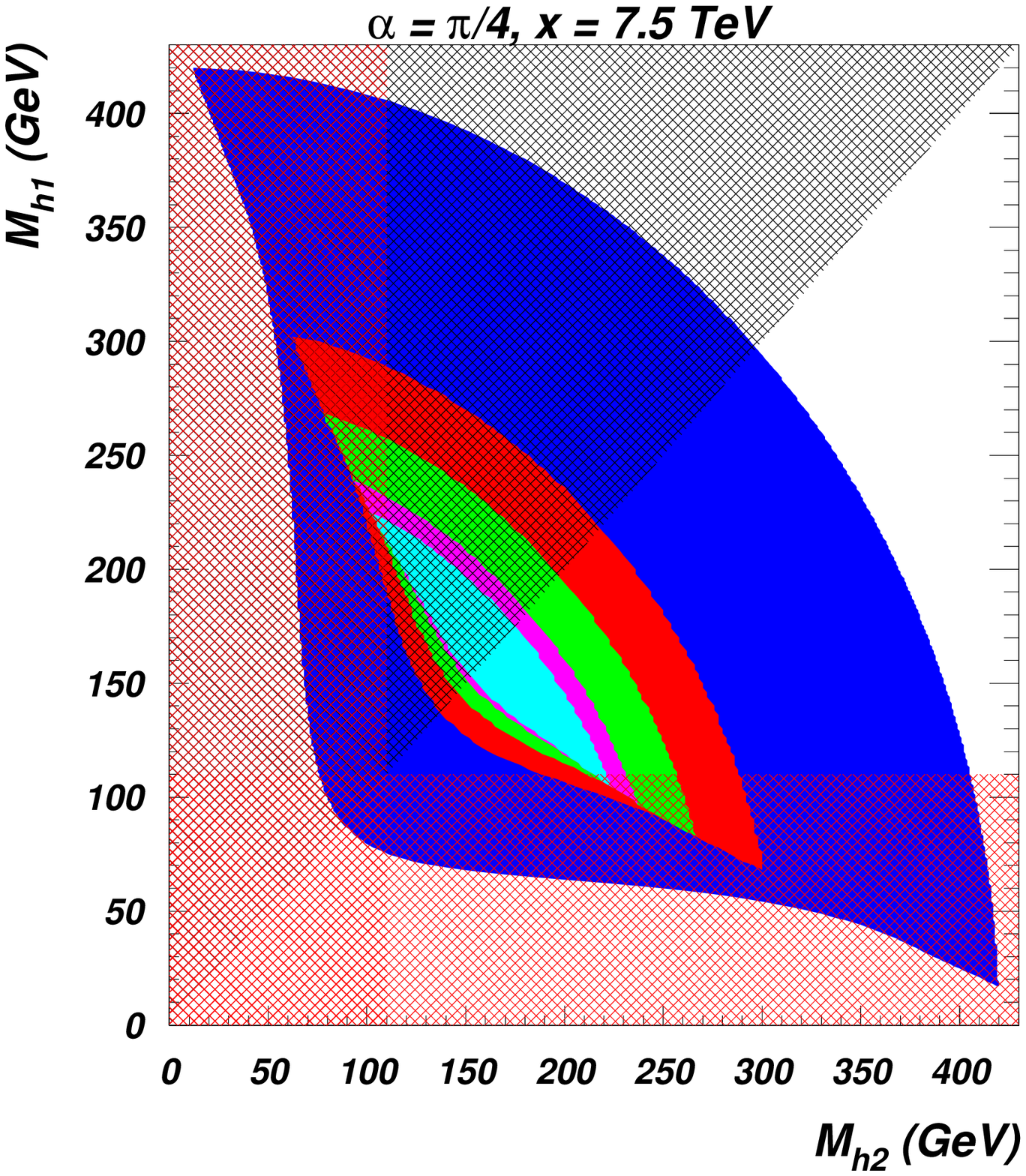}} 
  \subfigure[]{
  \label{mh1_mh2_api3}
  \includegraphics[angle=0,width=0.48\textwidth
  ]{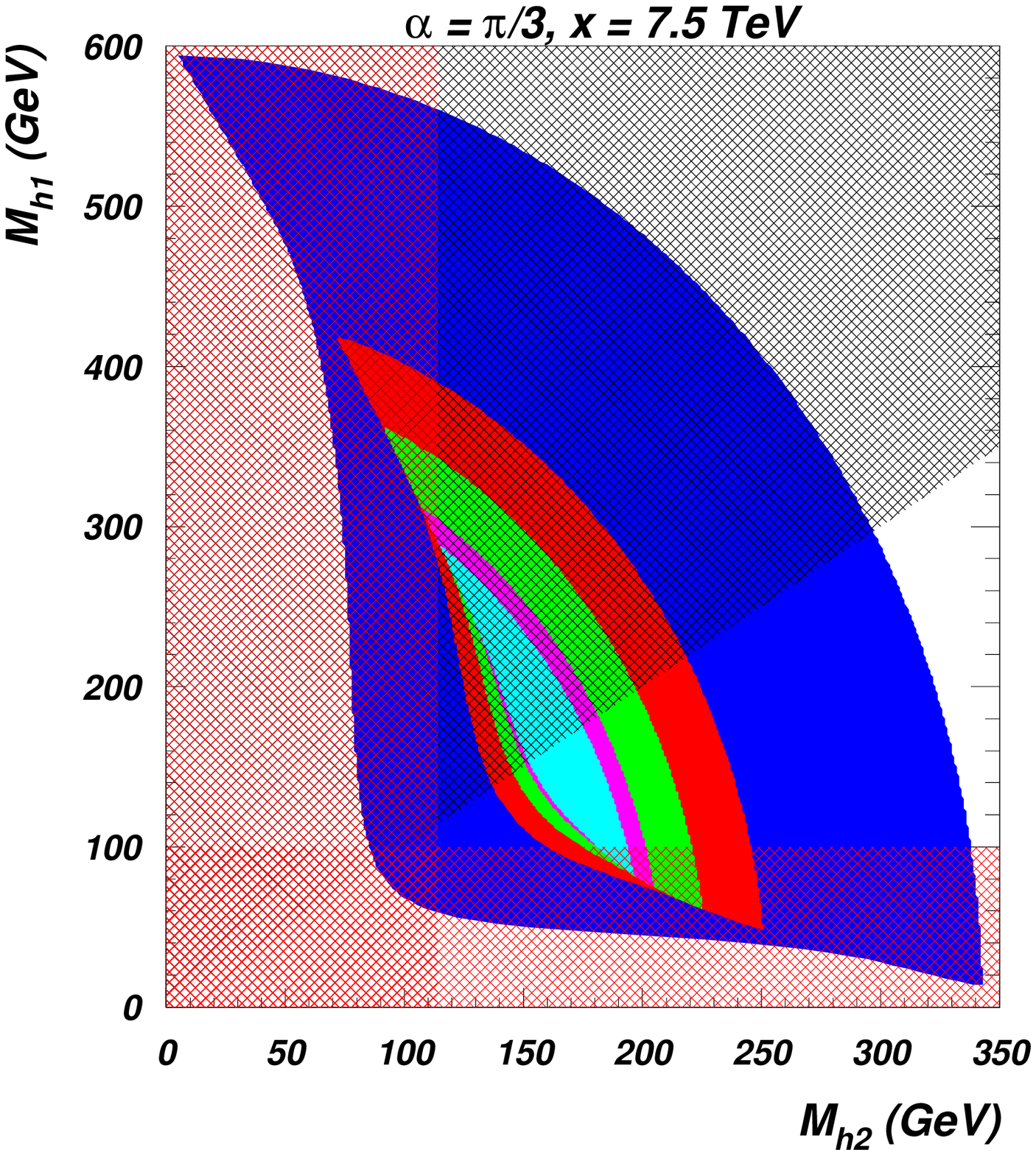}} 
  \caption[Theoretical constraints on the Higgs boson sector -
  Triviality and vacuum stability (1)]{Allowed values in the $M_{h_1}$
    vs. $M_{h_2}$ space in the
    $B-L$ model by eqs.~(\ref{cond_1}) and (\ref{cond_2}), for
    (\ref{mh1_mh2_a0}) $\alpha =0$, (\ref{mh1_mh2_a01}) $\alpha =0.1$,
    (\ref{mh1_mh2_api4}) $\alpha =\pi /4$ and (\ref{mh1_mh2_api3})
    $\alpha =\pi /3$. Colours refer to different values of $Q/$GeV:
    blue ($10^{3}$), red ($10^{7}$), green ($10^{10}$), purple
    ($10^{15}$) and cyan ($10^{19}$). The shaded black region is
    forbidden by our convention $M_{h_2} > M_{h_1}$, while the shaded
    red region refers to the values of of the scalar masses forbidden
    by LEP. Here: $x=7.5$ TeV, $M_{\nu_h}=200$ GeV.  \label{mh1_mh2}}
\end{figure}

Figure~\ref{mh1_mh2} shows the allowed region in the parameter space
$M_{h_1}$-$M_{h_2}$ for increasing values of the mixing angle
$\alpha$, for fixed $VEV$ $x=7.5$ TeV and heavy neutrino masses
$M_{\nu_h}=200$ GeV, corresponding to Yukawa couplings whose effect on
the $RGE$ running can be considered negligible.

For $\alpha=0$, the allowed values for $M_{h_1}$ are the $SM$ ones and
the extended scalar sector is completely decoupled. The allowed space
is therefore the simple direct product of the two, as we can see in
figure~\ref{mh1_mh2_a0}. When there is no mixing, the bounds that we get
for the new heavy scalar are quite loose, allowing a several TeV range
for $M_{h_2}$, still depending on the scale of validity of the theory. We
observe no significant lower bounds (i.e., $M_{h_2}>0.5$ GeV), as the
right-handed Majorana neutrino Yukawa couplings are negligible.

As we increase the value for the angle, the allowed space deforms
towards smaller values of $M_{h_1}$. If for very small scales $Q$ of
validity of the theory such masses have already been excluded by LEP,
for big enough values of $Q$, at a small angle as $\alpha=0.1$, the
presence of a heavier boson allows the model to survive up to higher
scales for smaller $h_1$ masses if compared to the $SM$ (in which just
$h_1$ would exist). Correspondingly, the constraints on $M_{h_2}$
become tighter. Moving to bigger values of the angle, the mixing
between $h_1$ and $h_2$ grows up to its maximum, at $\alpha = \pi /4$,
where $h_1$ and $h_2$ both contain an equal amount of doublet and
singlet scalars. The situation is therefore perfectly symmetric, as
one can see from figure~\ref{mh1_mh2_api4}. Finally, in
figure~\ref{mh1_mh2_api3}, we see that the bounds on $M_{h_2}$ are
getting tighter, approaching the $SM$ ones, and those for $M_{h_1}$ are
relaxing. That is, for values of the angle $\pi/4 < \alpha < \pi /2$,
the situation is qualitatively not changed, but now $h_2$ is the
$SM$-like Higgs boson. Visually, one can get the allowed regions at a
given angle $\pi/2 - \alpha$ by simply taking the specular figure
about the $M_{h_1}=M_{h_2}$ line of the plot for the given angle
$\alpha$.

\begin{figure}[!t]
  \subfigure[]{ 
  \label{mh2_a_mh1-100}
  \includegraphics[angle=0,width=0.48\textwidth
  ]{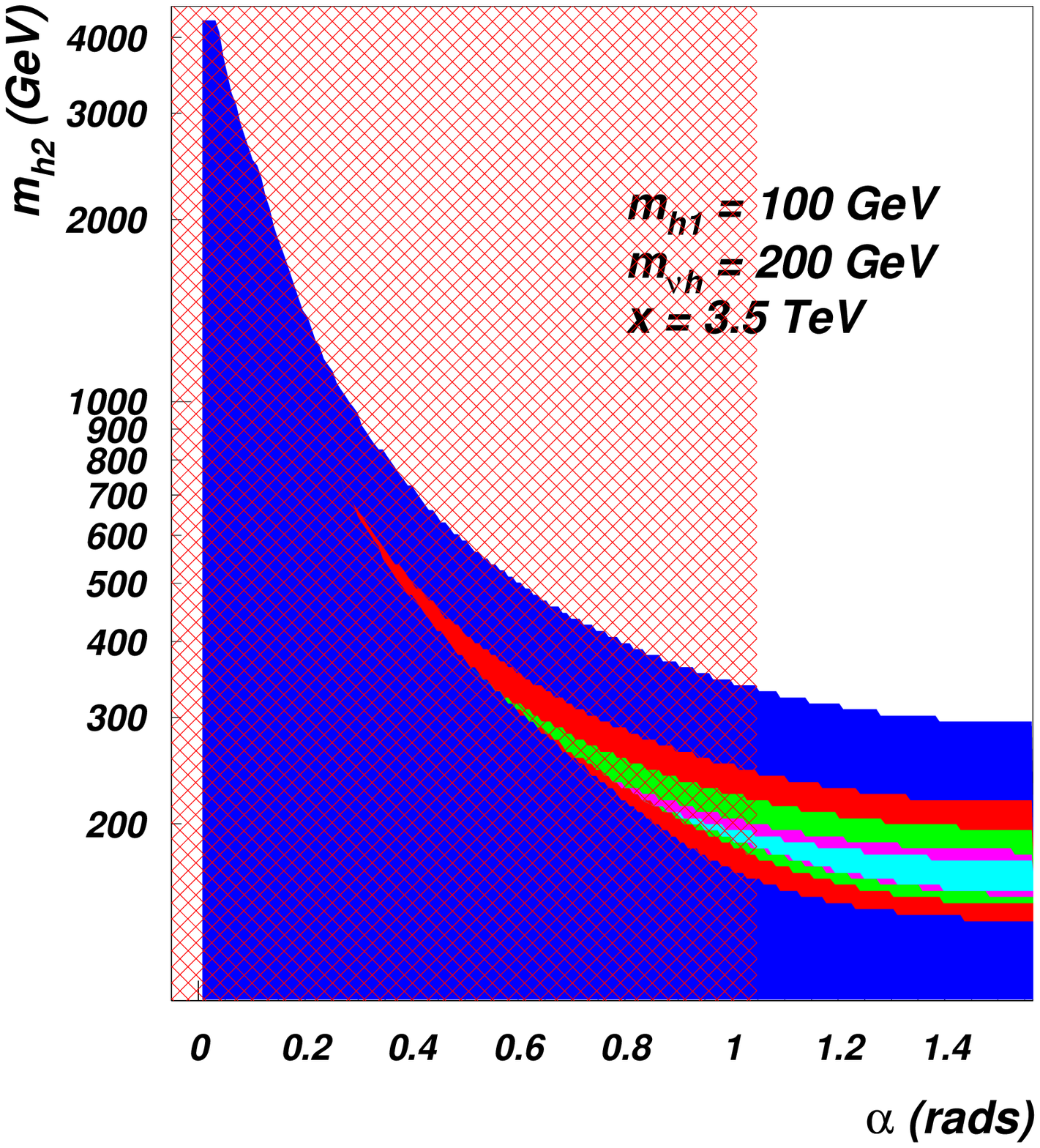}} 
  \subfigure[]{
  \label{mh2_a_mh1-120}
  \includegraphics[angle=0,width=0.48\textwidth
  ]{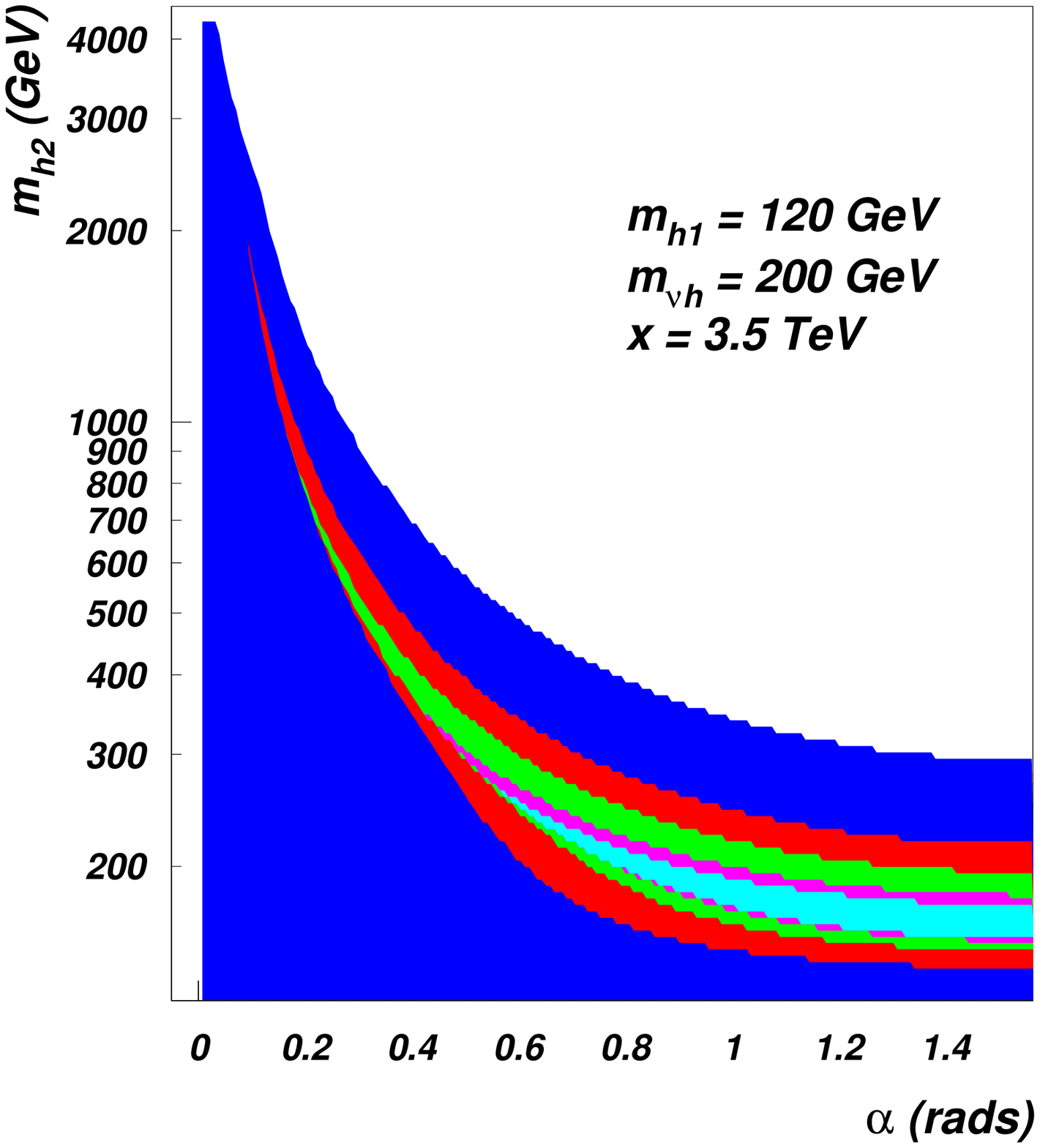}} 
\\
  \subfigure[]{
  \label{mh2_a_mh1-160}
  \includegraphics[angle=0,width=0.48\textwidth
  ]{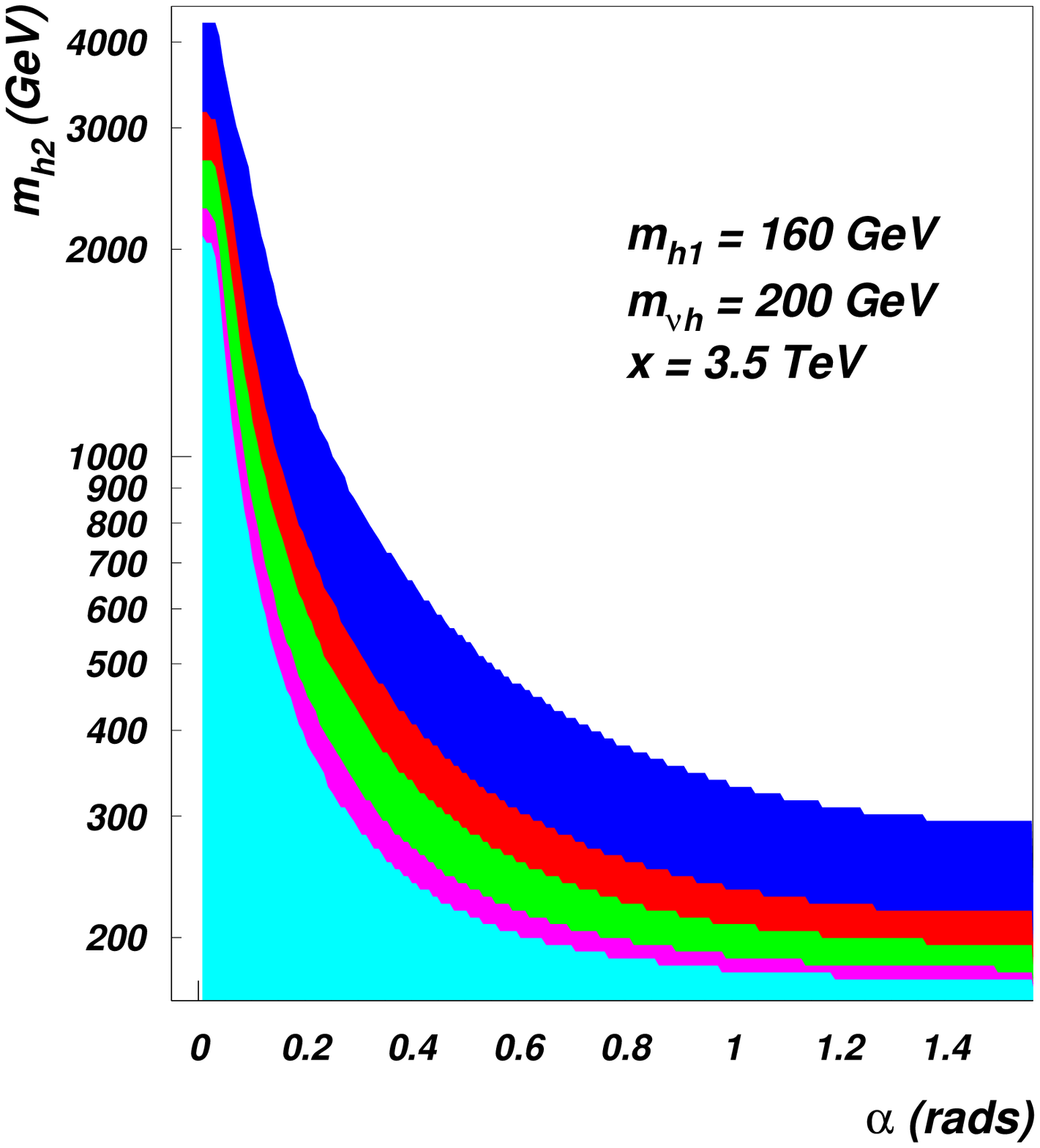}} 
  \subfigure[]{
  \label{mh2_a_mh1-180}
  \includegraphics[angle=0,width=0.48\textwidth
  ]{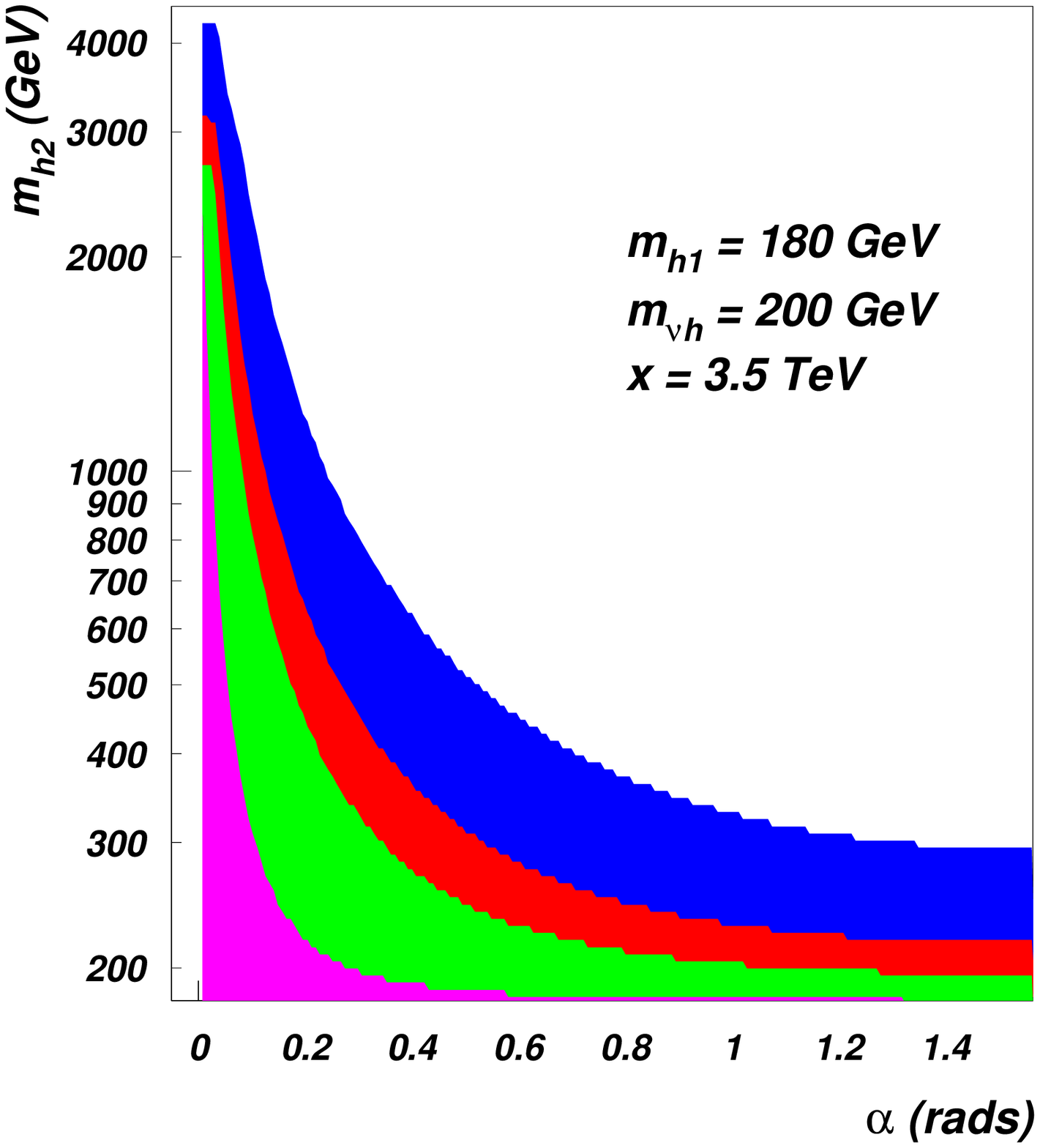}} 
  \caption[Theoretical constraints on the Higgs boson sector -
  Triviality and vacuum stability (2)]{Allowed values in the $M_{h_2}$
    vs. $\alpha$ space in the 
    $B-L$ model by eqs.~(\ref{cond_1}) and (\ref{cond_2}), for
    (\ref{mh2_a_mh1-100}) $M_{h_1}=100$ GeV, (\ref{mh2_a_mh1-100})
    $M_{h_1}=120$ GeV, (\ref{mh2_a_mh1-100}) $M_{h_1}=160$ GeV and
    (\ref{mh2_a_mh1-100}) $M_{h_1}=180$ GeV. Colours refer to
    different values of $Q/$GeV: blue ($10^{3}$), red ($10^{7}$),
    green ($10^{10}$), purple ($10^{15}$) and cyan ($10^{19}$). The
    plots already encode our convention $M_{h_2} > M_{h_1}$ and the
    shaded red region refers to the values of $\alpha$ forbidden by
    LEP. Here: $x=3.5$ TeV, $M_{\nu_h}=200$ GeV.  \label{mh2_alpha}} 
\end{figure}

Complementary to the previous study, we can now fix the light Higgs
mass at specific, experimentally interesting\footnote{The chosen
  values maximise the probability for the decays $h_1\rightarrow
  b\overline{b}$, $h_1\rightarrow \gamma \gamma$, $h_1\rightarrow
  W^+W^-$ and $h_1\rightarrow ZZ$, respectively.}, values, i.e.,
$M_{h_1} = 100$, $120$, $160$ and $180$ GeV, and show the allowed
region in the $M_{h_2}$ vs. $\alpha$ plane. This is done in
figure~\ref{mh2_alpha}.

From this figures it is clear that the transition of $h_2$ from the
new extra scalar to the $SM$-like Higgs boson as we scan on the
angle. As we increase $M_{h_1}$ (up to $M_{h_1} = 160$ GeV), a bigger
region in $M_{h_2}$ is allowed for the model to be valid up to the
Plank scale (the most inner regions, in cyan). Nonetheless, such a
region exists also for a value of the light Higgs mass excluded by LEP
for the $SM$, $M_{h_1} = 100$ GeV, but only for big values of the
mixing angle. No new regions (with respect to the $SM$) in which the
model can survive up to the Plank scale open for $M_{h_1} > 160$ GeV,
as the allowed space deforms towards smaller values of $M_{h_1}$. 

Combining the results of this Subsection with those on the previous one
(Subsection~\ref{subs:3-2-1}), we are now in a position to investigate
the production and decay phenomenology of both Higgs states of the
minimal $B-L$ model at present and future accelerators. This will be
realised in Chapter~\ref{chap:4}.

\subsection{Unitarity and triviality constraints on
  $g_1'$}\label{subs:3-2-3}

As well as constraining the Higgs sector parameters, the
perturbative unitarity and $RGE$-based techniques are also helpful to
constraining the $g_1'$ parameter.

For this reason, in this Subsection, we apply this methods to the
analysis of the $B-L$ gauge coupling.

Focusing on the computational aspects of this analysis, in
Subsection~\ref{subs:3-2-1} we have already seen that perturbative
unitarity violation at high energy occurs only in 
vector and Higgs bosons elastic scatterings, our 
interest is focused on the corresponding sectors.

We have also already made use of the Equivalence Theorem, which
guarantees that we can replace the gauge bosons and Higgses
interacting Lagrangian with the corresponding Goldstone and Higgs
bosons effective theory (further details of
this formalism can be found in Subsection~\ref{subs:3-2-1} and in
\cite{Basso:2010jt}.)


Moreover, since we want to focus on $g'_1$ limits,
we assume that the two Higgs bosons of the model have masses such that
no significant contribution to the spherical partial wave amplitude
will come from the scalar four-point and three-point functions.
Taking Higgs boson masses smaller than the unitarity limits is
therefore a way to exclude any other source of unitarity violation
different from the possible largeness of the $g'_1$ gauge coupling.

In the search for the $g'_1$ upper limits, we assume that we can
neglect all the gauge couplings in the covariant derivative but
$g_1'$, and the equation~(\ref{cov_der}) becomes:
\begin{equation}
D_{\mu}\simeq \partial _{\mu} + 2 i g_1'Z'_{\mu}\,.
\end{equation}

The inclusion of $g'_1$ in the covariant derivative gives rise to two
new Feynman rules with respect to the set described in
Appendix~\ref{appe:a}, i.e.:
\begin{eqnarray}
Z'h_1z' &=& -2 i g'_1\sin{\alpha}(p_{h_1}^\mu-p_{z'}^\mu), \\
Z'h_2z' &=& \phantom{-} 2 i g'_1 \cos{\alpha}
(p_{h_2}^\mu-p_{z'}^\mu),
\end{eqnarray}
where all the momenta are considered incoming and $z'$ is the
would-be-Goldstone boson associated with the new $Z'$ gauge field.

Now that the background is set, we focus on the techniques that
we have used to obtain the unitarity
bounds (Subsection~\ref{subs:3-2-1}) in combination with the $RGE$
analysis (Subsection~\ref{subs:3-2-2}).

Firstly, we recall that equations~(\ref{RGE_g2}) and the boundary
conditions where $g_1(EW)\simeq 0.36$ and $\widetilde{g}(EW)=0$ fully
fix the evolution of $g'_1$ with the scale.

In the search for the maximum $g'_1(EW)$ allowed by theoretical
constraints, we previously showed that the contour condition
\begin{eqnarray}\label{triviality_condition}
g'_1(\Lambda)\leq k,
\end{eqnarray}
also known as the triviality condition, is the assumption that enables
to solve the system of equations and gives the traditional upper bound on
$g'_1$ at the $EW$ scale (where it is usually assumed either $k=1$ or
$k=\sqrt{4\pi}$, calling for a coupling that preserves the
perturbative convergence of the theory).

Nevertheless, what is important to say is that this is an ``ad hoc''
assumption, while our aim is to extract the boundary condition by
perturbative unitarity arguments, showing that under certain
conditions it represents a stronger constraint on the domain of
$g'_1$.

While in the search for the Higgs boson mass bound it is widely
accepted to assume small values for the gauge couplings and large
Higgs boson masses, for our purpose we reverse such argument with the
same logic: we assume that the Higgs boson masses are compatible with
the unitarity limits and we study the two-to-two scattering amplitudes
of the whole scalar sector, pushing the largeness of $g'_1$ to the
perturbative limit.

By direct computation, also in this case it turns out that only $J=0$
(corresponding to the spherical partial wave contribution) leads to
some bound, and the only divergent 
contribution to the spherical amplitude is due to the size
of the coupling $g'_1$ in the intermediate $Z'$ vector boson exchange
contributions.
Hence, the only relevant channels are: $z'z'\rightarrow h_1 h_1$,
$z'z'\rightarrow h_1 h_2$, $z'z'\rightarrow h_2 h_2$.

As an example, we evaluate the $a_0$ partial wave amplitude for
$z'z'\rightarrow h_1 h_1$ scattering in the $s\gg M_{Z'}, M_{h_{1}}$
limit.

Firstly, we know that:
\begin{eqnarray}\label{mzpzph1h1}
M(s,\cos{\theta})\simeq (2 g'_1 \sin{\alpha})^2
\left(
1-\frac{4s}{s
\left(
1-\cos{\theta}
\right)+2M^2_{Z'}\cos{\theta}}
\right),
\end{eqnarray}
and by mean of the integration proposed in equation~(\ref{integral}),
we extract the $J=0$ partial wave:
\begin{eqnarray}\label{a_0z'z'h_1h_1}
a_0(z'z'\rightarrow h_1h_1)=\frac{(2 g'_1)^2}{16\pi}
\left(
1+2\log{\left( \frac{M_{Z'}^2}{s} \right)}
\right)\sin^2{\alpha}.
\end{eqnarray}

It is important to notice that the mass of the $Z'$ acts as a natural
regularisator that preserves both the amplitude and the spherical
partial wave from any collinear divergence.

Considering the three aforementioned scattering channels, their
spherical partial wave (in the
high energy limit $s\gg M_{Z'}, M_{h_{1,2}}$) is represented by the
following matrix: 
\begin{eqnarray}\label{a0_matrix}
a_0=f(g'_1,s;x)
\left(
\begin{tabular}{cccc}
$0$ & $\frac{1}{2}\sin^2{\alpha}$
& $-\frac{1}{\sqrt{2}} \sin{\alpha} \cos{\alpha} $ &
$\frac{1}{2}\cos^2{\alpha}$ \\
$\frac{1}{2}\sin^2{\alpha}$ & $0$ & $0$ & $0$ \\
$-\frac{1}{\sqrt{2}} \sin{\alpha} \cos{\alpha} $ & $0$ &
$0$ & $0$ \\
$\frac{1}{2}\cos^2{\alpha}$ & $0$ & $0$ & $0$ \\
\end{tabular}
\right),
\end{eqnarray}
where, according to equation~(\ref{Mzz'}),
\begin{eqnarray}\label{function}
f(g'_1,s;x)=\frac{(2 g'_1)^2}{16\pi}
\left(
1+2\log{\left( \frac{(2 g'_1 x)^2}{s} \right)}
\right),
\end{eqnarray}
and the elements of the matrix are related to the system consisting of
$\frac{1}{\sqrt{2}}z'z'$, $\frac{1}{\sqrt{2}}h_1h_1$, $h_1h_2$,
$\frac{1}{\sqrt{2}}h_2h_2$.

The most stringent unitarity bound on the $g'_1$ coupling is derived
from the requirement that the magnitude of the largest eigenvalue
combined with the function $f(g'_1,s;x)$ does not exceed
$1/2$ (see equation~(\ref{condition})).

If we diagonalise the matrix in equation~(\ref{a0_matrix}), we find
that the maximum eigenvalue\footnote{It is important to mention the
  fact that the maximum eigenvalue does not depend on $\alpha$, hence
  also the unitarity condition is $\alpha$-independent.} and the
corresponding eigenvector are:
\begin{eqnarray}\label{eigenvalue}
\frac{1}{2} \Rightarrow  \frac{1}{2}
\left( z'z' + h_1h_1\sin^2{\alpha}
- h_1h_2\sin{(2\alpha) + h_2h_2\cos^2{\alpha}}
\right).
\end{eqnarray}

Combining the informations of
equations~(\ref{function})-(\ref{eigenvalue}),
together with the perturbative unitarity condition in
equation~(\ref{condition}), we obtain:
\begin{eqnarray}\label{unitarity_condition}
|\textrm{Re}(a_0)|=\frac{(2 g'_1)^2}{32\pi}
\left|
1+2\log{\left( \frac{(2 g'_1x)^2}{s} \right)}
\right|\leq \frac{1}{2}.
\end{eqnarray}

In the last equation, $s$ represents the scale of energy squared at
which the scattering is consistent with perturbative unitarity,
i.e. $s=\Lambda^2$, where $\Lambda$ is the evolution energy scale
cut-off.

Finally, if we consider the contour of this inequality, we find
exactly the boundary condition that solve the set of differential
equations in~(\ref{RGE_g2}), giving us the upper limit for $g'_1$ at
the $EW$ scale.

\begin{figure}[!t]
  \begin{center}
        \includegraphics[angle=0,width=0.98\textwidth]{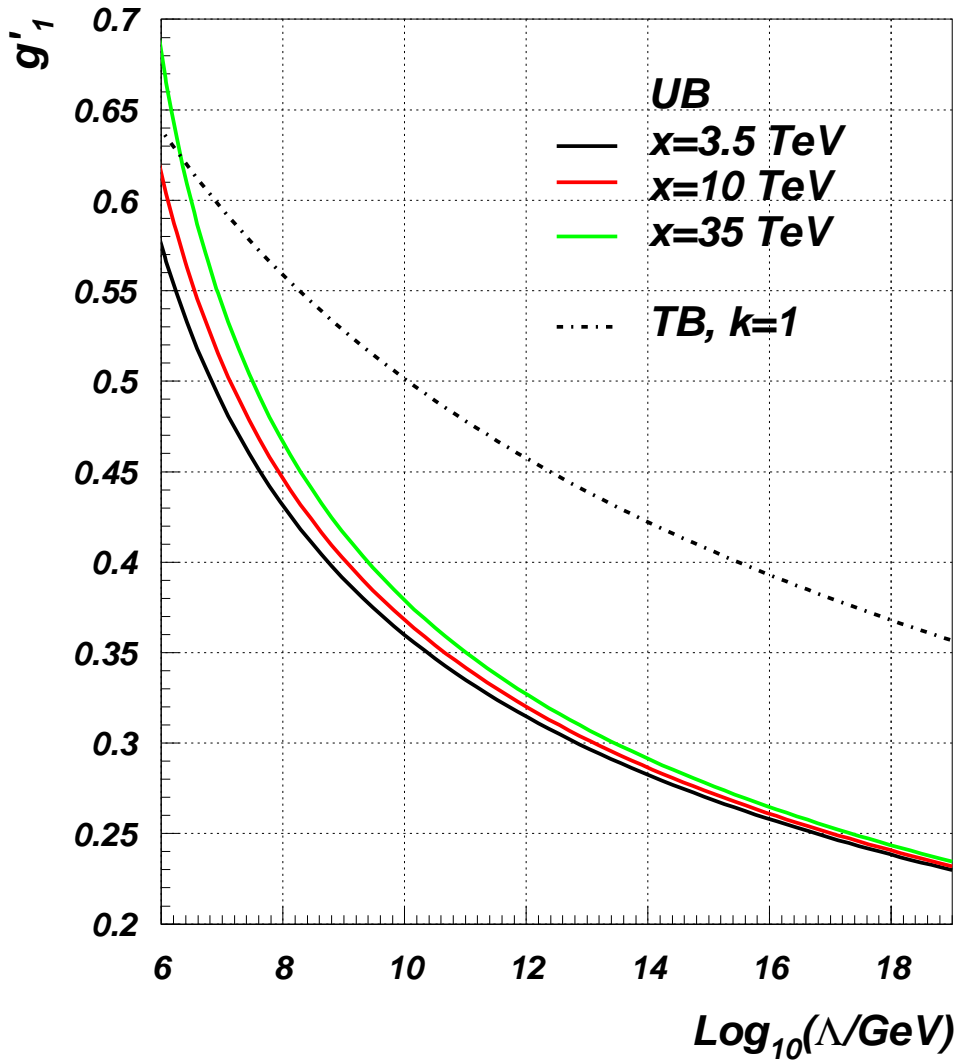}
  \end{center}
  \caption[Theoretical constraints on the Higgs boson sector -
  Unitarity and triviality constraints on $g_1'$]{Triviality (assuming
    $k=1$, dotted-dashed line) and
        unitarity (continuous and dashed lines) limits on the
        $g'_1$ coupling of
        the minimal $B-L$ model, plotted against the energy cut-off
        $\Lambda$ in $\log_{10}$-scale, for several choices of the
        singlet $VEV$ ($x=3.5$ TeV, black lines; $x=10$ TeV, red (dark
        grey) lines; $x=35$ TeV,
        green (light grey) lines).}
\label{g1pvsL}
\end{figure}

Evaluating the set of differential equations~(\ref{RGE_g2})
by means of the well-known Runge-Kutta algorithm, and imposing both
the unitarity and triviality condition as a two-point boundary value
with a simple shooting method\footnote{It consisted in varying the
  initial conditions in dichotomous-converging steps until the
  unitarity bound was fulfilled.}, we have made a comparison between
the two arguments for several values of $x$, the Higgs singlet $VEV$:
the results are plotted in figure~\ref{g1pvsL}.

By direct comparison of the two formulae in
equations~(\ref{triviality_condition})-(\ref{unitarity_condition}), it
is easy to see that the unitarity bounds become more important than
the triviality bounds when
\begin{eqnarray}
\frac{\Lambda}{x}\simeq
{\rm exp}\left(
\frac{16\pi+4}{16}
\right),
\end{eqnarray}
with the assumption that $M_{Z'} \sim x$ and $k=1$.

From this equation, it is straightforward to see that the ``ad hoc''
choice of the triviality parameter $k$ is crucial for establishing
which limit is the most stringent one.

If we then choose a value of the $VEV$ $x$ that is compatible with the
experimental limits and still in the TeV range, $x\in [3.5,35]$ TeV
(see equation~(\ref{LEP_bound})), we find that the unitarity bounds
are more stringent than the triviality ones when the energy scale is
greater than a critical value of $\Lambda_c\simeq 10^6$ GeV, and this
is consistent with the results in figure~\ref{g1pvsL}.

In order to summarise these results, in table~(\ref{g1p-up_bound}) we
present a comparison between the triviality and the unitarity bounds
for several energy scales and ($B-L$)-breaking $VEVs$.

\begin{table}[!ht]
\begin{center}
\begin{tabular}{|c|c|c|c|c|c|}
\hline
$Log_{10} (\Lambda/ \mbox{GeV}) $ & 7 & 10 & 15 & 19  \\
\hline
TB, $g_1'(k=1)$  & 0.595 & 0.501 & 0.407 & 0.357  \\ 
\hline
UB, $g_1'(x=3.5 \ \mbox{TeV})$  & 0.487 & 0.360 & 0.269
& 0.230  \\ 
\hline
UB, $g_1'(x=10 \ \mbox{TeV})$  & 0.510 & 0.368 & 0.273
& 0.232  \\ 
\hline
UB, $g_1'(x=35 \ \mbox{TeV})$  & 0.542 & 0.379 & 0.277
& 0.234  \\ 
\hline
\end{tabular}
\end{center}
\caption[Theoretical constraints on the Higgs boson sector -
  Unitarity and triviality constraints on $g_1'$]{Triviality bounds,
  equation~(\ref{triviality_condition}) with
  $k=1$, and unitarity bounds, equation~(\ref{unitarity_condition})
  with $x=3.5,10,35$ TeV, for $g'_1$ in the minimal $B-L$ model for
  several values of the energy scale $\Lambda$.
\label{g1p-up_bound}}
\end{table}

Though these results are scale dependent, we see that, if $\Lambda
\gg \Lambda_c$, our method basically refines the triviality bound by an
absolute value of $\simeq 0.1$, that represents a correction of (at
least) $20\%$ on the results that have recently appeared in the literature
(see \cite{Basso:2010jm}).

In conclusion, we have shown that, by combining perturbative unitarity
and $RGE$ methods, one can significantly constrain the $g'_1$ coupling
of the minimal $B-L$ extension of the $SM$, by imposing limits on its
upper value that are more stringent than standard triviality
bounds\footnote{Moreover, as unitarity is more constraining than
  triviality, the stability of the perturbative solution obtained
  through the former is already guaranteed by the latter.}.

\subsection{The fine-tuning constraint}\label{subs:3-2-4}

In this Subsection we discuss the ``naturalness'' aspects of the
model through the theoretical indication on the Higgs mass value that
comes from the ``fine-tuning'' problem: this is introduced by the
analysis of the one-loop radiative corrections to the Higgs
boson mass(es).

In the $SM$ framework, the dominant contributions to the quantum
corrections are
\begin{equation}\label{SM_ft}
M^2_H=(M^0_H)^2+\frac{3\Lambda^2}{8\pi^2v^2}(M_H^2+2M_W^2+M_Z^2-4M_t^2),
\end{equation}
where $\Lambda$ is the energy-scale cut-off and $M_H^0$ is the bare
mass contained in the unrenormalised Lagrangian. This induced
correction is quadratically divergent rather than the usual
logarithmic ones. If $\Lambda$ is large (for example of the order of
the grand unification scale), a fine adjustment between the bare
mass and the corrections is needed in order to have a physical Higgs
boson with a mass in the range of $EW$ symmetry breaking scale: this is
the ``naturalness'' problem of the $SM$.

However, following the Veltman conjecture (see \cite{Veltman:1980mj}),
if the following relation holds:
\begin{equation}\label{SM_veltman}
M_H^2\simeq 4M_t^2-2M_W^2-M_Z^2,
\end{equation}
not only the fine-tuning problem is smoothed, but we could have an
indirect indication of the $SM$ Higgs mass ($\sim 320$ GeV).

We apply this argument to the $B-L$ model: firstly, we have two Higgs
bosons and two related expressions for the mass radiative corrections:
\begin{eqnarray}\label{SM_veltman_1}
&\ &\Delta M_{h_1}^2 =\nonumber \\
&=& \frac{\Lambda^2}{64 M_{Z'}^2 \pi ^2 v^2} \left(3
M_{h_1}^2 \left(3 M_{Z'}^2+8 (g_1')^2 v^2\right)+M_{h_2}^2 \left(3
M_{Z'}^2+8 (g_1')^2 v^2\right) \right. \nonumber \\
&-&12 \left. \left(4 M_t^2 M_{Z'}^2+16
c_a^4 (g_1')^2 M_{\nu h}^2 v^2-M_{Z'}^2 \left(2 M_W^2+M_Z^2+4
v^2 \left((g_1')^2-2 c_a^2 s_a^2
(y^{\nu})^2\right)\right)\right)\right. \nonumber \\
&+& \left. 4 \left(M_{h_1}^2 \left(3 M_{Z'}^2-8 (g_1')^2
v^2\right)+3 \left(16 c_a^4 (g_1')^2 M_{\nu h}^2 v^2
\right. \right. \right. \nonumber
\\
&+&\left. \left. \left. M_{Z'}^2
\left(-4 M_t^2+2 M_W^2+M_Z^2-4 v^2 \left((g_1')^2+2 c_a^2
s_a^2 (y^{\nu})^2\right)\right)\right)\right) \cos{(2
  \alpha)}\right. \nonumber \\ 
&-& \left. 12 (g_1')
M_{Z'} v \left(M_{h_1}^2-M_{h_2}^2+16 \sqrt{2} c_a^3 M_{\nu h}
s_a v (y^{\nu})\right) \sin{(2 \alpha)}\right. \nonumber \\
&+&\left. (M_{h_1}^2-M_{h_2}^2)
\left(\left(3 M_{Z'}^2+8 (g_1')^2 v^2\right) \cos{(4 \alpha)}+2 (g_1')
M_{Z'} v \sin{(4 \alpha)}\right)\right),
\end{eqnarray}
and
\begin{eqnarray}\label{SM_veltman_2}
&\ &\Delta M_{h_2}^2 =\nonumber \\
&=& \frac{\Lambda^2}{64 M_{Z'}^2 \pi ^2 v^2} \left(M_{h_1}^2 \left(3
  M_{Z'}^2+8 (g_1')^2 v^2\right)+3 \left(M_{h_2}^2 \left(3 M_{Z'}^2+8
  (g_1')^2 v^2\right)\right. \right. \nonumber \\
&+& 4 \left. \left. \left(-16 c_a^4 (g_1')^2 M_{\nu h}^2
  v^2+M_{Z'}^2 \left(-4 M_t^2+2 M_W^2+M_Z^2+4 v^2
  \left((g_1')^2-2 c_a^2 s_a^2 (y^{\nu})^2\right)\right)\right)\right)
  \right. \nonumber \\
&-& 4 \left.
  \left(M_{h_2}^2 \left(3 M_{Z'}^2-8 (g_1')^2 v^2\right)+3 \left(16
  c_a^4 (g_1')^2 M_{\nu h}^2 v^2\right. \right. \right. \nonumber \\
&+& \left. \left. \left. M_{Z'}^2 \left(-4 M_t^2+2
  M_W^2+M_Z^2-4 v^2 \left((g_1')^2+2 c_a^2 s_a^2
  (y^{\nu})^2\right)\right)\right)\right) \cos{(2 \alpha)} \right. \nonumber \\
&+& \left. 12 (g_1') M_{Z'}
  v \left(-M_{h_1}^2+M_{h_2}^2+ 16 \sqrt{2} c_a^3 M_{\nu h} s_a v
  (y^{\nu})\right) \sin{(2 \alpha)}\right. \nonumber \\
&-&\left. (M_{h_1}^2-M_{h_2}^2) \left(\left(3
  M_{Z'}^2+8 (g_1')^2 v^2\right) \cos{(4 \alpha)}+2 (g_1') M_{Z'} v
  \sin{(4 \alpha)}\right)\right),
\end{eqnarray}
where we have chosen the neutrinos being degenerate ($M_{\nu
  h}^{1,2,3}\sim M_{\nu h}$, $y^{\nu}_{1,2,3}\sim y^{\nu}$) and their mixing
angles being averaged ($s_{a1,a2,a3}\sim s_a$).

By mean of the Veltman conjecture, we set the radiative one-loop
corrections to be small ($\sim 0$), obtaining a set of two equations
in two variables, $M_{h_1}$ and $M_{h_2}$, and the solution is:
\begin{eqnarray}
&\ & M_{h_1}^2 \simeq \nonumber \\
&\simeq &\left(3 \left((g_1') M_{Z'} v \left(12 
c_a^4 M_{\nu h}^2-8 M_t^2+4 M_W^2+2 M_Z^2-3 
M_{Z'}^2-16 c_a^2 s_a^2 v^2 (y^\nu)^2\right) \right. \right. \nonumber \\
&+& \left. \left. (g_1') 
M_{Z'} v \left(-12 c_a^4 M_{\nu h}^2-8 M_t^2+4 
M_W^2+2 M_Z^2+3 M_{Z'}^2-16 c_a^2 s_a^2 
v^2 (y^\nu)^2\right) \cos{(2 \alpha)}\right. \right. \nonumber \\
&+& \left. \left. 8 c_a^4 (g_1')^2 
M_{\nu h}^2 v^2 \sin{(2 \alpha)}+M_{Z'}^2 \left(-4 M_t^2+2 
M_W^2+M_Z^2-2 (g_1')^2 v^2\right) \sin{(2 \alpha)}\right. \right. \nonumber \\
&-&\left. \left. 8 
c_a^2 M_{Z'}^2 s_a^2 v^2 (y^\nu)^2 \sin{(2
  \alpha)}- 4 
\sqrt{2} c_a^3 M_{\nu h} s_a v (y^\nu) \left(2 \cos{(\alpha)} 
\sec{(2 \alpha)} \sin{(\alpha)} \times
\right. \right. \right. \nonumber \\ 
&\times &\left. \left. \left. \left(8 (g_1')^2 v^2 
\cos^2{(\alpha)}+3 M_{Z'}^2
\sin^2{(\alpha)}\right)+ 3(g_1') 
M_{Z'} v \sin{(2 \alpha)} \tan{(2 
  \alpha)}\right)\right)\right)\div \nonumber \\
&\div & \left(-12 (g_1') M_{Z'} v 
\cos{(2 \alpha)}+\left(-3 M_{Z'}^2+4 (g_1')^2 v^2\right) 
\sin{(2 \alpha)}\right),
\end{eqnarray}
and
\begin{eqnarray}
&\ &M_{h_2}^2 \simeq \nonumber \\
&\simeq &\left(3 \left((g_1') M_{Z'} v \left(-12
c_a^4 M_{\nu h}^2+8 M_t^2-4 M_W^2-2 M_Z^2+3
M_{Z'}^2+16 c_a^2 s_a^2 v^2 (y^\nu)^2\right) \right. \right. \nonumber \\
&+&\left. \left. (g_1')
M_{Z'} v \left(-12 c_a^4 M_{\nu h}^2-8 M_t^2+4
M_W^2+2 M_Z^2+3 M_{Z'}^2-16 c_a^2 s_a^2
v^2 (y^\nu)^2\right) \cos{(2 \alpha)}\right. \right. \nonumber \\
&+&8 \left. \left. c_a^4 (g_1')^2
M_{\nu h}^2 v^2 \sin{(2 \alpha)}+M_{Z'}^2 \left(-4 M_t^2+2
M_W^2+M_Z^2-2 (g_1')^2 v^2\right) \sin{(2 \alpha)} \right. \right. \nonumber \\
&-&\left. \left. 8
c_a^2 M_{Z'}^2 s_a^2 v^2 (y^\nu)^2 \sin{(2 \alpha)}+4
\sqrt{2} c_a^3 M_{\nu h} s_a v (y^\nu) \left(2 \cos{(\alpha)}
\sec{(2 \alpha)} \sin{(\alpha)} \times \right. \right. \right. \nonumber \\
&\times &\left. \left. \left. \left(3 M_{Z'}^2 \cos^2{(\alpha)}+8
(g_1')^2 v^2 \sin^2{(\alpha)}\right)-3 (g_1') M_{Z'} v
\sin{(2 \alpha)} \tan{(2 \alpha)}\right)\right)\right) \div \nonumber
\\
&\div &\left(-12
(g_1') M_{Z'} v \cos{(2 \alpha)}+\left(-3 M_{Z'}^2+4
(g_1')^2 v^2\right) \sin{(2 \alpha)}\right).
\end{eqnarray}

Indeed, it is important to say that the Veltman conjecture is only
indicative of the one-loop behaviour\footnote{Moreover, since too many
  free parameters enter in the game, no significative study could be
  performed in the $B-L$ context.}, and evaluating the stability of
the Veltman solution at two-loops is not among the purposes of this
work. However, it is interesting to notice that in the decoupled
scenario $\alpha=0$, the solution is:
\begin{eqnarray}\label{ft_sol_a0_h1}
M_{h_1}^2 &\simeq &4 M_t^2 -2 M_W^2 - M_Z^2 + 8 c_a^2s_a^2 v^2 (y^\nu)^2, \\
M_{h_2}^2 &\simeq &6 c_a^4 M_{\nu h}^2 - \frac{3
  M_{Z'}^2}{2}. \label{ft_sol_a0_h2}
\end{eqnarray}

We find the $SM$-like solution plus a non-significative contribution 
from heavy neutrinos (equation~(\ref{ft_sol_a0_h1})), while the second
solution is totally ruled by the interplay between the heavy neutrinos
and $Z'$ boson mass (equation~(\ref{ft_sol_a0_h2})).

Again, too many free-parameters are floating around, hence we
can only give an illustrative example of a choice of parameters
related to a reasonable amount of fine-tuning.

The role of heavy neutrinos in the first equation is absolutely
negligible, so as in the $SM$ case $M_{h_1}\simeq 320$ GeV represents
a good choice for the light Higgs; assuming a relatively heavy
right-handed neutrinos ($\sim 400$ GeV) and a relatively light $Z'$
boson ($\sim 700$ GeV), the Veltman conjecture implies that a theory
with a heavy Higgs boson with mass of $\sim 470$ GeV is the most
``natural''.

\chapter{Higgs phenomenology at colliders} 
\label{chap:4}
\lhead{Chapter 4. \emph{Higgs phenomenology at colliders}} 

This Chapter is devoted to the presentation of the results of our
phenomenological investigation of the $B-L$ Higgs sector.

Firstly, after a summary of our choice of parameters and
computational details, we present
the total decay width and the branching ratios ($BRs$) of the Higgs
bosons, in order to set the background.

Then, we focus on the LHC experimental
scenario, with emphasis on the possible peculiar signatures of the
model.

Finally, we will present a set of possible signature cross sections
that could represent a relevant test for the $SM$ break-down at future
LCs.

\section{Parameter space and implementation of the model in
  CalcHEP}\label{sect:4-1}

As spelled out in the previous Chapter, the independent physical
parameters of the Higgs sector of the scenario considered here are:
\begin{itemize}
\item $M_{h_1}$, $M_{h_2}$ and $\alpha$, the Higgs boson masses and
  mixing angle. We investigate this parameter space spanning
  over continuous intervals in the case of 
  the first two quantities while adopting discrete values for the
  third one (taking into account all the experimental and theoretical
  constraints presented in Chapter~\ref{chap:3}).
\end{itemize}

In order to explore efficiently the expanse of parameter space
pertaining to the minimal $B-L$ model, we introduce two extreme
conditions, which makes the model intuitive, though at the end it
should be borne in mind that intermediate solutions are
phenomenologically favoured. The two conditions are obtained by
setting:
\begin{enumerate}
\item $\alpha=0$, this is the ``decoupling limit'', with $h_1$
  behaving like the $SM$ Higgs.
\item $\alpha=\frac{\pi}{2}$, this is the so-called ``inversion
  limit'', with $h_2$ behaving like the $SM$ Higgs (though recall that
  this possibility is phenomenologically not viable, see
  Subsection~\ref{subs:3-1-1}).
\end{enumerate}

Furthermore, concerning the strength of Higgs interactions, some of
the salient phenomenological behaviours can be summarised as
follows\footnote{All of these results are a consequence of the way the
particles couple, i.e., the Feynman rules in Appendix~\ref{appe:b}.}:
\begin{itemize}
\item $SM$-like interactions scale with $\cos{\alpha}$($\sin{\alpha}$)
  for $h_1$($h_2$);
\item those involving the other new $B-L$ fields, like $Z'$ and heavy
  neutrinos, scale with the complementary angle, i.e., with
  $\sin{\alpha}$($\cos{\alpha}$) for $h_1$($h_2$);
\item triple (and quadruple) Higgs couplings are possible and can
  induce resonant behaviours, so that, e.g., the $h_2\rightarrow
  h_1\,h_1$ decay can become dominant if $M_{h_2}\sim 2M_{h_1}$.
\end{itemize}

Other than $M_{h_1}$, $M_{h_2}$ and $\alpha$, additional parameters
are the following:
\begin{itemize}
\item $g'_1$, the new $U(1)_{B-L}$ gauge  coupling. We will adopt
  discrete perturbative values for this quantity, taking into account
  the analysis of the allowed parameter space of
  Subsection~\ref{subs:3-2-3}.
\item $M_{Z'}$, the new gauge boson mass, considering the experimental
  limits presented in Subsection~\ref{subs:3-1-3}
  (equation~(\ref{LEP_bound})).
\item $M_{\nu _h}$, the heavy neutrino masses, are assumed to be
  degenerate, diagonal and relatively light.
\item $M_{\nu _l}$, the $SM$ (or light) neutrino masses; they have
  been conservatively taken to be $M_{\nu_l}=10^{-2}$ eV in order to
  fulfil the cosmological bound requirement presented in
  Subsection~\ref{subs:3-1-3}; they are assumed to be degenerate and
  diagonal.
\end{itemize}

The numerical analysis was performed with CalcHEP~\cite{Pukhov:2004ca}
with the model introduced through
LanHEP~\cite{Semenov:1996es}. Moreover, the implementation has been
enriched with the following plug-ins:
\begin{itemize}
\item the one-loop vertices $g-g-h_1(h_2)$\footnote{Finally, the $NLO$
  $QCD$ $k$-factor for the gluon fusion process
  \cite{Graudenz:1992pv,Spira:1995rr,Djouadi:2005gi} has been
  used. Regarding the other processes, we decided to not implement
  their $k$-factors since they are much smaller in comparison.},
  $\gamma -\gamma-h_1(h_2)$ and $\gamma -Z(Z')-h_1(h_2)$ via $W$ gauge
  bosons and heavy quarks (top, bottom and charm) have been
  implemented, adapting the formulae in \cite{Gunion:1989we}.
\item Running masses for top, bottom and charm quarks, evaluated at
  the Higgs boson mass: $Q= M_{h_1}(M_{h_2})$, depending on which
  scalar boson is involved in the interaction.
\item Running of the $QCD$ coupling constant, at two-loops with $5$
  active flavours.
\end{itemize}

\section{Branching ratios and total widths}\label{sect:4-2}

\begin{figure}[!t]
  \subfigure[]{ 
  \label{BR_h1}
  \includegraphics[angle=0,width=0.48\textwidth ]{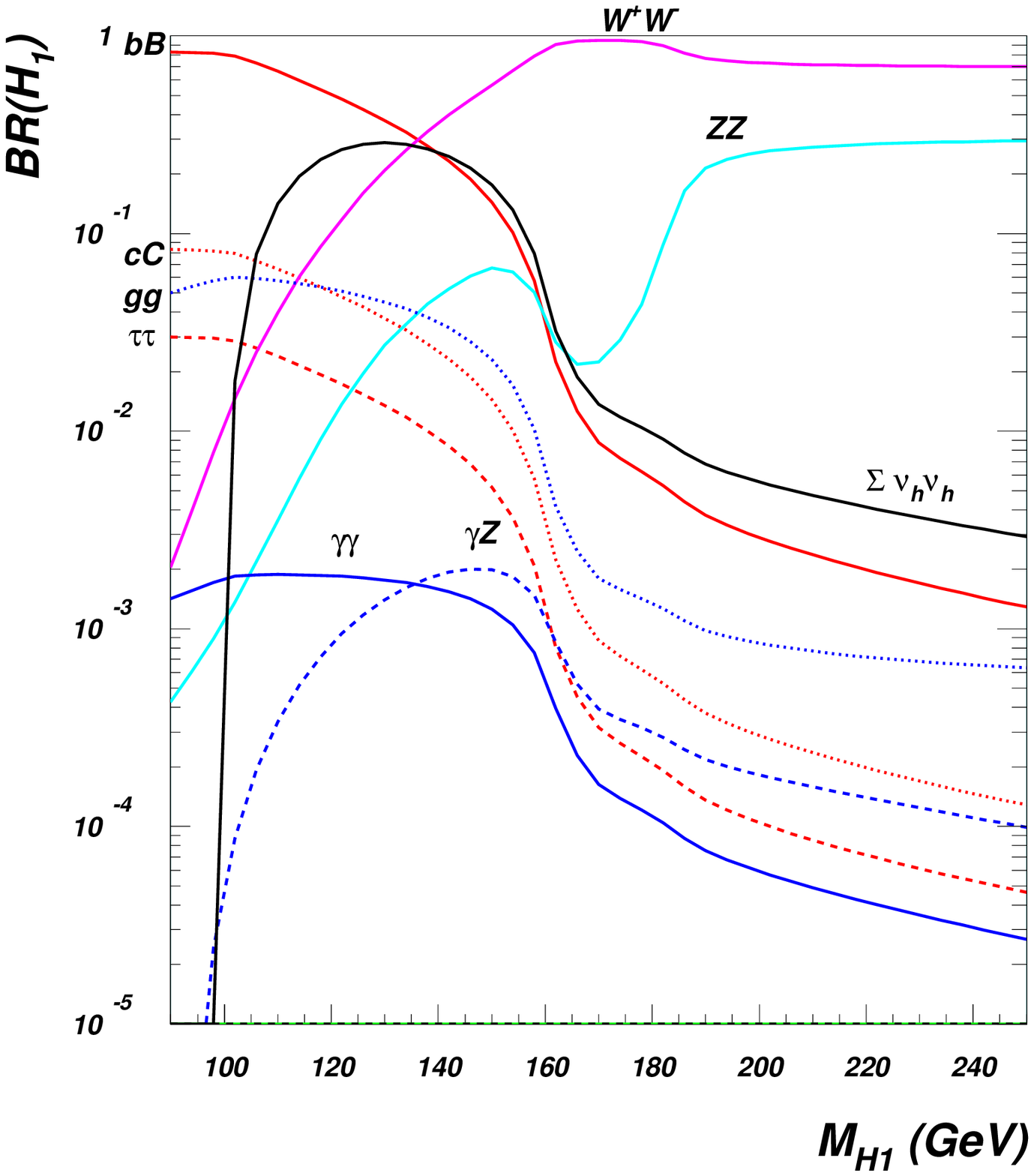}}
  \subfigure[]{
  \label{BR_h2}
  \includegraphics[angle=0,width=0.48\textwidth ]{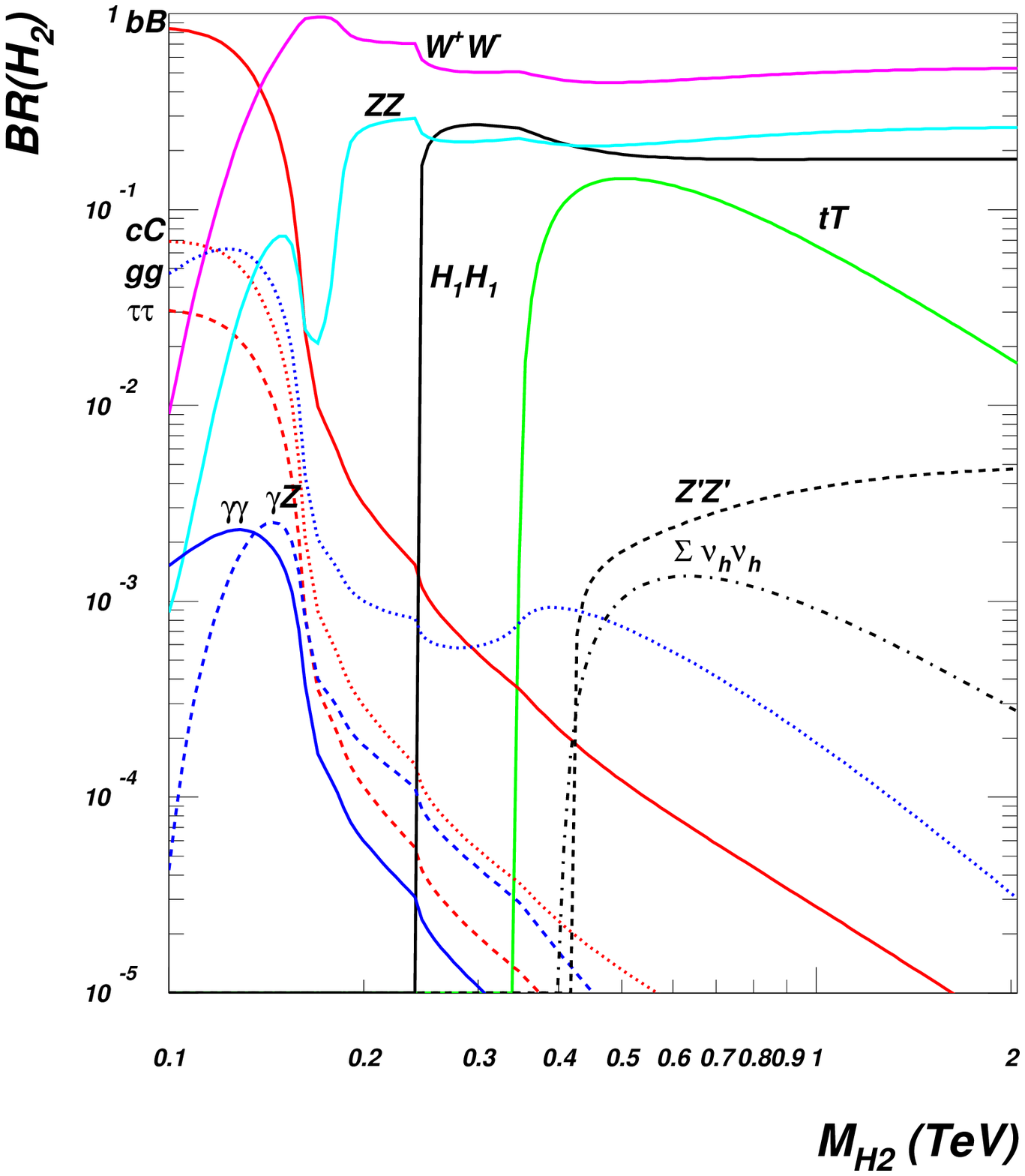}} \\
  \subfigure[]{ 
  \label{H1_TW}
  \includegraphics[angle=0,width=0.48\textwidth ]{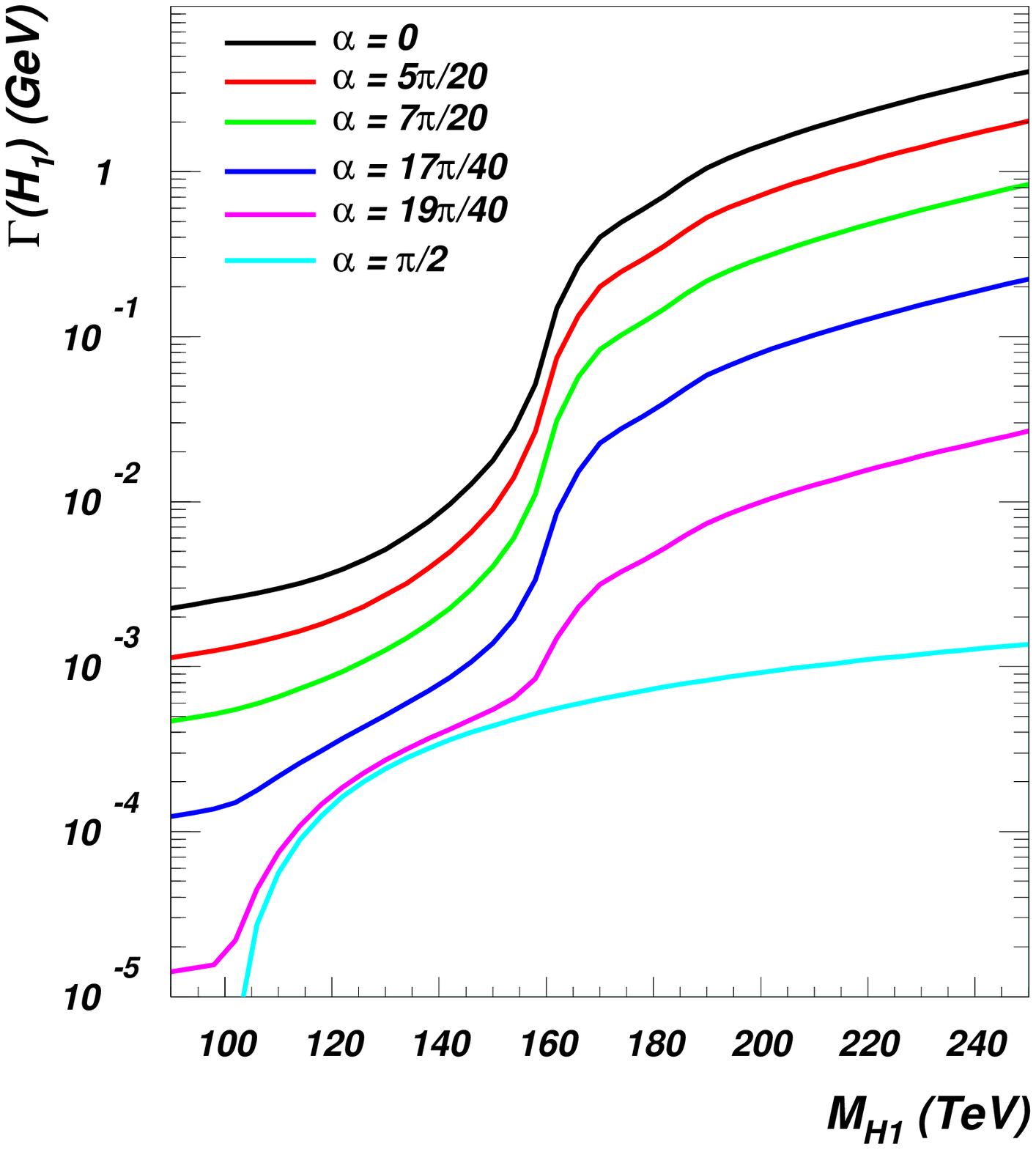}}
  \subfigure[]{
  \label{H2_TW}
  \includegraphics[angle=0,width=0.48\textwidth ]{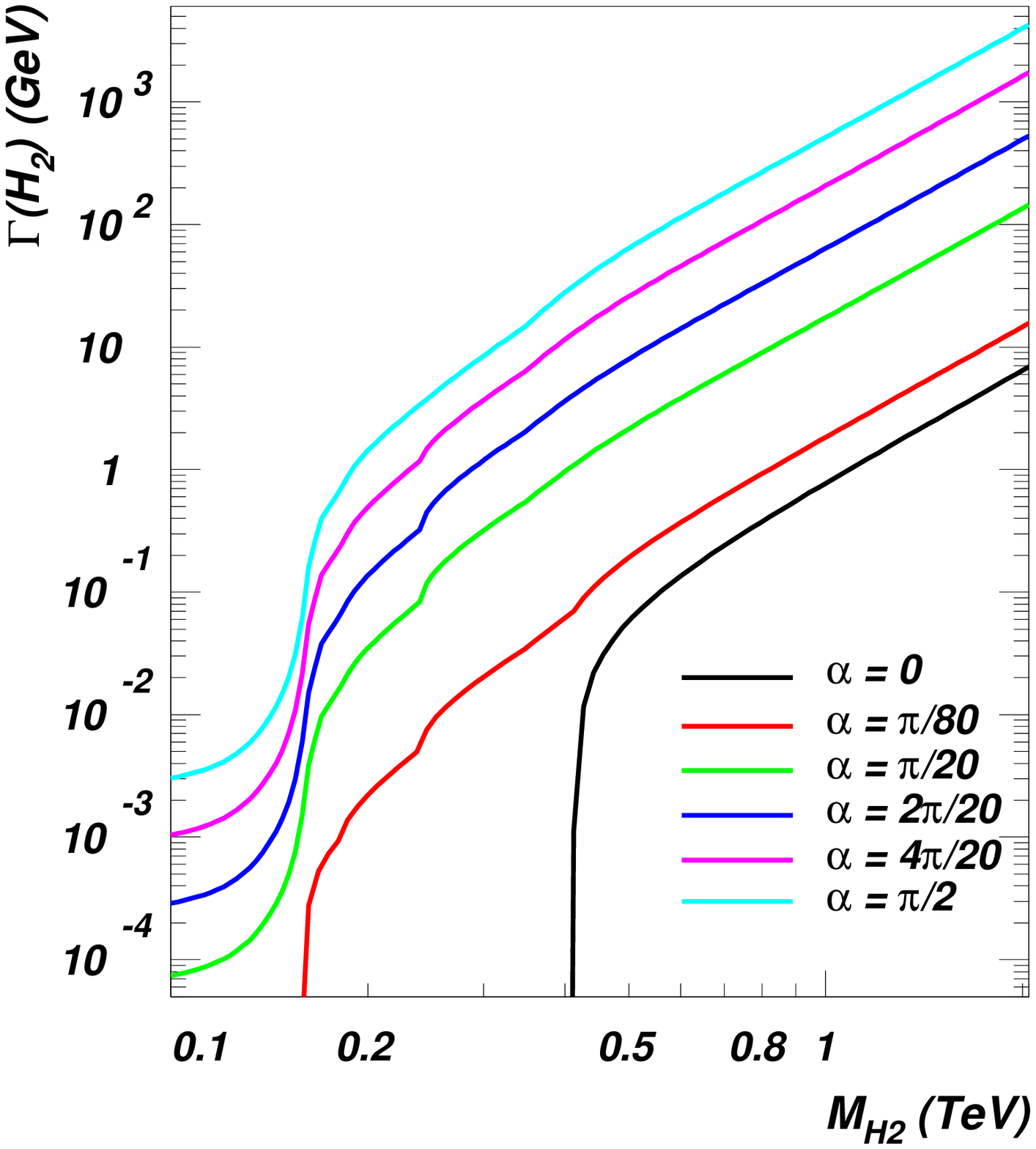}} 
  \caption[Branching ratios and total widths (1)]{(\ref{BR_h1})
    Branching ratios for $h_1$ for $\alpha
    =2\pi /5$ and $M_{\nu_h}=50$ GeV and (\ref{H1_TW}) $h_1$ total
    width for a choice of mixing angles and (\ref{BR_h2}) $BRs$ for
    $h_2$ for $\alpha =3\pi /20$ and $M_{h_1}=120$ GeV, $M_{Z'}=210$
    GeV and $M_{\nu_h}=200$ GeV and (\ref{H2_TW}) $h_2$ total width
    for a choice of mixing angles.}
  \label{Brs}
\end{figure}

Moving to the Higgs boson decays, figure~\ref{Brs} shows the $BRs$ for
both the Higgs bosons, $h_1$ and $h_2$, respectively. Only the
two-body decay channels are shown here.

Regarding the light Higgs boson, the only new particle it can decay
into is the heavy neutrino (we consider a very light $Z'$ boson
unlikely), if the channel is kinematically open. In
figure~\ref{BR_h1} we show this case, for a small heavy neutrino mass,
i.e., $M_{\nu_h}=50$ GeV, and we see that the relative $BR$ of this
channel can be rather important, as the decay into $b$-quark pairs or
into $W$ boson pairs, in the range of masses $110$ GeV $\leq M_{h_1}
\leq 150$ GeV. Such range happens to be critical in the $SM$ since here
the $SM$ Higgs boson passes from decaying dominantly into $b$-quark
pairs to a region in masses in which the decay into $W$ boson pairs is
the prevailing one. These two decay channels have completely different
signatures and discovery methods. The fact that the signal of
the Higgs boson decaying into $b$-quark pairs is many orders of
magnitude below the natural $QCD$ background spoils its sensitivity. In
the case of the $B-L$ model, the decay into heavy neutrino pairs is
therefore phenomenologically very important, besides being an
interesting feature of the $B-L$ model if $M_{\nu_h} < M_W$, as it
allows multileptons signatures of the light Higgs boson. Among them,
there is the decay of the Higgs boson into $3\ell$, $2j$ and $\met$
(that we have already studied for the $Z'$ case in
reference~\cite{Basso:2008iv}), into $4\ell$ and
$\met$ (as, again, already studied for the $Z'$ case in
reference~\cite{Huitu:2008gf}) or into $4\ell$ and $2j$ (as already
studied, when $\ell = \mu$, in the $4^{th}$ family extension of the $SM$
\cite{CuhadarDonszelmann:2008jp}). All these peculiar signatures allow
the Higgs boson signal to be studied in channels much cleaner than the
decay into $b$-quark pairs.

\begin{figure}[!t]
\centering
\subfigure[]{ 
  \label{H2_BR-a_H1}
  \includegraphics[angle=0,width=0.48\textwidth ]{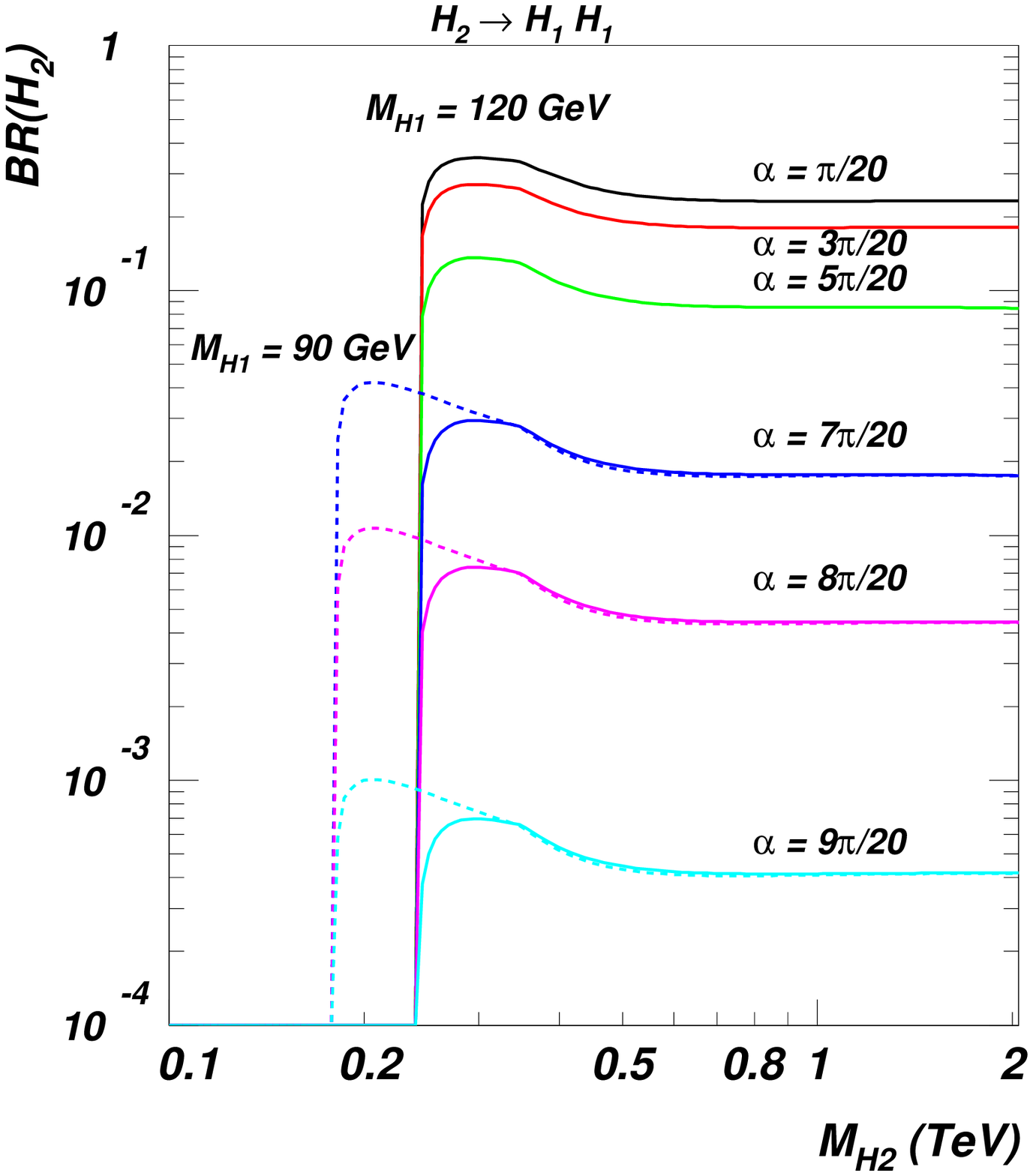}
}\\
  \subfigure[]{
  \label{H2_BR-a_Hnu}
  \includegraphics[angle=0,width=0.48\textwidth ]{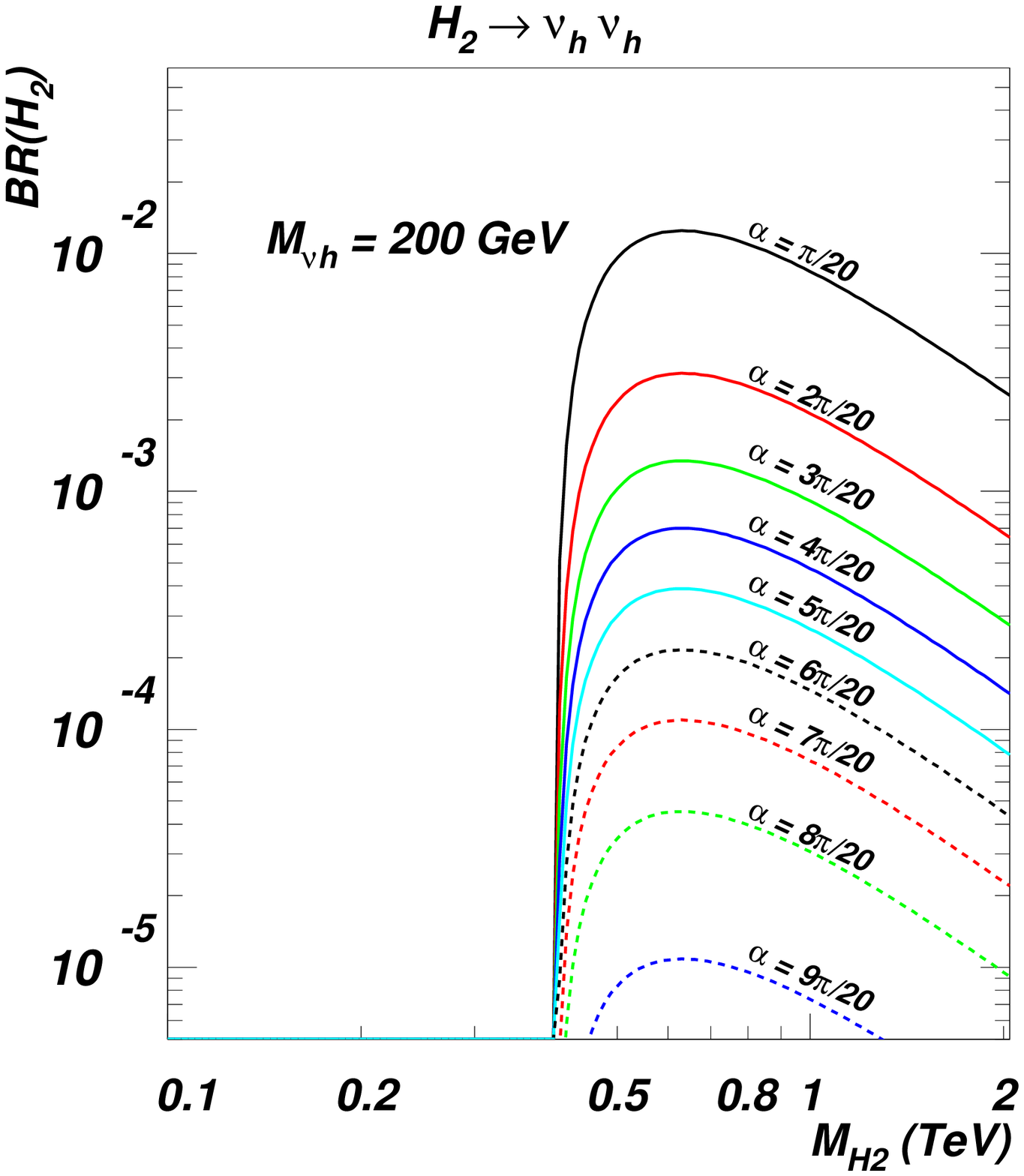}}  
  \subfigure[]{
  \label{H2_BR-a_Zp}
  \includegraphics[angle=0,width=0.48\textwidth ]{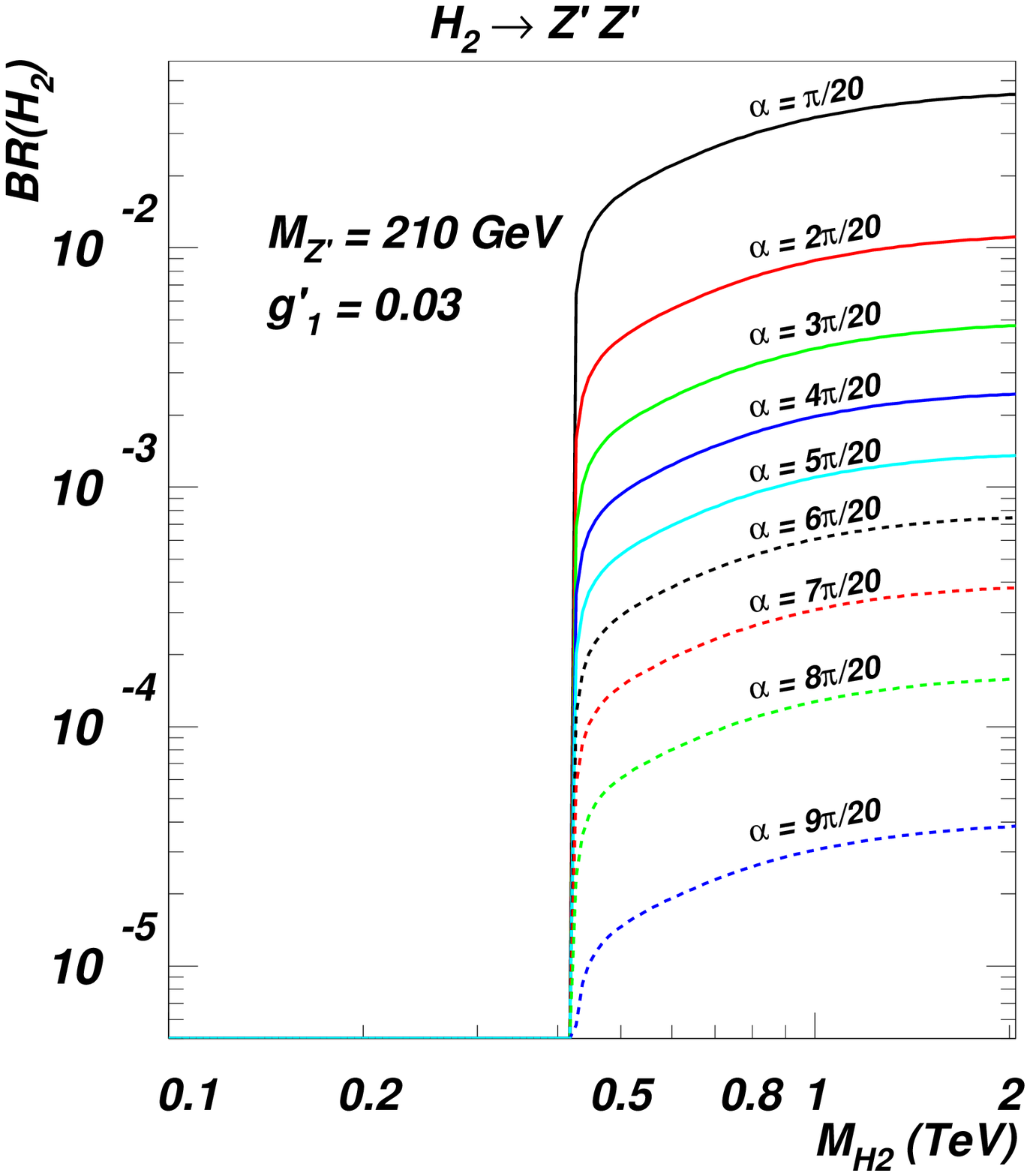}}  
  \caption[Branching ratios and total widths (2)]{Dependence on the
    mixing angle $\alpha$ of 
    (\ref{H2_BR-a_H1}) $BR(h_2\rightarrow h_1\, h_1)$,  of
    (\ref{H2_BR-a_Hnu}) $BR(h_2\rightarrow \nu _h\, \nu _h)$ and of
    (\ref{H2_BR-a_Zp}) $BR(h_2\rightarrow Z'\, Z')$.}
  \label{Br-alpha}
\end{figure}

In the case of the heavy Higgs boson, further decay channels are
possible in the $B-L$ model, if kinematically open. The heavy Higgs
boson can decay in pairs of the light Higgs boson ($h_2 \rightarrow
h_1\, h_1$) or even in triplets ($h_2 \rightarrow h_1\, h_1\, h_1$),
in pairs of heavy neutrinos and $Z'$ bosons. Even for a small value of
the angle, figure~\ref{BR_h2} shows that the decay of a heavy Higgs
boson into pairs of the light one can be quite sizeable, at the level
of the decay into $SM$ $Z$ bosons for $M_{h_1} = 120$ GeV. It is
important to note that this channel does not have a simple dependence
on the mixing angle $\alpha$, as we can see in figure~\ref{Br-alpha}.

The $BRs$ of the heavy Higgs boson decaying into $Z'$ boson pairs and
heavy neutrino pairs decrease as the mixing angle increases, getting
to their maxima (comparable to the $W$ and $Z$ ones) for a vanishing
$\alpha$, for which the production cross section is however
negligible. As usual, and also clear from figure~\ref{BR_h2}, the
decay of the heavy Higgs boson into $Z'$ gauge bosons is
always bigger than the decay into pairs of fermions (the heavy
neutrinos, even when summed over the generations as plotted), when
they have comparable masses (here, $M_{Z'}=210$ GeV and $M_{\nu _h} =
200$ GeV).

\begin{figure}[!t]
\centering
  \subfigure[]{
  \label{H2-Br-3H1}
  \includegraphics[angle=0,width=0.48\textwidth
  ]{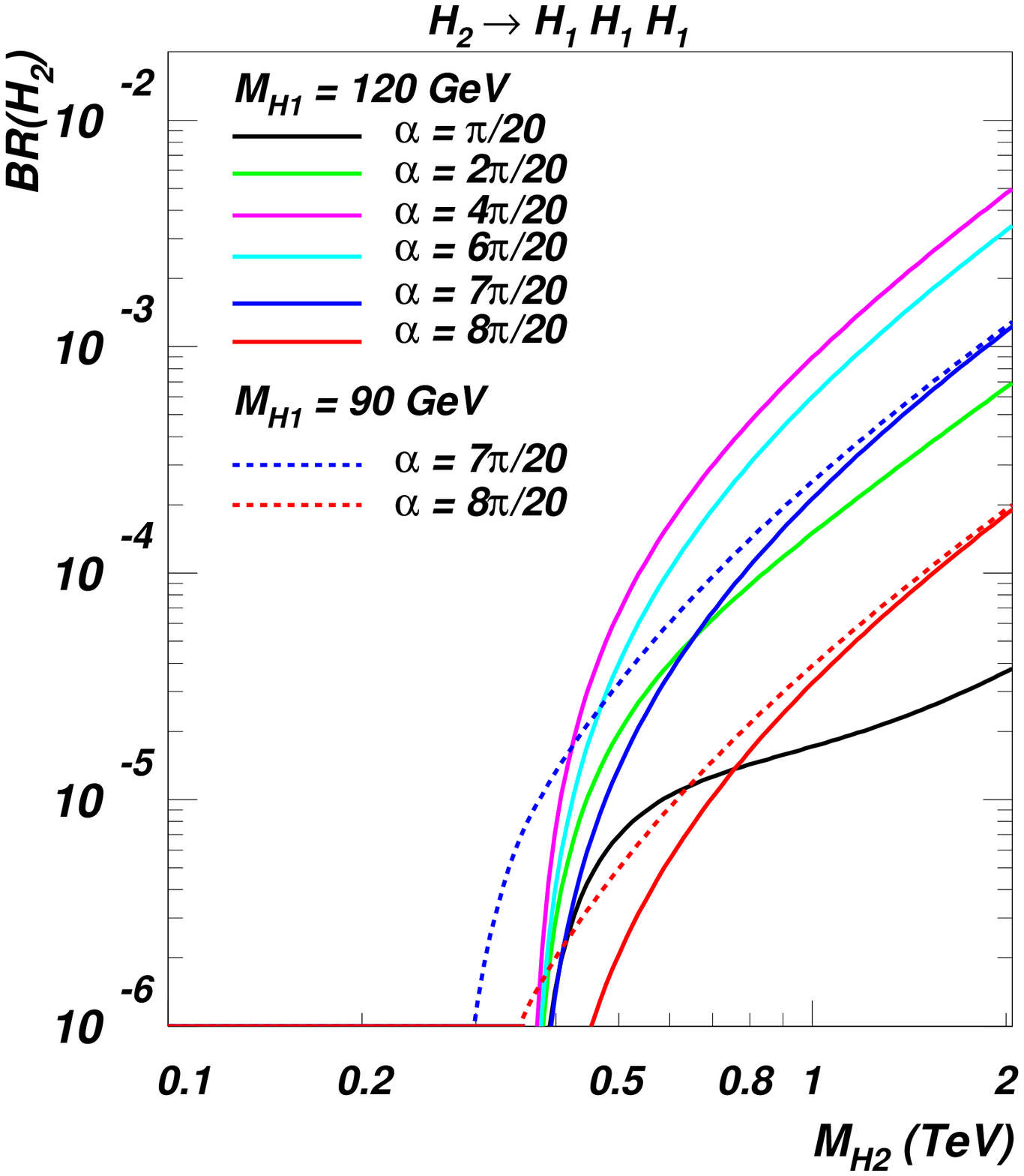}}\\ 
  \subfigure[]{
  \label{H2-Br-2Vh1}
  \includegraphics[angle=0,width=0.48\textwidth ]{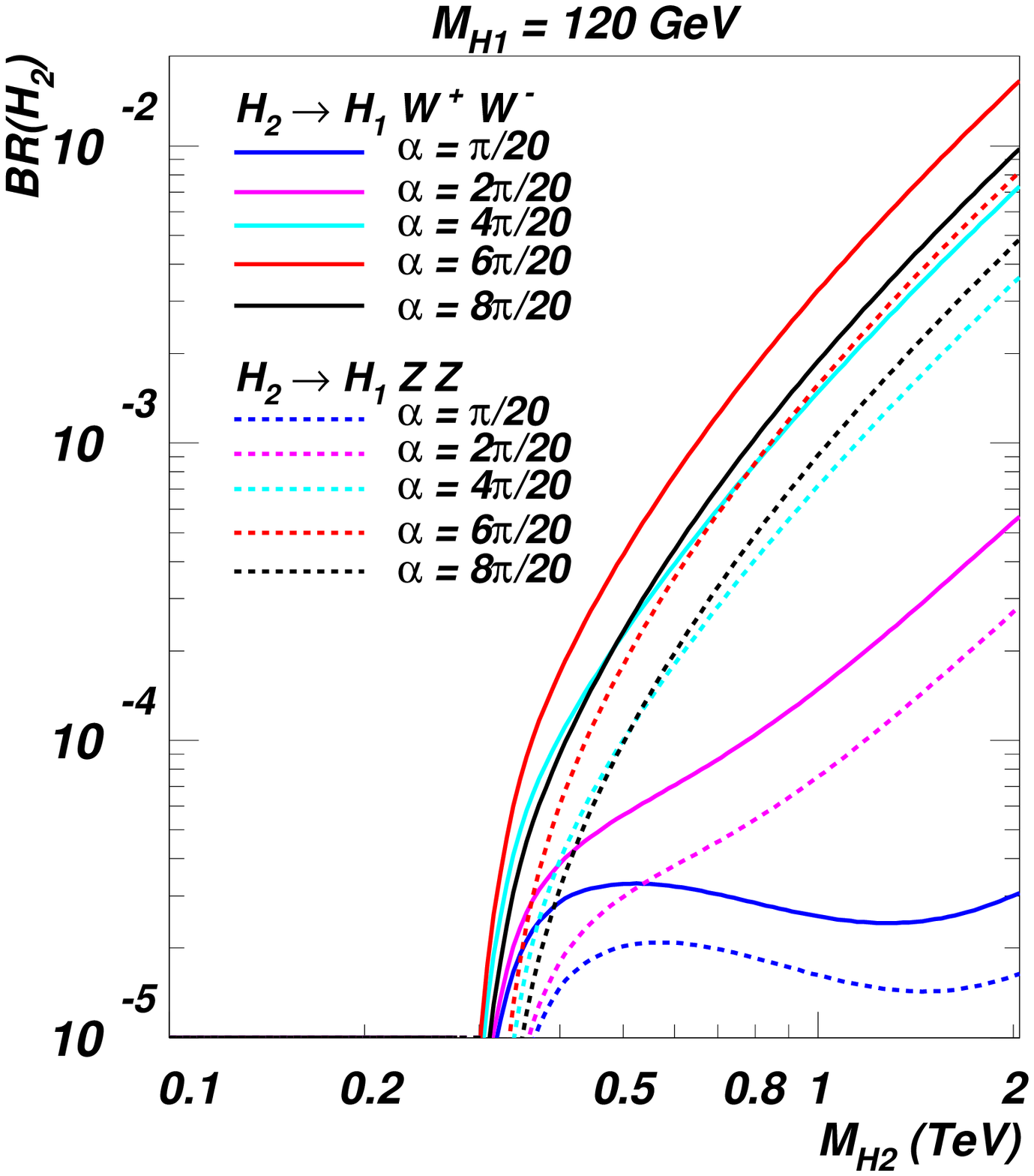}}
  \subfigure[]{
  \label{H2-Br-2Vh1-90}
  \includegraphics[angle=0,width=0.48\textwidth ]{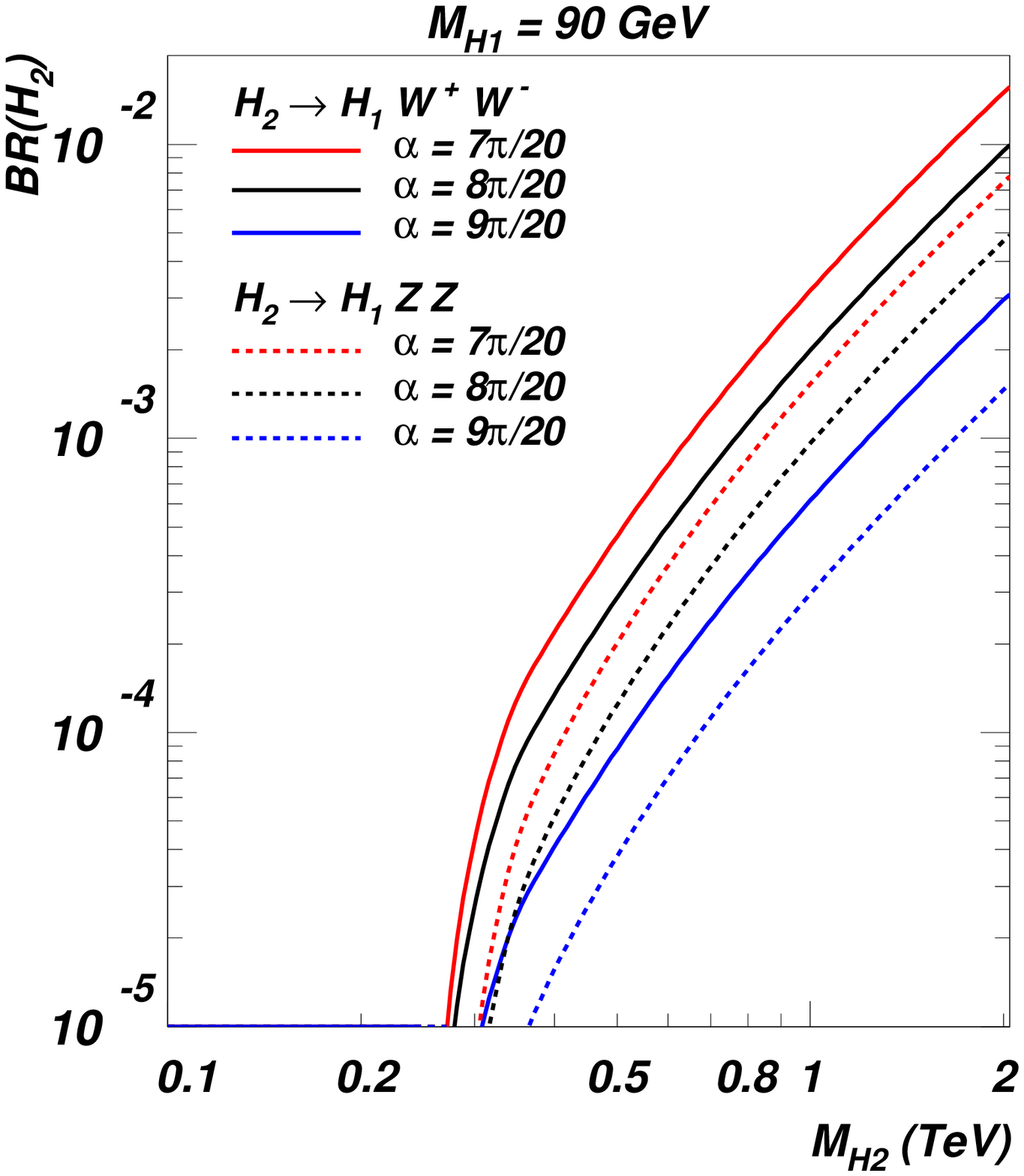}}
  \caption[Branching ratios and total widths (3)]{Dependence on the
    mixing angle $\alpha$ of the three body
    decays (\ref{H2-Br-3H1}) $BR(h_2\rightarrow h_1\, h_1\, h_1)$ and
    (\ref{H2-Br-2Vh1}) $BR(h_2\rightarrow h_1\, V\, V)$ ($V=W^\pm,Z$)
    for $M_{h_1} = 120$ GeV and (\ref{H2-Br-2Vh1-90}) for $M_{h_1} =
    90$ GeV, respectively.}
  \label{Br-3b}
\end{figure}

The other standard decays of both the light and the heavy Higgs bosons
are not modified substantially in the $B-L$ model (i.e., the decay of
the Higgs boson to $W$ boson pair is always dominant when
kinematically open,  while before that the decay into $b$-quarks is
the prevailing one; further, radiative decays, such as Higgs boson
decays into pairs of photons, peak at around $120$ GeV). Only when
other new channels open, the standard decay channels alter
accordingly. This rather common picture could be altered when the
mixing angle $\alpha$ approaches $\pi/2$, but such situation is
phenomenologically not viable (see Subsection~\ref{subs:3-1-1}).

Figures~\ref{H1_TW} and \ref{H2_TW} show the total widths for $h_1$
and $h_2$, respectively. In the first case, few thresholds are clearly
recognisable, as the heavy neutrino one at $100$ GeV (for angles very
close to $\pi /2$ only), the $WW$ and the $ZZ$ ones. Over the mass
range considered ($90$ GeV $< M_{h_1} < 250$ GeV), the particle's
width is very small until the $WW$ threshold, less than $1-10$ MeV,
rising steeply to few GeV for higher $h_1$ masses and small angles
(i.e., for a $SM$-like light Higgs boson). As we increase the mixing
angle, the couplings of the light Higgs boson to $SM$ particles is
reduced, and so its total width.

On the contrary, as we increase $\alpha$, the $h_2$ total width
increases, as clear from figure~\ref{H2_TW}. Also in this case, few
thresholds are recognisable, as the usual $WW$ and $ZZ$ gauge boson
ones, the light Higgs boson one (at $240$ GeV) and the $t\bar{t}$ one
(only for big angles, i.e., when $h_2$ is the $SM$-like Higgs
boson). When the mixing angle is small, the $h_2$ total width stays
below $1$ GeV all the way up to $M_{h_2} \sim 300\div 500$ GeV, rising
as the mass increases towards values for which $\Gamma_{h_2} \sim
M_{h_2}\sim 1$ TeV and $h_2$ loses the meaning of resonant state, only
for angles very close to $\pi /2$. Instead, if the angle is small,
i.e., less than $\pi /10$, the ratio of width over mass is less than
$10 \%$ and the heavy Higgs boson is a well defined particle. In the
decoupling regime, i.e., when $\alpha =0$, the only particles $h_2$
couples to are the $Z'$ and the heavy neutrinos. The width is
therefore dominated by the decay into them and is tiny, as clear from
figure~\ref{H2_TW}.

As already mentioned, figure~\ref{Br-alpha} shows the dependence on
the mixing angle $\alpha$ of the $BRs$ of $h_2$ into pairs of non-$SM$
particles. In particular, we consider the decays $h_2 \rightarrow
h_1\, h_1$ (for two different $h_1$ masses, $M_{h_1}=90$ GeV and
$M_{h_1}=120$ GeV, only for the allowed values of $\alpha$), $h_2
\rightarrow \nu _h\, \nu _h$ and $h_2 \rightarrow Z'\, Z'$ (not
influenced by $M_{h_1}$). As discussed in
Section~\ref{sect:4-1}, the interaction of the heavy Higgs
boson with $SM$ (or non-$SM$) particles has an overall $\sin{\alpha}$
(or $\cos{\alpha}$, respectively) dependence. Nonetheless, the $BRs$
in figure~\ref{Br-alpha} depend also on the total width, that for
$\alpha > \pi/4$ is dominated by the $h_2 \rightarrow W^+ W^-$
decay. Hence, when the angle assumes big values, the angle dependence
of the $h_2$ $BRs$ into heavy neutrino pairs and into $Z'$ boson pairs
follow a simple $\cot{\alpha}$ behaviour. Regarding $h_2 \rightarrow
h_1\, h_1$, its $BR$ is complicated by the fact that the contribution of
this process to the total width is not negligible when the mixing
angle is small, i.e., $\alpha < \pi/4$. In general, this channel
vanishes when $\alpha \rightarrow 0$, and it gets to its maximum, of
around $10\% \div 30 \%$ of the total width, as $\alpha$ takes a
non-trivial value, being almost constant with the angle if it is small
enough.

The heavy Higgs boson can be relatively massive and the tree-level
three-body decays are interesting decay modes too. Besides being clear
$BSM$ signatures, they are crucial to test the theory beyond the
observation of any scalar particle: its self-interactions and the
quartic interactions with the vector bosons could be tested directly
in these decay modes. In the $B-L$ model with no $Z-Z'$ mixing, the
quartic interactions that can be tested as $h_2$ decay modes, if the
respective channels are kinematically open, are: $h_2 \rightarrow
h_1\, h_1\, h_1$, $h_2 \rightarrow h_1\, W^+\, W^-$ and $h_2
\rightarrow h_1\, Z\, Z$, as shown in figure~\ref{Br-3b}, again for
$M_{h_1} = 90$~GeV and $120$~GeV. Although possible, $h_2 \rightarrow
h_1\, Z'\, Z'$ is negligible always, even if the $Z'$ boson is light
enough to allow the decay. For $M_{Z'}=210$ GeV, $BR(h_2 \rightarrow
h_1\, Z'\, Z') \lesssim 10^{-5}$ for $M_{h_2} < 2$ TeV.

The $BRs$ for both the $h_2 \rightarrow h_1\, h_1\, h_1$ and the $h_2
\rightarrow h_1\, V\, V$ ($V=W^\pm,\, Z$) channels are maximised
roughly when the mixing between the two scalars is maximum, i.e., when
$\alpha \sim \pi /4$, regardless of $M_{h_1}$. The former channel,
that is interesting because would produce three light Higgs bosons
simultaneously, can contribute at most at $10^{-3}$ of the total width
for $h_2$, as we are neglecting values of $M_{h_2}$ and $\alpha$ for
which $\Gamma_{h_2} \sim M_{h_2}$ (see figure~\ref{H2_TW}). For
instance, for $M_{h_2} = 800$ GeV, $\alpha$ needs to be less the $\pi
/5$ to have a reasonable small width-over-mass ratio ($\sim 10\%$),
and $BR(h_2 \rightarrow h_1\, h_1\, h_1) \leq 0.6\cdot 10^{-3}$. The
situation is similar for the latter channel, involving pairs of $SM$
gauge bosons. Again, for $M_{h_2} = 800$ GeV and $\alpha = \pi /5$,
$BR(h_2 \rightarrow h_1\, W^+\, W^-)=2BR(h_2 \rightarrow h_1\,
Z\, Z) = 10^{-3}$ for $m_{h_1} = 120$ GeV. For $M_{h_1} = 90$ GeV, the
mixing angle is constrained to be bigger than $7\pi /20$. For these
values and the same $M_{h_2}$ as before, such $BRs$ are doubled.

\section{Higgs bosons at the LHC}\label{sect:4-3}

In this Section we present our results for the analysis of the scalar
sector of the minimal $B-L$ model at the LHC. We shortly introduce the
scheduled working plan at the accelerator. Then, we present
cross sections at $\sqrt{s}=7$ and $14$
TeV for the two Higgs bosons. Finally, we will focus on some
phenomenologically viable signatures and their event rates.

\subsection{The LHC scheduled working plan}\label{subs:4-3-1}

The scheduled programme is planned to be the following:
\begin{itemize}
\item $7$ TeV is total energy of the two proton beams (energy in the
  hadronic center of mass) and $1$ fb$^{-1}$ is the scheduled
  integrated luminosity ($\sim 1-2$ operational years in the
  time scale). This is what we label as ``early discovery scenario''.
\item $14$ TeV is total energy of the two proton beams (energy in the
  hadronic center of mass) and $300$ fb$^{-1}$ is the scheduled
  integrated luminosity ($\sim 10$ operational years in the
  time scale). This is what we label as ``full luminosity scenario''.
\end{itemize}

\subsection{Standard production mechanisms}\label{subs:4-3-2}

\begin{figure}[!t]
  \subfigure[]{ 
  \label{xs_h1_7}
  \includegraphics[angle=0,width=0.48\textwidth ]{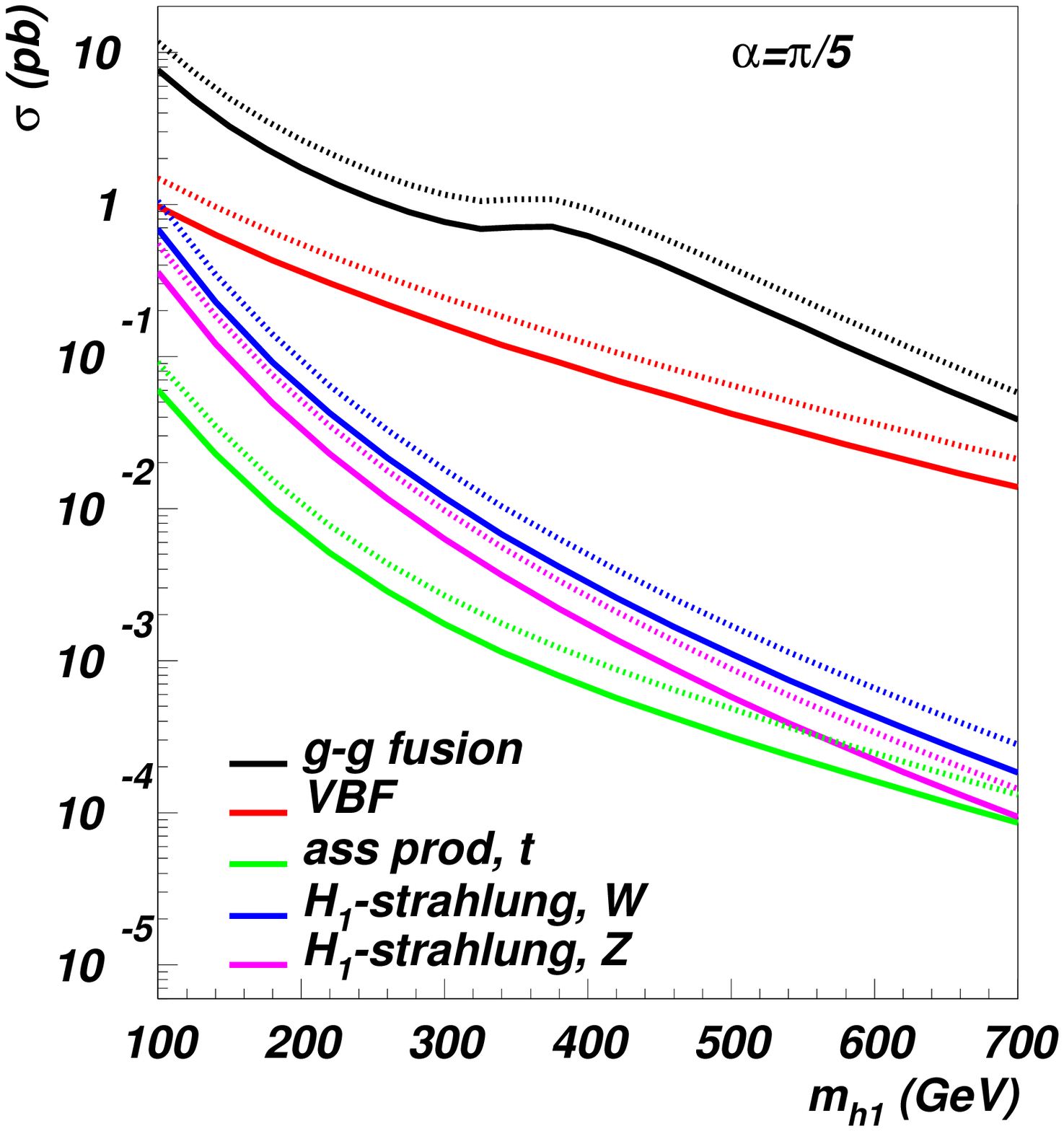}}
  \subfigure[]{
  \label{xs_h2_7}
  \includegraphics[angle=0,width=0.48\textwidth ]{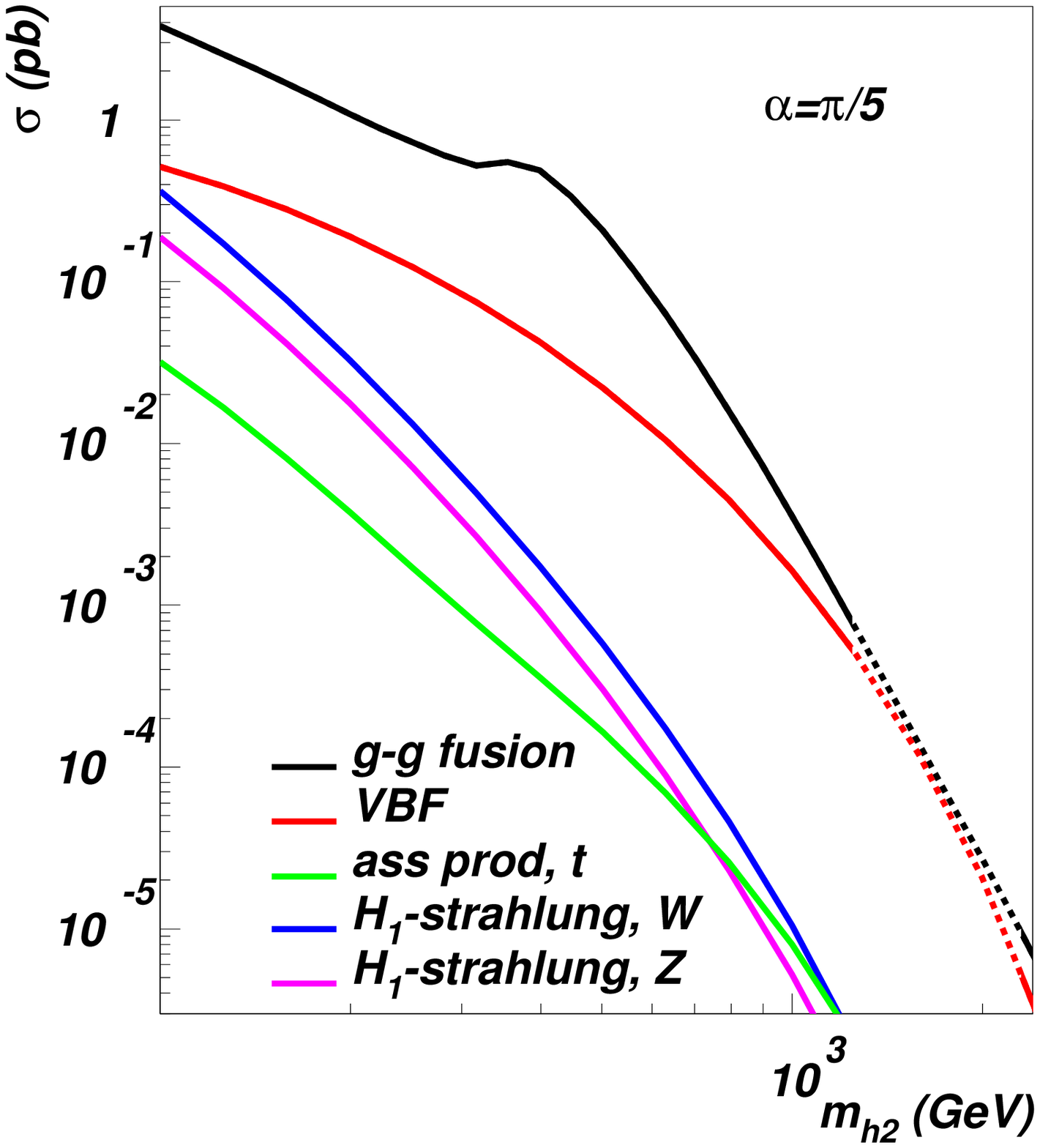}}\\
  \subfigure[]{
  \label{xs_h1_14}
  \includegraphics[angle=0,width=0.48\textwidth
  ]{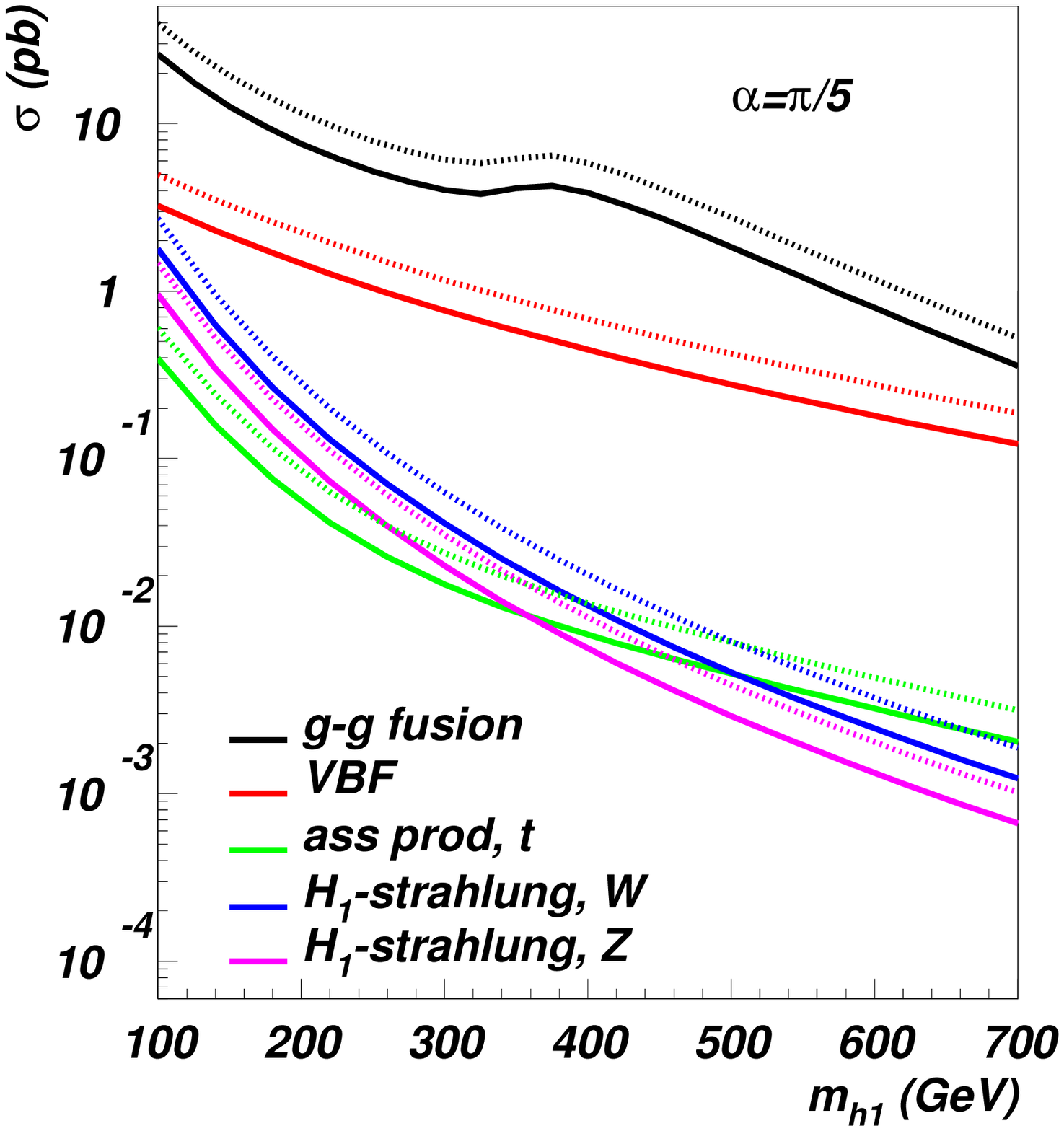}} 
  \subfigure[]{
  \label{xs_h2_14}
  \includegraphics[angle=0,width=0.48\textwidth ]{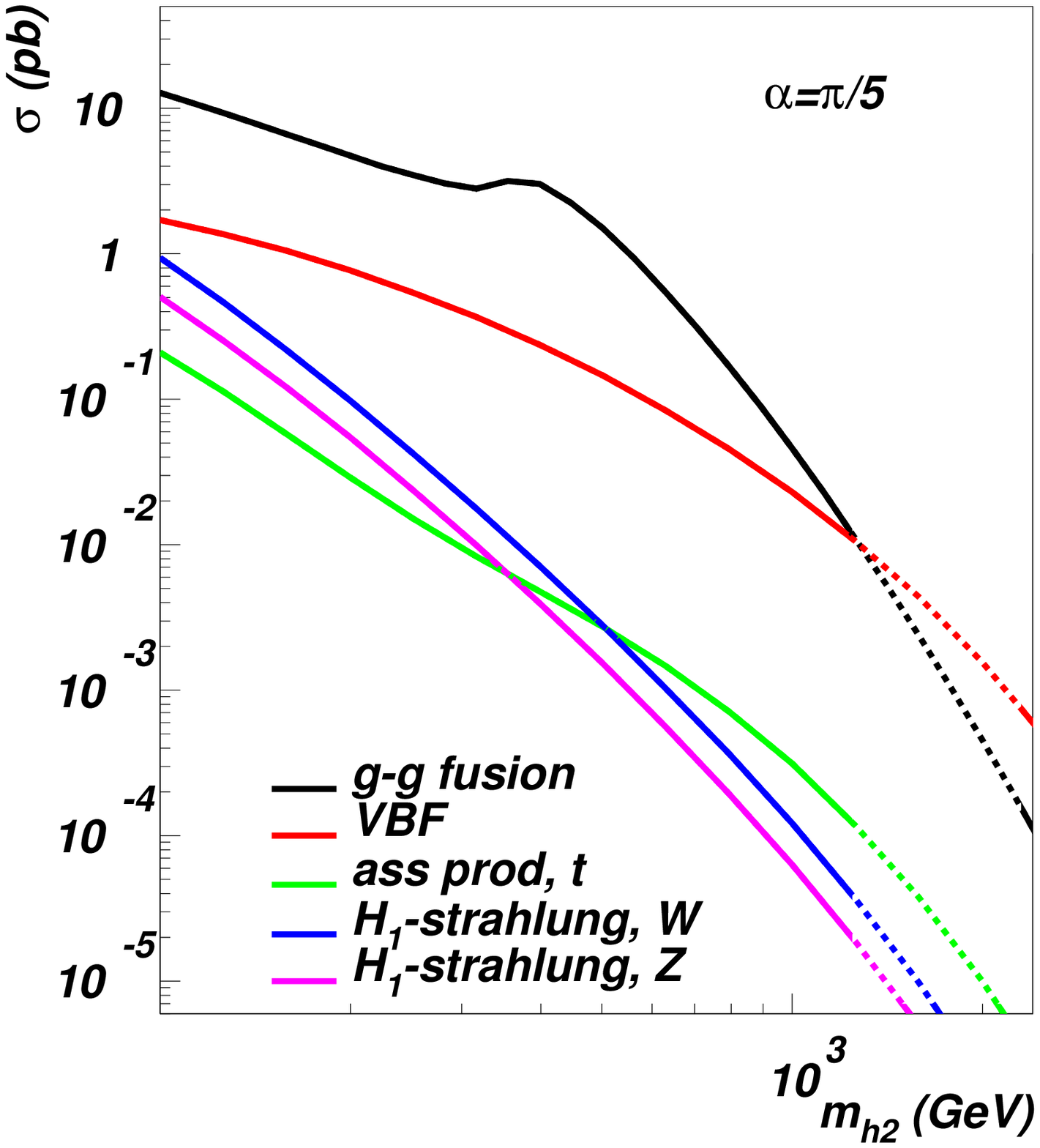}}  
  \caption[Higgs at LHC - Standard production
    mechanisms]{Cross sections in the minimal $B-L$ model for $h_1$ at
    the LHC (\ref{xs_h1_7}) at $\sqrt{s}=7$ TeV and (\ref{xs_h1_14}) 
    at $\sqrt{s}=14$ TeV, and for $h_2$ (\ref{xs_h2_7}) at
    $\sqrt{s}=7$ TeV and (\ref{xs_h2_14}) at $\sqrt{s}=14$ TeV. Dashed
    lines in figures~\ref{xs_h1_7} and \ref{xs_h1_14} refer to
    $\alpha = 0$. The dotted part of the lines in
    figure~\ref{xs_h2_14} refer to $h_2$ masses excluded by
    Unitarity (see Subsection~\ref{subs:3-2-1}).}
  \label{Xs}
\end{figure}

In figure~\ref{Xs} we present the cross sections for the most relevant
production mechanisms, i.e., the usual $SM$ processes such as
gluon-gluon fusion, vector-boson fusion, $t\overline{t}$ associated
production and Higgs-strahlung. For reference, we show in dashed lines
the $SM$ case (only for $h_1$), that corresponds to $\alpha =0$.

Comparing figure~\ref{xs_h1_14} to figure~\ref{xs_h1_7}, there is a
factor two enhancement passing from a $\sqrt{s}=7$ TeV to a
$\sqrt{s}=14$ TeV centre-of-mass energy at the LHC. 

The cross sections are a smooth function of the mixing angle $\alpha$,
so, as expected, every sub-channel has a cross section that scales with
$\cos{\alpha}$ ($\sin{\alpha}$), respectively for $h_1$ ($h_2$). As a
general rule, the cross section for $h_1$ at an angle $\alpha$ is
equal to that one of $h_2$ for $\pi /2 -\alpha$. In particular, the
maximum cross section for $h_2$ (i.e., when $\alpha =\pi/2$) 
coincides with the cross section of $h_1$ for $\alpha =0$.

We notice that these results are in agreement with the ones that have
been discussed in
\cite{BahatTreidel:2006kx,Barger:2007im,Bhattacharyya:2007pb} in the
context of a scalar singlet extension of the $SM$, having
the latter the same Higgs production phenomenology.
Moreover, as already showed in \cite{BahatTreidel:2006kx}, also in the
minimal $B-L$ context an high value of the mixing angle could lead
to important consequences for Higgs boson discovery at the LHC: $h_1$
production could be suppressed below
an observable rate at $\sqrt{s}=7$ TeV and heavy Higgs
boson production could be favoured, with peculiar final states clearly
beyond the $SM$, or
even hide the production of both (if no more than $1$ fb$^{-1}$ of
data is accumulated). Instead, at $\sqrt{s}=14$ TeV we expect that at
least one Higgs boson will be observed, either the light one or the
heavy one, or indeed both, thus shedding light on the scalar sector of
the $B-L$ extension of the $SM$ discussed in this work.

\subsection{Non-standard production mechanisms}\label{subs:4-3-3}

\begin{figure}[!t]
  \subfigure[]{ 
  \label{xs_Zp-h1}
  \includegraphics[angle=0,width=0.48\textwidth ]{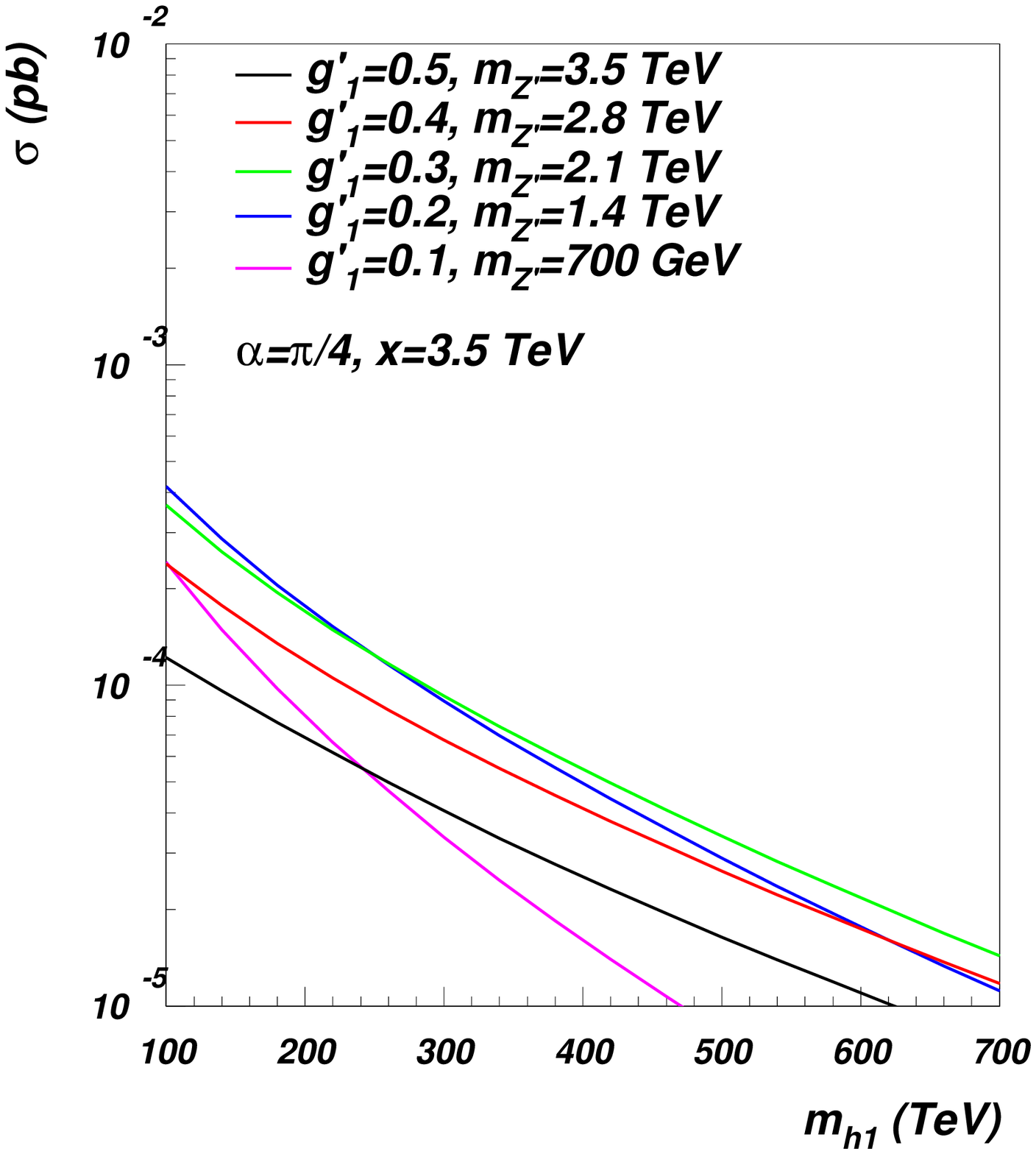}
}
  \subfigure[]{
  \label{xs_Zp-h2}
  \includegraphics[angle=0,width=0.48\textwidth ]{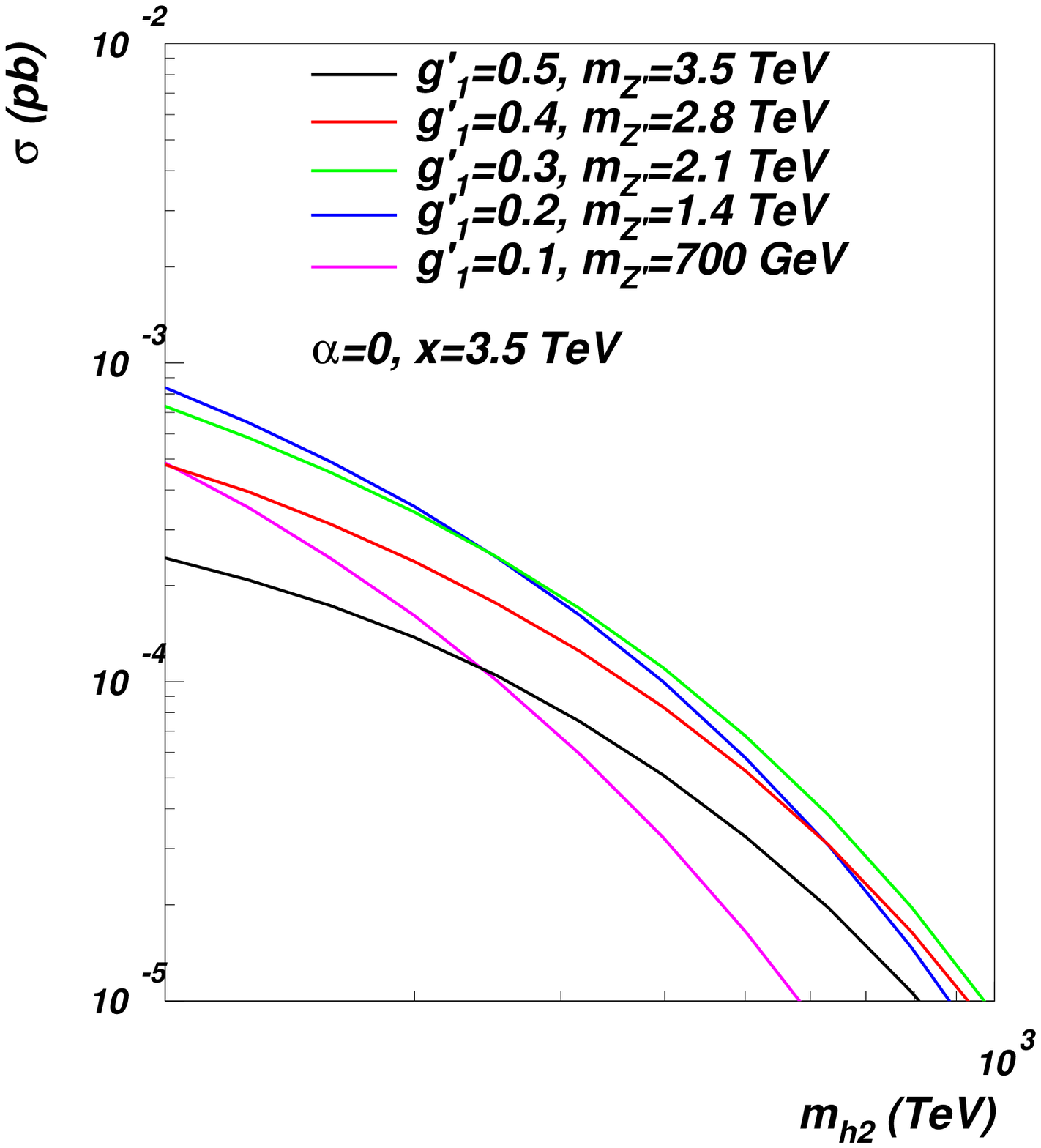}
}
  \caption[Higgs at LHC - Non-standard production
    mechanisms (1)]{Cross sections in the minimal $B-L$ model for the
    associated 
    production with the $Z'_{B-L}$ boson (\ref{xs_Zp-h1}) of $h_1$ at
    $\alpha = \pi /4$ and (\ref{xs_Zp-h2}) of $h_2$ at $\alpha = 0$.}
  \label{strah-Xs}
\end{figure}

\begin{figure}[!t]
\centering
  \subfigure[]{
  \label{xs_h1nn}
  \includegraphics[angle=0,width=0.48\textwidth ]{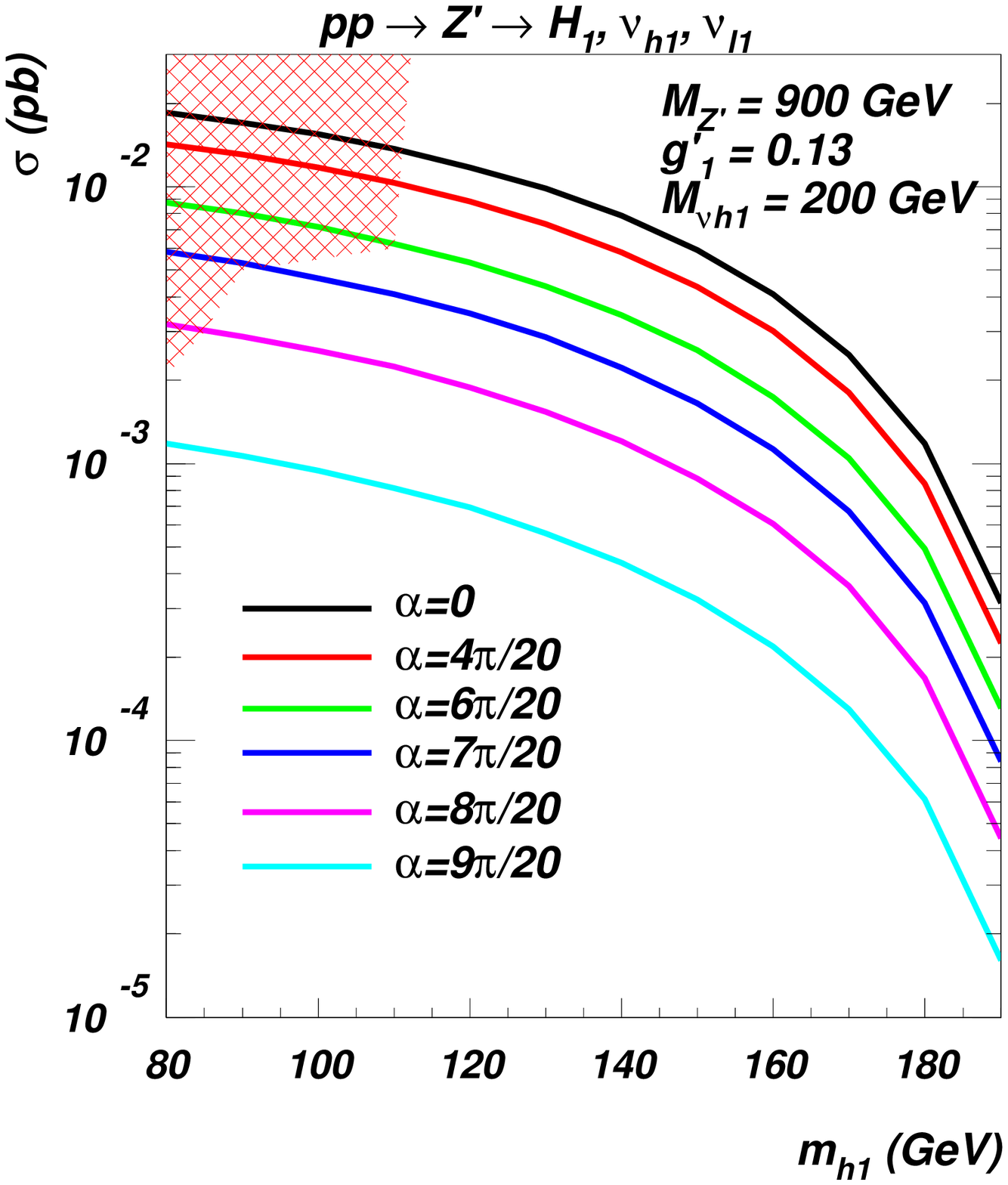}}  \\
  \subfigure[]{
  \label{xs_h1A}
  \includegraphics[angle=0,width=0.48\textwidth
  ]{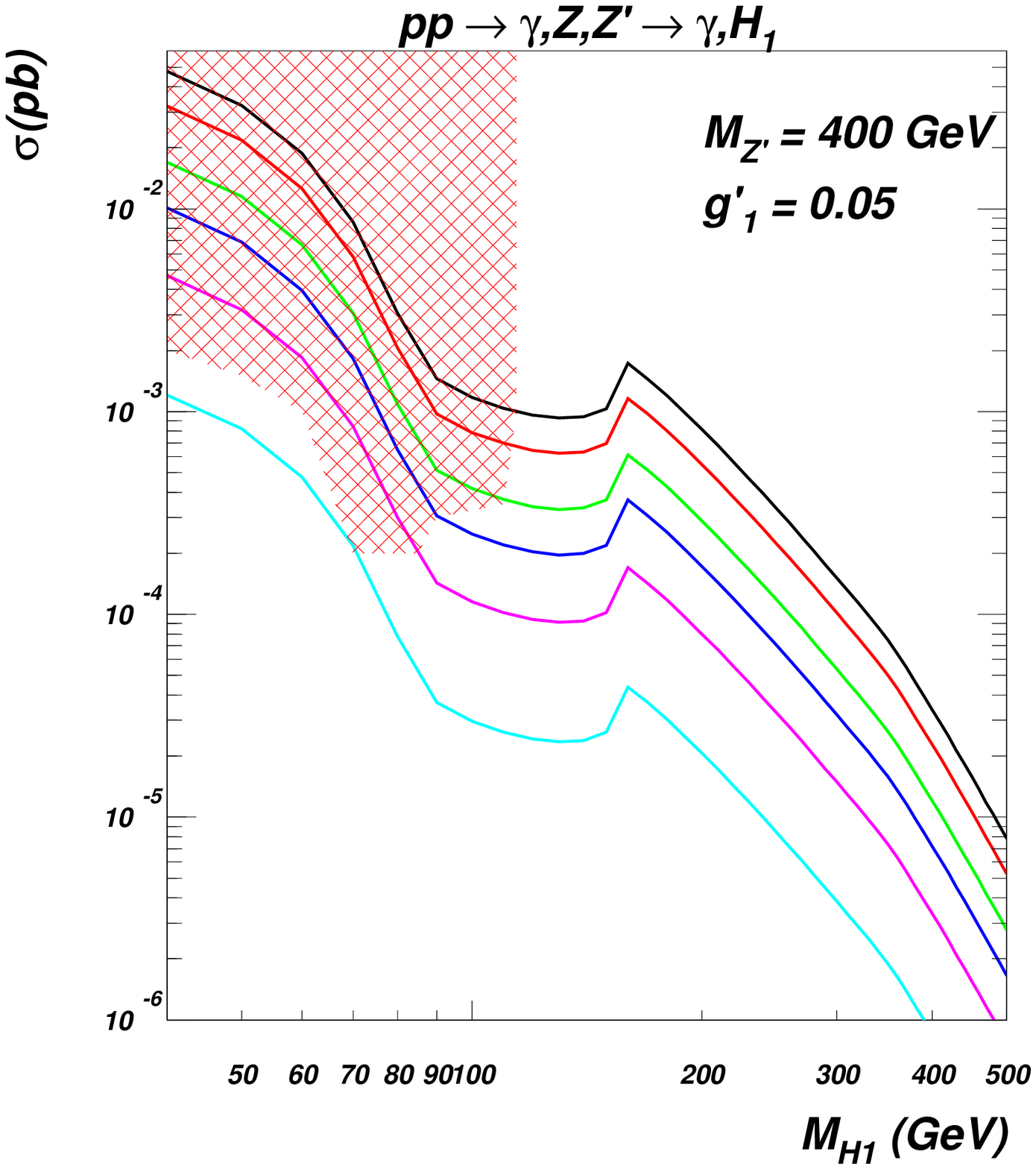}}
  \subfigure[]{
  \label{xs_h1A_VBF}
  \includegraphics[angle=0,width=0.48\textwidth ]{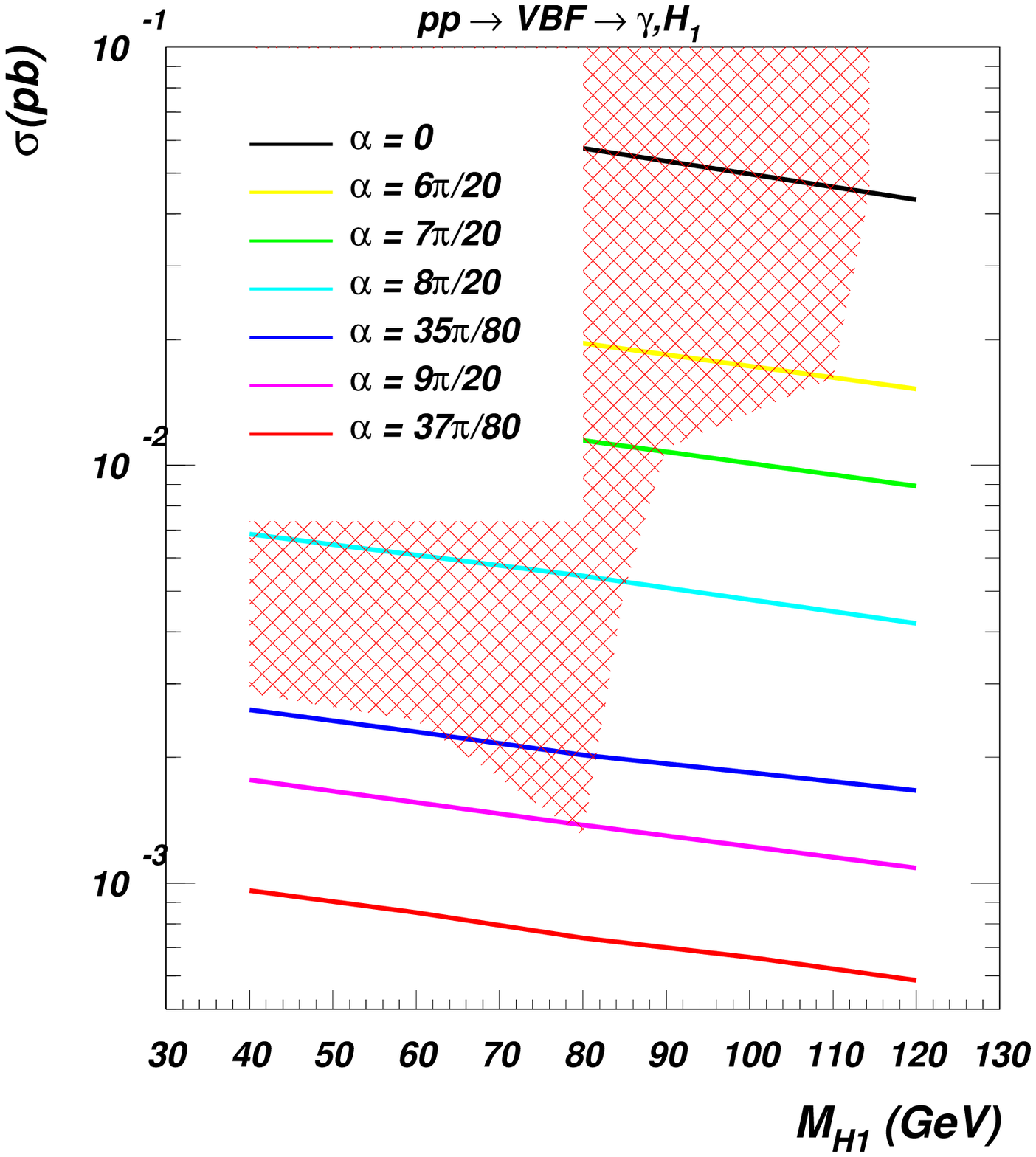}} 
  \caption[Higgs at LHC - Non-standard production
    mechanisms (2)]{Cross sections in the minimal $B-L$ model for the
    associated production of $h_1$ (\ref{xs_h1nn}) with one heavy and
    one light neutrinos, (\ref{xs_h1A}) with a photon via $\gamma$,
    $Z$ and $Z'$ bosons exchange (same legend as in
    figure~\ref{xs_h1nn} applies here) and (\ref{xs_h1A_VBF}) in the
    vector-boson fusion, all at $\sqrt{s}=14$ TeV. The red shading is
    the region excluded by LEP constraints
    (Subsection~\ref{subs:3-1-2}).}
  \label{non-std-Xs}
\end{figure}

All the new particles in the $B-L$ model interact with the scalar
sector, so novel production mechanisms can arise considering the
exchange of new intermediate particles. Among the new production
mechanisms, the associated production of the scalar boson with the
$Z'$ boson and the decay of a heavy neutrino into a Higgs boson are
certainly the most promising, depending on the specific masses. Notice
also that the viable parameter space, that allows a Higgs mass lighter
than the $SM$ limit of $114.4$ GeV for certain $\alpha - M_{h_2}$
configurations, enables us to investigate also production mechanisms
that in the $SM$ are subleading, as the associated production of a Higgs
boson with a photon. Figures~\ref{strah-Xs} and \ref{non-std-Xs} show
the cross sections for the non-standard production mechanisms, for
$\sqrt{s}=14$ TeV and several values of $\alpha$.

Figures~\ref{xs_Zp-h1} and \ref{xs_Zp-h2} show the cross sections for
associated production with the $Z'$ boson of $h_1$ and of $h_2$,
respectively, for several combinations of $Z'$ boson masses and $g'_1$
couplings. The process is
\begin{equation}
q\,\overline{q}\rightarrow Z'^{\ast} \rightarrow Z'\, h_{1(2)}\, ,
\end{equation}
and it is dominated by the $Z'$ boson's production cross sections (see
\cite{Basso:2008iv,Basso:2010pe}). 
Although never dominant (always below $1$ fb), this channel is the
only viable mechanism to produce $h_2$ in the decoupling scenario,
i.e., $\alpha =0$.

In figure~\ref{non-std-Xs} we plot the cross sections of the other
non-standard production mechanisms against the light Higgs mass, for
several choices of parameters (as explicitly indicated in the labels).
We superimposed the red-shadowed region in order to avoid any value of
the cross section that has been already excluded by LEP constraints
(see Subsection~\ref{subs:3-1-2}), mapping each
value of the boundary cross section as produced by the related maximum
value allowed for the 
light Higgs mass $M_{h_1}$ (at fixed mixing angle $\alpha$).

First of the showed plots is the decay of a heavy neutrino into a
Higgs boson. The whole process chain is
\begin{equation}
q\,\overline{q}\rightarrow Z' \rightarrow \nu _h\, \nu _h \rightarrow
\nu _h\, \nu _l\, h_{1(2)}\, ,
\end{equation}
and it requires to pair produce heavy neutrinos, again via the $Z'$
boson (see \cite{Basso:2008iv,Perez:2009mu} for a detailed analysis of
the $pp\rightarrow Z' \rightarrow \nu_h \nu_h$ process and other
aspects of $Z'$ and heavy
neutrinos phenomenology in the minimal $B-L$ model). Although rather
involved, this mechanism has the advantage that
the whole decay chain can be of on-shell particles, besides the
peculiar final state of a Higgs boson and a heavy neutrino. For a
choice of the parameters that roughly maximises this mechanism
($M_{Z'}=900$ GeV, $g'_1=0.13$ and $M_{\nu _h}=200$ GeV),
figure~\ref{xs_h1nn} shows that the cross sections for the production
of the light Higgs boson (when only one generation of heavy neutrinos
is considered) are above $10$ fb for $M_{h_1} < 130$ GeV (and small
values of $\alpha$), dropping steeply when the light Higgs boson mass
approaches the kinematical limit for the heavy neutrino to decay into
it. Assuming the transformation $\alpha \rightarrow \pi/2 -
\alpha$, the production of the heavy Higgs boson via this mechanism
shows analogous features.

Next, figures~\ref{xs_h1A} and \ref{xs_h1A_VBF} shows the associated
production of the light Higgs boson with a photon. The processes are,
respectively,
\begin{equation}\label{H1_A_GGfus}
q\,\overline{q}\rightarrow \gamma / Z / Z' \rightarrow \gamma \, h_{1}
\end{equation}
via the $SM$ neutral gauge bosons ($\gamma$ and $Z$) and the new $Z'$ boson, and
\begin{equation}\label{H1_A_VBF}
q\,q'\rightarrow \gamma \, h_{1}\, q'' \, q''' \, ,
\end{equation}
through vector-boson fusion (only $W$ and $Z$ bosons).

In the first instance, we notice that the $Z'$ sub-channel in
equation~(\ref{H1_A_GGfus}) is always negligible,
as there is no $Z'-W-W$ interaction and the $V-h-\gamma$ effective
vertex is only via a $t$-quark loop (an order of magnitude lower than
the $V-h-\gamma$ effective vertex via a $W$ boson loop)
\cite{Gunion:1989we}. What is relevant in these two channels is that
the light Higgs
boson mass can be considerably smaller than the LEP limit
(they are valid for the $SM$, or equivalently when $\alpha=0$ in the
$B-L$ model). Hence, the phase space factor can enhance the mechanism
of equation~(\ref{H1_A_GGfus}) for small masses, up to the level of
$1$ fb for $M_{h_1} < 60$ GeV (and suitable values of the mixing angle
$\alpha$, depending on the experimental and theoretical limits, see
\cite{Basso:2010jt,Basso:2010jm} for a complete tratement of the
allowed parameter space of the Higgs sector of the minimal $B-L$
model). Moreover, it has recently been
observed that the associated production with a photon in the 
vector-boson fusion channel could be useful for low Higgs boson masses
in order to trigger
events in which the Higgs boson decays into $b$-quark pairs
\cite{Asner:2010ve}. Complementary to that, the process in
equation~(\ref{H1_A_GGfus}) can also be of similar interest, with the
advantage that the photon will always be back-to-back relative to the
$b$-quark pair. For comparison, figures~\ref{xs_h1A} and
\ref{xs_h1A_VBF} show the cross section for these
processes\footnote{In order to produce figure~\ref{xs_h1A_VBF}, we
  included the following cuts: $P_t ^{\gamma,\mbox{jet}} > 15$ GeV,
  $|\eta ^{\gamma}| < 3$ and $|\eta ^{\mbox{jet}}|~<~5.5$, where
  ``jet'' refers to the actual final state, though we use partons here
  to emulate it  
  \cite{Asner:2010ve}.}. Certainly, for a ${h_1}$ boson heavier than
the $SM$ limit, vector-boson fusion is the dominant process for
associated production of $h_1$ with a photon, and this is also true
for $M_{h_1} > 60$~GeV. However, for light Higgs boson masses lower
than $60$~GeV, the two mechanisms of eqs.~(\ref{H1_A_GGfus}) and
(\ref{H1_A_VBF}) become equally competitive, up to the level of
$\mathcal{O}(1)$~fb each, for suitable values of the mixing angle
$\alpha$.

\subsection{Event Rates}\label{subs:4-3-4}

In this Section we combine the results from the Higgs boson cross
sections and those from the $BR$ analysis in order to perform a
detailed study of typical event rates for some Higgs signatures which
are specific to the $B-L$ model.

Before all else, we remind the terminology previously introduced in
Subsection~\ref{subs:4-3-1}: we will generally refer to an ``early
discovery
scenario'' by considering an energy in the hadronic $CM$ of
$\sqrt{s}=7$ TeV and an integrated luminosity of 
$\int{L}=1$~fb$^{-1}$ (according to the official schedule, this is
what is expected to be collected after the first couple of years of
LHC running) and to a ``full luminosity scenario'' by considering an
energy in the hadronic $CM$ of $\sqrt{s}=14$ TeV and an integrated
luminosity of $\int L~=300$~fb$^{-1}$ (according to the official
schedule, this is what is expected to be realistically collected at
the higher energy stage).

As we shall see by combining the production cross sections
and the decay $BRs$ already presented, the two different scenarios
open different possibilities for the detection of peculiar signatures
of the model: in the ``early discovery scenario'' there is a clear
possibility to detect a light Higgs state yielding heavy neutrino
pairs while the ``full luminosity scenario'' affords the possibility
of numerous discovery mechanisms (in addition to the previous
mechanism, for the heavy Higgs state one also has decays of the latter
into $Z'$ boson and light Higgs boson pairs).

\begin{figure}[!t]
  \subfigure[]{ 
  \label{gg-h1-nunu50}
  \includegraphics[angle=0,width=0.48\textwidth ]{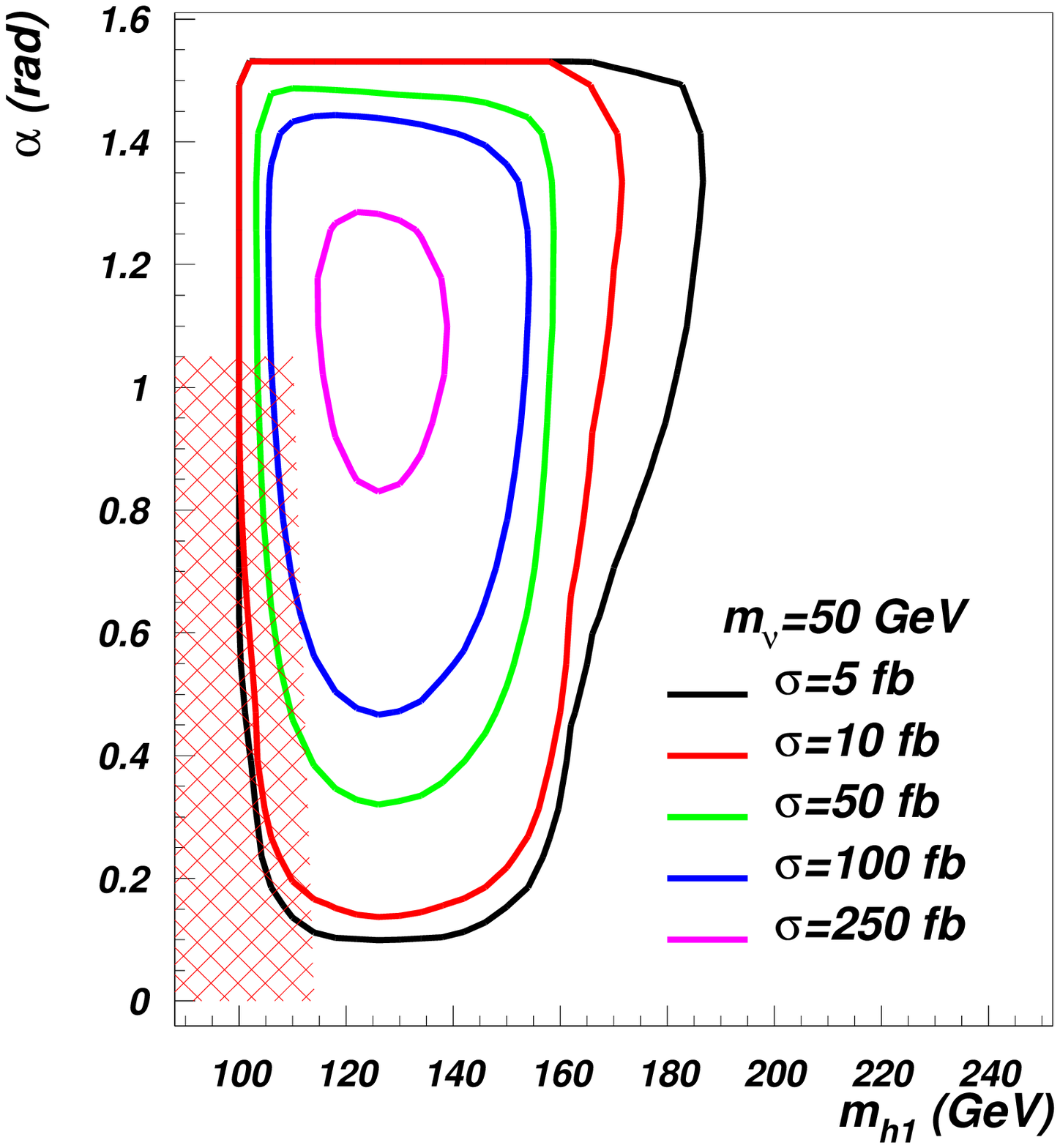}
}
  \subfigure[]{
  \label{gg-h1-nunu60}
  \includegraphics[angle=0,width=0.48\textwidth ]{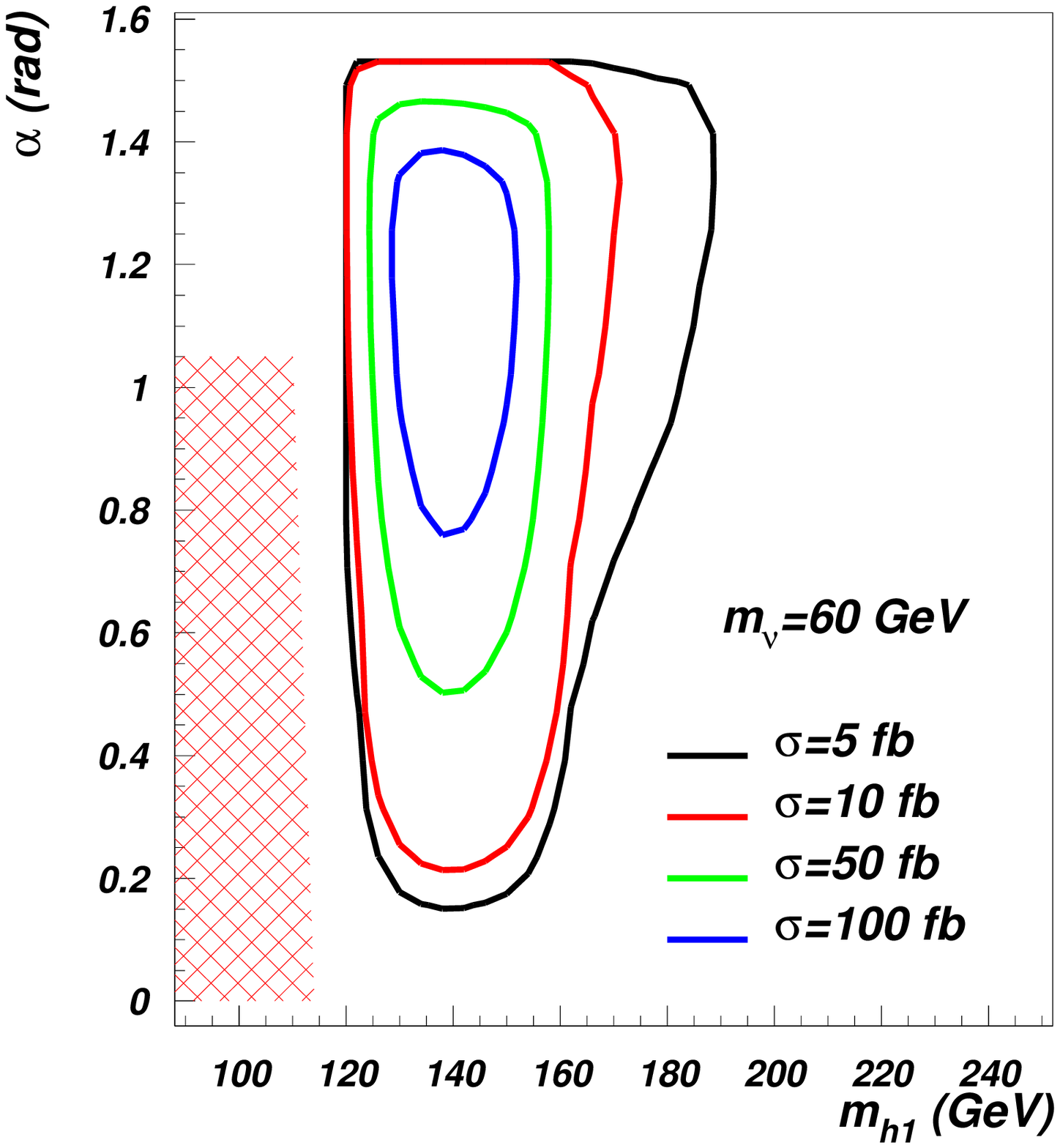}
}
  \caption[Higgs at LHC - Event rates (1)]{Cross section times $BR$
    contour plot for the $B-L$ process $pp\rightarrow h_1\rightarrow
    \nu_h \nu_h$ at the LHC with $\sqrt{s}=7$ TeV, plotted against
    $M_{h_1}$-$\alpha$, with $M_{\nu_h}=50$ GeV 
    (\ref{gg-h1-nunu50}) and $M_{\nu_h}=60$ GeV
    (\ref{gg-h1-nunu60}). Several values of
    cross section times $BR$ have been considered: $\sigma=5$ fb
    (black line), $\sigma=10$ fb (red line), $\sigma=50$ fb (green
    line), $\sigma=100$ fb (blue line) and $\sigma=250$ fb (violet
    line). The red-shadowed region is excluded by the LEP
    experiments.}
  \label{ggnunu50-60}
\end{figure}

Firstly, we focus on the ``early discovery scenario'': in this
experimental configuration, the most important $B-L$ distinctive process is
represented by heavy neutrino pair production via a light Higgs boson,
through the channel 
$pp\rightarrow h_1 \rightarrow \nu_h\nu_h$.
In figure~\ref{ggnunu50-60} we show the explicit results for
the $pp\rightarrow h_1\rightarrow \nu_h \nu_h$ process at the LHC with
$\sqrt{s}=7$ TeV, for $M_{\nu_h}=50$ GeV (figure~\ref{gg-h1-nunu50}) and
$M_{\nu_h}=60$ GeV (figure~\ref{gg-h1-nunu60}), obtained by combining the
light Higgs boson production cross section via gluon-gluon fusion only 
(since it represents the main contribution) and the $BR$ of the light
Higgs boson to heavy neutrino pairs. The obtained rate is projected in
the $M_{h_1}$-$\alpha$ plane and several values of the
cross section times $BR$ have been considered: $\sigma=5$, $10$, $50$,
$100$ and $250$ fb. The red-shadowed region takes into account the
exclusion limits established by the LEP experiments.

\begin{figure}[!t]
  \subfigure[]{ 
  \label{gg-h2-h1h1120}
  \includegraphics[angle=0,width=0.48\textwidth ]{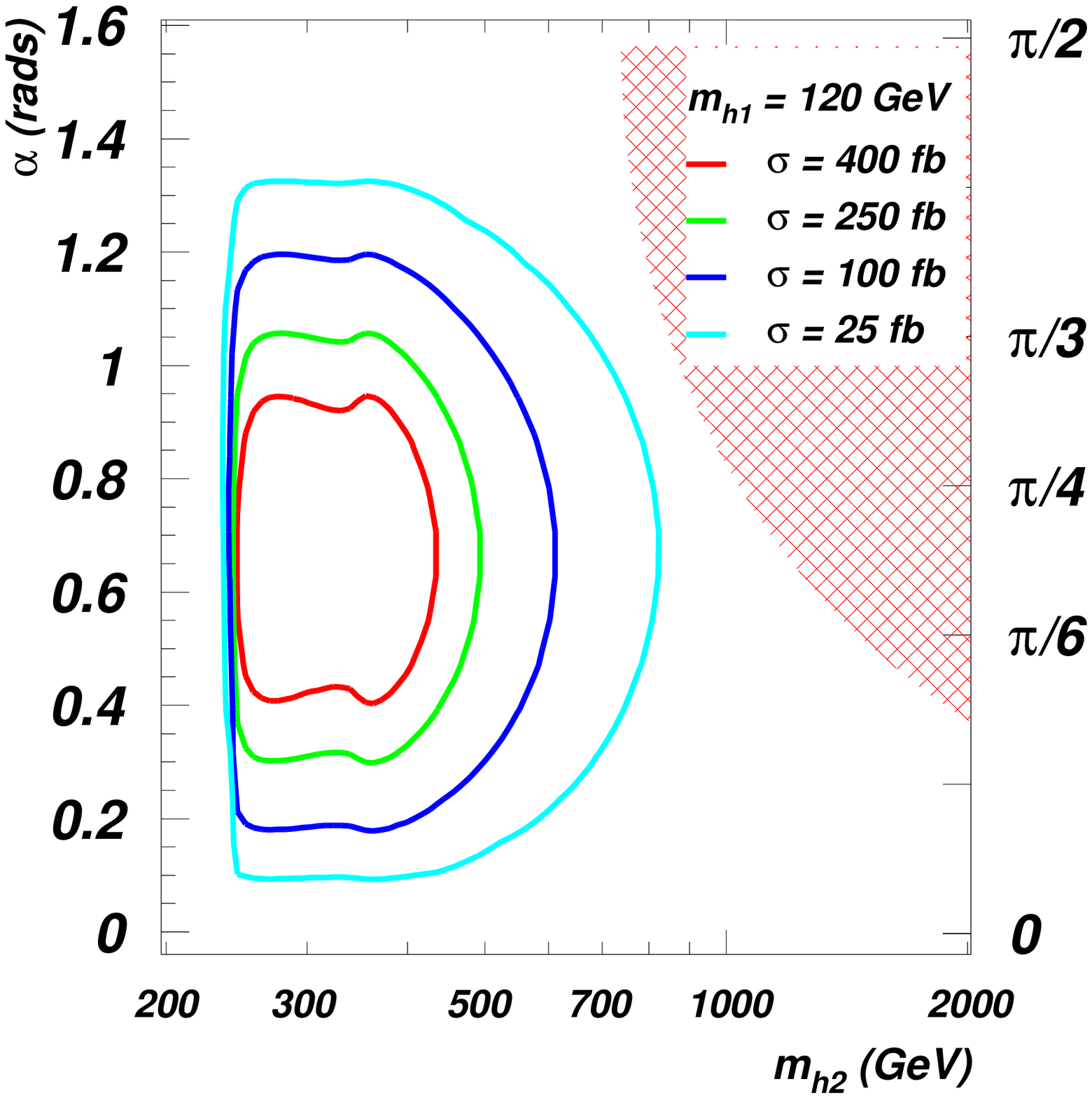}
}
  \subfigure[]{
  \label{gg-h2-h1h1240}
  \includegraphics[angle=0,width=0.48\textwidth ]{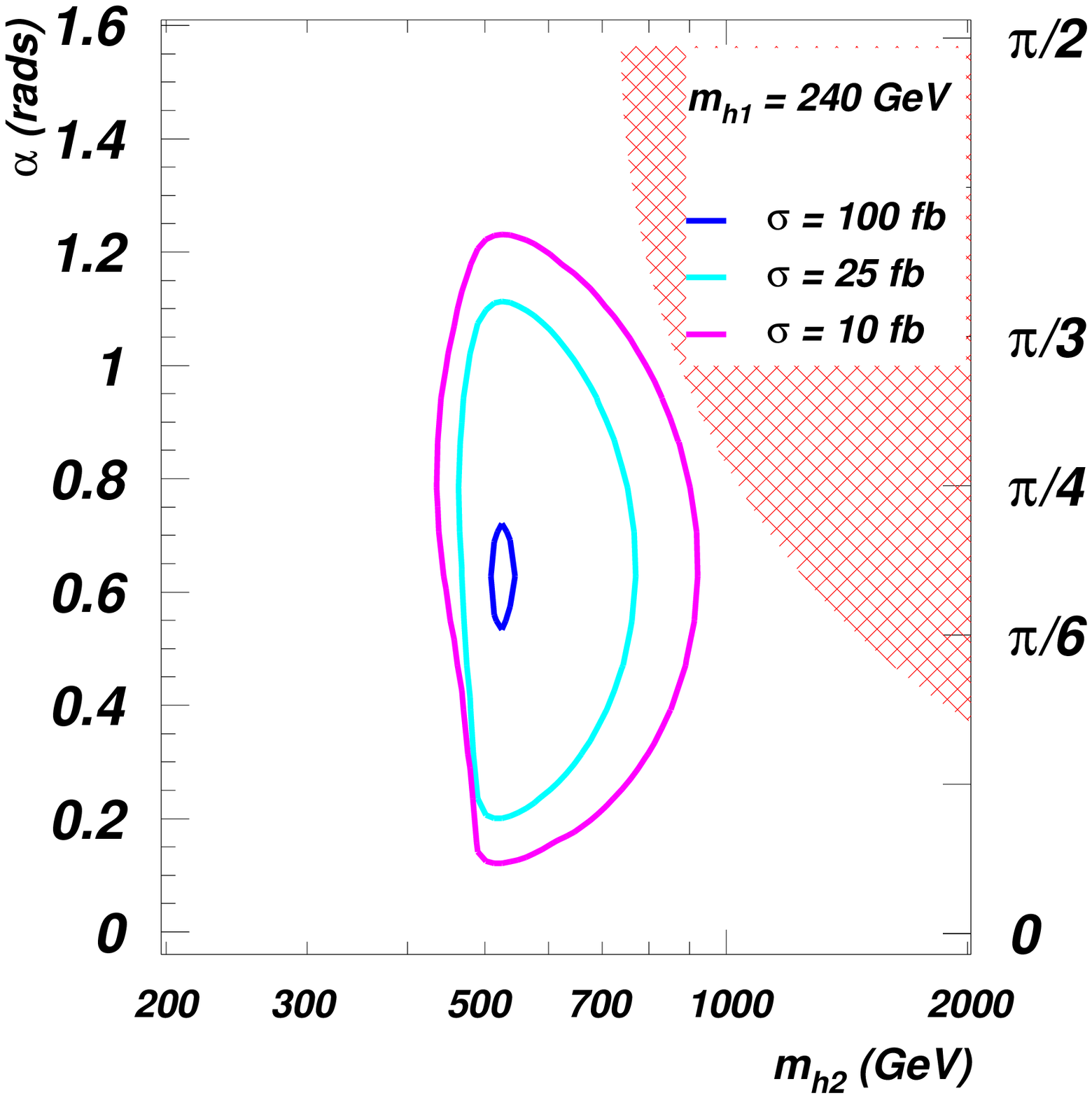}
}
  \caption[Higgs at LHC - Event rates (2)]{Cross section times $BR$
    contour plot for the $B-L$ process $pp\rightarrow h_2\rightarrow
    h_1 h_1$ at the LHC with $\sqrt{s}=14$ TeV, plotted against
    $M_{h_2}$-$\alpha$, with $M_{h_1}=120$ GeV (\ref{gg-h2-h1h1120})
    and $M_{h_1}=240$ GeV (\ref{gg-h2-h1h1240}). Several values of
    cross section times $BR$ have been considered: $\sigma=10$ fb
    (violet line), $\sigma=25$ fb (light-blue line), $\sigma=100$ fb
    (blue line), $\sigma=250$ fb (green line) and $\sigma=400$ fb (red
    line). The red-shadowed region is excluded by unitarity
    constraints.}
  \label{ggh1h1120-240}
\end{figure}

Even considering a low-luminosity scenario
(i.e., $\int L\simeq~1$~fb$^{-1}$), there is a
noticeable allowed parameter space for which the rate of such events is
considerably large: in the case of $M_{\nu_h}=50$ GeV, when the
integrated luminosity reaches $\int L=1$~fb$^{-1}$, we estimated a
collection of $\sim 10$ heavy neutrino pairs from 
the light Higgs boson production and decay for $100$ GeV$<M_{h_1}<170$ GeV and
$0.05\pi<\alpha<0.48\pi$, that scales up to $\sim 10^2$ events for
$110$ GeV$<M_{h_1}<155$ GeV and $0.16\pi<\alpha<0.46\pi$. In the case
of $M_{\nu_h}=60$ GeV, we estimated a
collection of $\sim 10$ heavy neutrino pairs from Higgs 
production for $120$ GeV$<M_{h_1}<170$ GeV and
$0.06\pi<\alpha<0.48\pi$, that scales up to $\sim 10^2$ events for
$125$ GeV$<M_{h_1}<150$ GeV and $0.25\pi<\alpha<0.44\pi$.

If we consider instead the ``full luminosity scenario'',
there are several important distinctive signatures: $pp\rightarrow h_2
\rightarrow h_1h_1$, $pp\rightarrow h_2 \rightarrow Z'Z'$ and
$pp\rightarrow h_2 \rightarrow \nu_h\nu_h$.
In figure~\ref{ggh1h1120-240} we show the results for
light Higgs boson pair production from heavy Higgs boson decays at the
LHC with $\sqrt{s}=14$ TeV for $M_{h_1}=120$ GeV
(figure~\ref{gg-h2-h1h1120}) and  
$M_{h_1}=240$ GeV (figure~\ref{gg-h2-h1h1240}). Again, if we project
the rates on the bi-dimensional $M_{h_2}$-$\alpha$ plane, we can
select the contours that relate the cross section times $BR$ to some
peculiar values. 

Considering an integrated luminosity of $300$~fb$^{-1}$, we can
relate $\sigma=25(250)$ fb to $7500(75000)$ events, hence for both
choices of the light Higgs mass the $\alpha$-$M_{h_2}$ parameter space
offers an abundant portion in which the event rate is noticeable for
light Higgs boson pair production from heavy Higgs boson decays: when
$M_{h_1}=120$ GeV the process is accessible almost over the entire
parameter space, with a cross section peak of $400$ fb in the
$240$ GeV$<M_{h_2}<400$ GeV and $0.13\pi<\alpha<0.30\pi$ intervals,
while in the $M_{h_1}=240$ GeV case the significant parameter space is
still large, even if slightly decreased, with a cross section peak of
$25$ fb in the $480$ GeV$<M_{h_2}<800$ GeV and $0.06\pi<\alpha<0.32\pi$
region.

\begin{figure}[!t]
  \subfigure[]{ 
  \label{gg-h2-zpzp210}
  \includegraphics[angle=0,width=0.48\textwidth ]{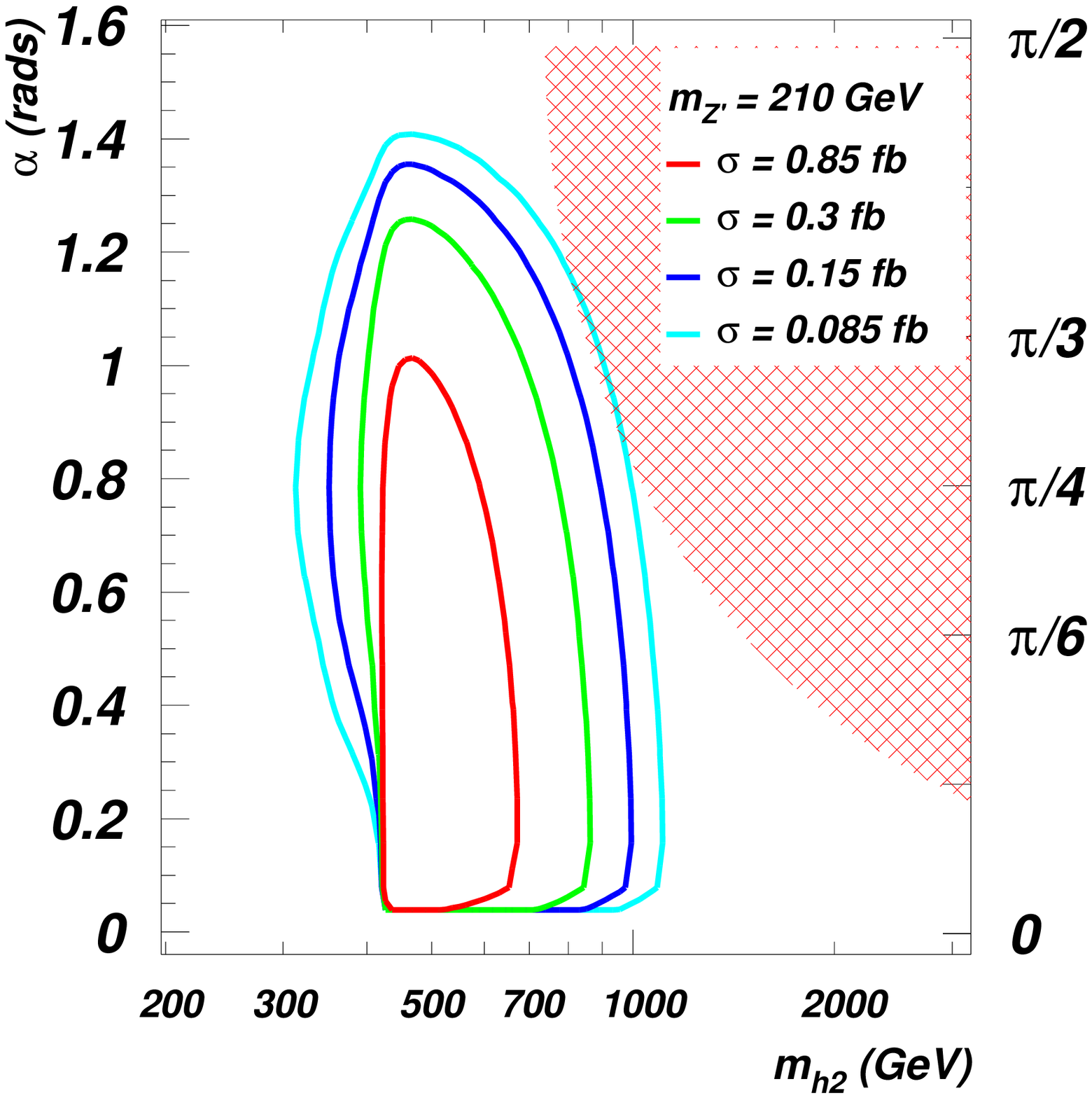}
}
  \subfigure[]{
  \label{gg-h2-zpzp280}
  \includegraphics[angle=0,width=0.48\textwidth ]{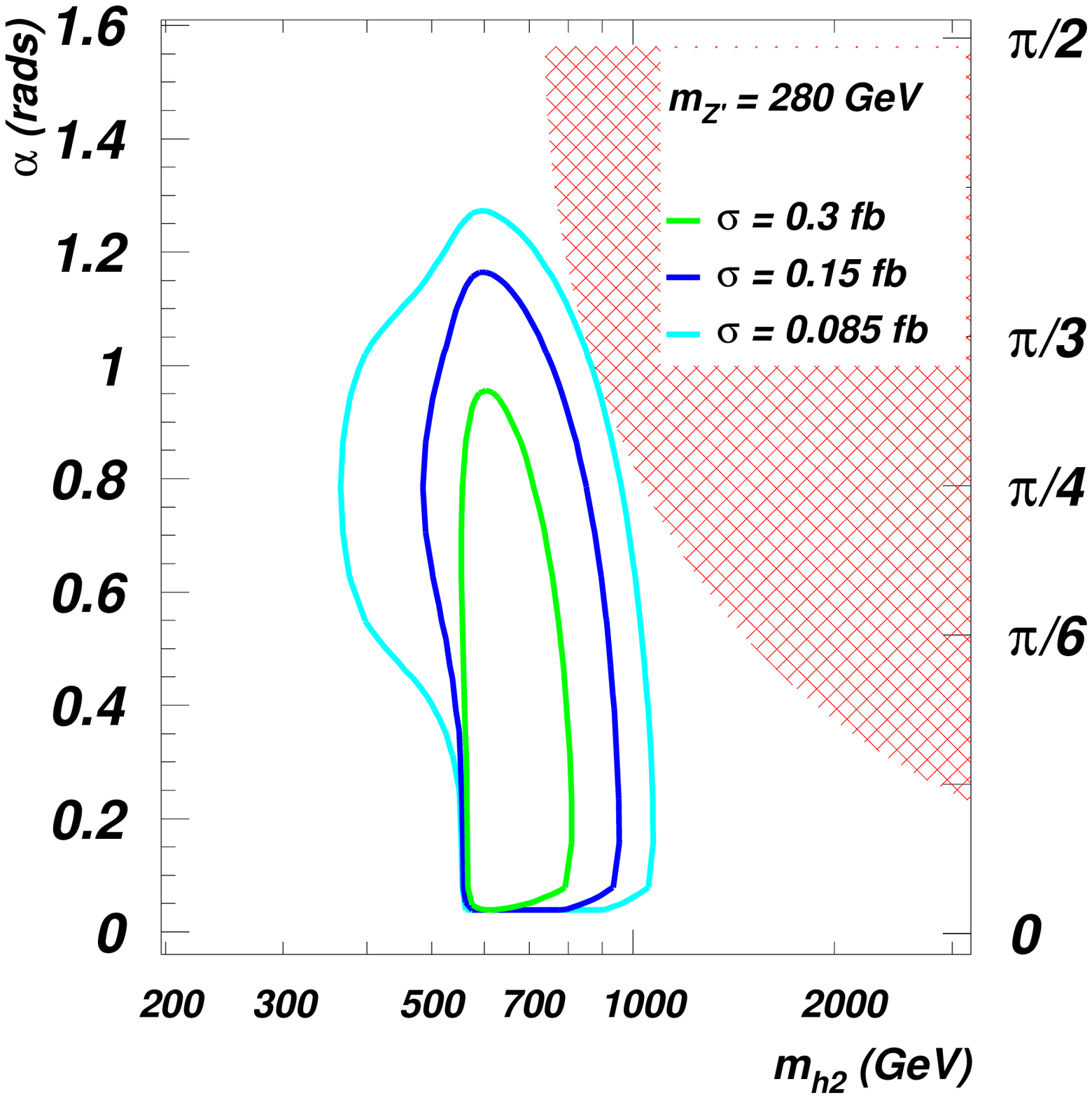}
}
  \caption[Higgs at LHC - Event rates (3)]{Cross section times $BR$
    contour plot for the $B-L$ process $pp\rightarrow h_2\rightarrow
    Z' Z'$ at the LHC with $\sqrt{s}=14$ TeV, plotted against
    $M_{h_2}$-$\alpha$, with $M_{Z'}=210$ GeV (\ref{gg-h2-zpzp210})
    and $M_{Z'}=280$ GeV (\ref{gg-h2-zpzp280}). Several values of
    cross section times $BR$ have been considered: $\sigma=0.085$ fb
    (light-blue line), $\sigma=0.15$ fb (blue line), $\sigma=0.3$ fb
    (green line), $\sigma=0.85$ fb (red line). The red-shadowed region
    is excluded by unitarity constraints.}
  \label{ggzpzp210-280}
\end{figure}

In figure~\ref{ggzpzp210-280} we show the results for
$Z'$ boson pair production from heavy Higgs boson decays at the LHC with
$\sqrt{s}=14$ TeV for $M_{Z'}=210$ GeV (figure~\ref{gg-h2-zpzp210}) and
$M_{Z'}=280$ GeV (figure~\ref{gg-h2-zpzp280}). Again, if we project
the rates on the bi-dimensional $M_{h_2}$-$\alpha$ plane, we can
select the contours that relate the cross section times $BR$ to some
peculiar values. Here, we have that $\sigma=0.085(0.85)$ fb
corresponds to $25(250)$ events, hence for both choices of $Z'$ mass
the $\alpha$-$M_{h_2}$ parameter space offers an abundant portion in
which the event rate could be interesting for $Z'$ boson pair
production from heavy Higgs boson decays: for $M_{Z'}=210$ GeV the
process has a peak of $0.85$ fb in the $420$ GeV$<M_{h_2}<650$ GeV and
$0.03\pi<\alpha<0.25\pi$ region, while if $M_{Z'}=280$ GeV a
noticeable parameter space is still potentially accessible with a rate
peak of $0.3$ fb ($100$ events) in the $560$ GeV$<M_{h_2}<800$ GeV and
$0.03\pi<\alpha<0.19\pi$ region.

\begin{figure}[!t]
  \subfigure[]{ 
  \label{gg-h2-nunu150}
  \includegraphics[angle=0,width=0.48\textwidth ]{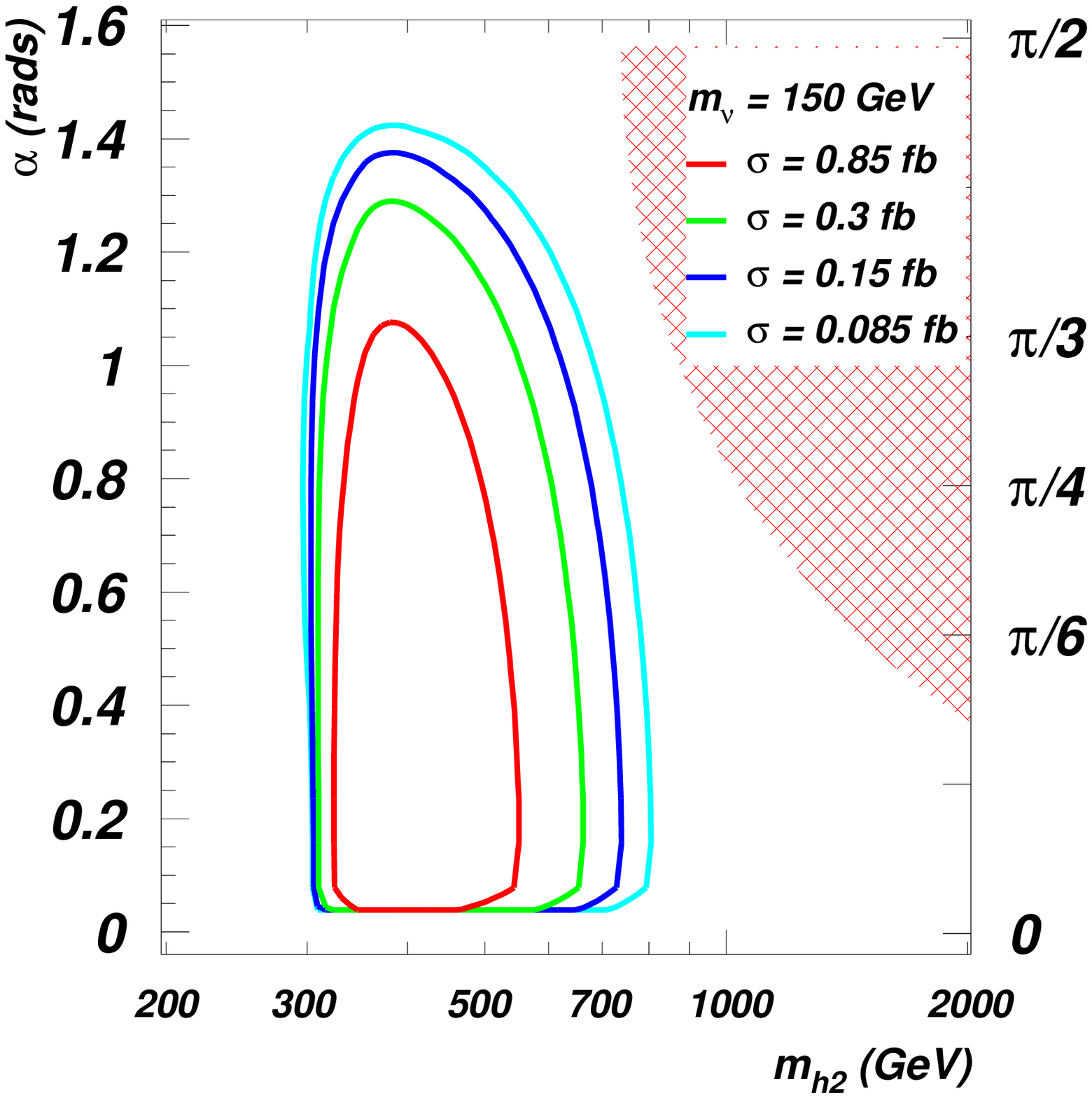}
}
  \subfigure[]{
  \label{gg-h2-nunu200}
  \includegraphics[angle=0,width=0.48\textwidth ]{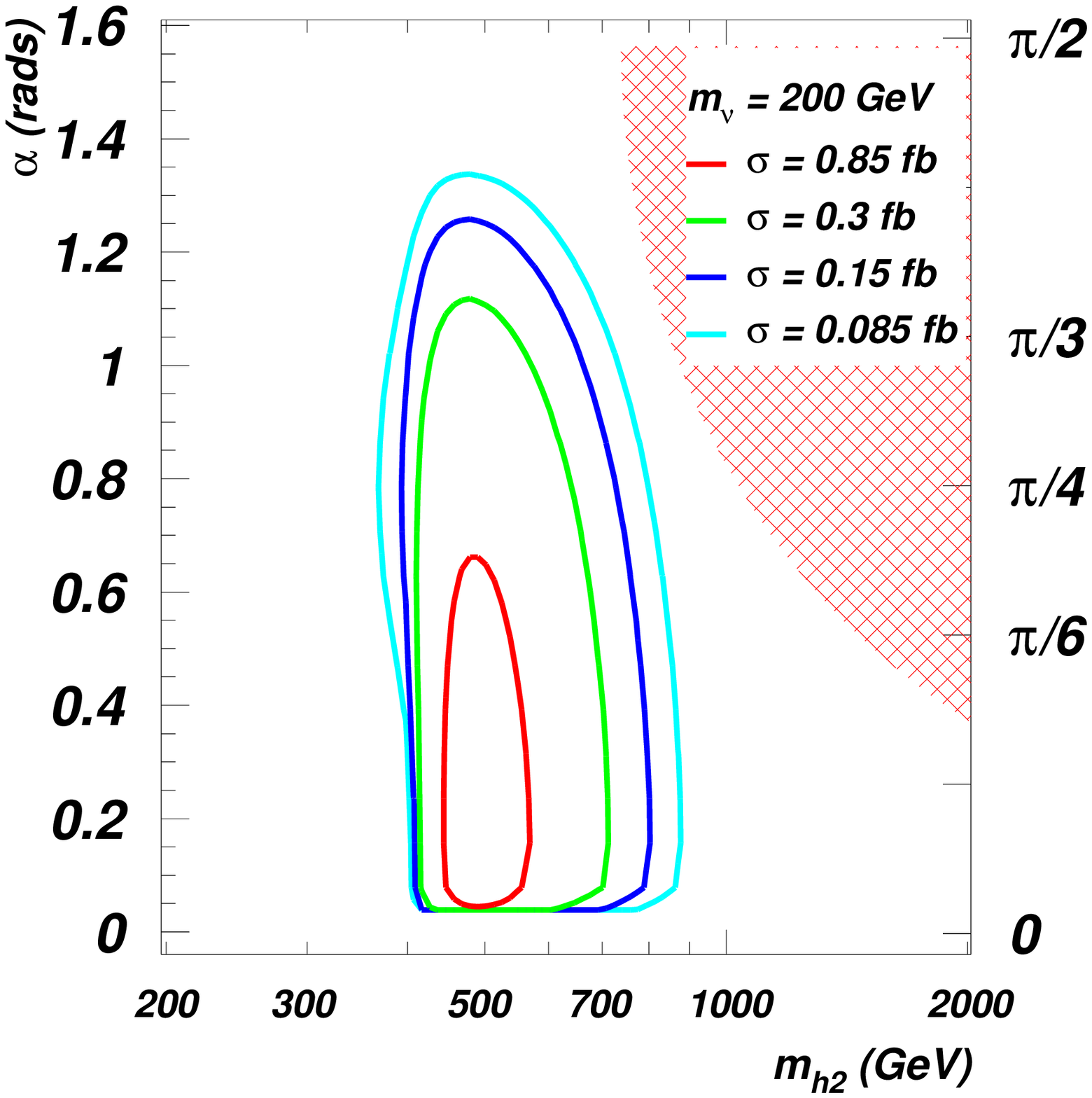}
}
  \caption[Higgs at LHC - Event rates (4)]{Cross section times $BR$
    contour plot for the $B-L$ process $pp\rightarrow h_2\rightarrow
    \nu_{h}\nu_{h}$ at the LHC with $\sqrt{s}=14$ TeV, plotted against
    $M_{h_2}$-$\alpha$, with $M_{\nu_{h}}=150$ GeV
    (\ref{gg-h2-nunu150}) and $M_{\nu_{h}}=200$ GeV
    (\ref{gg-h2-nunu200}). Several values of cross section times $BR$
    have been considered: $\sigma=0.085$ fb (light-blue line),
    $\sigma=0.15$ fb (blue line), $\sigma=0.3$ fb (green line),
    $\sigma=0.85$ fb (red line). The red-shadowed region is excluded
    by unitarity constraints.}
  \label{ggnunu150-200}
\end{figure}

In analogy with the previous two cases, in figure~\ref{ggnunu150-200}
we show the results for heavy neutrino pair production at the LHC with
$\sqrt{s}=14$ TeV plus $M_{\nu_h}=150$ GeV
(figure~\ref{gg-h2-nunu150}) and $M_{\nu_h}=200$ GeV
(figure~\ref{gg-h2-nunu200}). The usual contour plot displays a
sizable
event rate in the $\alpha$-$M_{h_2}$ parameter space for both choices
of the $\nu_h$ mass. For example, when $M_{\nu_h}=150$ GeV we find a
cross section times $BR$ peak of $0.85$ fb ($\sim 250$ events) in the
$320$ GeV$<M_{h_2}<520$ GeV and $0.03\pi<\alpha<0.33\pi$ region, while
if $M_{\nu_h}=200$ GeV we find a peak of $0.85$ fb in the $450$
GeV$<M_{h_2}<550$ GeV and $0.03\pi<\alpha<0.21\pi$ region.

\section{Higgs bosons at future linear colliders}\label{sect:4-4}

In this Section we present our results for the scalar sector of the
minimal $B-L$ model at future LCs. In general, sub-TeV $CM$ 
energies ($\sqrt{s}=500$ GeV) will be suitable for an ILC, multi-TeV
$CM$ energies ($\sqrt{s}=3$ TeV) will be appropriate for CLIC while the
case $1$ TeV may be appropriate to both. In all cases, results will be
shown for some discrete choices of the $Z'$ mass and of the scalar
mixing angle~$\alpha$. Their values have been chosen in each plot to
highlight some relevant phenomenological aspects. Concerning single
Higgs production, we will distinguish the standard production
mechanisms (via $SM$ gauge bosons) from the novel mechanisms present
in the model under discussion (emphasising in particular the role of
the $Z'$ gauge boson). Finally, plots for double Higgs production will
also be presented.

\subsection{The future linear collider running
  proposals}\label{subs:4-4-1}

Although there are not official approvals of either ILC or CLIC yet,
we already know what are the energy parameters of the two proposed
machines:
\begin{itemize}
\item ILC: $500$ GeV is the planned initial energy, with the open
  possibility of an upgrade to $1$ TeV, and $500$ fb$^{-1}$ of
  integrated luminosity at fixed energy; moreover, the possibility is
  planned to span over the energy range up the maximum energy ($500$
  GeV or $1$ TeV) at $10$ fb$^{-1}$ of integrated luminosity.
\item CLIC: $3$ TeV is the planned initial energy, with the open
  possibility of an upgrade to $5$ TeV; there is no scheduled
  integrated luminosity yet, then we conservatively assume that it
  will correspond to the ILC prototype, i.e. $500$ fb$^{-1}$.
\end{itemize}

The standard Initial State Radiation ($ISR$)
functions are implemented, according to the formulae in
\cite{calchep_man,Jadach:1988gb,Basso:2009hf}.

Finally, though the beam-strahlung parameters have only been set for
the ILC prototype, in this work we assume that the same set of values
holds for the CLIC framework, hence we will take into account these
values as they appear in the ``ILC Reference Design Report'' (see
\cite{:2007sg}). We list them in table~\ref{beam}.

\begin{table}[!htbp]
\begin{center}
\begin{tabular}{|c|c|c|}
\hline
\  & Nominal value & Unit  \\
\hline
Bunch population & $2$ & $\times 10^{10}$ \\
\hline
RMS bunch lenght & $300$ & $\mu$m \\
\hline
RMS horizontal beam size & $640$ & nm \\
\hline
RMS vertical beam size & $5.7$ & nm \\
\hline
\end{tabular}
\end{center}
\caption[Higgs at future linear colliders - Future linear collider
  prototypes]{Nominal values of beam parameters at the ILC (see
  \cite{:2007sg}).}
\label{beam}
\end{table}

\subsection{Standard single-Higgs production mechanisms}\label{subs:4-4-2}

\begin{figure}[!t]
\centering
  \subfigure[]{ 
  \label{LC_h1_500}
  \includegraphics[angle=0,width=0.48\textwidth ]{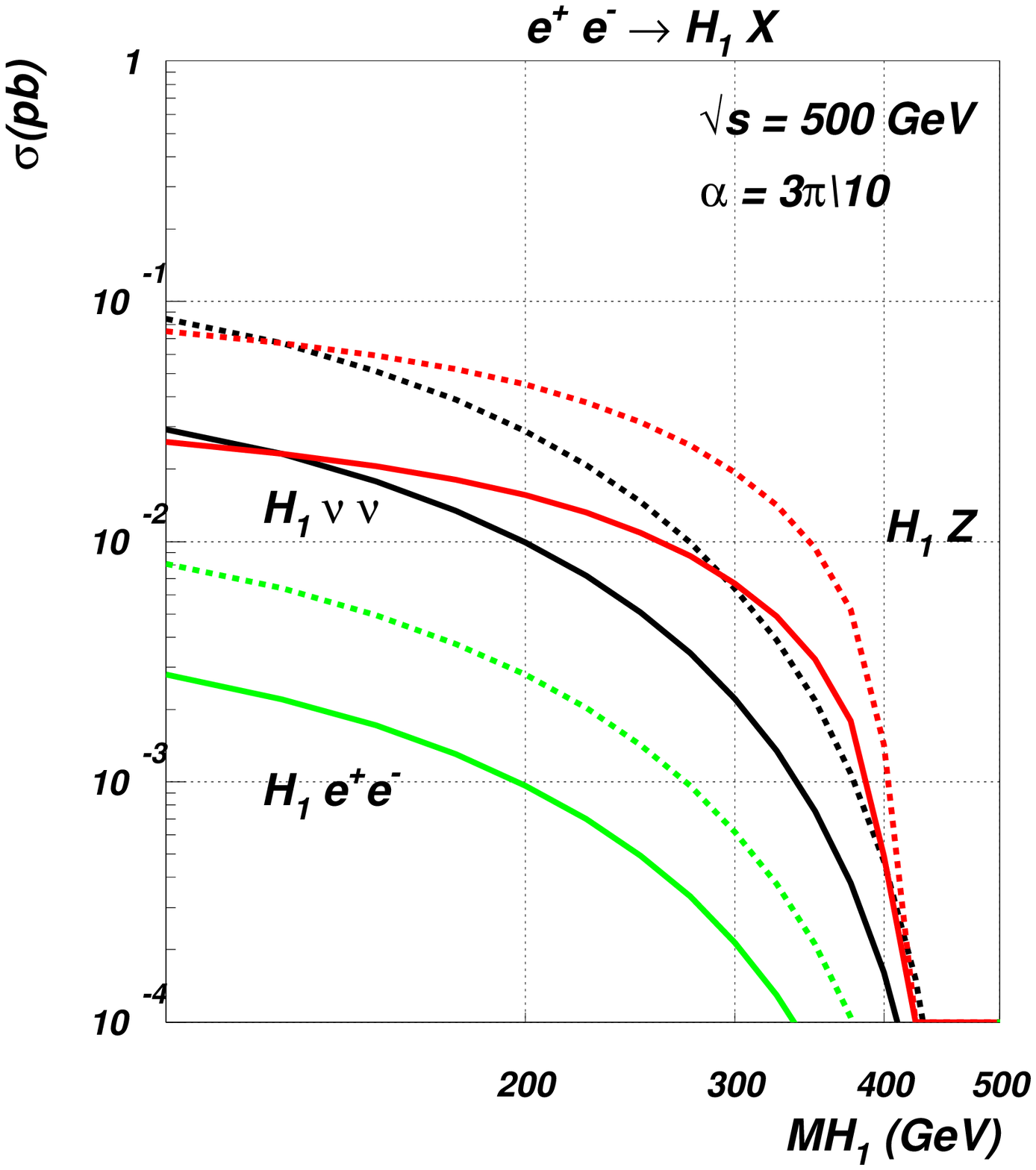}}
  \\
  \subfigure[]{
  \label{LC_h1_1000}
  \includegraphics[angle=0,width=0.48\textwidth ]{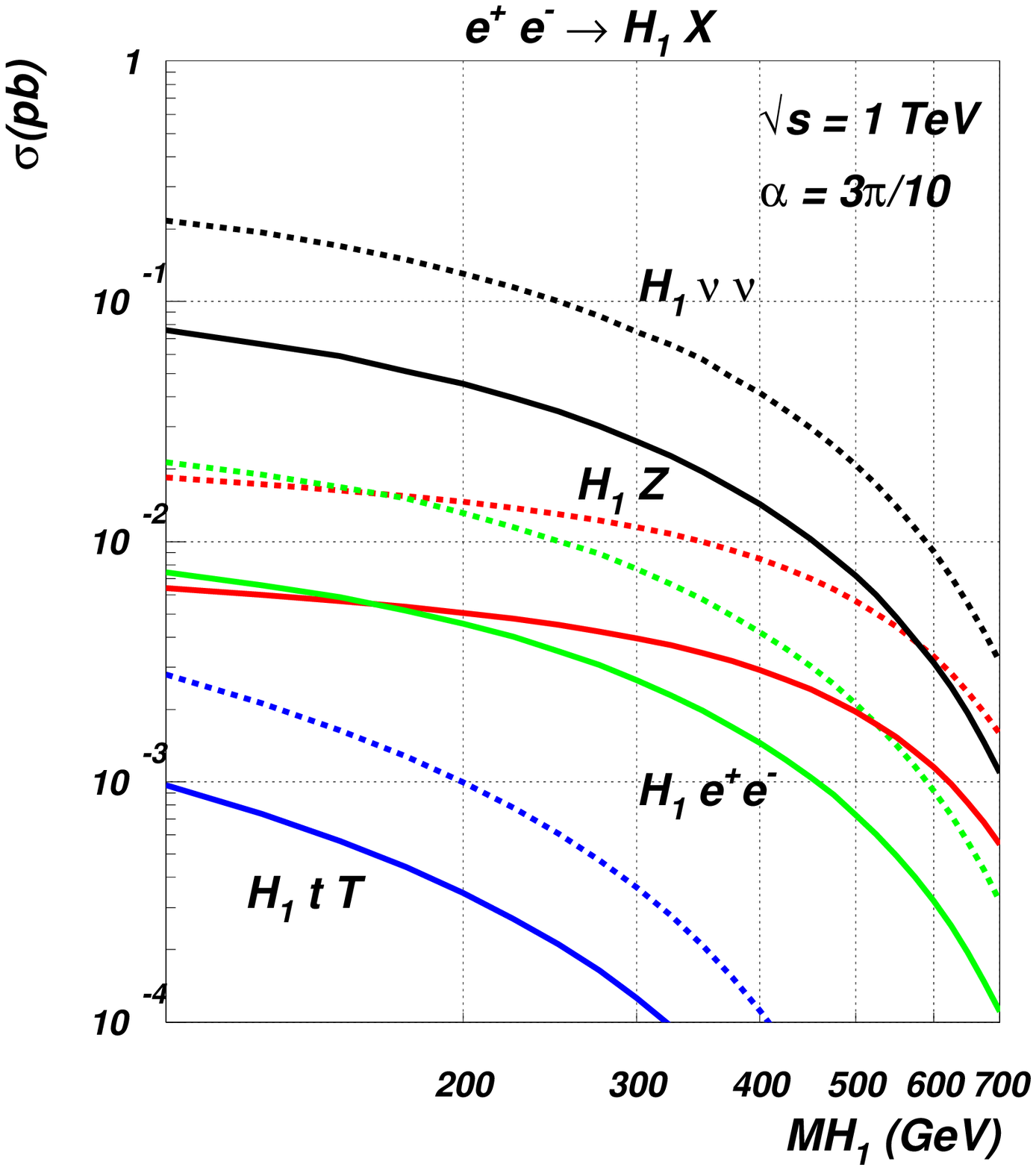}}
  \subfigure[]{
  \label{LC_h1_3000}
  \includegraphics[angle=0,width=0.48\textwidth ]{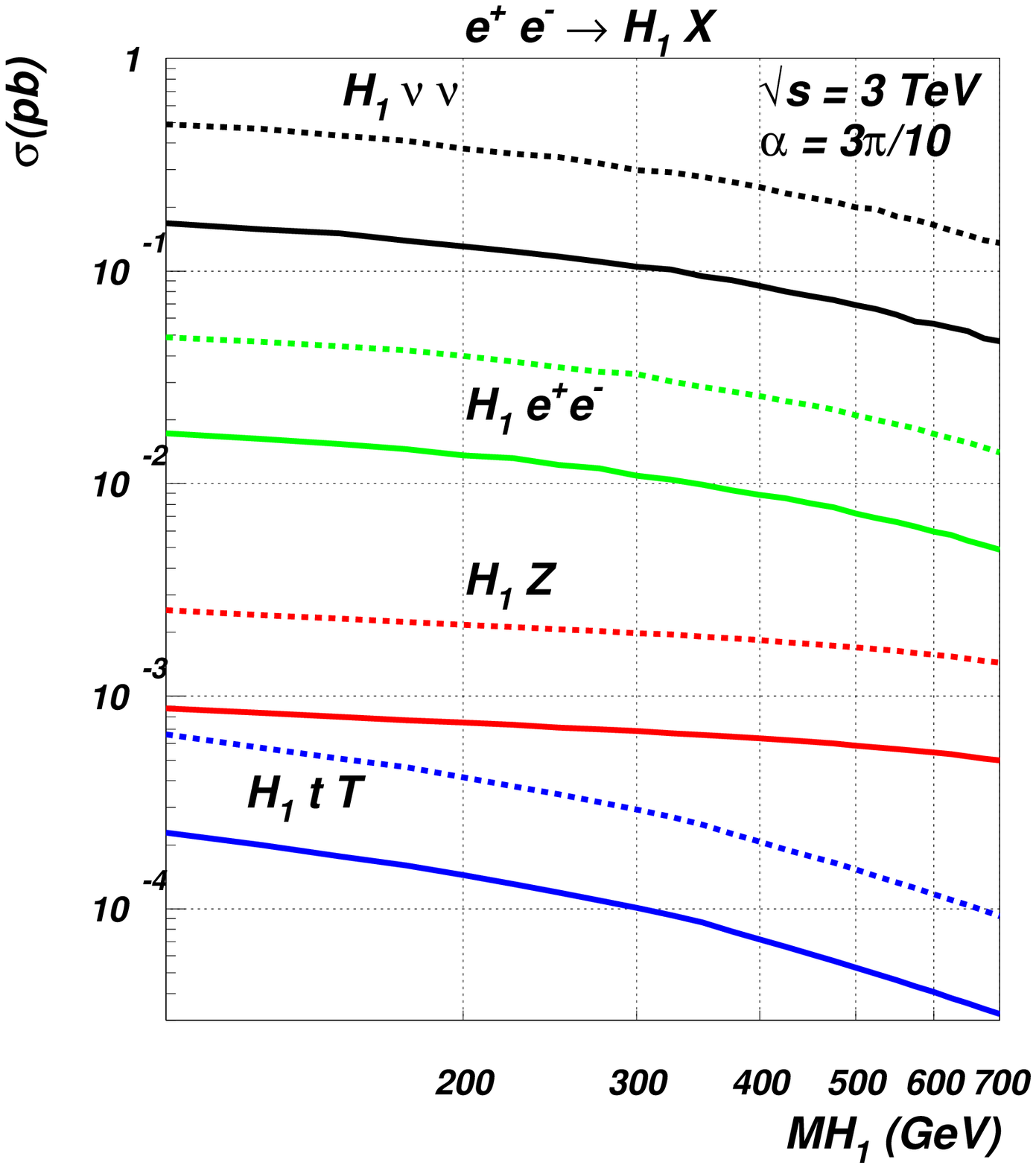}}
  \caption[Higgs at future linear colliders - Standard single-Higgs
    production mechanisms]{Cross sections for the process standard
    Higgs boson production mechanism as a function of the mass at the
    LC for $\alpha = 3\pi/10$ for (\ref{LC_h1_500}) $h_1$ and at
    $\sqrt{s}=500$ GeV, for (\ref{LC_h1_1000}) $h_1$ at $\sqrt{s}=1$
    TeV and for (\ref{LC_h1_3000}) $h_1$ at $\sqrt{s}=3$ TeV. The
    dashed lines refer to $\alpha =0$. \label{ILC_stand_prod}}
\end{figure}

Figure~\ref{ILC_stand_prod} shows the cross sections for the standard
production mechanisms of a single Higgs boson: vector boson fusion
($V=W^\pm,Z)$, the strahlung from the $Z$ boson and the
associated production with a $t$-quark pair. These standard production
mechanisms are modulated by a $\cos{\alpha}$($\sin{\alpha}$) prefactor
in the vertices when considering $h_1$($h_2$), as generally true for
all the scalar singlet extensions of the $SM$. Hence, we will not
spend too much time in discussing them.

As well known \cite{Djouadi:2005gi}, for low energies, Higgs-strahlung
from a $Z$ boson is competitive with $W$-boson fusion, that becomes the
main mechanism as we increase the $CM$ energy. At
$\sqrt{s}=1$ TeV, the cross sections for a light Higgs boson vary from
$\mathcal{O}(0.1)$ pb for $M_{h_1}=100$ GeV to $\mathcal{O}(10)$ fb
for $M_{h_1}=600$ GeV, in the decoupling regime (i.e., for $\alpha
=0$, that corresponds to recovering the $SM$ Higgs boson). All the
cross sections are then scaled by a factor $\cos{\alpha}$
$(\sin{\alpha})$ when considering $h_1$ $(h_2)$. At this value of $CM$
energy then, the associated production with a $t$-quark pair reaches
its highest value (for the $CM$ energies we are plotting), i.e.,
$\mathcal{O}(1)$ fb for $M_{h_1} \leq 200$ GeV, depending on the
mixing angle.

Not surprisingly, as we increase the $CM$ energy towards $\sqrt{s}=3$
TeV, $W$-boson fusion increases considerably, staying around fractions
of pb for several masses and angles, for both Higgs bosons, while the
Higgs-strahlung from the $Z$ boson plunges towards cross sections of
the order of few fb.

The associated production with a $t$-quark pair in the $SM$ scenario
is the least effective production mechanism, with cross sections of
few fb at most. However, we will show in the following Subsection that
this mode can be enhanced by the presence of the $Z'$ boson.

\subsection{Non-standard single-Higgs production
  mechanisms}\label{subs:4-4-3}

In this Section we will discuss the novel mechanisms to produce a
Higgs boson (both the light one or the heavy one) in the minimal
$B-L$ model. All the new features arise from having a $Z'$ that
interacts with both the
scalar and fermion sectors. We recall here another important feature:
the $Z'$ boson is dominantly coupled to leptons
\cite{Basso:2008iv}. In fact:
\begin{eqnarray} 
\sum _\ell BR(Z' \rightarrow \ell \ell) &\sim& \frac{3}{4}\, ;\\
\sum _q BR(Z' \rightarrow q \overline{q}) &\sim& \frac{1}{4}\, ;
\end{eqnarray}
and in particular, $BR(Z' \rightarrow e ^+ e ^-) \simeq 15\%$, which
makes a lepton collider the most suitable environment for testing this
model.

Again, the $Z'$ mass and $g'_1$ gauge coupling values have been
chosen to respect the constraints coming from LEP
and Tevatron (Subsection~\ref{subs:3-1-3}).
%
%

\begin{figure}[!t]
  \subfigure[]{ 
  \label{H1Zp_1TeV}
  \includegraphics[angle=0,width=0.48\textwidth ]{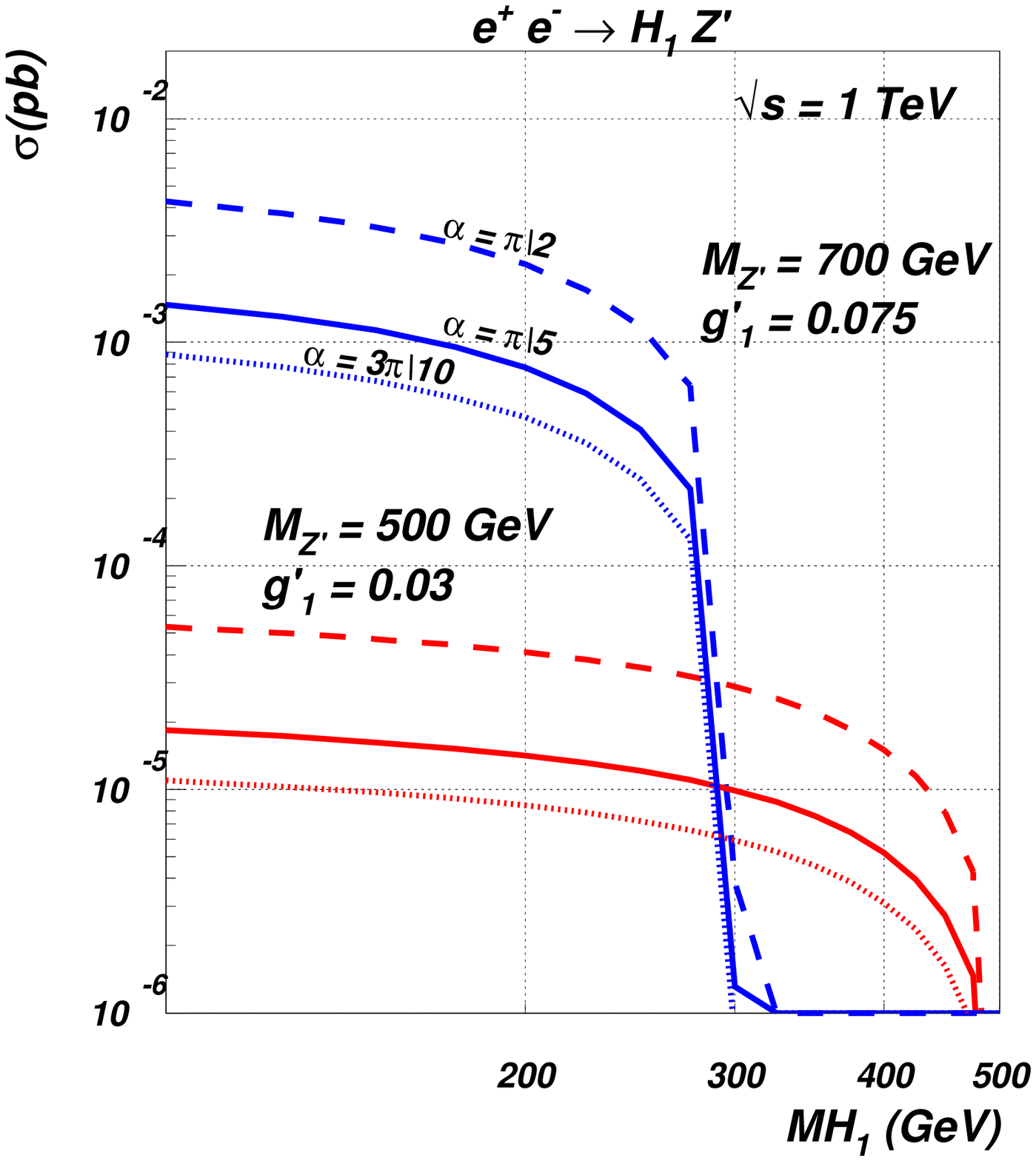}}
  \subfigure[]{
  \label{H2Zp_1TeV}
  \includegraphics[angle=0,width=0.48\textwidth ]{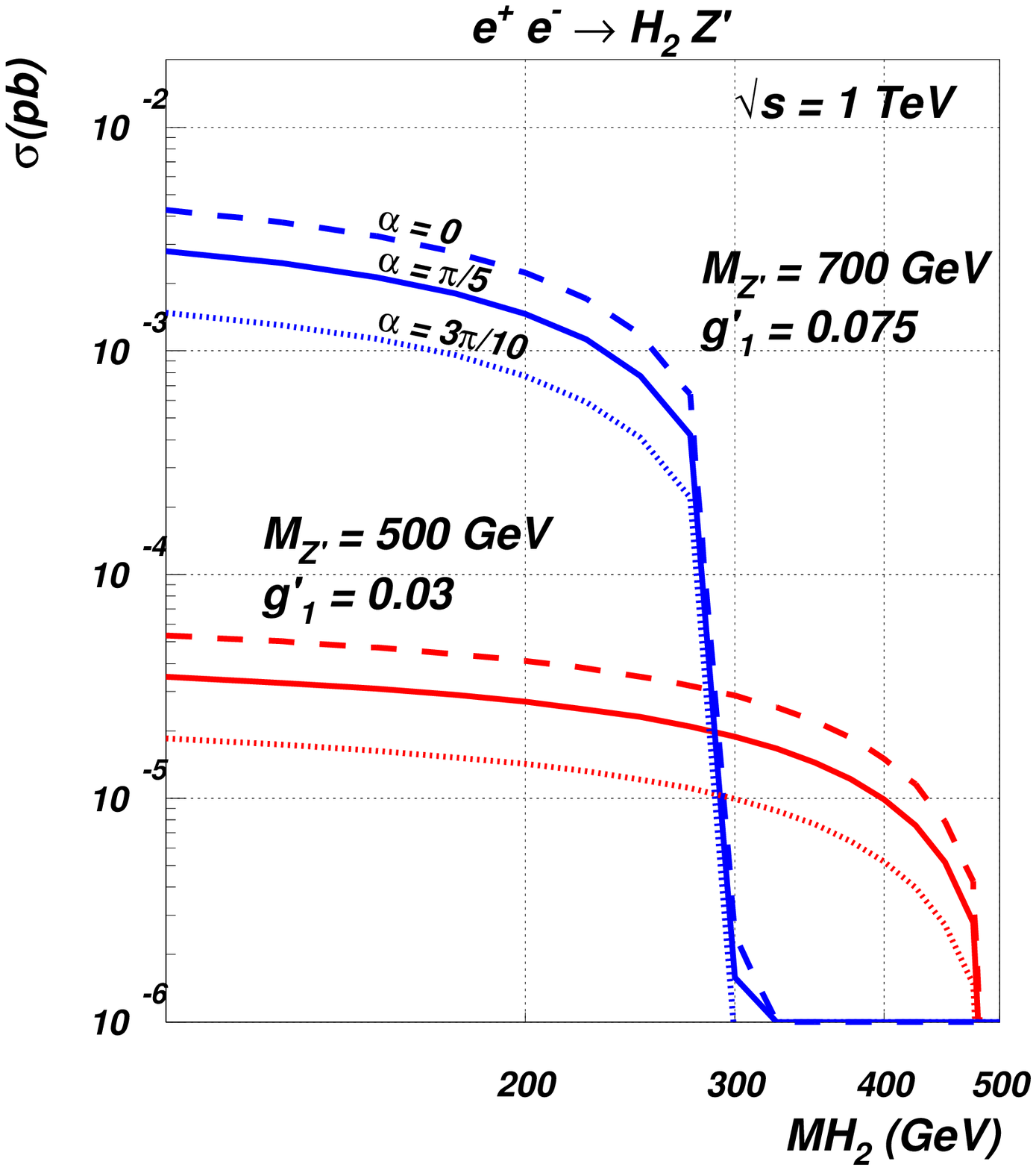}}\\
  \subfigure[]{ 
  \label{H1Zp_3TeV}
  \includegraphics[angle=0,width=0.48\textwidth ]{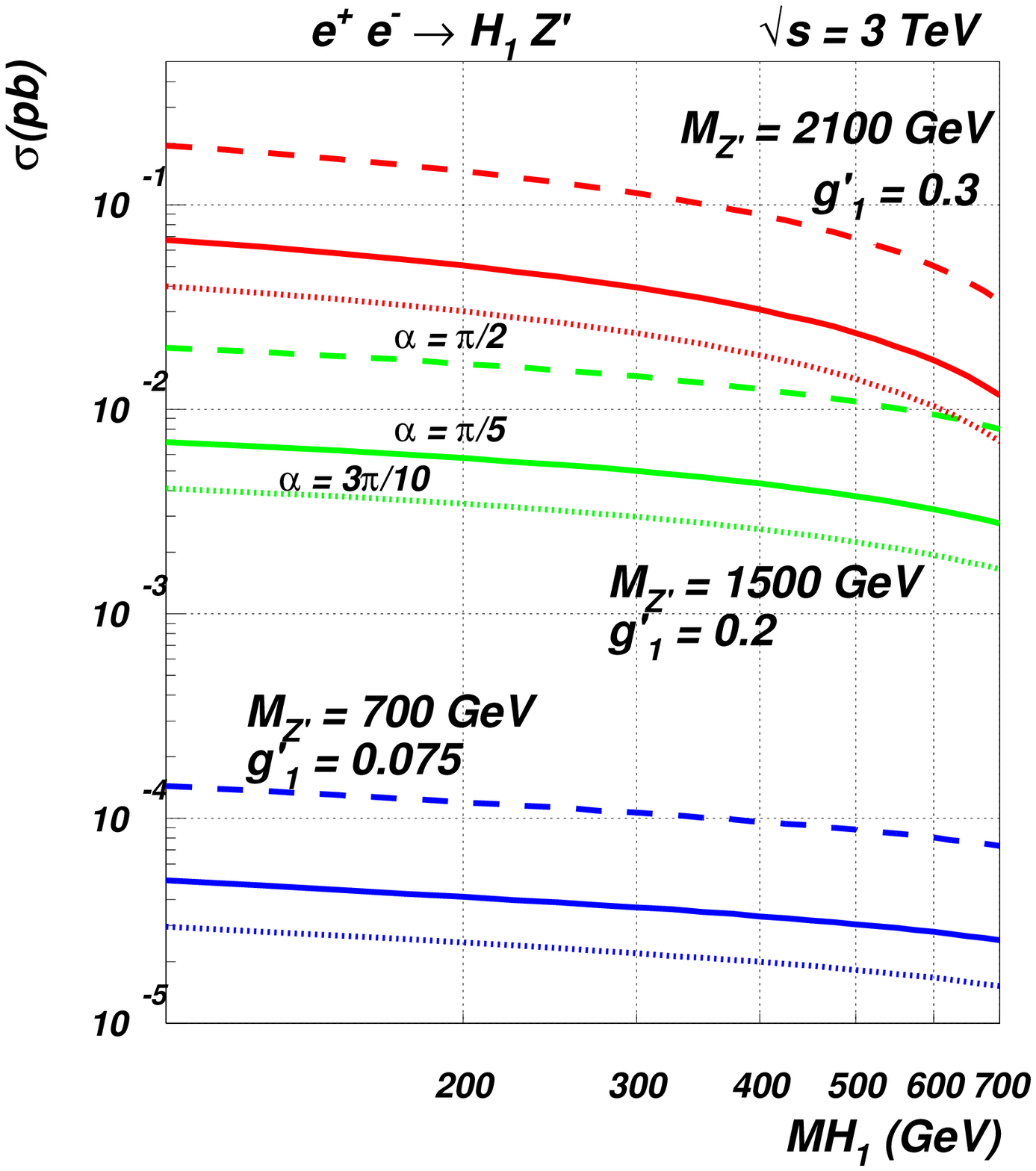}}
  \subfigure[]{
  \label{H2Zp_3TeV}
  \includegraphics[angle=0,width=0.48\textwidth ]{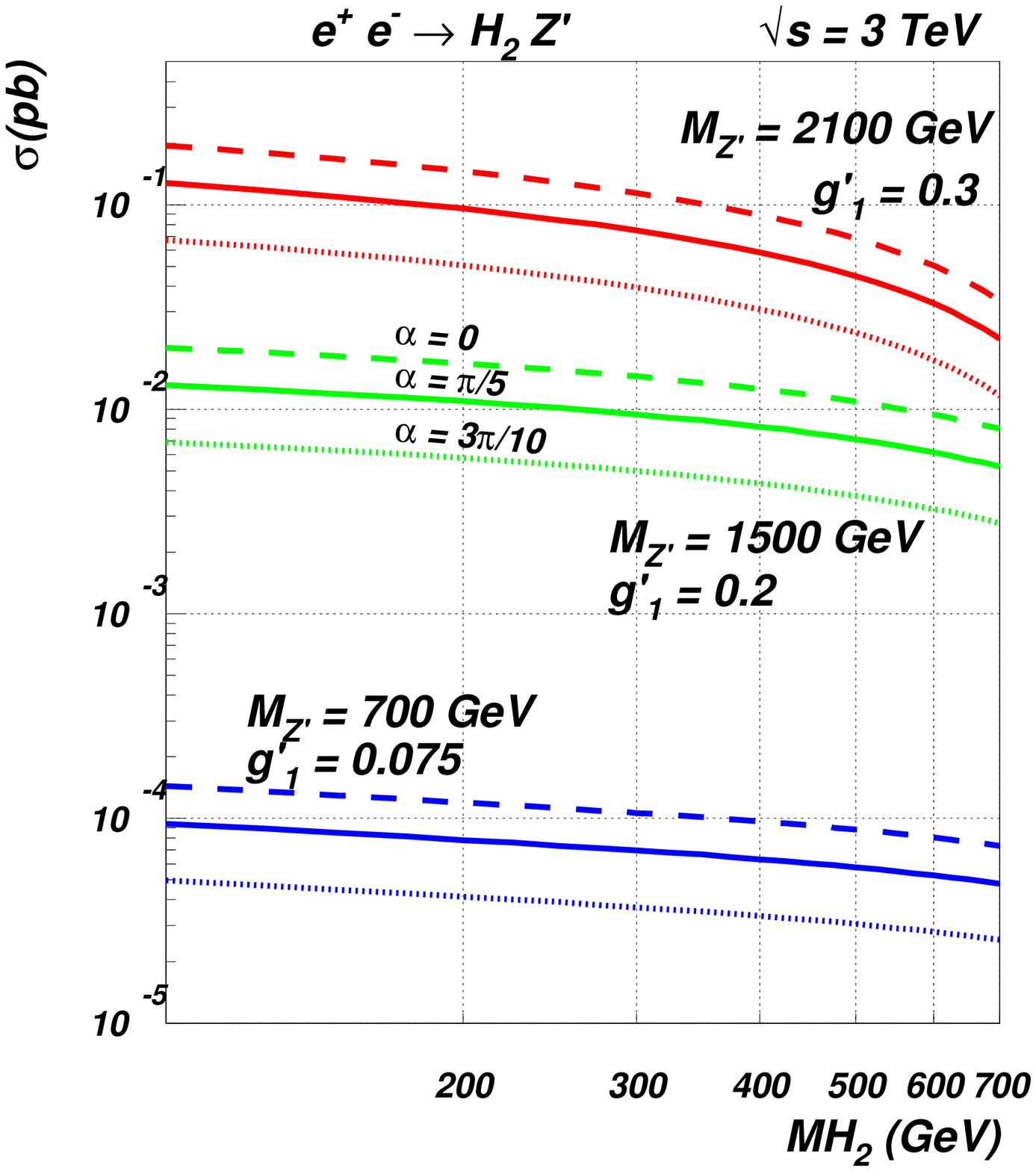}}  
  \caption[Higgs at future linear colliders - Non-standard
    single-Higgs production mechanisms (1)]{Cross sections for the
    process $e^+e^-\rightarrow Z^{'\ast} \rightarrow h_{1(2)} Z'$
    (\ref{H1Zp_1TeV}) for $h_1$ and (\ref{H2Zp_1TeV}) for $h_2$ at the
    LC at $\sqrt{s}=1$ TeV and (\ref{H1Zp_3TeV}) for $h_1$ and
    (\ref{H2Zp_3TeV}) for $h_2$ at the LC at $\sqrt{s}=3$
    TeV. \label{ILC_Zp_strah}}
\end{figure}

We start by showing the cross section for the associated production of
a Higgs boson and a $Z'$ boson, as in figure~\ref{ILC_Zp_strah}. Due
to the stringent bounds on the $Z'$ boson mass and coupling to
fermions, a sub-TeV $CM$ energy collider is not capable of benefiting
from this production mechanism, especially because of the naive
kinematic limitation in the final state phase space. In other words,
there is not enough energy to produce a $Z'$ and a Higgs boson,
if both are on-shell. This is clear in figures~\ref{H1Zp_1TeV} and
\ref{H2Zp_1TeV}, where a light $Z'$ boson (with mass of $500$ GeV)
gives cross sections below $0.1$ fb. For a $Z'$ boson of $700$ GeV
mass instead, the cross sections can be of the order of few fb, only
for Higgs masses below $300$ GeV, the kinematical limit. These results
are similar to the LHC, in which the $Z'$ strahlung process has
cross sections below $1$ fb (below $0.1$ fb for $M_{h_2} > 400$ GeV)
\cite{Basso:2010yz}, even if the LC is expected to accumulate roughly
an order of magnitude more integrated luminosity than the LHC.

The situation is considerably improved for a multi-TeV collider, not
anymore limited in kinematics. As shown in figures~\ref{H1Zp_3TeV} and~\ref{H2Zp_3TeV},  a Higgs boson can be produced in association with a
$Z'$ boson of $1.5$ TeV mass with cross sections of $\sim 10$ fb in
the whole range of the scalar masses considered, rising to
$\mathcal{O}(100)$ fb if $M_{Z'}=2.1$ TeV is considered (and a
suitable value for $g'_1$ is chosen). Although such configuration could
suffer from kinematical limitations for the Higgs boson mass to be
produced, the cross sections when the scalar mass is close to $700$
GeV (the maximum value considered here) are still above those when a
$Z'$ boson of $1.5$ TeV mass is considered, regardless of the Higgs
boson. It is crucial to note that this is the only production
mechanism that can potentially lead to the discovery of the heavy
Higgs boson in the decoupling limit, i.e., for $\alpha \to 0$. As
previously stated, the strahlung from $Z'$ is not suitable for
the LHC, making a multi-TeV LC possibly the ultimate
chance for its discovery. 
%
%

\begin{figure}[!t]
  \subfigure[]{ 
  \label{H1ZptT_1TeV}
  \includegraphics[angle=0,width=0.48\textwidth ]{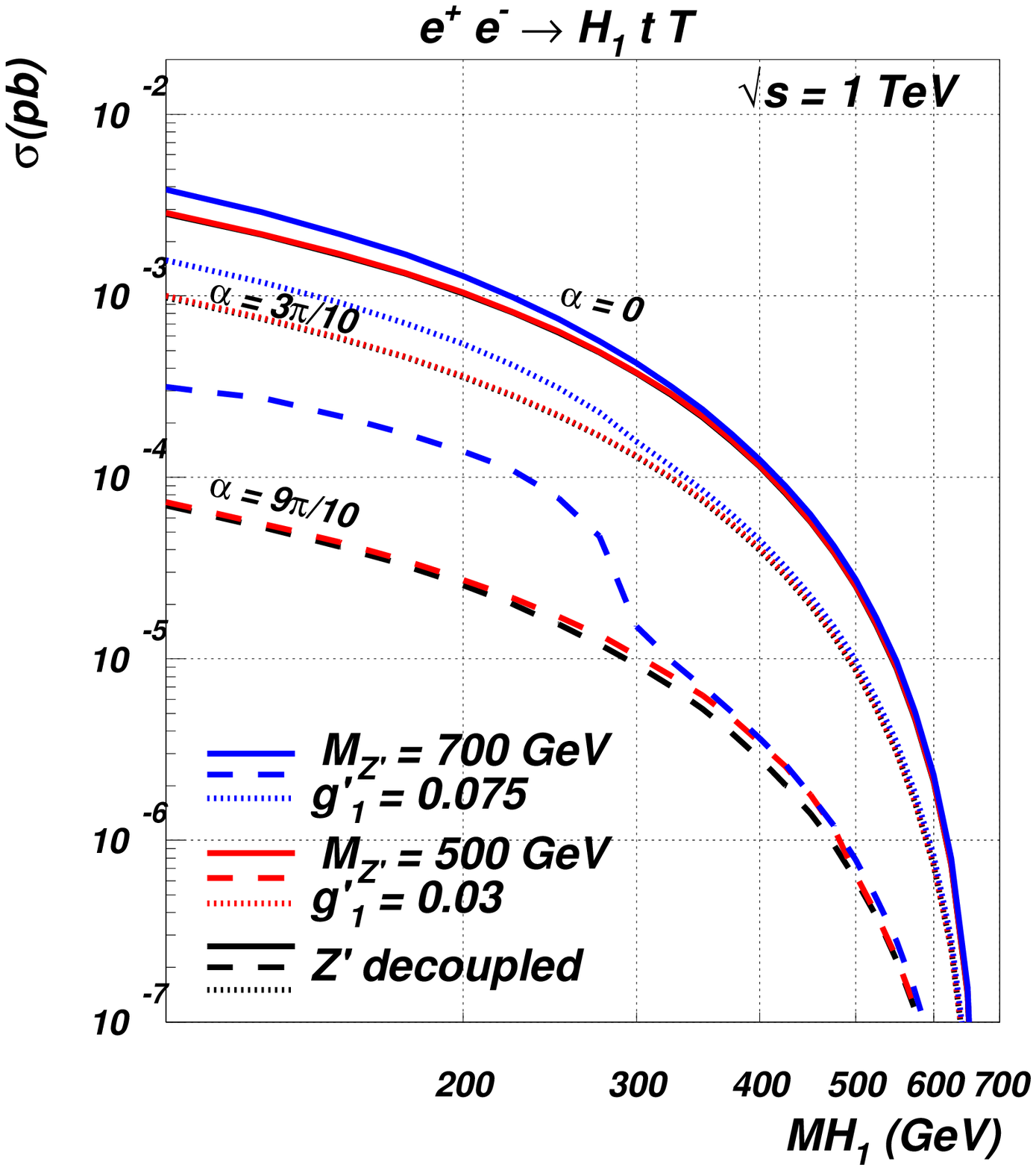}}
  \subfigure[]{
  \label{H2ZptT_1TeV}
  \includegraphics[angle=0,width=0.48\textwidth ]{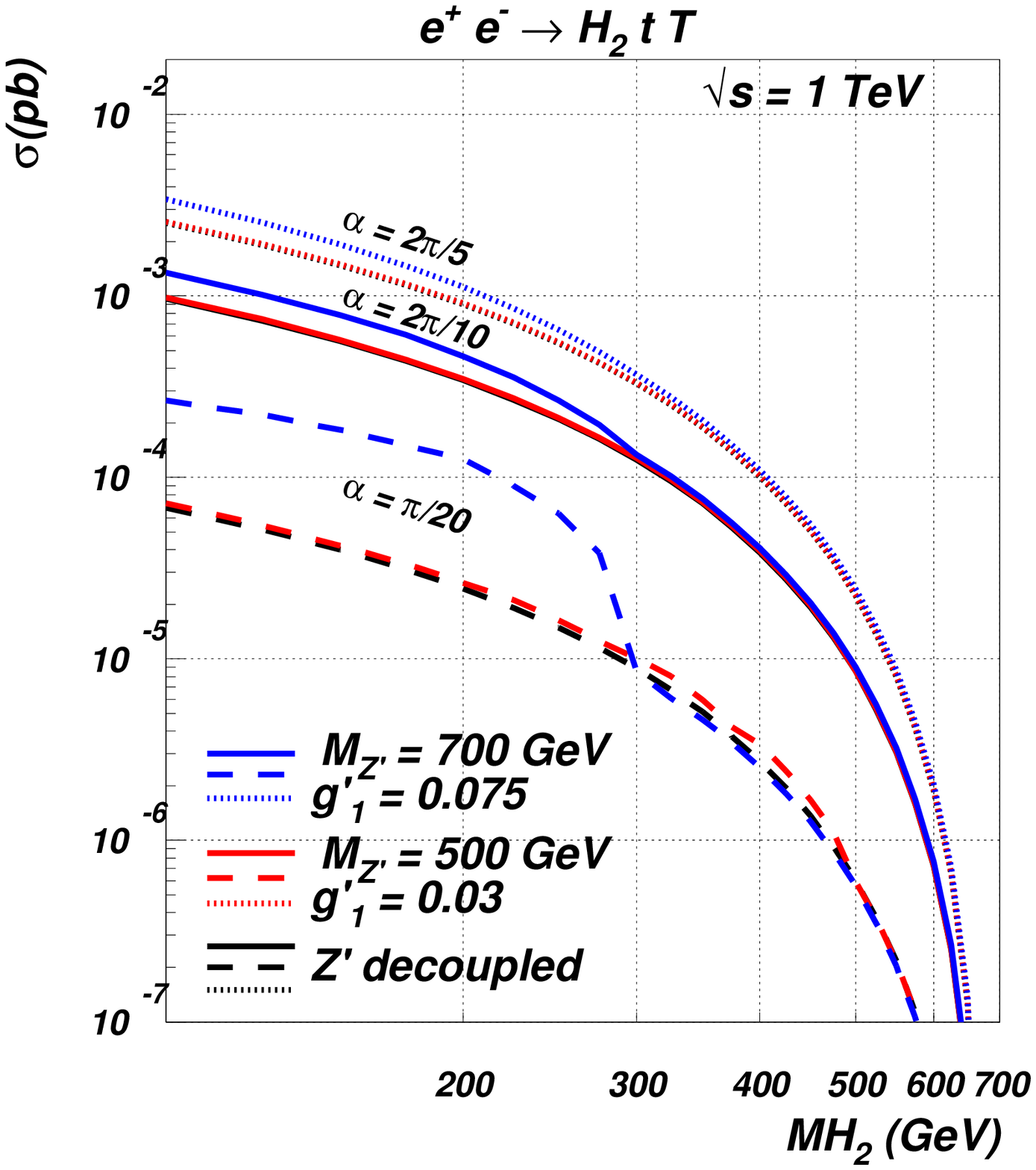}}
  \\
  \subfigure[]{ 
  \label{H1ZptT_3TeV}
  \includegraphics[angle=0,width=0.48\textwidth ]{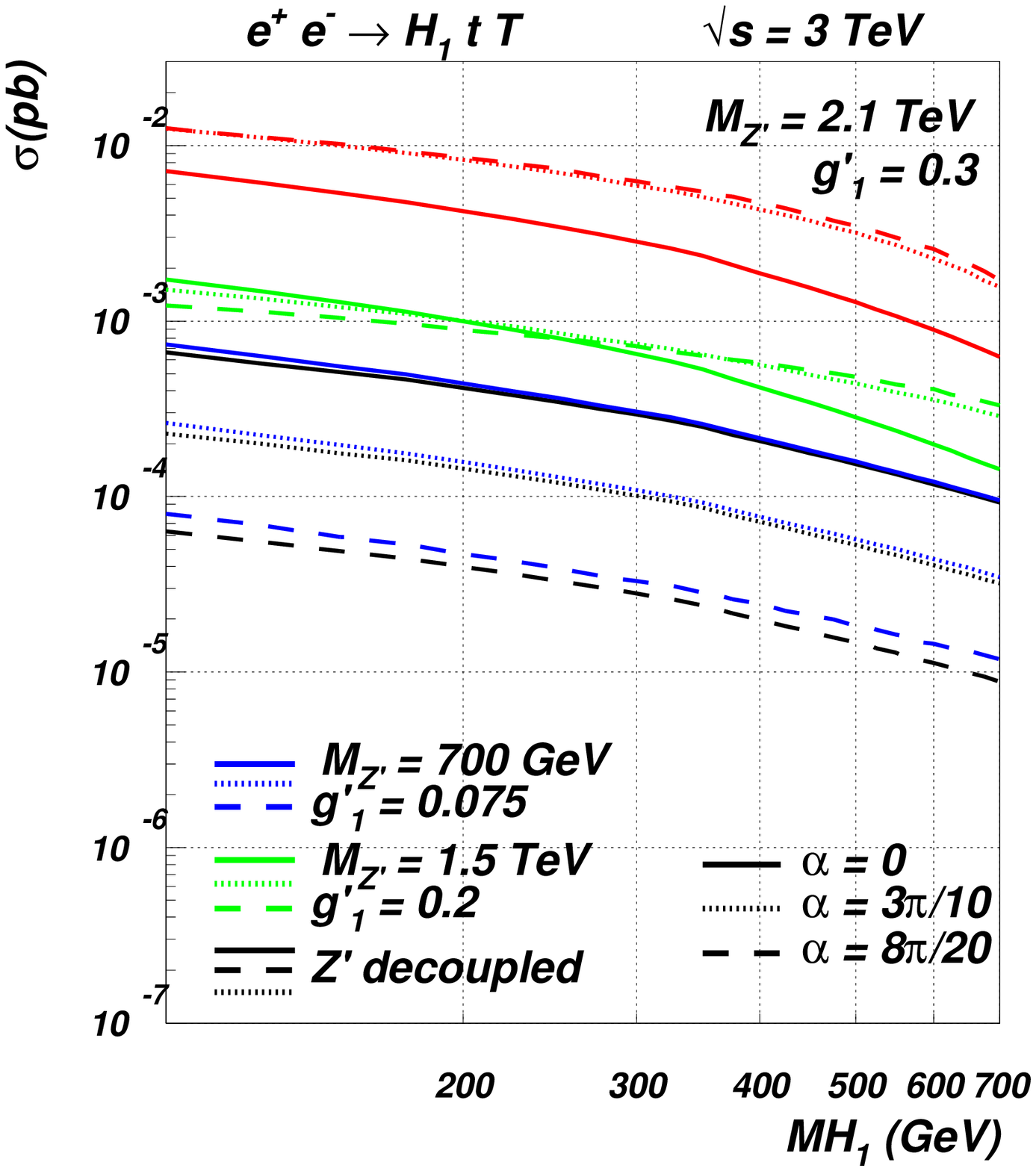}}
  \subfigure[]{
  \label{H2ZptT_3TeV}
  \includegraphics[angle=0,width=0.48\textwidth ]{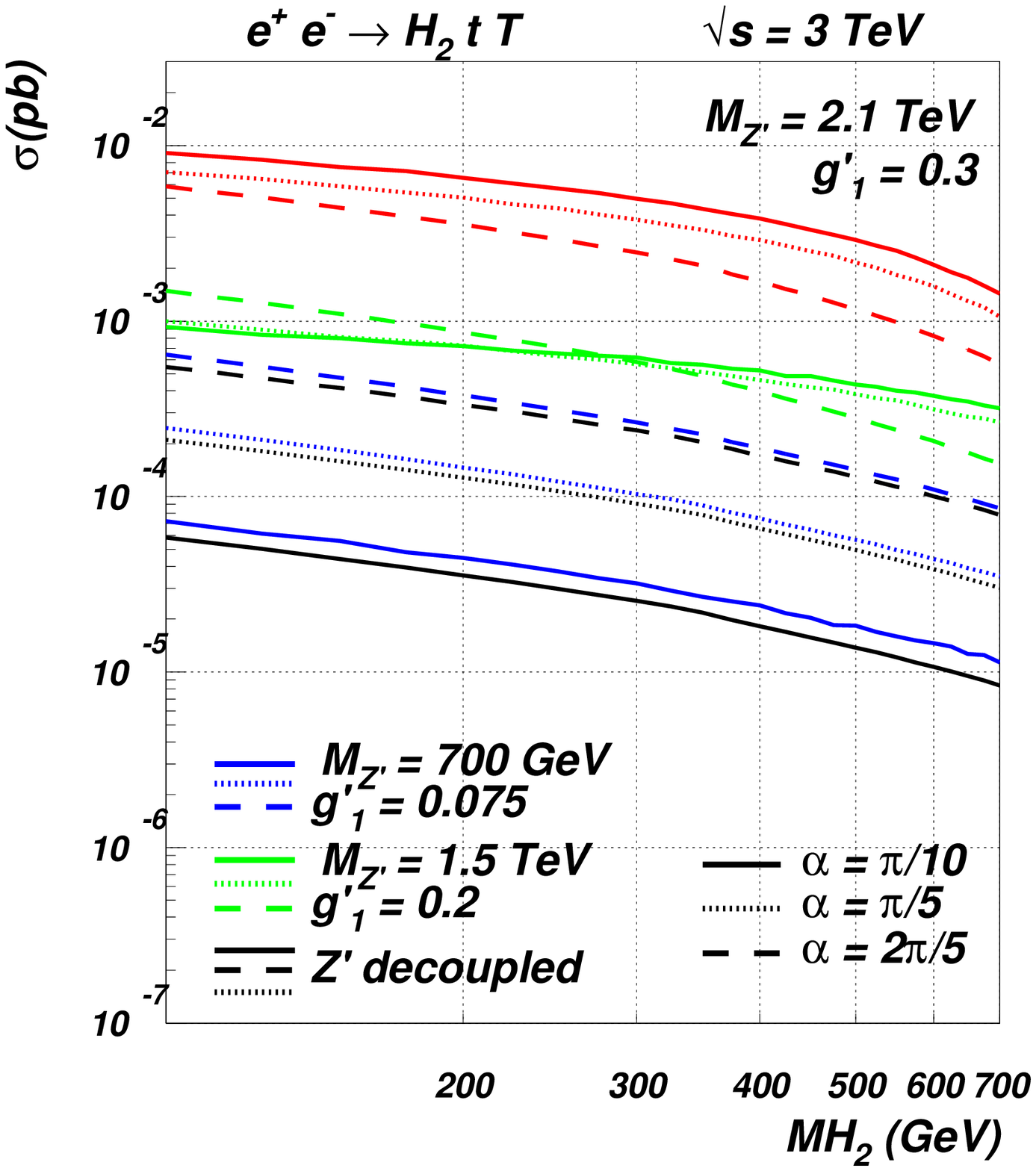}}
  \\
  \caption[Higgs at future linear colliders - Non-standard
    single-Higgs production mechanisms (2)]{Cross sections for the
    process $e^+e^-\rightarrow h_{1(2)} t \overline{t}$
    (\ref{H1ZptT_1TeV}) for $h_1$ and (\ref{H2ZptT_1TeV}) for $h_2$,
    at the LC at $\sqrt{s}=1$ TeV, (\ref{H1ZptT_3TeV}) for $h_1$ and
    (\ref{H2ZptT_3TeV}) for $h_2$, at $\sqrt{s}=3$ TeV, for several
    angles and $M_{Z'}$.
 \label{ILC_tT}}
\end{figure}

As anticipated, the associated production with a $t$-quark pair can be
enhanced exploiting the $Z'$ boson. In figures~\ref{H1ZptT_1TeV}
and~\ref{H2ZptT_1TeV} are shown the cross sections for the associated 
production of a Higgs boson and a pair of $t$-quarks, for $M_{Z'}=500$
GeV, $700$ GeV and in the  case of a much heavier $Z'$ boson, hence
decoupled, for $\sqrt{s}=1$ TeV. As known, the Higgs boson in this
channel can be radiated both by a $t$-(anti)quark or by the vector
boson, even though the fraction of events with a $Z$ boson is
negligible with respect to the $t$-quark pair produced by a
photon. Therefore, in the $SM$, the measure of the Higgs coupling to
the $t$-quark is possible, though difficult because of the small cross
sections \cite{Baer:1999ge}. We are therefore left to evaluate the
relative contribution of the $Z'$ boson, to check whether the same
situation holds.

It is first interesting to note that, in the decoupling limit (i.e.,
for vanishing scalar mixing angle $\alpha$), $h_1$ does not couple
directly to the $Z'$ boson. Nonetheless, the $Z'$ boson can decay to
$t$-quark pair, one of which then radiates the light Higgs boson. This
channel has the same final state than the $SM$ ones, and will
therefore increase the total number of events, as clear from
figure~\ref{H1ZptT_1TeV}. Hence, the chances of measuring the
($SM$-like) Higgs boson to $t$-quark coupling are improved in this
case, only slightly for $\sqrt{s}=1$~TeV but quite considerably for
$\sqrt{s}=3$~TeV and a few TeV $Z'$ boson mass.

As we increase the scalar mixing angle, the relative contribution of
the $Z'$ boson increases, although the total cross sections for
$\sqrt{s}=1$ TeV fall below the fraction of fb, making it even harder
to be observed. The situation is opposite for the multi-TeV $CM$ energy
case (figures~\ref{H1ZptT_3TeV} and~\ref{H2ZptT_3TeV}), in which the
$Z'$ boson is produced abundantly and it can enhance the Higgs boson
associated production with a $t$-quark pair. In this case, anyway, it
is not true anymore that the majority of the events are those in which
the Higgs boson is radiated by a $t$-quark: the Higgs-strahlung from
the $Z'$ boson is now an important channel, as clear from
figures~\ref{H1Zp_3TeV} and~\ref{H2Zp_3TeV} and from the fact that,
for low $Z'$ masses, the total cross section is smaller as we start
increasing the angle (due to the reduced coupling to the $t$-quark),
while for TeV $Z'$ boson masses it always increases. If the $Z'$ boson
mass is below the maximum $CM$ energy of the collider, the fraction of
strahlung events off the $Z'$ boson can be reduced by tuning the $CM$
energy and sitting at the peak of the $Z'$ boson itself. In this case,
the vast majority of $Z'$ bosons are produced on-shell, enhancing the
total cross sections and the portion of events in which the $Z'$ boson
decays into a $t$-quark pair, one of which will then radiate the Higgs
boson. The possibility of sitting at the peak of the $Z'$ boson is
therefore very important phenomenologically, as it allows the Higgs
coupling to the $t$-quark to be measured much more precisely than in
the $SM$, as the cross section in the minimal $B-L$ model for this
channel can 
rise up to $10\div 100$~fb, depending on the Higgs boson mass and
mixing angle. Notice that for $h_1$ the angle has to be small (i.e.,
less than $\pi/5$) to allow for this measurement, as only in this case
$h_1$ couples more to the $t$-quark than to the $Z'$ boson, though
$Z'$ strahlung events are still important (the ratio of the two
subchannels is in fact $<5 \%$). This situation is exactly specular
when the heavy Higgs boson is considered: when the $CM$ energy is
maximal, the associated production with a $t$-quark pair has good
cross sections but it does not allow a direct measurement of the Higgs
boson to $t$-quark coupling, that instead is possible for big angles
($\alpha \gtrsim 3\pi/10$) when sitting at the $Z'$ boson peak. Notice
that, in this configuration, the total cross section is independent of
the $Z'$ boson mass, if the $t$-pair and the Higgs boson can be
produced all on-shell. Otherwise, the cross sections are suppressed by
the phase space.
%
%

\begin{figure}[!t]
  \subfigure[]{ 
  \label{LC_CM-1}
  \includegraphics[angle=0,width=0.48\textwidth ]{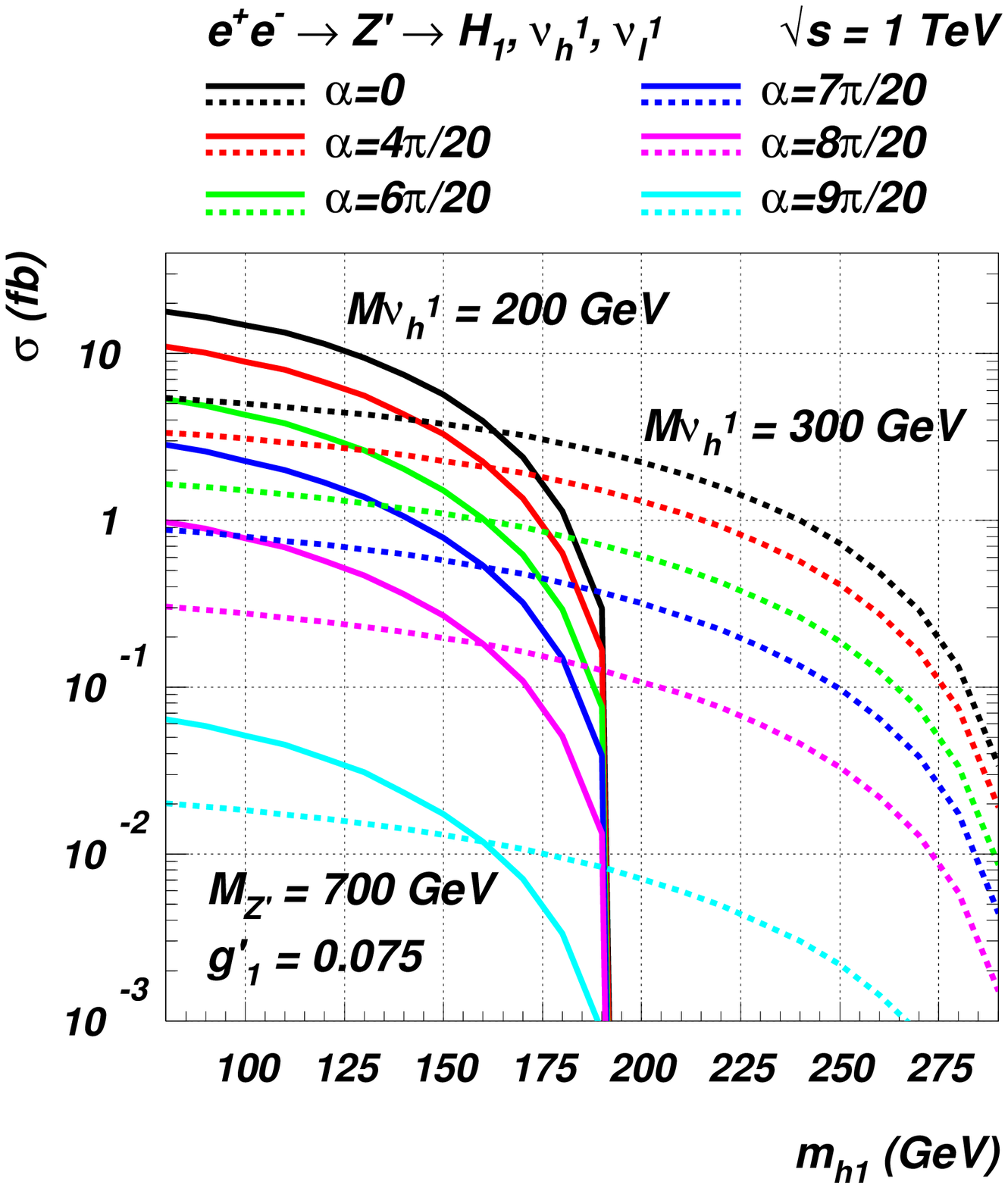}}
  \subfigure[]{
  \label{LC_CM-MZp}
  \includegraphics[angle=0,width=0.48\textwidth ]{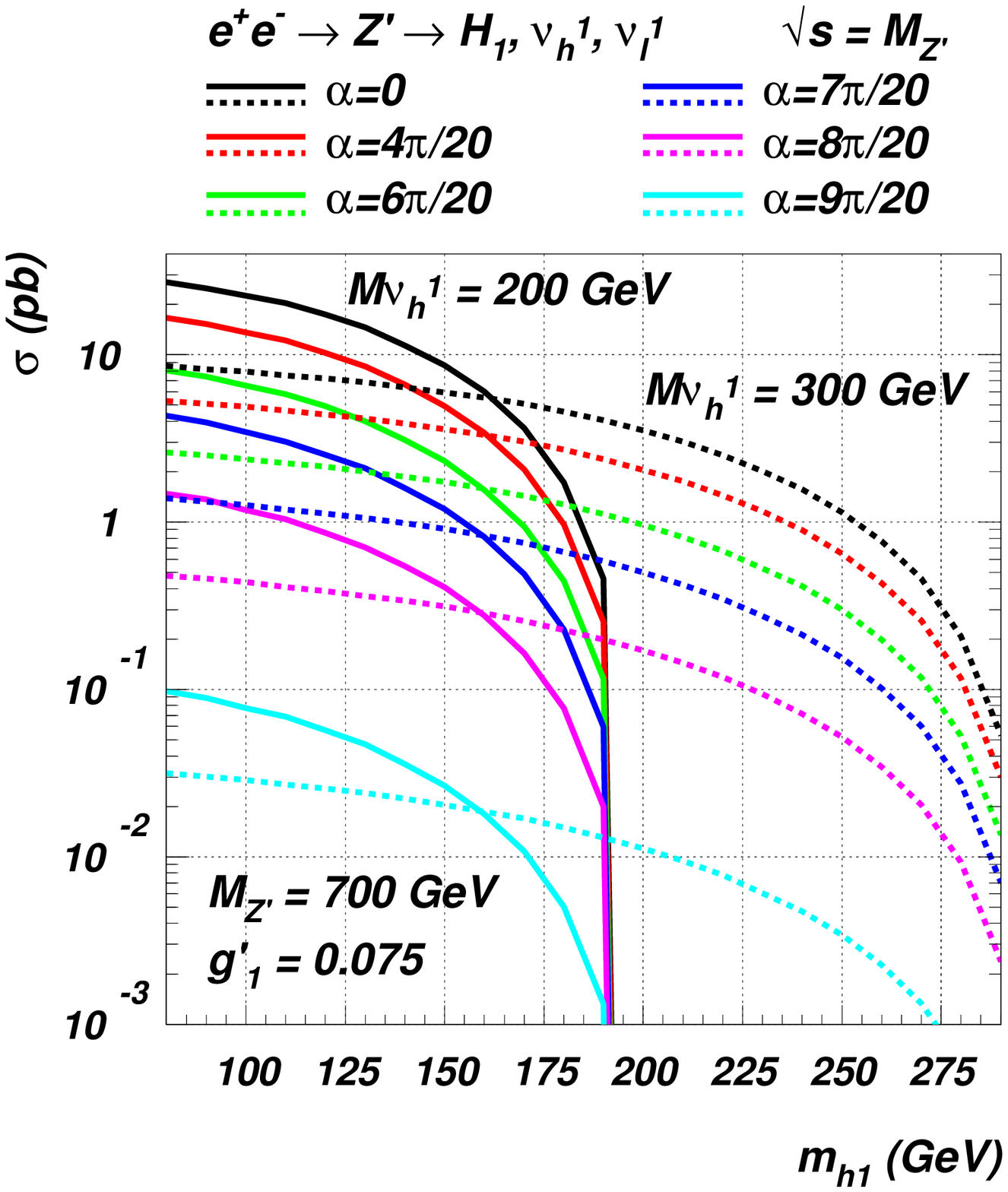}}
  \caption[Higgs at future linear colliders - Non-standard
    single-Higgs production mechanisms (3)]{Cross sections for the
    associated production of the 
    light higgs boson and one heavy and one light first generation
    neutrinos (via $Z'\rightarrow \nu _h \nu _h$) at the LC
    (\ref{LC_CM-1})  for $\sqrt{s}=1$ TeV and (\ref{LC_CM-MZp})  for
    $\sqrt{s}\equiv M_{Z'}$. \label{ILC_neutrino}}
\end{figure}

Next, a possibility already highlighted in
\cite{Basso:2008iv,Basso:2010yz} for the LHC is using the heavy
neutrino as a source of light Higgs bosons. Besides to provide a
further production mechanism and being a very peculiar feature of the
minimal $B-L$ model, it also allows for a direct measure for the Higgs
boson 
to heavy neutrinos coupling when the decay of the Higgs to neutrino
pairs is kinematically forbidden. Though, in \cite{Basso:2010yz} is
showed that it gives low cross sections at the LHC,
making it hard to probe. In contrast, a LC is a more suitable
environment to test this mechanism. One reason is that the $Z'$
couples dominantly to leptons, as already intimated. Further, the
possibility of tuning the $CM$ energy and sitting exactly on the $Z'$
peak will enhance the $Z'$ production cross section by a factor of
roughly $10^3$. Another key factor is that the $BR$ of
a heavy neutrino into a light Higgs boson and a light neutrino is
$\sim 20\%$ (at the very most, see \cite{Basso:2008iv}), when
kinematically 
allowed. This mechanism is not suitable for the heavy scalar though:
since it is heavier than the light one, for sure one would observe the
latter first. 
Altogether, for a $Z'$ boson of $700$~GeV mass, figure~\ref{LC_CM-1}
shows the cross sections for the production of a (first generation
only) heavy neutrino pair and the subsequent decay of one of them into
a light Higgs boson, for two different masses of the heavy neutrino,
at $\sqrt{s}=1$ TeV. At this stage, the mechanism is giving
$\mathcal{O}(1\div 10)$~fb cross sections for a heavy neutrino of
$200$ GeV mass, decreasing to $\mathcal{O}(1)$ fb when a mass of $300$
GeV is considered, for a good range in the mixing
angle. Figure~\ref{LC_CM-MZp} shows the full potentiality of this
model at a LC: by sitting on the $Z'$ peak, the heavy neutrino pair
production is enhanced by a factor $\sim 10^3$, giving cross sections
well above the pb range for a large portion of the allowed parameter
space, and staying above $10$ fb whatever the mixing angle. When
kinematically allowed
though, this peculiar mechanism really carries the hallmark of the
minimal $B-L$ model and it does not depend dramatically on the $Z'$
mass, if
below the maximum $CM$ energy of the collider.
%
%

\begin{figure}[!t]
  \subfigure[]{ 
  \label{LC_intH1-500}
  \includegraphics[angle=0,width=0.48\textwidth
  ]{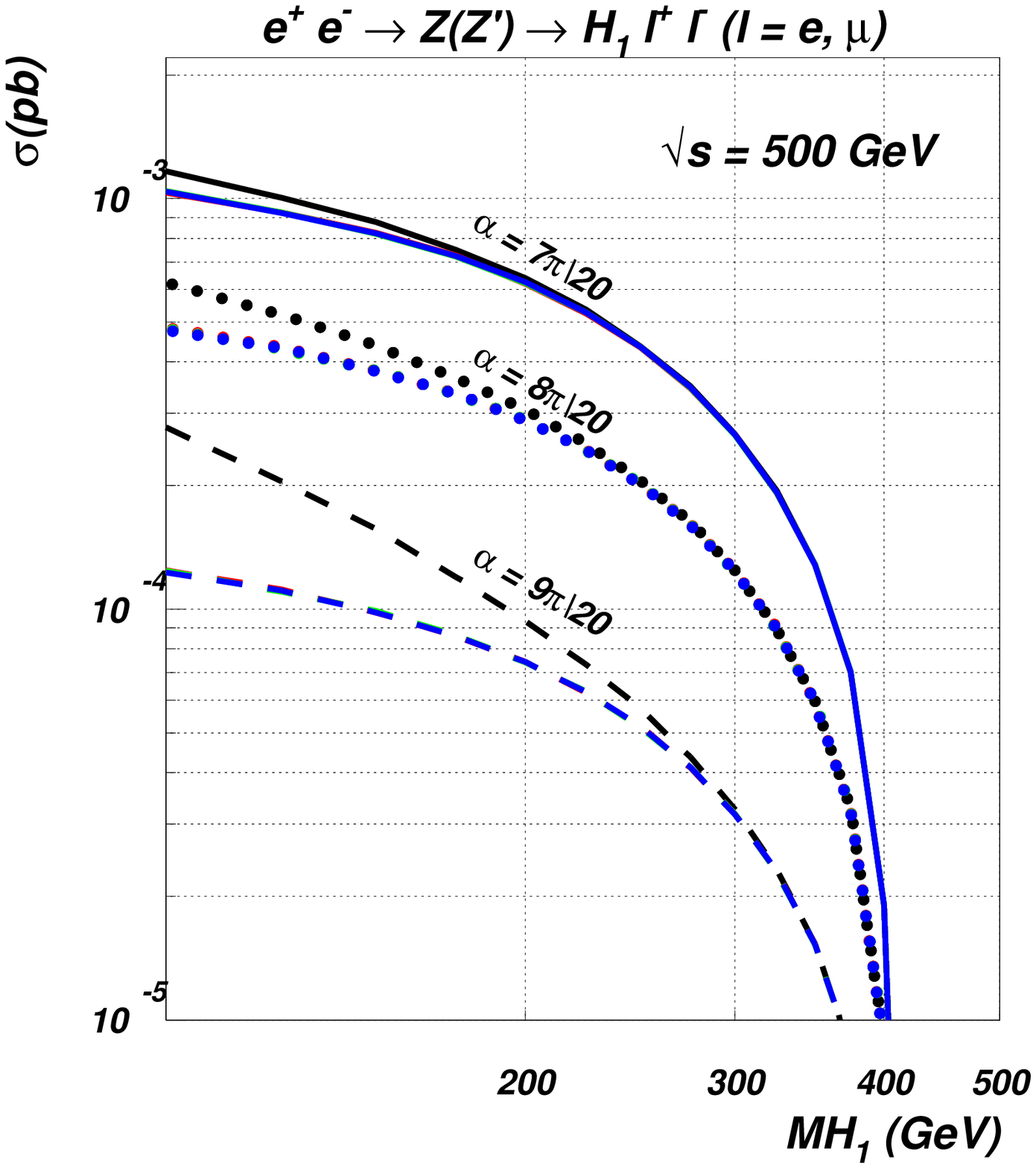}} 
  \subfigure[]{
  \label{LC_intH2-1000}
  \includegraphics[angle=0,width=0.48\textwidth
  ]{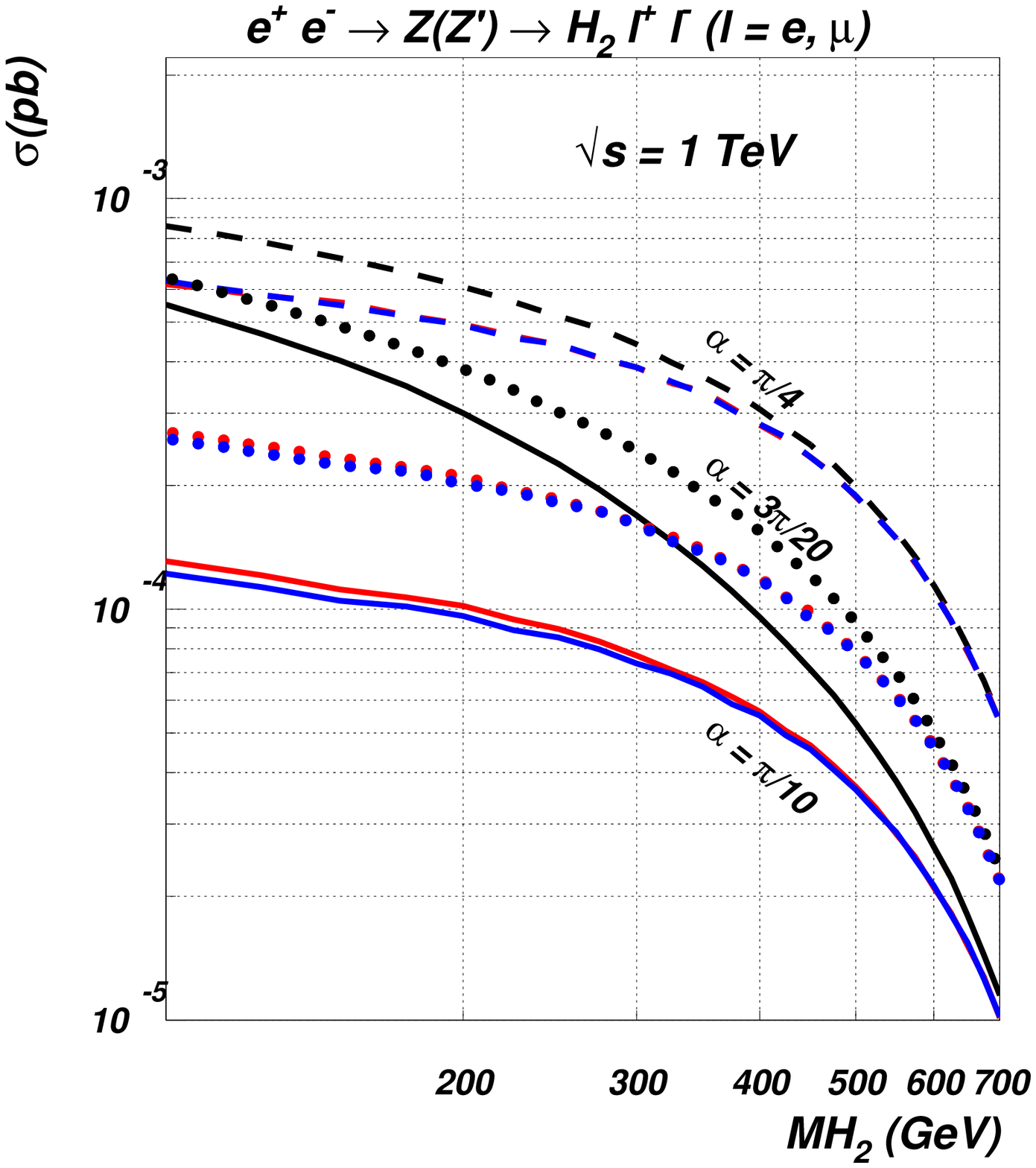}} 
  \caption[Higgs at future linear colliders - Non-standard
    single-Higgs production mechanisms (4)]{Cross section for $e^+e^-
    \rightarrow Z(Z')^\ast 
    \rightarrow h_1 \ell^+ \ell^-$ $(\ell
    =e,\mu$). (\ref{LC_intH1-500}) Black line is for $M_{Z'}=420$ GeV,
    other lines for $M_{Z'}=700,\, 1500,\, 7000$ GeV; $\sqrt{s}=500$
    GeV. (\ref{LC_intH2-1000}) Black line is for $M_{Z'}=1050$ GeV,
    other lines for $M_{Z'}=1400,\, 3500$ GeV; $\sqrt{s}=1$
    TeV.  \label{ILC_Interf}} 
\end{figure}

Finally, the interference between the $Z$ and the $Z'$
could play an important role in the scalar sector, besides the
$t$-quark Yukawa coupling measure presented above. As well known, and
remarked upon in \cite{Basso:2009hf}, the negative
interference between the neutral gauge bosons can be substantial. One
could then look for information about a further neutral vector boson
also by looking at the interference when a Higgs boson is radiated
from the $Z$ bosons. To highlight this effect, in this model it
is possible to select just the leptonic decay modes of the vector
bosons, reducing the predominance of the $Z$ boson. Nonetheless,
as shown in figure~\ref{ILC_Interf}, such effects are minimal when the
$Z'$ boson mass is above the $CM$ energy of the LC.

Other subleading processes for Higgs boson production consist of
single production in association with two vector bosons and of double
Higgs boson production. In these contexts, many new mechanisms arise
in this model, especially exploiting the $Z'$ boson and because a
resonant $h_2 \rightarrow h_1 h_1$ process is allowed.

\subsection{Single scalar production in association with a pair of
  vector bosons}\label{subs:4-4-4}

\begin{figure}[!t]
\centering
  \subfigure[]{
  \label{LC_h1VV_500}
  \includegraphics[angle=0,width=0.48\textwidth ]{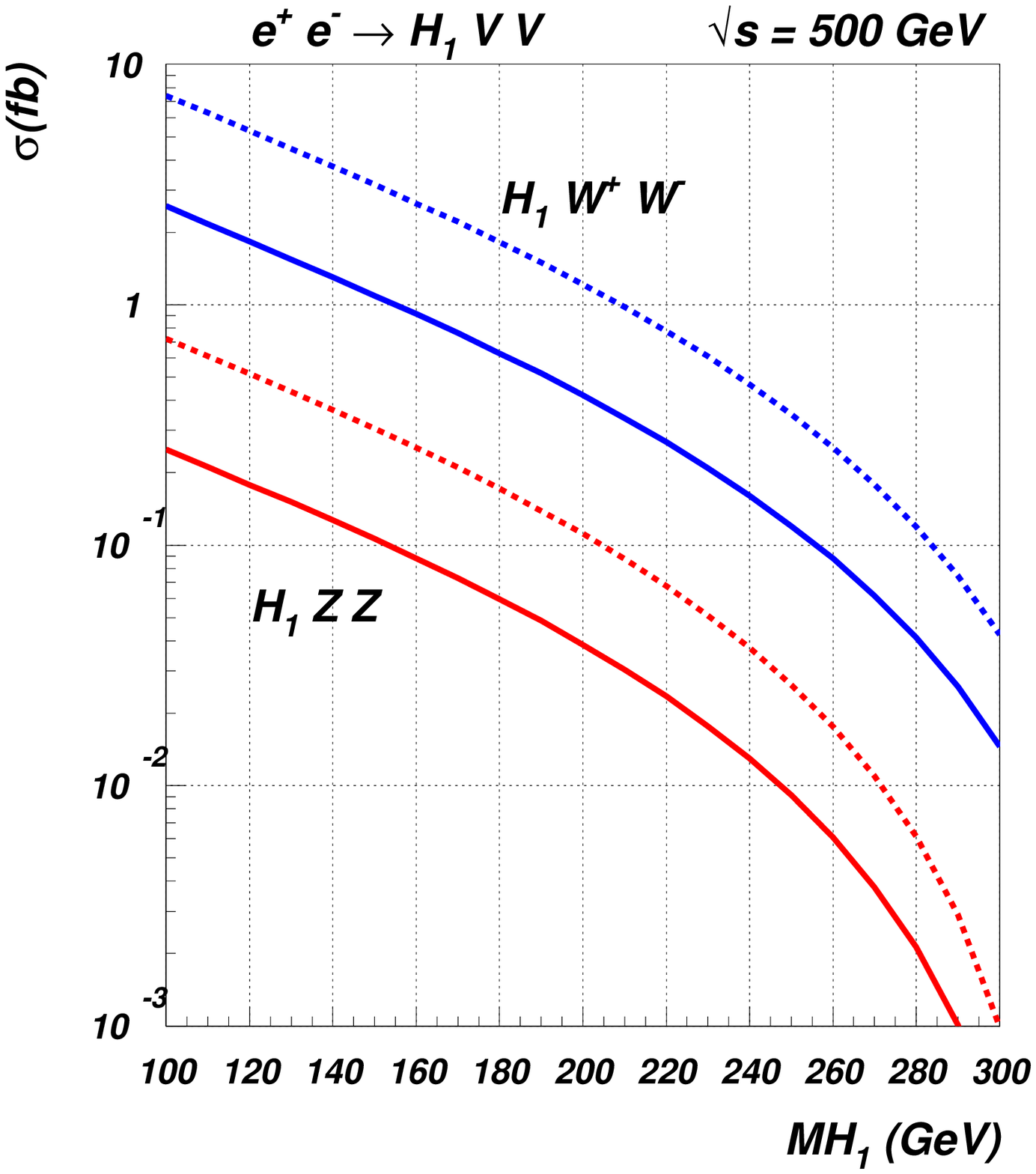}}
\\
  \subfigure[]{
  \label{LC_h1VV_1}
  \includegraphics[angle=0,width=0.48\textwidth ]{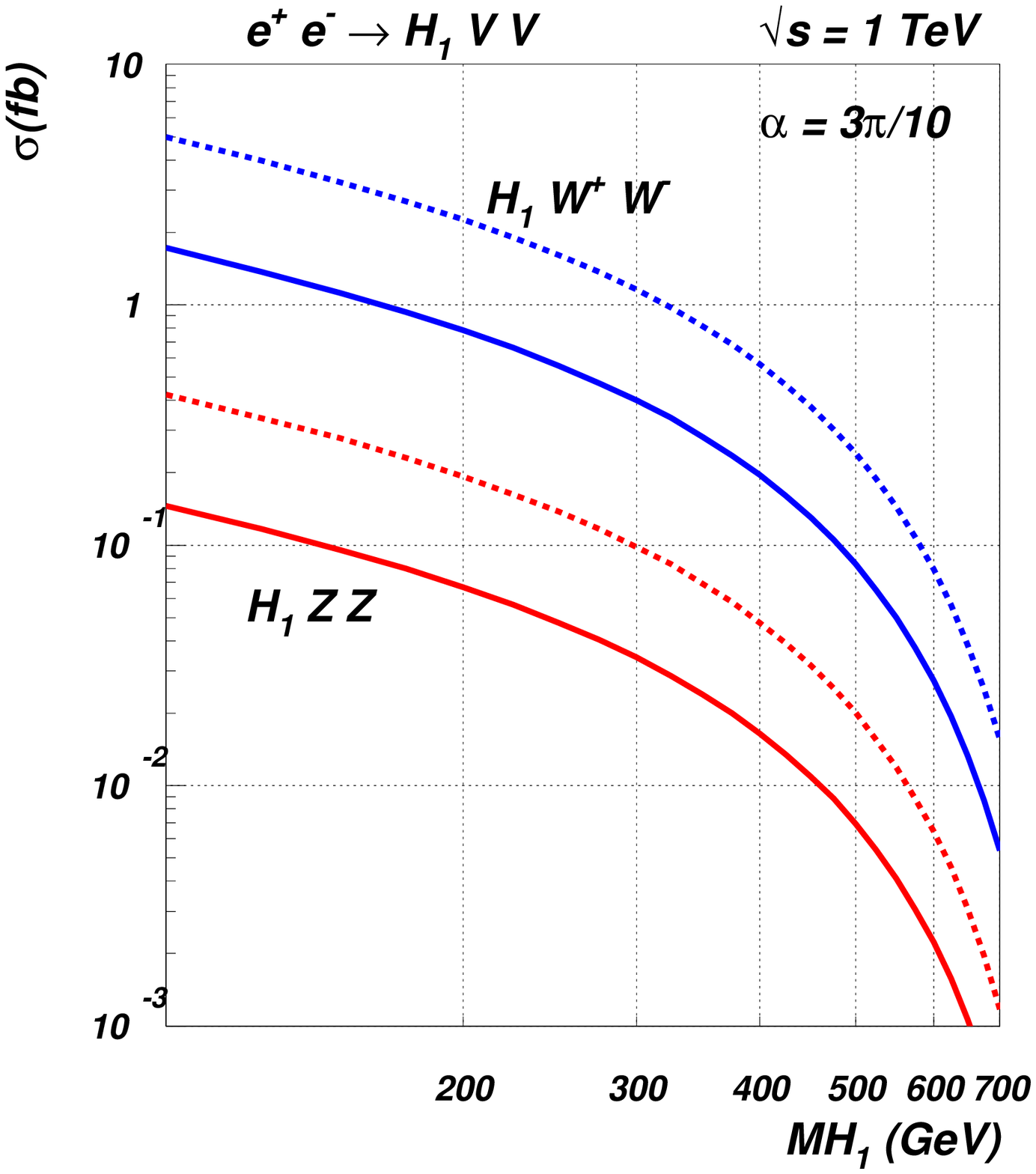}}
  \subfigure[]{
  \label{LC_h1VV_3}
  \includegraphics[angle=0,width=0.48\textwidth ]{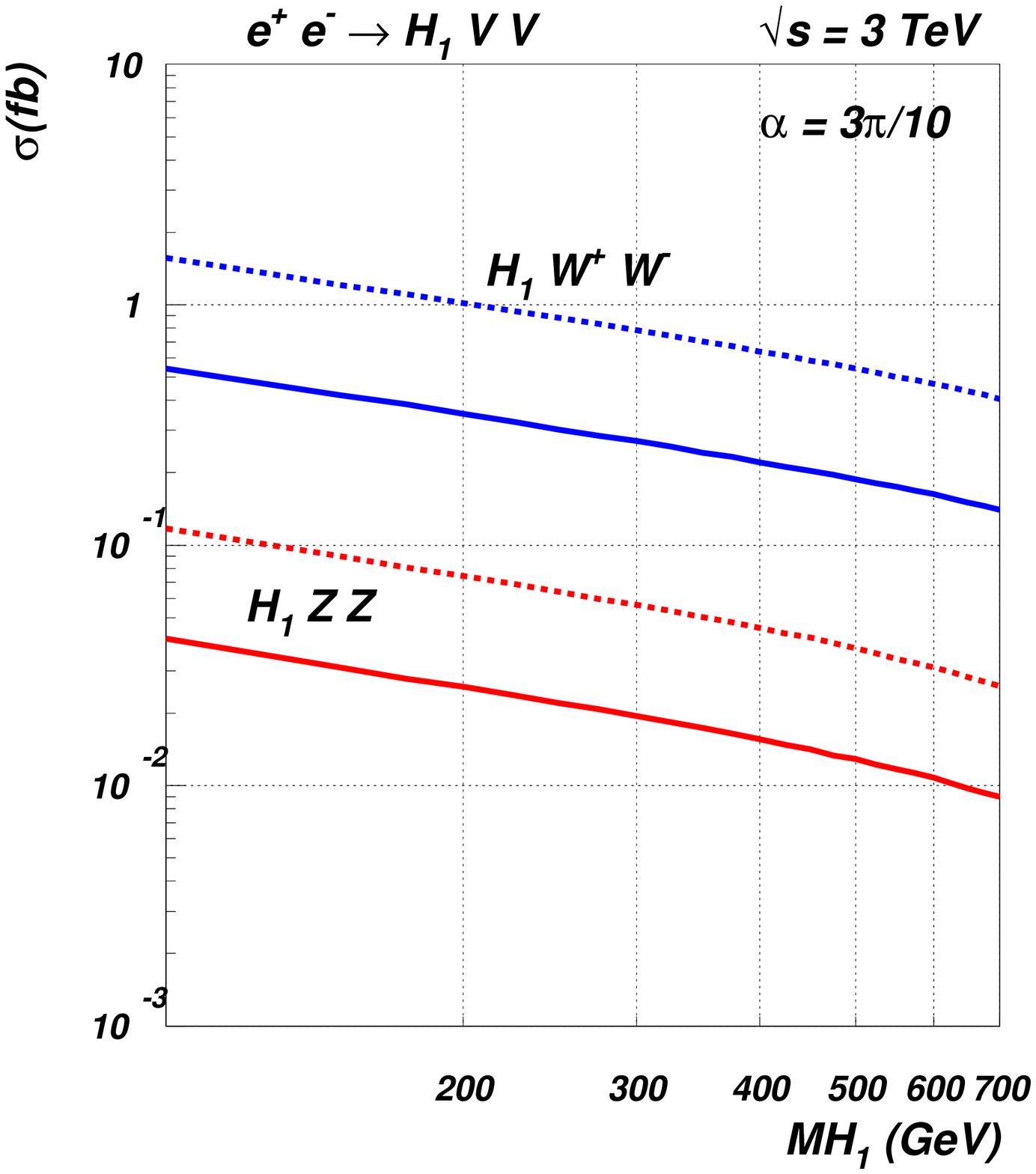}}
  \caption[Higgs at future linear colliders - Single scalar production
    in association to a pair of vector bosons (1)]{Cross sections for
    the light Higgs boson production
    with two vector bosons ($V=W^\pm,Z$) as a function of the mass at
    the LC for $\alpha = 3\pi/10$ (\ref{LC_h1VV_500}) for $h_1$ at
    $\sqrt{s}=500$ GeV, (\ref{LC_h1VV_1}) for $h_1$ at $\sqrt{s}=1$
    TeV and (\ref{LC_h1VV_3}) for $h_1$ at $\sqrt{s}=3$ TeV. The
    dashed lines refer to $\alpha =0$.
\label{LC-HVV}}
\end{figure}

The $SM$ gauge boson pair production has really large cross sections
at the LC, therefore the radiation of a Higgs boson could still have
observable rates. Also, once the Higgs boson has been seen in the main
production mechanisms of Section~\ref{subs:4-4-2}, these
subprocesses could be useful to test the quartic coupling to the $SM$
gauge bosons. Figure~\ref{LC-HVV} shows the case for $\sqrt{s}=0.5,1$
and $3$ TeV $CM$ energies. We neglect here final state photons: although
the cross section of channels comprising the photon could be
comparable to the $WW$ subchannel (see, e.g.,
\cite{Djouadi:2005gi}), the absence of a direct coupling to
the scalar bosons ensures that these channels would not provide
further informations than the other processes that have been
considered.

We then see that the $WW$ channel is roughly an order of magnitude
higher than the $ZZ$ one and that, for low Higgs boson masses, the
cross sections decrease as we increase the $CM$ energy. However, a
larger $CM$ energy allows the production of more massive scalars and to
avoid kinematic limitations. So that, if $\sqrt{s}=1$ TeV is
preferable to test these mechanisms for Higgs boson masses between
$100$ and $300$ GeV (with comparable or higher cross sections to the
case for $\sqrt{s}=500$ GeV), the $\sqrt{s}=3$ TeV configuration is
essential for masses above $300$ GeV, for both the light and heavy
Higgs boson. Unless very high (and disfavoured) values of the mixing
angle, the $h_1WW$ channel has cross section above $0.1$ fb for the
whole range of scalar masses considered. The opposite is true for the
heavy 
Higgs boson, for which only big values of $\alpha$ allow this channel
to be above $0.1$ fb. The case for the $ZZ$ channel is different:
since its cross sections are rather small, it has chances of being
detected, staying above $0.1$ fb, only for $\sqrt{s}=500$ GeV and
$\sqrt{s}=1$ TeV. For $\sqrt{s}=3$ TeV its observation requires very
high statistics.

\begin{figure}[!t]
  \subfigure[]{ 
  \label{LC_ZZpH1_3}
  \includegraphics[angle=0,width=0.48\textwidth ]{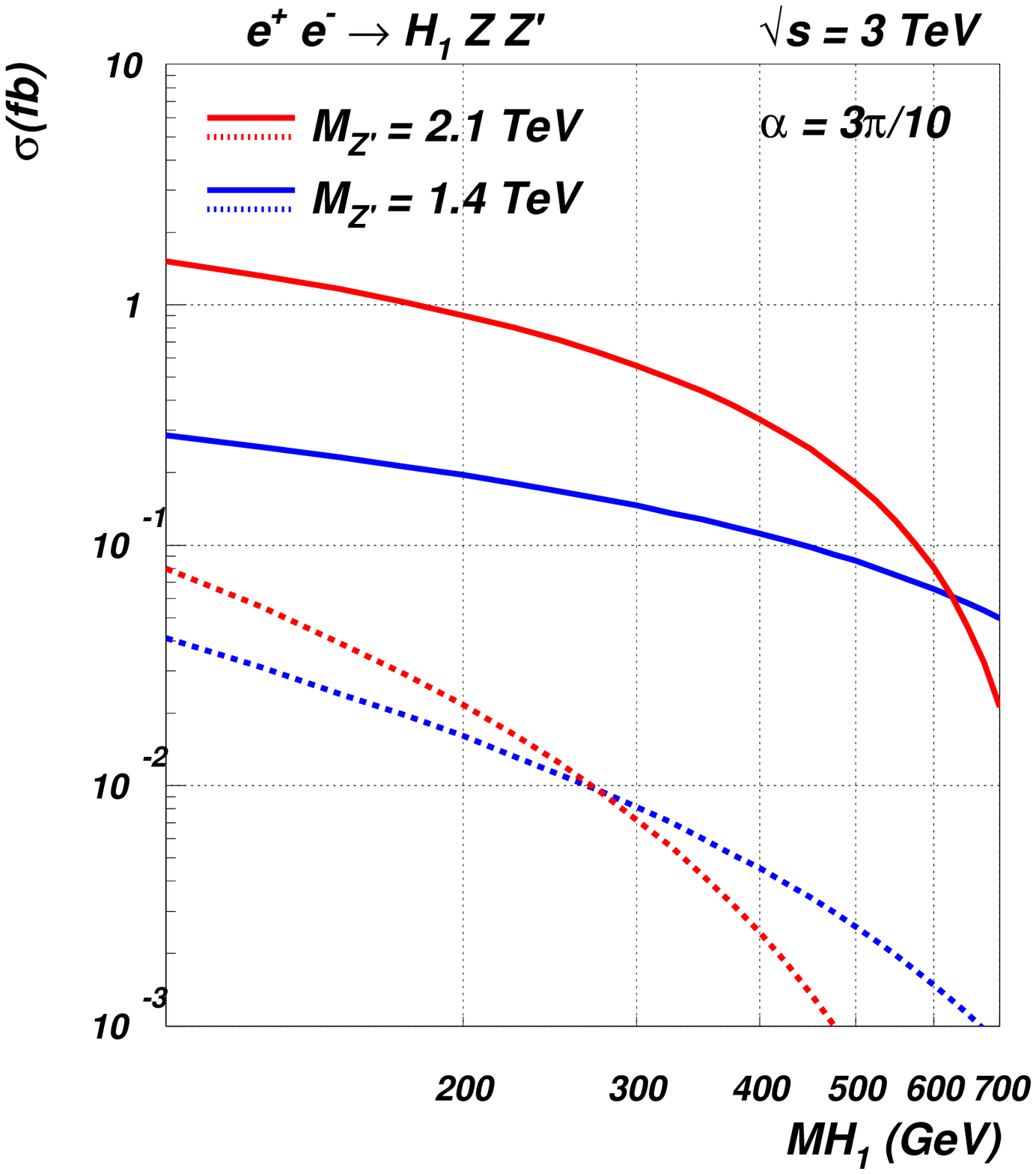}}
  \subfigure[]{
  \label{LC_ZZpH2_3}
  \includegraphics[angle=0,width=0.48\textwidth ]{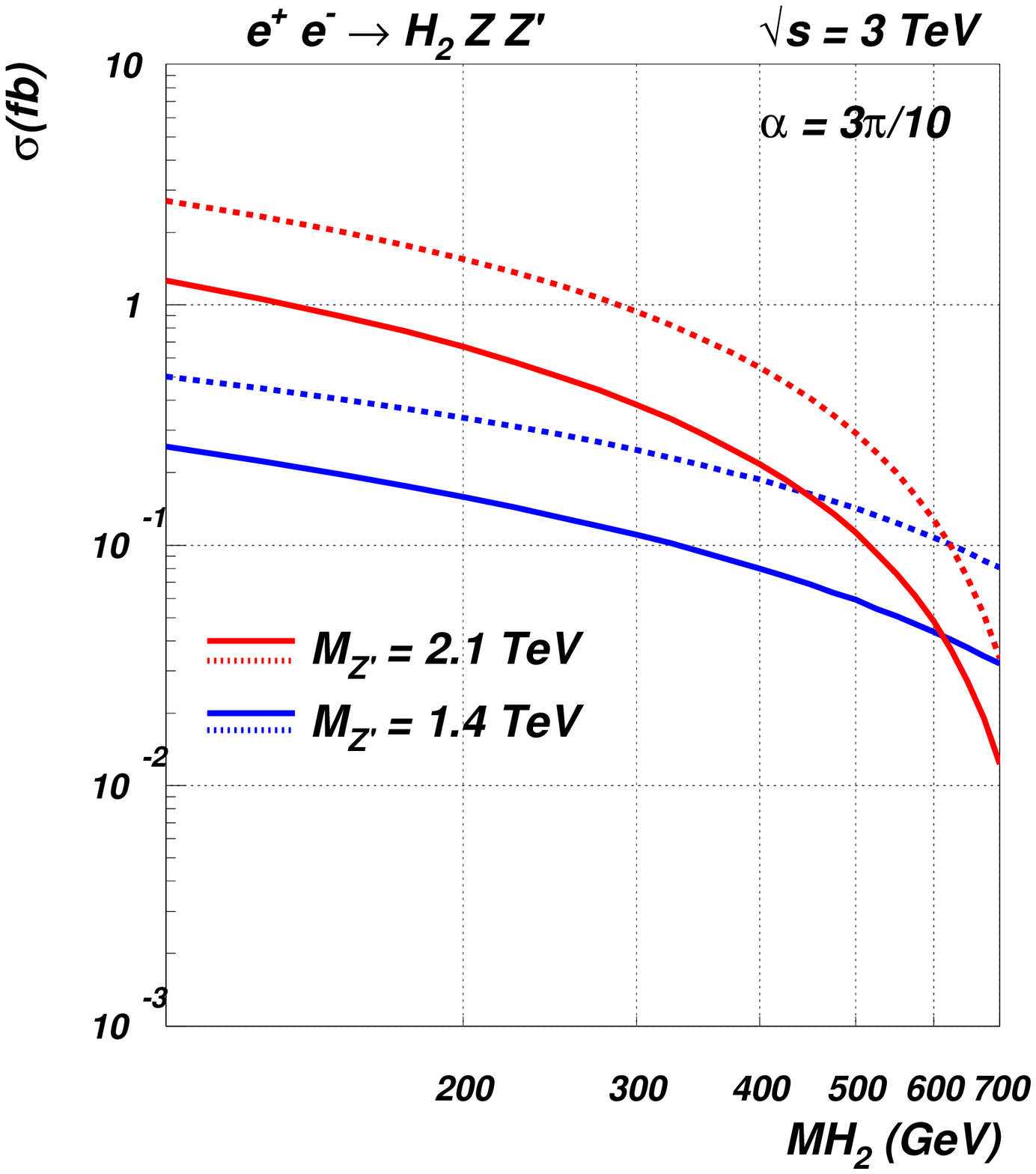}}
  \caption[Higgs at future linear colliders - Single scalar production
    in association to a pair of vector bosons (2)]{Cross sections for
    the Higgs boson production with a
    $Z$ and a $Z'$ boson as a function of the scalar mass at the LC
    (\ref{LC_ZZpH1_3}) for $h_1$ and (\ref{LC_ZZpH2_3}) for $h_2$, at
    $\sqrt{s}=3$~TeV. The dashed lines refer to $\alpha = 0$. 
  \label{LC_HZZp}} 
\end{figure}

The cross sections for the case of one $Z'$ boson in the final state
could be important and comparable to the $WW$ channel. Also,
this particular channel is useful to test the absence of a tree-level
$h-Z-Z'$ coupling. Figure~\ref{LC_HZZp} shows the cross sections for
$\sqrt{s}=3$~TeV for two values of the $Z'$ mass, $M_{Z'}=1.4$ and
$2.1$ TeV, and suitable $g'_1$ coupling. The heavier the $Z'$ boson,
the higher the cross sections, until kinematical limitations occur. In
fact, the cross sections for $M_{Z'}=2.1$~TeV are always above those
for $M_{Z'}=1.4$~TeV for scalar masses below $600$~GeV, for which the
process with the lighter $Z'$ boson overtakes. It is important to note
that the behaviour of these processes with the scalar mixing angle is
opposite to the previous case. Hence, for $h_1$ and for small values
of the angle, the associated production with a pair of $SM$ gauge
bosons is favoured, while the process with the $Z'$ boson is favoured
for big angles. For $h_2$ it is again the opposite. For intermediate
angles, instead, both processes can have small but observable rates,
between $0.1$ and $\mathcal{O}(1)$ fb, for both Higgs bosons and in
the whole range of considered masses.

Finally, notice that the case for $\alpha=0$ represents the radiative
correction to the $Z$($Z'$) strahlung of $h_1$($h_2$) for a $Z'$($Z$)
boson emission.

\subsection{Double scalar production mechanisms}\label{subs:4-4-5}

The observation of a Higgs boson pair is crucial to measure parameters
of the scalar Lagrangian entering directly in the trilinear and
quartic self-couplings
\cite{Castanier:2001sf,Baur:2002rb,Baur:2009uw}, although it requires
high statistics and large $CM$ energy. Remarkable in this sense is the
possible complementarity between the LHC and LCs, as shown in
\cite{Baur:2003gpa,Plehn:2005nk}.

In the minimal $B-L$ model, the $h_2 \rightarrow h_1 h_1$ process is
also present for a large portion of the parameter space, contrary to
the $MSSM$ case, for instance (where it is important only for very low
values of $\tan{\beta}$, region that has been constrained at LEP
\cite{Schael:2006cr}). On the one side, light Higgs boson pair
production is enhanced by this channel, especially when it is
resonant. On the other side, also the $h_2-h_1-h_1$ coupling is
directly testable. Moreover, the $Z'$ boson can give further scope to
this, providing an extra mechanism for Higgs pair production, both
without and through heavy Higgs boson production.

\begin{figure}[!t]
  \subfigure[]{ 
  \label{LC_H1H1_1}
  \includegraphics[angle=0,width=0.48\textwidth ]{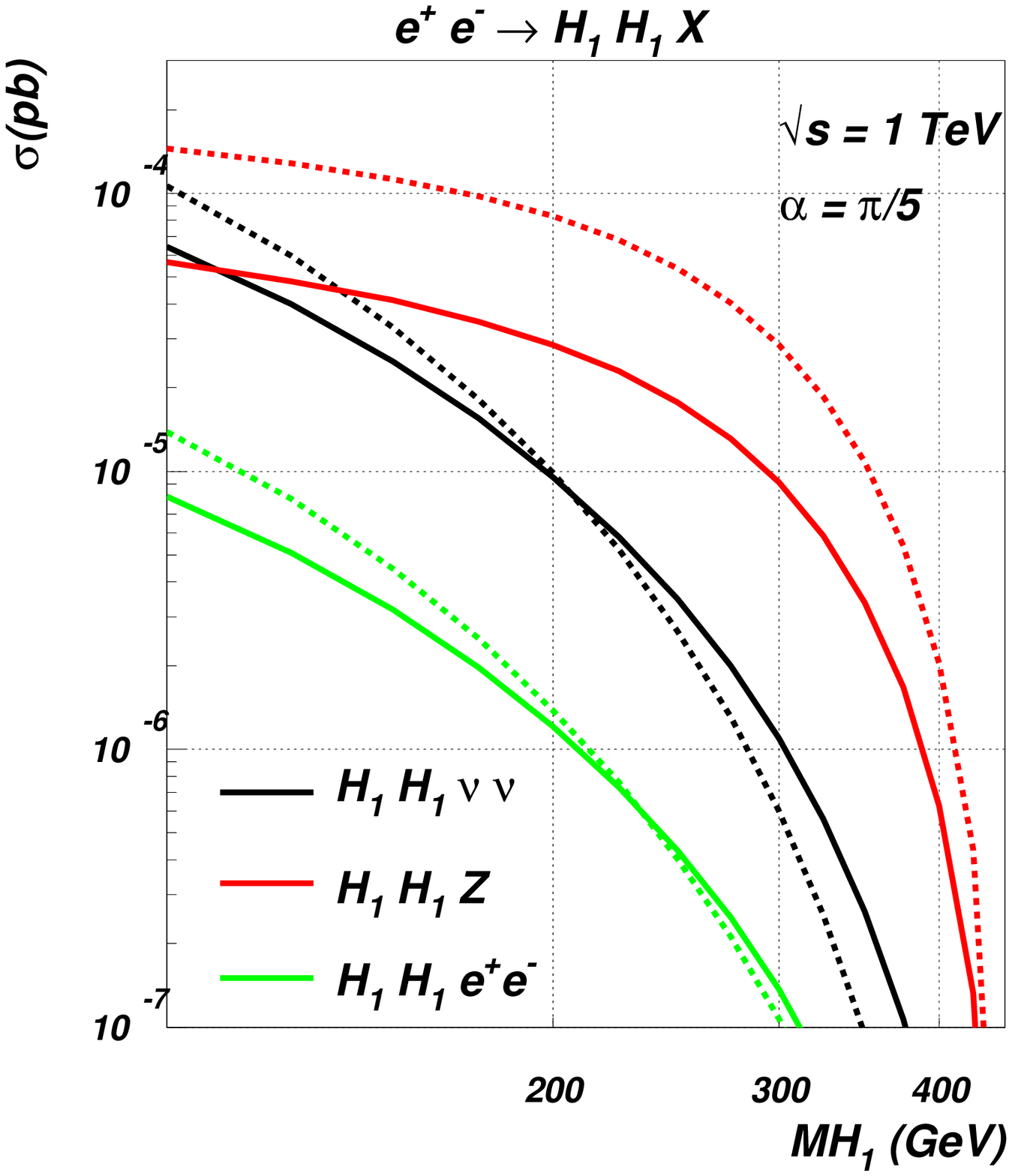}}
  \subfigure[]{
  \label{LC_H2-H1H1_1}
  \includegraphics[angle=0,width=0.48\textwidth ]{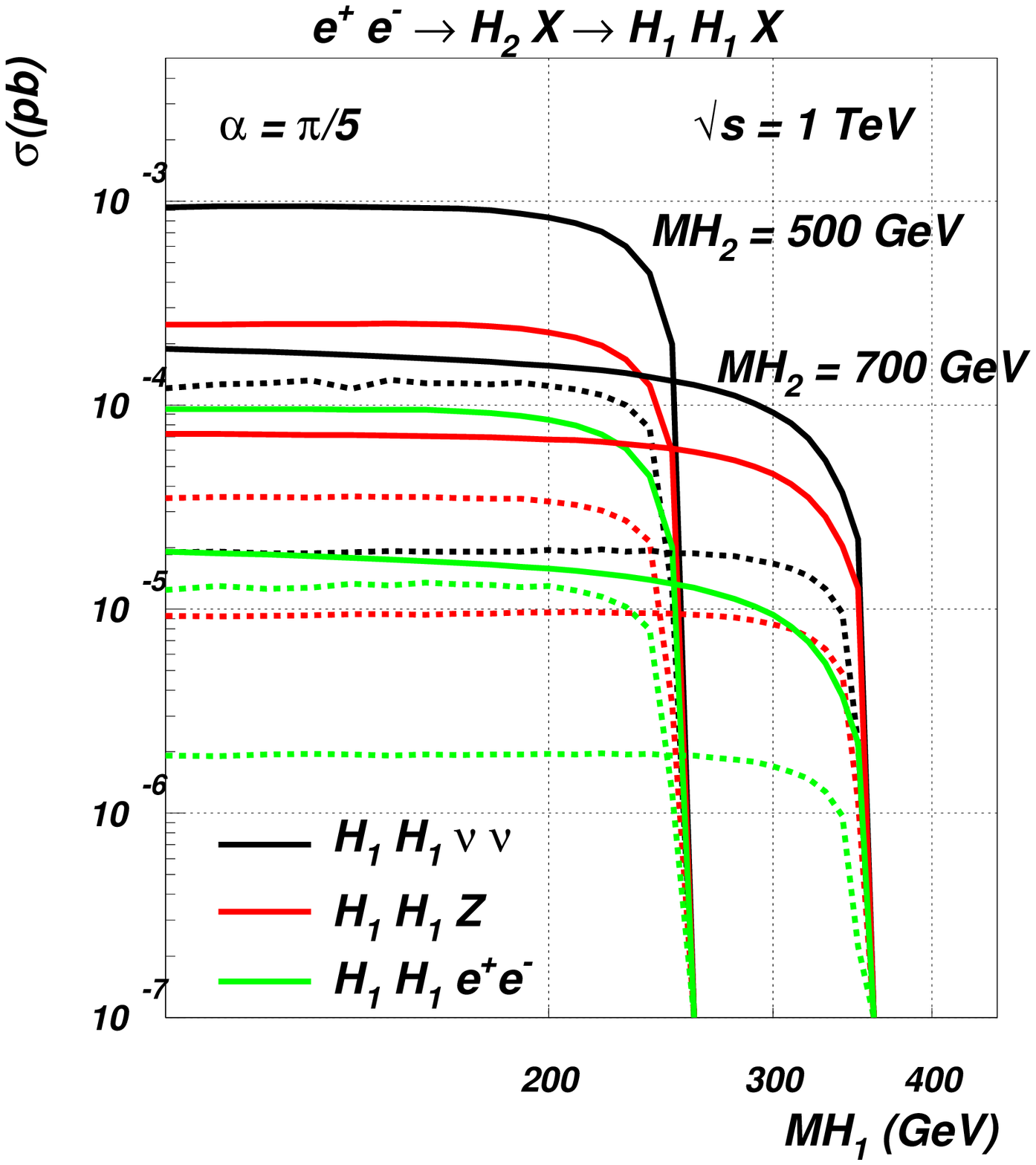}}\\
  \subfigure[]{ 
  \label{LC_H1H1_3}
  \includegraphics[angle=0,width=0.48\textwidth ]{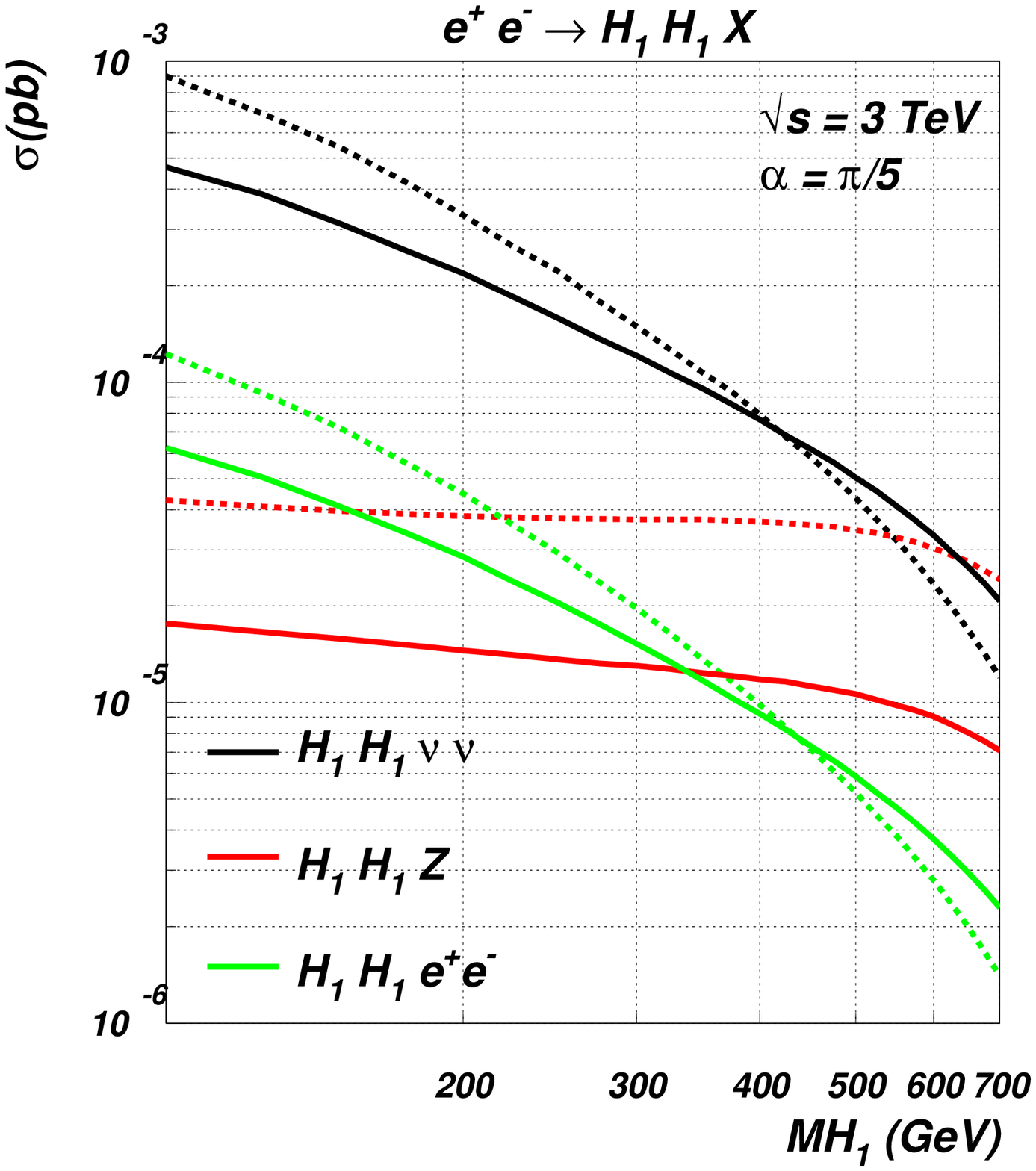}}
  \subfigure[]{
  \label{LC_H2-H1H1_3}
  \includegraphics[angle=0,width=0.48\textwidth ]{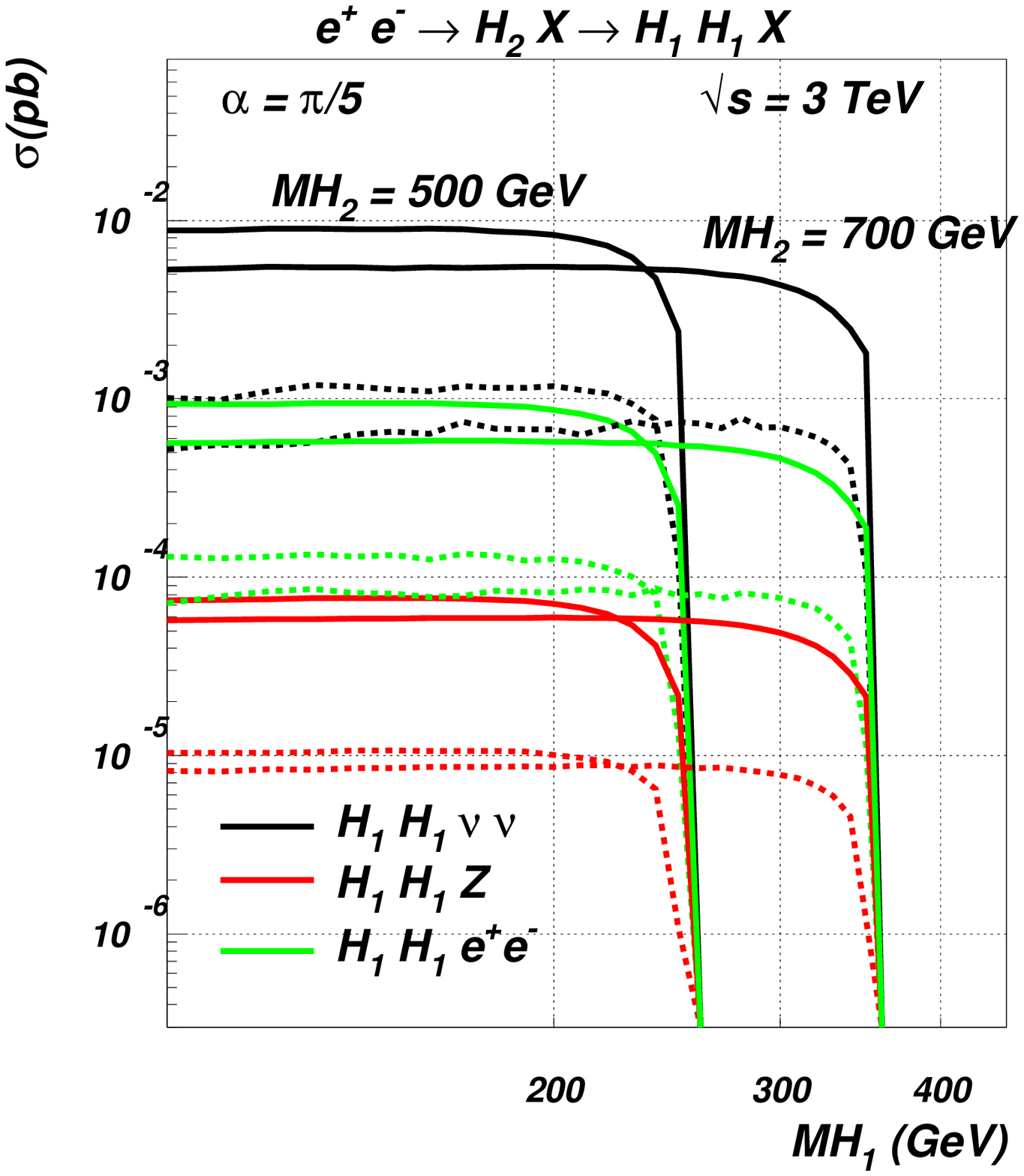}}
  \caption[Higgs at future linear colliders - Double scalar production
    mechanisms (1)]{Cross sections for the double light higgs boson
    production at the LC (\ref{LC_H1H1_1}) alone and
    (\ref{LC_H2-H1H1_1}) via $h_2$, for $\sqrt{s}=1$ TeV and
    (\ref{LC_H1H1_3}) alone and (\ref{LC_H2-H1H1_3}) via $h_2$, for
    $\sqrt{s}=3$ TeV. The dashed lines in figures~\ref{LC_H1H1_1}
    and~\ref{LC_H1H1_3} refer to $\alpha =0$, while in
    figures~\ref{LC_H2-H1H1_1} and~\ref{LC_H2-H1H1_3} refer to
    $\alpha =\pi/20$.
\label{ILC_Standard_double}} 
\end{figure}

Figure~\ref{ILC_Standard_double} shows the standard production
mechanisms of a pair of light Higgs bosons. In the $SM$ case (or when
we neglect $h_2$), they are the same mechanisms discussed in
Subsection~\ref{subs:4-4-2} when a further Higgs boson is
attached. Cross sections for these processes are always below $0.1(1)$
fb at $\sqrt{s}=1(3)$ TeV, and above $0.1$ fb at $\sqrt{s}=3$ TeV only
for the $W$-fusion process and for $M_{h_1} \lesssim 350$ GeV, as
clear from figures~\ref{LC_H1H1_1} and~\ref{LC_H1H1_3}, respectively.

When instead the light Higgs boson pair is originated by the decay of
the heavy Higgs boson, cross sections can be of
$\mathcal{O}(1)-\mathcal{O}(10)$ fb at $\sqrt{s}=1-3$ TeV, in the
$W$-fusion channel (at most, when the mixing is maximal, i.e., $\alpha
\approx \pi/4$, and we chose $M_{h_2} = 500$ GeV.). If we choose a
higher value for $M_{h_2}$, a more massive $h_1$ boson can be pair
produced, but with a sensibly lower cross sections: for $M_{h_2}=700$
GeV, they are roughly a factor $5(2)$ smaller than for $M_{h_2} = 500$
GeV at $\sqrt{s}=1(3)$ TeV. Notice that the cross sections are
constant with $M_{h_1}$ as long as the $h_2 \rightarrow h_1 h_1$ decay
is allowed. This is a consequence of having chosen a specific value
for $M_{h_2}$ and that $BR(h_2 \rightarrow h_1 h_1) \approx 20\% $ is
approximately constant for $M_{h_1} > M_W,\, M_Z$
\cite{Basso:2010yz}. Other channels give smaller cross sections: the
$Z$-fusion channel is always an order of magnitude below the
$W$-fusion one, while the strahlung from the $Z$ boson channel
gives at most $\sim 0.2 (0.08)$ fb for $M_{h_2} = 500$ GeV and $\sim
0.08(0.06)$ fb for $M_{h_2} = 700$ GeV at $\sqrt{s}=1(3)$ TeV,
respectively.

%
%
\begin{figure}[!t]
\centering
  \subfigure[]{
  \label{H2-H1H1Zp_1}
  \includegraphics[angle=0,width=0.48\textwidth
  ]{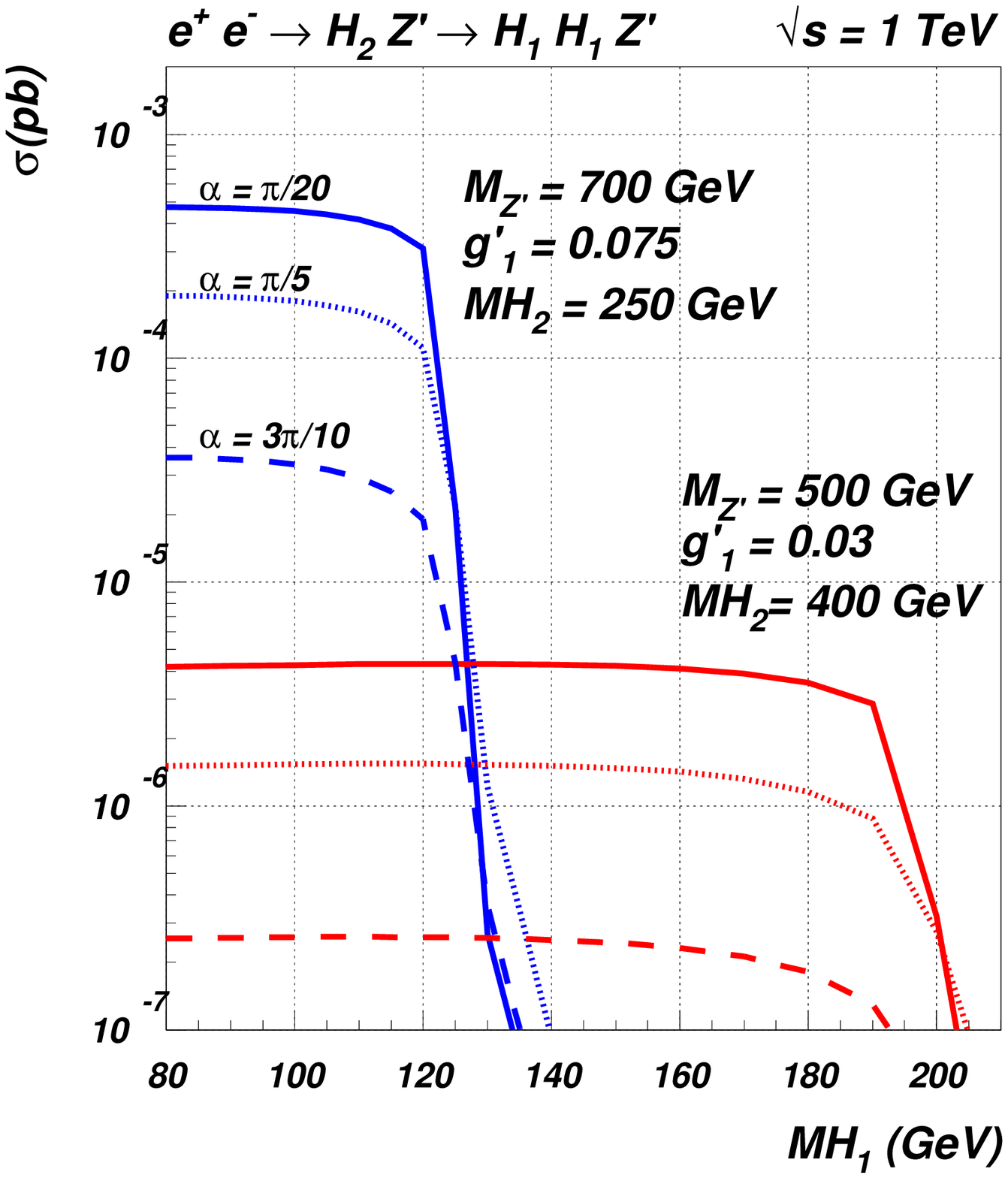}}\\ 
  \subfigure[]{ 
  \label{H1H1Zp_3}
  \includegraphics[angle=0,width=0.48\textwidth
  ]{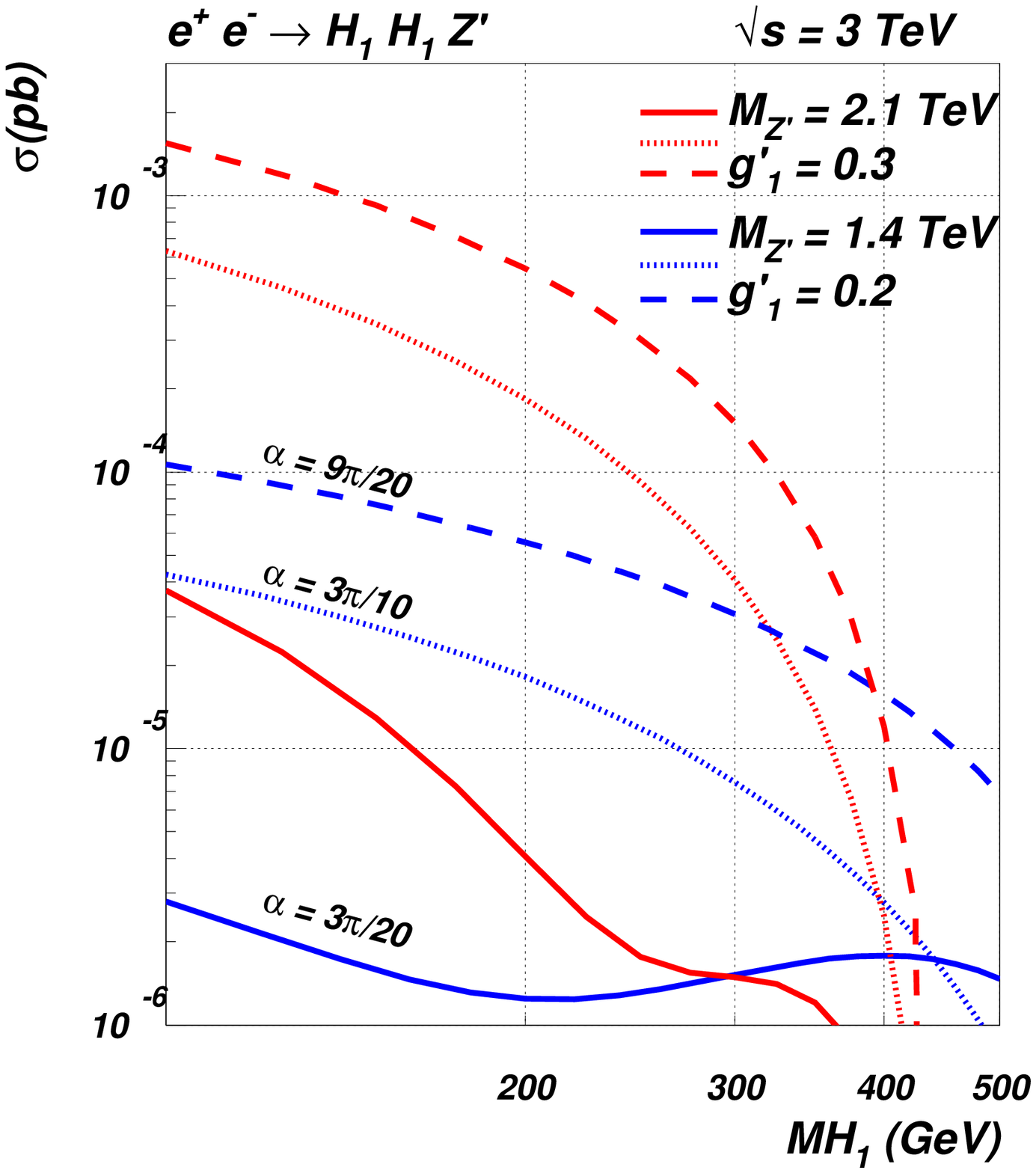}} 
  \subfigure[]{
  \label{H2-H1H1Zp_3}
  \includegraphics[angle=0,width=0.48\textwidth
  ]{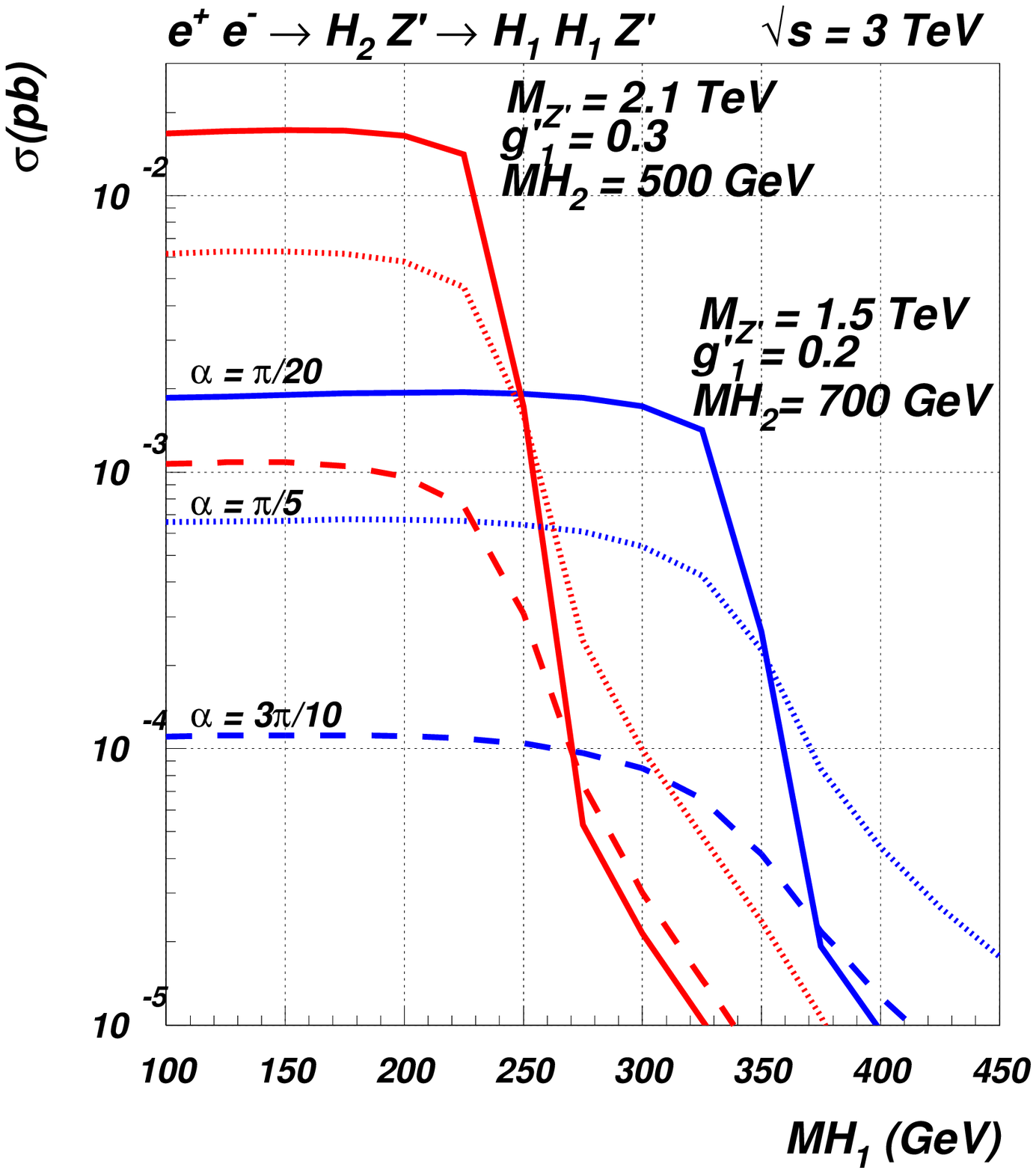}}   
  \caption[Higgs at future linear colliders - Double scalar production
    mechanisms (2)]{Cross sections for the process $e^+e^-\rightarrow H_2
    Z' \rightarrow H_1 H_1 Z'$ (\ref{H2-H1H1Zp_1})  for $\sqrt{s}=1$
    TeV and (\ref{H2-H1H1Zp_3}) for $\sqrt{s}=3$ TeV, for suitable
    values of $M_{H_2}$ and for the process $e^+e^-\rightarrow H_1 H_1
    Z'$ (\ref{H1H1Zp_3}) at $\sqrt{s}=3$ TeV, several values of the
    angle and of $M_{Z'}$.
  \label{H1H1Zp_H2}}
\end{figure}

As anticipated, the $Z'$ boson in the minimal $B-L$ model can give
further 
scope to produce also a pair of light Higgs bosons at a LC, both
directly (without or through $h_2$) and indirectly (pair producing
heavy neutrinos). 

Figure~\ref{H1H1Zp_H2} shows double Higgs strahlung from the $Z'$
boson (for $\sqrt{s}=3$~TeV only) and the case in which $h_2$ is
radiated from the $Z'$ boson and it subsequently decays into a light
Higgs boson pair. The double Higgs-strahlung from the $Z'$ boson at
$\sqrt{s}=1$ TeV has negligible cross sections, below $10^{-3}$ fb,
especially because of kinematic limitations, and therefore we neglect
it here. The cross sections for double Higgs-strahlung at
$\sqrt{s}=3$~TeV are presented in figure~\ref{H1H1Zp_3}, where we see
that, for $M_{Z'}=2.1$~TeV (and $g'_1=0.3$), a pair of light Higgs
bosons can be produced with cross section $\gtrsim 0.1$~fb for
$m_{h_1} \lesssim 300$ GeV and for big values of the scalar mixing
angle (roughly bigger than $\pi/4$). The situation improves
if we consider the Higgs-strahlung of $h_2$ from the $Z'$ boson and its
subsequent decay into $h_1$ pairs. Notice that this channel
reduces\footnote{This is true when both the $Z'$ boson and the heavy
  Higgs boson are on-shell. When $h_2$ is an off-shell intermediate
  state, the cross sections for light Higgs pair production via
  $h_2$ increases as we increase the value of the mixing angle.} as we
increase the value of the mixing angle, vanishing in the decoupling
regimes (both for $\alpha \equiv 0$ and $\pi/2$). At $\sqrt{s}=1$~TeV
this process is still limited by the kinematics: the higher the $Z'$
boson mass the higher the cross sections and the smaller the
producible $h_2$ mass. For $M_{Z'}=700$ GeV (and suitable values for
the $g'_1$ coupling), the light Higgs boson can be pair produced
through $h_2$ with cross sections bigger than $0.1$ fb (for $\alpha <
\pi /4$, up to $4$ fb) through a heavy Higgs boson of $250$ GeV, hence
for $h_1$ masses up to $120$ GeV only. To extend the range in
$M_{h_1}$, a higher mass for the heavy Higgs boson has to be
considered, needing a smaller $Z'$ boson mass: the cross sections in
this case become unobservable, below $10^{-2}$ fb. If the collider $CM$
energy is increased though, heavier $h_1$'s can be pair produced
through the heavy Higgs boson, in association with a much heavier $Z'$
boson, with bigger cross sections. Figure~\ref{H2-H1H1Zp_3} shows
that, for $M_{Z'}=2.1$ TeV, a heavy Higgs boson with $500$ GeV mass
can pair produce the light Higgs boson with cross sections well above
the fb level up to $M_{h_1} = 200$ GeV, reaching $\mathcal{O}(10)$ fb
for small (but not negligible) values of the mixing angle (i.e., $\pi
/20<\alpha <\pi/5$). If a $Z'$ boson of $1.5$ TeV mass is considered,
there are no more kinematical limitations for the producible $h_2$
boson and, in the case of $M_{h_2}=700$ GeV, an even heavier $h_1$ can
be pair produced, up to masses of $350$ GeV with cross sections bigger
than $0.1$ fb and $\mathcal{O}(1)$ fb for small (but not negligible)
values of the mixing angle (i.e., for the same values of the previous
case).

%
%
\begin{figure}[!t]
  \subfigure[]{ 
  \label{LC_H1H1-CM-1}
  \includegraphics[angle=0,width=0.48\textwidth
  ]{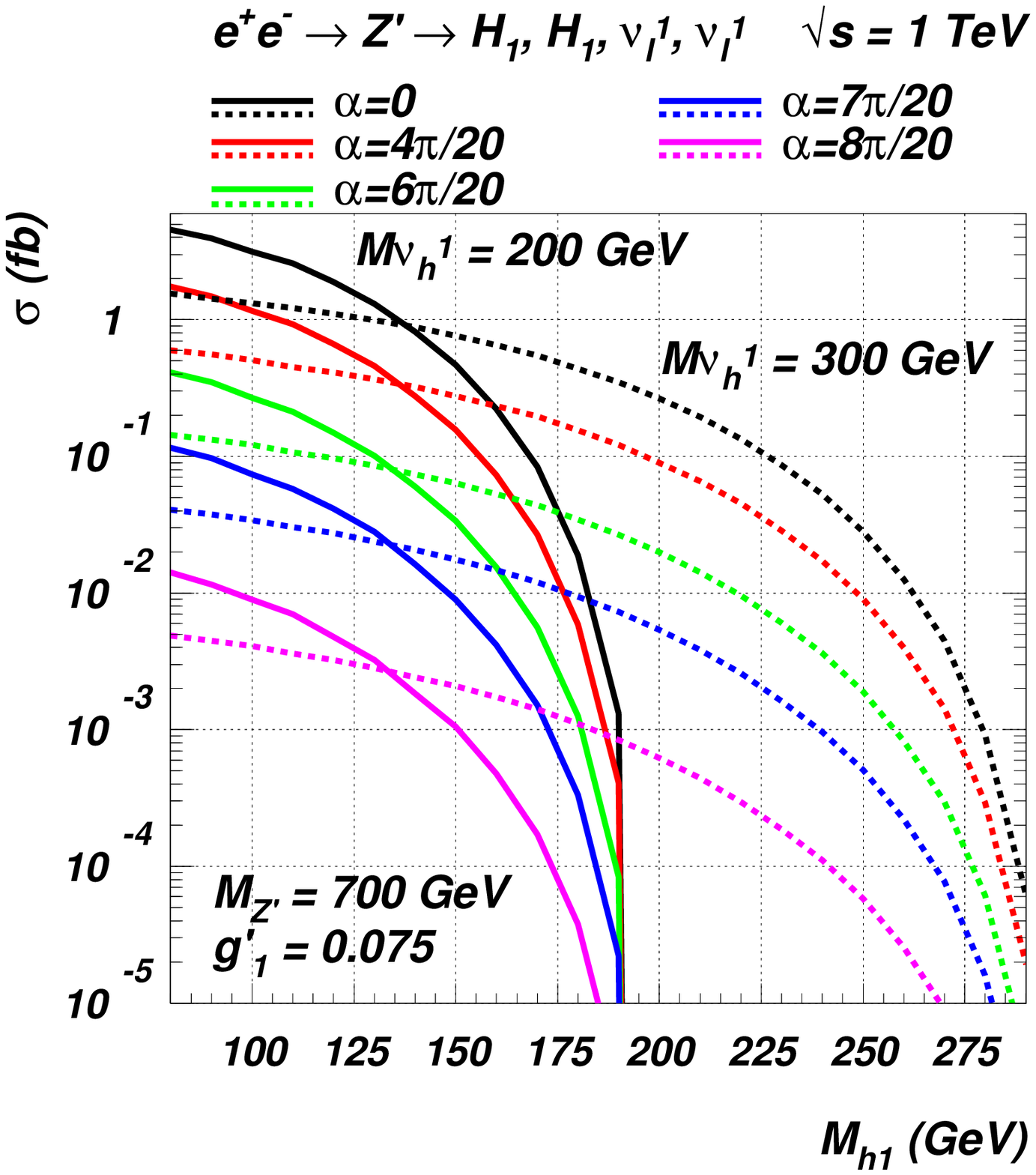}} 
  \subfigure[]{
  \label{LC_H1H1-CM-MZp}
  \includegraphics[angle=0,width=0.48\textwidth
  ]{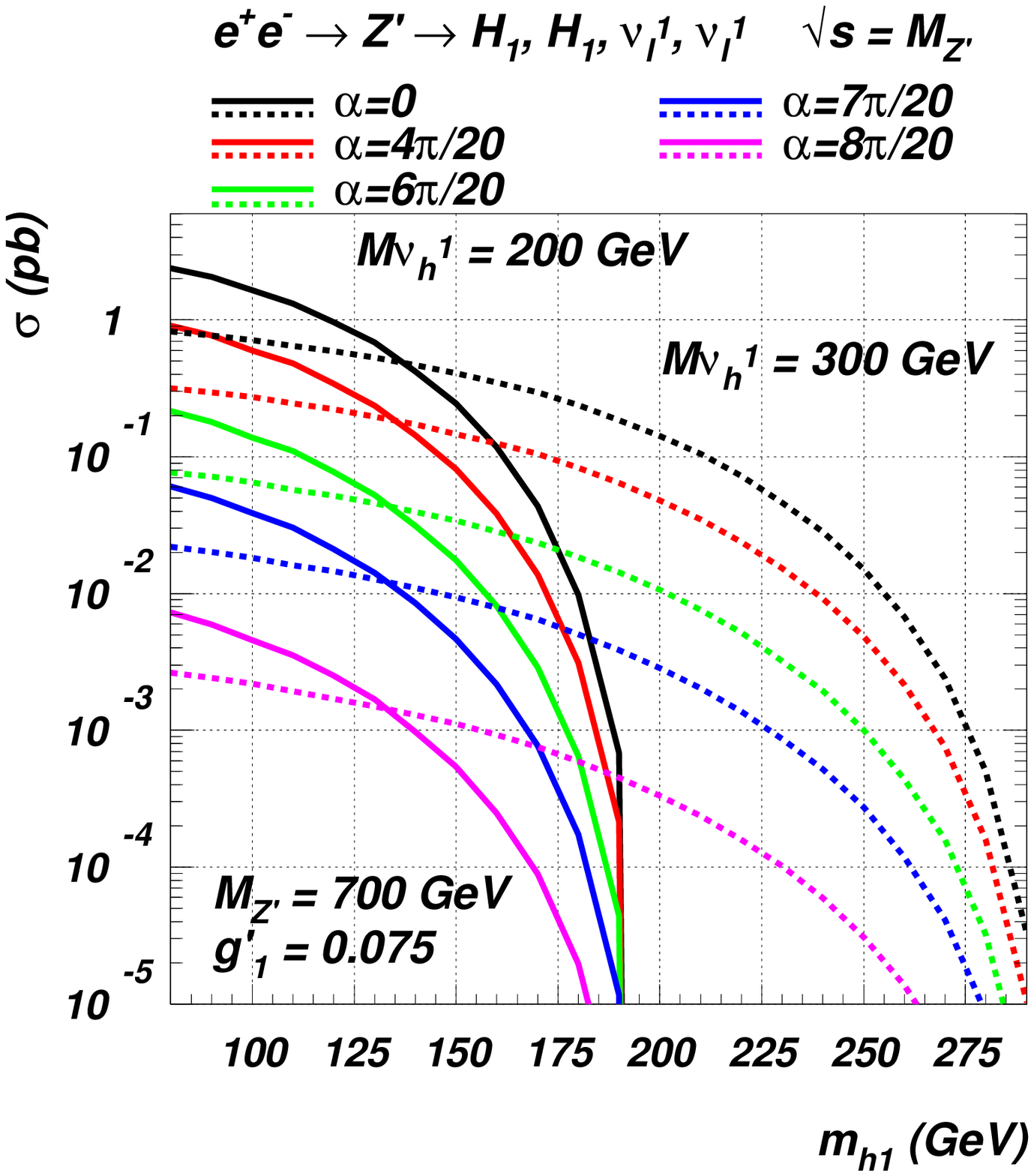}} 
  \caption[Higgs at future linear colliders - Double scalar production
    mechanisms (3)]{Cross sections for the associated production of two
    light higgs bosons and two light first generation neutrinos (via
    $Z'\rightarrow \nu _h \nu _h$) at the LC (\ref{LC_H1H1-CM-1})  for
    $\sqrt{s}=1$ TeV and (\ref{LC_H1H1-CM-MZp})  for $\sqrt{s}\equiv
    M_{Z'}$.}
\label{ILC_double_neutrino}
\end{figure}

The high cross sections of figure~\ref{ILC_neutrino} (and the fact
that $BR(\nu _h \rightarrow h_1 \nu _l) \approx 20\%$) allows one to
consider the case in which both heavy neutrinos decay into a light
Higgs boson each. In figure~\ref{ILC_double_neutrino} we show this
case. Once again, the possibility of tuning the $CM$ energy of the LC
to sit at the $Z'$ boson peak is crucial to test this mechanism. Without
it, the cross sections would be about $0.1 \div 1$~fb for small values
of the scalar mixing angle only: for instance, for $\alpha = 6\pi
/20$, light Higgs boson pair production through heavy neutrino pair
production (via the $Z'$ boson, one generation only) are above $0.1$
fb for $M_{h_1} < 125$ GeV, for both heavy neutrino masses
chosen. When instead the $CM$ is tuned to the $Z'$ boson peak, the cross
sections are enhanced and well above the fb level whatever the value
for the mixing angle, for $h_1$ masses kinematically allowed, reaching
a few pb (or fractions of pb) for small mixing angles and $M_{h_1}$
values.

\begin{figure}[!t]
  \subfigure[]{
  \label{LC_H1H2-300_3}
  \includegraphics[angle=0,width=0.48\textwidth ]{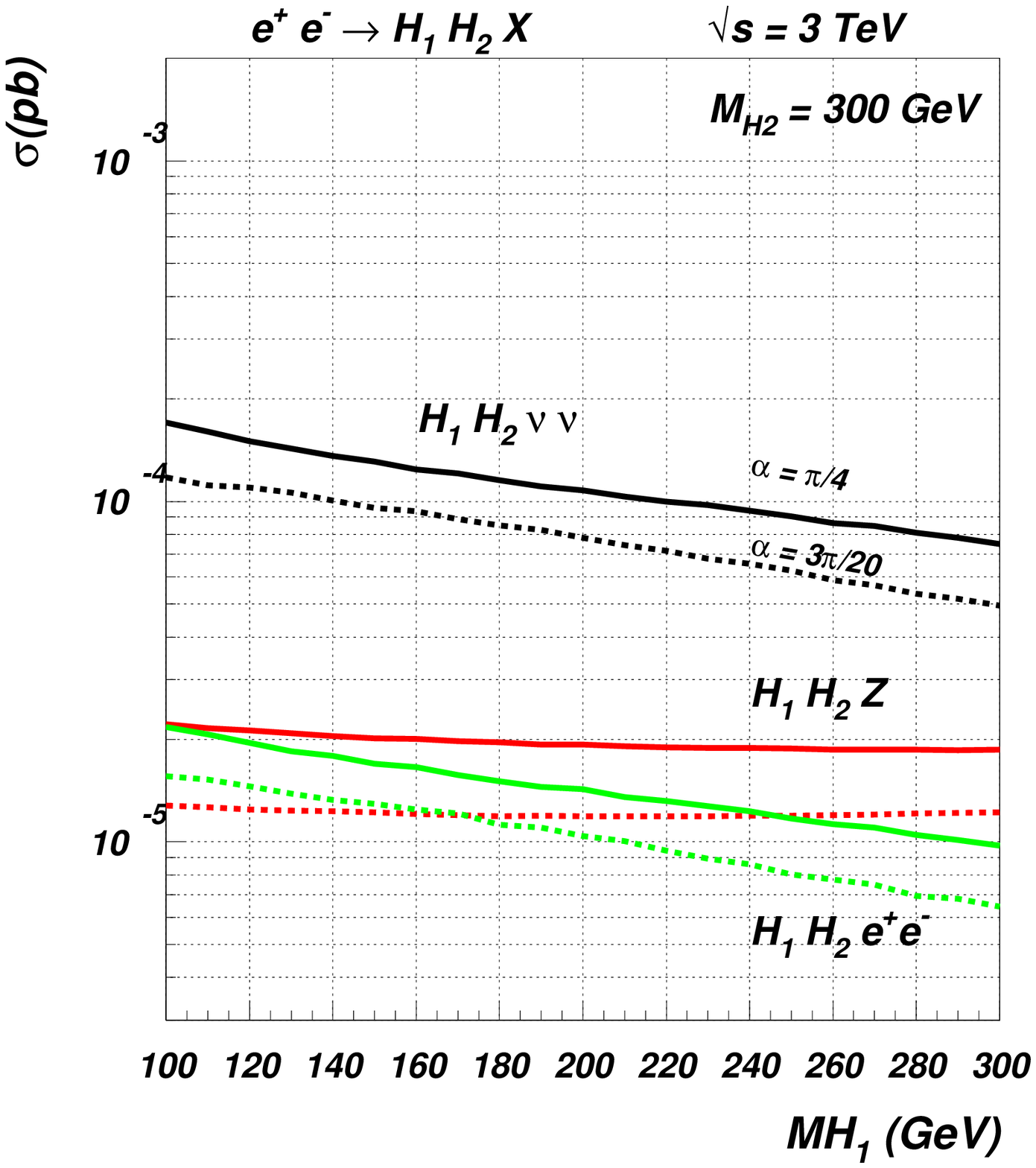}}
  \subfigure[]{
  \label{LC_H1H2-500_3}
  \includegraphics[angle=0,width=0.48\textwidth ]{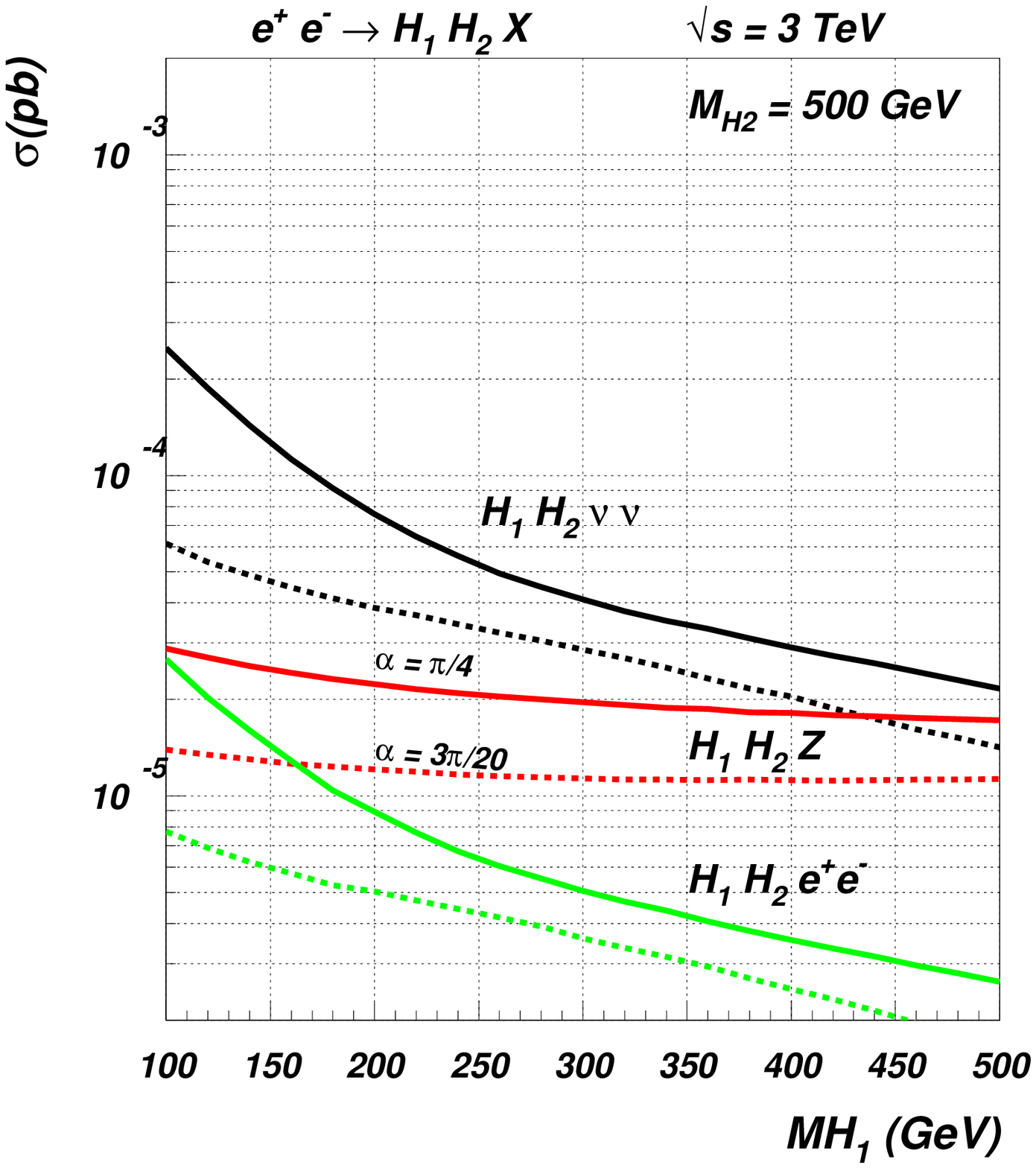}}
  \\
  \subfigure[]{
  \label{LC_H1H2Zp-300_3}
  \includegraphics[angle=0,width=0.48\textwidth
  ]{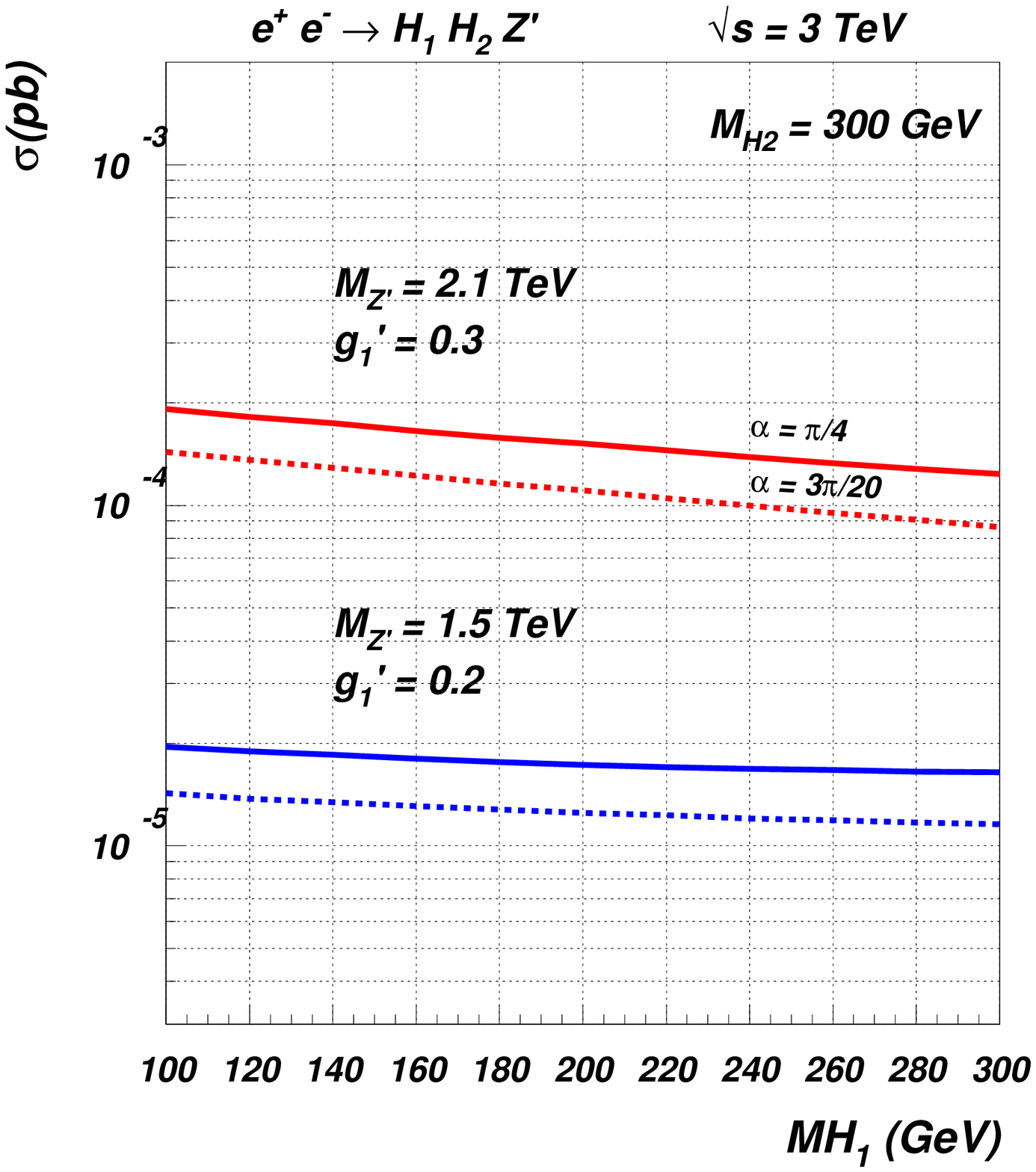}} 
  \subfigure[]{
  \label{LC_H1H2Zp-500_3}
  \includegraphics[angle=0,width=0.48\textwidth
  ]{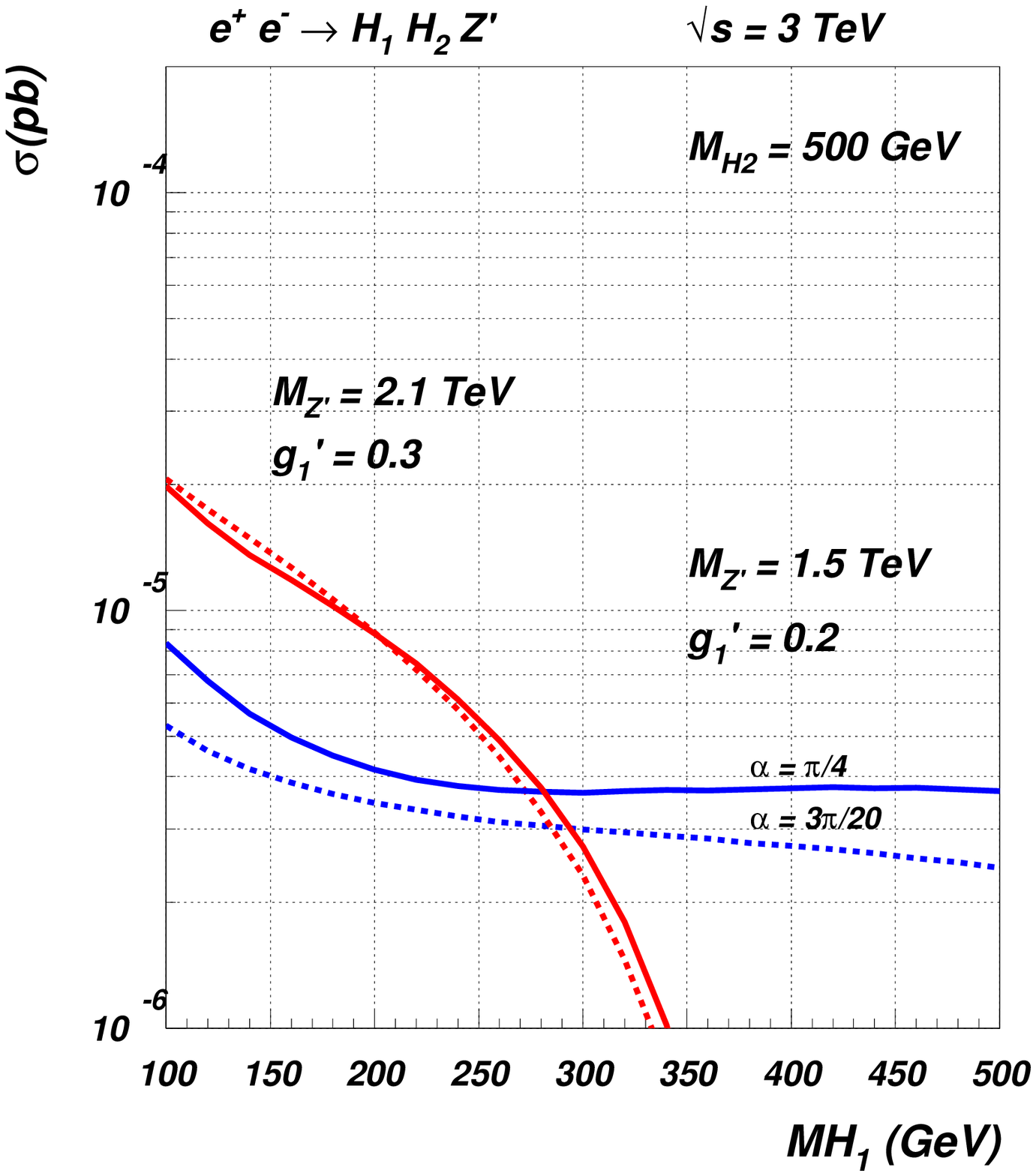}} 
  \caption[Higgs at future linear colliders - Double scalar production
    mechanisms (4)]{Cross sections for the associated production of the two
    higgs bosons at the LC through the standard production mechanisms
    (\ref{LC_H1H2-300_3}) for $M_{h_2}=300$ GeV and
    (\ref{LC_H1H2-500_3}) for $M_{h_2}=500$ GeV, for $\sqrt{s}=3$ TeV,
    and in association with a $Z'$ boson (\ref{LC_H1H2Zp-300_3}) for
    $M_{h_2}=300$ GeV and  (\ref{LC_H1H2Zp-500_3}) for $M_{h_2}=500$
    GeV, for $\sqrt{s}=3$ TeV and several $Z'$
    masses.}
\label{ILC_H1H2} 
\end{figure}

Finally, the two Higgs bosons could be produced together, as shown in
figure~\ref{ILC_H1H2}. Although subleading, this mechanism is peculiar
for several reasons: it requires both Higgs bosons to be
(simultaneously) significantly coupled to the gauge bosons, it has a
very complicated dependence upon $\alpha$, due to the trilinear and
quartic scalar self-couplings, that makes it not invariant under
$\alpha \rightarrow \frac{\pi}{2} - \alpha$ (being $\alpha$ the scalar
mixing angle)\footnote{The behaviour of the trilinear and quartic
  self-interaction couplings with the scalar mixing angle is not a
  simple trigonometric function, as one can verify by checking the
  Feynman rules listed in Appendix~\ref{appe:b}.} and it is maximum
when the mixing is maximal, i.e., for $\alpha = \pi/4$.
These processes could be important to reconstruct the scalar potential
and the whole set of self-interaction couplings, together with the
$h_2\rightarrow h_1 h_1$ decay.

If at $\sqrt{s}=1$ TeV the cross sections for this process are always
below $0.1$ fb, at $\sqrt{s}=3$ TeV the $W$-fusion channel can produce
the two Higgs bosons with cross sections of fractions of fb, up to
$0.08(0.02)\div 0.2(0.3)$ fb, for $M_{h_1}<M_{h_2} = 300(500)$
GeV. The double Higgs-strahlung from a $Z'$ boson of $2.1$ TeV mass
(and $g'_1=0.3$) has also comparable cross sections, of
$\mathcal{O}(0.1)$~fb for $M_{h_1}<M_{h_2} = 300$~GeV only, for values
of the mixing angle close to maximal. Notice that the cross sections
for this process scale approximately with $\sin{2 \alpha}$, whatever
production mechanism is considered. The mixing angle can be measured
from other processes and used as an input for these channels, provided
that also both scalar masses have been measured elsewhere. If so, the
deviation of the cross sections from the naive ones (when the two
Higgs bosons are produced independently, i.e., neglecting the
self-interactions, that would be exactly proportional to $\sin{2
  \alpha}$) will give further indications about the self-interaction
couplings. Very high statistics is required for such a study, barely
within the potentiality of the next generation of LCs.

\newpage

\ 

\chapter{Conclusions} 
\label{chap:5}
\lhead{Chapter 5. \emph{Conclusions}} 

In this Thesis we have investigated the phenomenology of the Higgs
sector of the minimal $B-L$ model at present and future colliders.

The model is realised by mean of a minimal extension of the $SM$,
i.e., by gauging the broken\footnote{At the TeV energy scale.}
$U(1)_{B-L}$ symmetry in addiction to the $SM$ gauge
symmetry\footnote{Assuming no-mixing between the two gauge groups at
  tree-level.}, hence realising the following symmetry pattern:
\begin{equation}
SU(3)_C \times SU(2)_L \times U(1)_Y \times U(1)_{B-L}.
\end{equation}

This extension is triple-minimal: in the gauge boson sector (adding
one gauge boson, $Z'$), in the Higgs sector (adding one scalar boson,
$h_2$) and in the fermion sector (adding one heavy right-handed
neutrino per generation).

In Chapter~\ref{chap:1} we have presented the main motivations that
call for the $U(1)_{B-L}$ extension, in particular the
phenomenological necessity of giving mass to the light neutrinos (via 
the ``see-saw'' mechanism) through a TeV-scale symmetry breaking. We
have also established two possible experimental environments in which
this extension could be studied: the LHC (a currently operative
collider) and the ILC/CLIC (two future linear collider proposal).

In Chapter~\ref{chap:2} we have presented the formal aspects of the
minimal $B-L$ model, focusing on the details that are needed to
implement the model Lagrangian in any of the public softwares that are
devoted to Feynman Rules generation (such as LanHEP, FeynRules, etc.).

In Chapter~\ref{chap:3} (based on the results appeared in
\cite{Basso:2010jt,Basso:2010jm,Basso:2010hk}) we have presented a
detailed study of the Higgs sector parameter space. In particular, we
have focused on the analysis of the parameter space, with emphasis on
the experimental and theoretical exclusion methods, in order to set
the basis for the phenomenological analysis of the Higgs sector at
colliders.

As for the theoretical methods, we have presented a full analysis on
unitarity bounds in the Higgs sector of the minimal $B-L$ model.
Using the equivalence theorem, we have evaluated the spherical partial
wave amplitude of all possible two-to-two scatterings in the scalar
Lagrangian at an infinite energy, identifying the $zz \rightarrow zz$
and $z'z'\rightarrow z'z'$ processes as the most relevant scattering
channels for this analysis ($z^{(')}$ is the would-be Goldstone boson
of the $Z^{(')}$ vector boson). Then, we have shown that these two
channels impose an upper bound on the two Higgs masses: the light one
cannot exceed the $SM$ bound while the limit on the heavy one is
established by the singlet Higgs $VEV$, whose value is presently
constrained by LEP and could shortly be extracted by experiment
following a possible discovery of a $Z'$. We also studied how the
discovery of a light Higgs boson at the LHC could impact on the heavy
Higgs mass bounds in the minimal $B-L$ model and we discovered that
the lighter the $h_1$ mass the more loose is the bound on $M_{h_2}$,
except in the low-mixing region ($\alpha \rightarrow 0$) of the Higgs
parameter space, in which the knowledge of the $x$ $VEV$ is again
fundamental.

Then, we have investigated the triviality and vacuum stability
conditions of the minimal $B-L$ model with a particular view to define
the phenomenologically viable regions of the parameter space of the
scalar sector, by computing all relevant $RGEs$ at the one-loop level
in presence of all available experimental constraints. The $RGE$
dependence on the Higgs masses and couplings (including mixings) has
been studied in detail for discrete choices of the singlet Higgs field
$VEV$, in order to make a fruitful comparison with the unitarity case.

Thereafter, we have shown that, by combining perturbative unitarity and
$RGE$ methods, one can significantly constrain the $g'_1$ coupling of
the minimal $B-L$ extension of the $SM$, by imposing limits on its
upper value that are more stringent than standard triviality
bounds.

Finally, we have made an illustrative study of the
``fine-tuning'' induced by the quantum one-loop corrections of the
Higgs boson masses, giving the correct solution for the Veltman
conjecture in the minimal $B-L$ context.

In Chapter~\ref{chap:4} (based on the results appeared in
\cite{Basso:2010yz,Basso:2010si}), we have studied in detail the
phenomenology of the Higgs sector of the minimal $B-L$ model at
colliders.

After a short discussion about the implementation of the model in
CalcHEP, we have presented the Higgs bosons
branching ratios and total widths for some interesting points of the
parameter space.

Then, we have investigated both the foreseen
energy stages of the LHC (and corresponding luminosities). While
virtually all relevant production and decay processes of the two Higgs
states of the model have been investigated, we have eventually paid
particular attention  to those that are peculiar to the described
$B-L$ scenario. The phenomenological analysis has been carried out in
presence of all available theoretical and experimental constraints and
by exploiting numerical programs at the parton level. While many Higgs
signatures already existing in the $SM$ could be replicated in the
case of its $B-L$ version, in either of the two Higgs states of the
latter (depending on their mixing), it is more important to notice
that several novel Higgs processes could act as hallmarks of the
minimal $B-L$ model. These include Higgs production via gluon-gluon
fusion, in either the light or heavy Higgs state, the former produced
at the lower energy stage of the CERN collider and decaying in two
heavy neutrinos and the latter produced at the higher energy stage of
such a machine and decaying not only in heavy neutrino pairs but also
in $Z'$ and light Higgs ones. For each of these signatures we have in
fact found parameter space regions where the event rates are sizable
and potentially amenable to discovery. Our results have laid the basis
for the phenomenological exploitation of the Higgs sector of the
minimal $B-L$ model at the LHC.

Finally, we have studied the potential of
future LCs in establishing the structure of the Higgs sector of the
minimal $B-L$ model. We have considered both an ILC and CLIC. The
scope of either machine in this respect is substantial as a large
variety of Higgs production processes are accessible. The latter
include both single and double Higgs boson channels, at times produced
in association with heavy particles, both $SM$ ($W$ and $Z$ bosons
and $t$ (anti)quarks) and $B-L$ ones ($Z'$ boson and
$\nu_h$ neutrinos), thus eventually yielding very peculiar signatures
at detector level. This variety of accessible Higgs production
processes potentially allows future LCs to accurately pin down the
structure of the $B-L$ Higgs sector, including not only the masses and
couplings of both Higgs states pertaining to this scenario, but also
trilinear and quartic self-couplings between the two scalar bosons
themselves. On this score, the interplay and complementarity of
measures at the LHC and at LCs is fundamental. 
The extension in the gauge sector, with a $Z'$
boson dominantly coupled to leptons, is fundamental to distinguish
this model from the classic scalar extensions of the $SM$. Although
the scalar Lagrangian is rather a simple one, we showed that the new
signatures and production mechanisms led by the $Z'$ are quite
peculiar and not shared with any extension of the $SM$ that keeps its
gauge content minimal. Finally, the fermion sector can also have very
important consequences for the $B-L$ scalar sector discovery and
identification, allowing for peculiar Higgs bosons decay patterns.

In conclusion, we firmly believe that the analysis that we have
presented in this Thesis could lay the basis for the phenomenological
exploitation of the Higgs sector of the minimal $B-L$ model at
present and future colliders, representing a ``today's challenge'' for
the former's working schedule and a pressing motivation for the
latter's approval.

For the sake of completeness, it is of fundamental importance to point
out the fact that a significative number of open issues is still under
investigation: in the first place, one of the next step of our
research is to study the phenomenology of the minimal $B-L$ at future
photon-photon colliders. Secondly, a whole set of
next-to-leading-order corrections should be evaluated both at the LHC
and LCs. Thereafter, it should be important to investigate the impact
of the mixing between the two $U(1)$ groups, realising the non-minimal
$B-L$ model. Finally, one of the most important open issues lies in
the supersymmetrisation of the model, for the purpose of making
another step of our bottom-up approach, toward the inclusion of the
minimal $B-L$ model in a bigger theoretical grand unification
picture.




\addtocontents{toc}{\vspace{2em}} 

\appendix 

\chapter{The scalar potential}
\label{appe:a}
\lhead{Appendix A. \emph{The scalar potential}}

In this Appendix, we rewrite the interaction part of
equation~(\ref{potential}) in terms of mass eigenstates, separating
four-point and three-point functions and classifying them by the
nature of the involved fields.

The part of the interacting potential that contains four-point
functions involving only would-be Goldstone bosons is:
\begin{eqnarray}\label{4-goldstone}
&\ &V_{4,g}=\nonumber \\
&-&\frac{\pi \alpha_W  \left(M_{h_1}^2 \cos^2{\alpha} + M_{h_2}^2 \sin^2{\alpha}
\right)}{8 M_W^2}(w^+w^-+z^2)^2
\nonumber \\
&-&\frac{(g'_1)^2 \left(M_{h_1}^2\sin^2{\alpha}
+M_{h_2}^2 \cos^2{\alpha} \right)}{2 M_{Z'}^2}(z')^4 \nonumber \\
&-&\frac{\sqrt{\pi \alpha_W}  g'_1  \left( M_{h_2}^2 - M_{h_1}^2 \right)
\sin{(2\alpha)}}{4 M_W M_{Z'}}(w^+w^-+z^2)(z')^2.
\end{eqnarray}

\newpage

The part of the interacting potential that contains four-point
functions involving both would-be Goldstone and Higgs bosons is:
\begin{eqnarray}\label{4-mixed}
&\ &V_{4,hg}= \nonumber \\
&-&\frac{\sqrt{\pi \alpha _W}\cos{\alpha}}{4 M_W^2 M_{Z'}}\big[ 2 
g'_1 \left( M_{h_2}^2 - M_{h_1}^2 \right) M_W \sin^3{\alpha}
 \nonumber \\
&+& \sqrt{\pi \alpha _W}  \left( 
M_{h_1}^2 \cos^2{\alpha} + M_{h_2}^2 \sin^2{\alpha} \right) M_{Z'} \cos{\alpha}
 \big]h^2_1(w^+w^-+z^2) \nonumber \\
&-&\frac{\sqrt{\pi \alpha _W}\sin{\alpha}}{4 M_W^2 M_{Z'}}\big[ 2 
g'_1  \left( M_{h_2}^2 - M_{h_1}^2 \right)M_W \cos^3{\alpha}
 \nonumber \\
&+& \sqrt{\pi \alpha _W}  \left( 
M_{h_1}^2 \cos^2{\alpha} + M_{h_2}^2 \sin^2{\alpha} \right) M_{Z'} \sin{\alpha}
 \big]h^2_2(w^+w^-+z^2) \nonumber \\
&-&\frac{\sqrt{\pi \alpha_W}\sin{(2\alpha)}}{4 M_W^2 M_{Z'}}\big[
g'_1  \left( M_{h_2}^2 - M_{h_1}^2 \right)M_W \sin{(2\alpha)}
 \nonumber \\
&+& \sqrt{\pi \alpha _W}  \left( 
M_{h_1}^2 \cos^2{\alpha} + M_{h_2}^2 \sin^2{\alpha} \right) M_{Z'} 
 \big]h_1h_2(w^+w^-+z^2) \nonumber \\
&-&\frac{g'_1 \sin{\alpha}}{2 M_W
M^2_{Z'}}\big[ -\sqrt{\pi \alpha_W}  \left( M_{h_2}^2 - M_{h_1}^2 \right)
M_{Z'} \cos^3{\alpha} \nonumber \\
&+& 2 g'_1 \left( 
M_{h_1}^2 \sin^2{\alpha} + M_{h_2}^2 \cos^2{\alpha} \right) M_{W} \sin{\alpha}
 \big]h^2_1(z')^2 \nonumber \\
&-&\frac{g'_1 \cos{\alpha}}{2 M_W
M^2_{Z'}}\big[ -\sqrt{\pi \alpha_W}  \left( M_{h_2}^2 -
M_{h_1}^2 \right) M_{Z'} \sin^3{\alpha}
\nonumber \\
&+& 2 g'_1 \left( 
M_{h_1}^2 \sin^2{\alpha} + M_{h_2}^2 \cos^2{\alpha} \right) M_{W} \cos{\alpha}
 \big]h^2_2(z')^2 \nonumber \\
&-&\frac{g'_1\sin{(2\alpha)}}{4 M_W M^2_{Z'}}\big[
\sqrt{\pi \alpha_W}  \left( M_{h_2}^2 - M_{h_1}^2 \right) M_{Z'} \sin{(2\alpha)}
\nonumber \\
&-& 4 g'_1 \left( 
M_{h_1}^2 \sin^2{\alpha} + M_{h_2}^2 \cos^2{\alpha} \right) M_{W} 
 \big]h_1h_2(z')^2.
\end{eqnarray}

\newpage

The part of the interacting potential that contains four-point
functions involving only Higgs bosons is:
\begin{eqnarray}\label{4-higgs}
&\ &V_{4,h}= \nonumber \\
&-&\frac{1}{16} \Bigg[ \frac{8  (g'_1)^2 \left(
M_{h_1}^2 \sin^2{\alpha} +
M_{h_2}^2 \cos^2{\alpha} \right) \sin^4{\alpha}}{M_{Z'}^2} \nonumber \\
&+&\frac{\sqrt{\pi \alpha_W}  g'_1  \left( M_{h_2}^2 - M_{h_1}^2 \right)
\sin^3{(2\alpha)}}{M_W M_{Z'}} \nonumber \\
&+& \frac{2 \pi \alpha_W  \left(
M_{h_1}^2\cos^2{\alpha} + M_{h_2}^2 \sin^2{\alpha}\right) \cos^4{\alpha}}{M_W^2}\Bigg]
h^4_1 \nonumber \\
&-&\frac{\sin{(2\alpha)}}{4 M_W^2 M_{Z'}^2} \left( 2 g'_1 M_W \sin{\alpha}
+\sqrt{\pi \alpha_W} 
M_{Z'} \cos{\alpha} \right) \times \nonumber \\
&\times &\Big[ -2 
g'_1 \left(M_{h_1}^2\sin^2{\alpha} + M_{h_2}^2 \cos^2{\alpha} \right)
M_W \sin{\alpha}
\nonumber \\
&+& \sqrt{\pi \alpha_W} \left(M_{h_1}^2\cos^2{\alpha} + M_{h_2}^2 \sin^2{\alpha} \right)
M_{Z'} \cos{\alpha} \Big]h_1^3h_2
\nonumber \\
&-&\frac{\sin{(2\alpha)}}{16 M_W^2
M_{Z'}^2} \Big[ 12  (g'_1)^2 \left( 
M_{h_1}^2 \sin^2{\alpha} + M_{h_2}^2 \cos^2{\alpha} \right)
M_W^2 \sin{(2\alpha)} \nonumber \\
&+&\sqrt{\pi \alpha_W} g'_1  \left( M_{h_2}^2 - M_{h_1}^2 \right) M_W
M_{Z'} (1 + 3 \cos{(4\alpha)}) \nonumber \\
&+& 3 \pi \alpha_W  \left(
M_{h_1}^2 \cos^2{\alpha} + M_{h_2}^2 \sin^2{\alpha} \right)
M_{Z'}^2 \sin{(2\alpha)} \Big] h_1^2h_2^2 \nonumber \\
&-&\frac{\sin{(2\alpha)}}{4 M_W^2 M_{Z'}^2}
\left( 2 g'_1 M_W \cos{\alpha}
+\sqrt{\pi \alpha_W} 
M_{Z'} \sin{\alpha} \right) \times \nonumber \\
&\times &\Big[-2 
g'_1 \left(M_{h_1}^2\sin^2{\alpha} + M_{h_2}^2 \cos^2{\alpha} \right)
M_W \cos{\alpha}
\nonumber \\
&+& \sqrt{\pi \alpha_W} \left(M_{h_1}^2\cos^2{\alpha} + M_{h_2}^2 \sin^2{\alpha} \right)
M_{Z'} \sin{\alpha} \Big]h_1h_2^3
\nonumber \\
&-&\frac{1}{16} \Bigg[ \frac{8  (g'_1)^2 \left(
M_{h_1}^2 \sin^2{\alpha} +
M_{h_2}^2 \cos^2{\alpha} \right)\cos^4{\alpha}}{M_{Z'}^2} \nonumber \\
&+&\frac{\sqrt{\pi \alpha_W}  g'_1  \left( M_{h_2}^2 - M_{h_1}^2 \right)
\sin^3{(2\alpha)}}{M_W M_{Z'}} \nonumber \\
&+& \frac{2 \pi \alpha_W  \left(
M_{h_1}^2\cos^2{\alpha} + M_{h_2}^2 \sin^2{\alpha}\right) \sin^4{\alpha}}{M_W^2}\Bigg]
h^4_2.
\end{eqnarray}

\newpage

The part of the interacting potential that contains three-point
functions involving both would-be Goldstone and Higgs bosons is:
\begin{eqnarray}\label{3-mixed}
&\ &V_{3,hg}= \nonumber \\
&-&\frac{\sqrt{\pi \alpha_W} M_{h_1}^2 \cos{\alpha}}{2
M_W}h_1(w^+w^-+z^2) \nonumber \\
&-&\frac{\sqrt{\pi \alpha_W} M_{h_2}^2 \sin{\alpha}}{2
M_W}h_2(w^+w^-+z^2) \nonumber \\
&+&\frac{ g'_1 M_{h_1}^2 \sin{\alpha}}{M_{Z'}} h_1(z')^2 -\frac{ g'_1
M_{h_2}^2 \cos{\alpha}}{M_{Z'}} h_2(z')^2.
\end{eqnarray}

The part of the interacting potential that contains three-point
functions involving only Higgs bosons is:
\begin{eqnarray}\label{3-higgs}
&\ & V_{3,h}= \nonumber \\
&-& \frac{M_{h_1}^2}{2}\Bigg( 
-\frac{2 g'_1 \sin^3{\alpha}}{M_{Z'}}
+\frac{\sqrt{\pi \alpha_W}\cos^3{\alpha}}{M_W}
\Bigg) h_1^3 \nonumber \\
&-&\frac{\sin{(2\alpha)}}{4 M_W M_{Z'}}
( 2 M_{h_1}^2 + M_{h_2}^2 ) \Big( 2 g'_1 M_W \sin{\alpha}
+ \sqrt{\pi \alpha_W} M_{Z'} \cos{\alpha} \Big) h_1^2h_2 \nonumber \\
&-& \frac{\sin{(2\alpha)}}{4 M_W M_{Z'}}
( M_{h_1}^2 + 2 M_{h_2}^2 ) \Big( -2 g'_1 M_W \cos{\alpha}
+ \sqrt{\pi \alpha_W} M_{Z'} \sin{\alpha}
  \Big) h_1h_2^2 \nonumber \\
&-& \frac{M_{h_2}^2}{2} \Bigg( 
\frac{2 g'_1 \cos^3{\alpha}}{M_{Z'}}
+\frac{\sqrt{\pi \alpha_W}\sin^3{\alpha}}{M_W}
\Bigg) h_2^3.
\end{eqnarray}


\chapter{The minimal $B-L$ Feynman rules}
\label{appe:b}
\lhead{Appendix B. \emph{The minimal $B-L$ Feynman rules}}

In this Appendix we list the Feynman rules of the minimal $B-L$ model
in the Feynman-gauge; the labelling of the fields is straightforward,
and follows the notation that has been introduced in
Chapter~\ref{chap:2}.

We remark upon the fact that the following list of Feynman rules
has been generated by means of the LanHEP package.

All the vertices must be coupled to a phase ``$i$''.

All the momenta appearing in the vertices are incoming.

\begin{itemize}
\item $e$ is the electric charge.
\item $s_w$($c_w$) $\Rightarrow$ $\sin{\theta_W}$($\cos{\theta_W}$).
\item $s_\alpha$($c_\alpha$) $\Rightarrow$ $\sin{\alpha}$($\cos{\alpha}$).
\item $V$ is the $CKM$ matrix (see
  \cite{Cabibbo:1963yz,Kobayashi:1973fv}).
\item $s_{\alpha i}$($c_{\alpha i}$) is the sinus(cosinus) of the
  ``see-saw'' mixing of the $i^{th}$ neutrino generation (no mixing
  between generations has been considered).
\end{itemize}

\newpage

\begin{center}

\begin{tabular}{|l|l|} \hline
Fields in the vertex & Variational derivative of Lagrangian by fields \\ \hline
${A}_{\mu }$ \phantom{-} $W^+_{\nu }$ \phantom{-} $W^-_{\rho }$ \phantom{-}  &
	$- e\big[(p_2^\rho-p_1^\rho) g^{\mu \nu} -(p_2^\mu-p_3^\mu) g^{\nu \rho} +(p_1^\nu-p_3^\nu) g^{\mu \rho} \big]$\\[2mm]
${A}_{\mu }$ \phantom{-} $W^+_{\nu }$ \phantom{-} $w^-$ \phantom{-}  &
	$ i   e   M_W  g^{\mu \nu} $\\[2mm]
${A}_{\mu }$ \phantom{-} $w^+$ \phantom{-} $W^-_{\nu }$ \phantom{-}  &
	$- i   e   M_W  g^{\mu \nu} $\\[2mm]
${A}_{\mu }$ \phantom{-} $w^+$ \phantom{-} $w^-$ \phantom{-}  &
	$ e\big(p_3^\mu -p_2^\mu \big)$\\[2mm]
$\bar{C}^A$ \phantom{-} $C^{W^+}$ \phantom{-} $W^-_{\mu }$ \phantom{-}  &
	$- e  p_1^\mu $\\[2mm]
$\bar{C}^A{}_{}$ \phantom{-} $C^{W^-}$ \phantom{-} $W^+_{\mu }$ \phantom{-}  &
	$ e  p_1^\mu $\\[2mm]
$\bar{b}{}_{a p }$ \phantom{-} $b{}_{b q }$ \phantom{-} ${A}_{\mu }$ \phantom{-}  &
	$\frac{1}{3} e\delta_{p q} \gamma_{a c}^\mu   \delta_{c b} $\\
\hline
\end{tabular}

\begin{tabular}{|l|l|} \hline
Fields in the vertex & Variational derivative of Lagrangian by fields \\ \hline
$\bar{c}{}_{a p }$ \phantom{-} $c{}_{b q }$ \phantom{-} ${Z}_{\mu }$ \phantom{-}  &
	$-\frac{1}{6}\frac{ e}{ c_w   s_w}\delta_{p q} \gamma_{a c}^\mu \big( (3-4 s_w^2)  {(1-\gamma^5)_{c b}\over 2} -4 s_w^2   {(1+\gamma^5)_{c b}\over 2} \big)$\\[2mm]
$\bar{c}{}_{a p }$ \phantom{-} $c{}_{b q }$ \phantom{-} $z$ \phantom{-}  &
	$\frac{1}{2}\frac{ i   e   M_c}{ M_W   s_w}\delta_{p q}   \gamma_{a b}^5 $\\[2mm]
$\bar{c}{}_{a p }$ \phantom{-} $c{}_{b q }$ \phantom{-} ${Z'}_{\mu }$ \phantom{-}  &
	$-\frac{1}{3} g^{\prime}_1\delta_{p q} \gamma_{a c}^\mu   \delta_{c b} $\\[2mm]
$\bar{c}{}_{a p }$ \phantom{-} $d{}_{b q }$ \phantom{-} $W^+_{\mu }$ \phantom{-}  &
	$-\frac{1}{2}\frac{ e  \sqrt{2}   V_{cd}}{ s_w}  \delta_{p q} \gamma_{a c}^\mu {(1-\gamma^5)_{c b}\over 2} $\\[2mm]
$\bar{c}{}_{a p }$ \phantom{-} $d{}_{b q }$ \phantom{-} $w^+$ \phantom{-}  &
	$-\frac{1}{2}\frac{ i   e   M_c  \sqrt{2}   V_{cd}}{ M_W   s_w}  \delta_{p q} {(1-\gamma^5)_{a b}\over 2} $\\[2mm]
$\bar{c}{}_{a p }$ \phantom{-} $s{}_{b q }$ \phantom{-} $W^+_{\mu }$ \phantom{-}  &
	$-\frac{1}{2}\frac{ e  \sqrt{2}   V_{cs}}{ s_w}  \delta_{p q} \gamma_{a c}^\mu {(1-\gamma^5)_{c b}\over 2} $\\[2mm]
$\bar{c}{}_{a p }$ \phantom{-} $s{}_{b q }$ \phantom{-} $w^+$ \phantom{-}  &
	$\frac{1}{2}\frac{ i   e  \sqrt{2}   V_{cs}}{ M_W   s_w}\delta_{p q} \big( M_s  {(1+\gamma^5)_{a b}\over 2} - M_c  {(1-\gamma^5)_{a b}\over 2} \big)$\\
\hline
\end{tabular}

\begin{tabular}{|l|l|} \hline
Fields in the vertex & Variational derivative of Lagrangian by fields \\ \hline
$\bar{e}{}_{a }$ \phantom{-} ${\nu h}^1_{b }$ \phantom{-} $w^-$ \phantom{-}  &
	$-\frac{1}{2}\frac{ i}{ M_W   s_w}\big( s_{a1}   e   M_e  \sqrt{2}  {(1-\gamma^5)_{a b}\over 2} -2 s_w   c_{a1}   M_W   y^{\nu}_1  {(1+\gamma^5)_{a b}\over 2} \big)$\\[2mm]
${G}_{\mu p }$ \phantom{-} ${G}_{\nu q }$ \phantom{-} ${G}_{\rho r }$ \phantom{-}  &
	$ g_sf_{p q r} \big[(p_3^\nu-p_1^\nu) g^{\mu \rho} -(p_3^\mu-p_2^\mu) g^{\nu \rho} +(p_1^\rho-p_2^\rho) g^{\mu \nu} \big]$\\[2mm]
$\bar{C}^G_{p }$ \phantom{-} $C^G_{q }$ \phantom{-} ${G}_{\mu r }$ \phantom{-}  &
	$ g_s  p_2^\mu f_{p q r} $\\[2mm]
${H}_{1}$ \phantom{-} ${H}_{1}$ \phantom{-} ${H}_{1}$ \phantom{-}  &
	$-3\frac{1}{ e}\big(4 c_{\alpha}^3  s_w M_W \lambda _1-2 s_{\alpha}^3  e \lambda _2 x- c_{\alpha}^2  s_{\alpha} e \lambda _3 x$ \\[2mm]
  & $+2 s_w s_{\alpha}^2  c_{\alpha} M_W \lambda _3\big)$\\[2mm]
${H}_{1}$ \phantom{-} ${H}_{1}$ \phantom{-} ${H}_{2}$ \phantom{-}  &
	$-\frac{1}{ e}\big(12 c_{\alpha}^2  s_w s_{\alpha} M_W \lambda _1+6 s_{\alpha}^2  c_{\alpha} e \lambda _2 x+ (1-3 s_{\alpha}^2) c_{\alpha} e \lambda _3 x$ \\[2mm]
  & $-2 (2-3 s_{\alpha}^2) s_w s_{\alpha} M_W \lambda _3\big)$\\
\hline
\end{tabular}

\begin{tabular}{|l|l|} \hline
Fields in the vertex & Variational derivative of Lagrangian by fields \\ \hline
${H}_{2}$ \phantom{-} ${Z}_{\mu }$ \phantom{-} ${Z}_{\nu }$ \phantom{-}  &
	$\frac{ e   M_W   s_{\alpha}}{ c_w^2    s_w}g^{\mu \nu} $\\[2mm]
${H}_{2}$ \phantom{-} ${Z}_{\mu }$ \phantom{-} $z$ \phantom{-}  &
	$\frac{1}{2}\frac{ i   e   s_{\alpha}}{ c_w   s_w}\big(p_3^\mu -p_1^\mu \big)$\\[2mm]
${H}_{2}$ \phantom{-} $z$ \phantom{-} $z$ \phantom{-}  &
	$-\frac{1}{ e}\big(4 s_w s_{\alpha} M_W \lambda _1+ c_{\alpha} e \lambda _3 x\big)$\\[2mm]
${H}_{2}$ \phantom{-} ${Z'}_{\mu }$ \phantom{-} ${Z'}_{\nu }$ \phantom{-}  &
	$8 c_{\alpha}   g^{\prime}_1{}^2    x  g^{\mu \nu} $\\[2mm]
${H}_{2}$ \phantom{-} ${Z'}_{\mu }$ \phantom{-} ${z'}_{}$ \phantom{-}  &
	$2 i   c_{\alpha}   g^{\prime}_1\big(p_1^\mu -p_3^\mu \big)$\\[2mm]
${H}_{2}$ \phantom{-} ${z'}_{}$ \phantom{-} ${z'}_{}$ \phantom{-}  &
	$-2\frac{1}{ e}\big( c_{\alpha} e \lambda _2 x+ s_w s_{\alpha} M_W \lambda _3\big)$\\[2mm]
$\bar{\tau}{}_{a }$ \phantom{-} $\tau{}_{b }$ \phantom{-} ${A}_{\mu }$ \phantom{-}  &
	$ e\gamma_{a c}^\mu   \delta_{c b} $\\
\hline
\end{tabular}

\begin{tabular}{|l|l|} \hline
Fields in the vertex & Variational derivative of Lagrangian by fields \\ \hline
$\bar{\mu}{}_{a }$ \phantom{-} ${\nu h}^2_{b }$ \phantom{-} $w^-$ \phantom{-}  &
	$-\frac{1}{2}\frac{ i}{ M_W   s_w}\big( s_{a2}   e   M_\mu  \sqrt{2}  {(1-\gamma^5)_{a b}\over 2} -2 s_w   c_{a2}   M_W   y^\nu_2  {(1+\gamma^5)_{a b}\over 2} \big)$\\[2mm]
$\bar{s}{}_{a p }$ \phantom{-} $c{}_{b q }$ \phantom{-} $W^-_{\mu }$ \phantom{-}  &
	$-\frac{1}{2}\frac{ e  \sqrt{2}   V_{cs}}{ s_w}  \delta_{p q} \gamma_{a c}^\mu {(1-\gamma^5)_{c b}\over 2} $\\[2mm]
$\bar{s}{}_{a p }$ \phantom{-} $c{}_{b q }$ \phantom{-} $w^-$ \phantom{-}  &
	$-\frac{1}{2}\frac{ i   e  \sqrt{2}   V_{cs}}{ M_W   s_w}\delta_{p q} \big( M_s  {(1-\gamma^5)_{a b}\over 2} - M_c  {(1+\gamma^5)_{a b}\over 2} \big)$\\[2mm]
$\bar{s}{}_{a p }$ \phantom{-} $s{}_{b q }$ \phantom{-} ${A}_{\mu }$ \phantom{-}  &
	$\frac{1}{3} e\delta_{p q} \gamma_{a c}^\mu   \delta_{c b} $\\[2mm]
$\bar{s}{}_{a p }$ \phantom{-} $s{}_{b q }$ \phantom{-} ${G}_{\mu r }$ \phantom{-}  &
	$ g_s  \lambda_{p q}^r \gamma_{a b}^\mu $\\[2mm]
$\bar{s}{}_{a p }$ \phantom{-} $s{}_{b q }$ \phantom{-} ${H}_{1}$ \phantom{-}  &
	$-\frac{1}{2}\frac{ c_{\alpha}   e   M_s}{ M_W   s_w}\delta_{p q}   \delta_{a b} $\\[2mm]
$\bar{s}{}_{a p }$ \phantom{-} $s{}_{b q }$ \phantom{-} ${H}_{2}$ \phantom{-}  &
	$-\frac{1}{2}\frac{ e   M_s   s_{\alpha}}{ M_W   s_w}\delta_{p q}   \delta_{a b} $\\
\hline
\end{tabular}

\begin{tabular}{|l|l|} \hline
Fields in the vertex & Variational derivative of Lagrangian by fields \\ \hline
$\bar{t}{}_{a p }$ \phantom{-} $t{}_{b q }$ \phantom{-} $z$ \phantom{-}  &
	$\frac{1}{2}\frac{ i   e   M_t}{ M_W   s_w}\delta_{p q}   \gamma_{a b}^5 $\\[2mm]
$\bar{t}{}_{a p }$ \phantom{-} $t{}_{b q }$ \phantom{-} ${Z'}_{\mu }$ \phantom{-}  &
	$-\frac{1}{3} g^{\prime}_1\delta_{p q} \gamma_{a c}^\mu   \delta_{c b} $\\[2mm]
$\bar{u}{}_{a p }$ \phantom{-} $b{}_{b q }$ \phantom{-} $W^+_{\mu }$ \phantom{-}  &
	$-\frac{1}{2}\frac{ e  \sqrt{2}   V_{ub}}{ s_w}  \delta_{p q} \gamma_{a c}^\mu {(1-\gamma^5)_{c b}\over 2} $\\[2mm]
$\bar{u}{}_{a p }$ \phantom{-} $b{}_{b q }$ \phantom{-} $w^+$ \phantom{-}  &
	$\frac{1}{2}\frac{ i   e   M_b  \sqrt{2}   V_{ub}}{ M_W   s_w}  \delta_{p q} {(1+\gamma^5)_{a b}\over 2} $\\[2mm]
$\bar{u}{}_{a p }$ \phantom{-} $d{}_{b q }$ \phantom{-} $W^+_{\mu }$ \phantom{-}  &
	$-\frac{1}{2}\frac{ e  \sqrt{2}   V_{ud}}{ s_w}  \delta_{p q} \gamma_{a c}^\mu {(1-\gamma^5)_{c b}\over 2} $\\[2mm]
$\bar{u}{}_{a p }$ \phantom{-} $s{}_{b q }$ \phantom{-} $W^+_{\mu }$ \phantom{-}  &
	$-\frac{1}{2}\frac{ e  \sqrt{2}   V_{us}}{ s_w}  \delta_{p q} \gamma_{a c}^\mu {(1-\gamma^5)_{c b}\over 2} $\\[2mm]
$\bar{u}{}_{a p }$ \phantom{-} $s{}_{b q }$ \phantom{-} $w^+$ \phantom{-}  &
	$\frac{1}{2}\frac{ i   e   M_s  \sqrt{2}   V_{us}}{ M_W   s_w}  \delta_{p q} {(1+\gamma^5)_{a b}\over 2} $\\
\hline
\end{tabular}

\begin{tabular}{|l|l|} \hline
Fields in the vertex & Variational derivative of Lagrangian by fields \\ \hline
$w^+$ \phantom{-} $w^-$ \phantom{-} ${Z}_{\mu }$ \phantom{-}  &
	$\frac{1}{2}\frac{ (1-2 s_w^2)   e}{ c_w   s_w}\big(p_2^\mu -p_1^\mu \big)$\\[2mm]
$\bar{C}^{W^-}{}_{}$ \phantom{-} $C^{Z}{}_{}$ \phantom{-} $W^+_{\mu }$ \phantom{-}  &
	$- e  p_1^\mu $\\[2mm]
$\bar{C}^{W^-}{}_{}$ \phantom{-} $C^{Z}{}_{}$ \phantom{-} $w^+$ \phantom{-}  &
	$ i   e   M_W$\\[2mm]
$\bar{C}^{W^-}{}_{}$ \phantom{-} $C^{W^+}{}_{}$ \phantom{-} ${A}_{\mu }$ \phantom{-}  &
	$ e  p_1^\mu $\\[2mm]
$\bar{C}^{W^-}{}_{}$ \phantom{-} $C^{W^+}{}_{}$ \phantom{-} ${H}_{1}$ \phantom{-}  &
	$-\frac{1}{2}\frac{ c_{\alpha}   e   M_W}{ s_w}$\\[2mm]
$\bar{C}^{W^-}{}_{}$ \phantom{-} $C^{W^+}{}_{}$ \phantom{-} ${H}_{2}$ \phantom{-}  &
	$-\frac{1}{2}\frac{ e   M_W   s_{\alpha}}{ s_w}$\\[2mm]
$\bar{C}^{W^-}{}_{}$ \phantom{-} $C^{W^+}{}_{}$ \phantom{-} ${Z}_{\mu }$ \phantom{-}  &
	$\frac{ c_w   e}{ s_w}  p_1^\mu $\\
\hline
\end{tabular}

\begin{tabular}{|l|l|} \hline
Fields in the vertex & Variational derivative of Lagrangian by fields \\ \hline
$\tau^c_{a }$ \phantom{-} ${\nu h}^3_{b }$ \phantom{-} $w^+$ \phantom{-}  &
	$\frac{1}{2}\frac{ i}{ M_W   s_w}\big( s_{a3}   e   M_\tau  \sqrt{2}  {(1+\gamma^5)_{a b}\over 2} -2 s_w   c_{a3}   M_W   y^\nu_3  {(1-\gamma^5)_{a b}\over 2} \big)$\\[2mm]
$\mu^c_{a }$ \phantom{-} ${\nu l}^2_{b }$ \phantom{-} $W^+_{\mu }$ \phantom{-}  &
	$\frac{1}{2}\frac{ c_{a2}   e  \sqrt{2}}{ s_w}  \gamma_{a c}^\mu {(1+\gamma^5)_{c b}\over 2} $\\[2mm]
$\mu^c_{a }$ \phantom{-} ${\nu l}^2_{b }$ \phantom{-} $w^+$ \phantom{-}  &
	$\frac{1}{2}\frac{ i}{ M_W   s_w}\big( c_{a2}   e   M_\mu  \sqrt{2}  {(1+\gamma^5)_{a b}\over 2} +2 s_w   s_{a2}   M_W   y^\nu_2  {(1-\gamma^5)_{a b}\over 2} \big)$\\[2mm]
$\mu^c_{a }$ \phantom{-} ${\nu h}^2_{b }$ \phantom{-} $W^+_{\mu }$ \phantom{-}  &
	$\frac{1}{2}\frac{ e   s_{a2}  \sqrt{2}}{ s_w}  \gamma_{a c}^\mu {(1+\gamma^5)_{c b}\over 2} $\\[2mm]
$\mu^c_{a }$ \phantom{-} ${\nu h}^2_{b }$ \phantom{-} $w^+$ \phantom{-}  &
	$\frac{1}{2}\frac{ i}{ M_W   s_w}\big( s_{a2}   e   M_\mu  \sqrt{2}  {(1+\gamma^5)_{a b}\over 2} -2 s_w   c_{a2}   M_W   y^\nu_2  {(1-\gamma^5)_{a b}\over 2} \big)$\\[2mm]
${\nu l}^1_{a }$ \phantom{-} ${\nu l}^1_{b }$ \phantom{-} ${H}_{1}$ \phantom{-}  &
	$\frac{ s_{a1}}{ x}\big( c_{\alpha}   c_{a1}  \sqrt{2}   x   y^{\nu}_1  \delta_{a b} + s_{\alpha}   s_{a1}   M_{\nu_1}  \delta_{a b} \big)$\\[2mm]
${\nu l}^1_{a }$ \phantom{-} ${\nu l}^1_{b }$ \phantom{-} ${H}_{2}$ \phantom{-}  &
	$\frac{ s_{a1}}{ x}\big( s_{\alpha}   c_{a1}  \sqrt{2}   x   y^{\nu}_1  \delta_{a b} - s_{a1}   c_{\alpha}   M_{\nu_1}  \delta_{a b} \big)$\\
\hline
\end{tabular}

\begin{tabular}{|l|l|} \hline
Fields in the vertex & Variational derivative of Lagrangian by fields \\ \hline
${\nu l}^2_{a }$ \phantom{-} ${\nu h}^2_{b }$ \phantom{-} ${Z}_{\mu }$ \phantom{-}  &
	$\frac{1}{2}\frac{ c_{a2}   e   s_{a2}}{ c_w   s_w}\gamma_{a c}^\mu   \gamma_{c b}^5 $\\[2mm]
${\nu l}^2_{a }$ \phantom{-} ${\nu h}^2_{b }$ \phantom{-} $z$ \phantom{-}  &
	$\frac{1}{2} i   (1-2 s_{a2} ^2)  \sqrt{2}   y^\nu_2  \gamma_{a b}^5 $\\[2mm]
${\nu l}^2_{a }$ \phantom{-} ${\nu h}^2_{b }$ \phantom{-} ${Z'}_{\mu }$ \phantom{-}  &
	$-2 c_{a2}   s_{a2}   g^{\prime}_1\gamma_{a c}^\mu   \gamma_{c b}^5 $\\[2mm]
${\nu l}^2_{a }$ \phantom{-} ${\nu h}^2_{b }$ \phantom{-} ${z'}_{}$ \phantom{-}  &
	$\frac{ i   c_{a2}   M_{\nu_2}   s_{a2}}{ x}  \gamma_{a b}^5 $\\[2mm]
${\nu l}^3_{a }$ \phantom{-} ${\nu l}^3_{b }$ \phantom{-} ${H}_{1}$ \phantom{-}  &
	$\frac{ s_{a3}}{ x}\big( c_{\alpha}   c_{a3}  \sqrt{2}   x   y^\nu_3  \delta_{a b} + s_{\alpha}   s_{a3}   M_{\nu_3}  \delta_{a b} \big)$\\[2mm]
${\nu l}^3_{a }$ \phantom{-} ${\nu l}^3_{b }$ \phantom{-} ${H}_{2}$ \phantom{-}  &
	$\frac{ s_{a3}}{ x}\big( s_{\alpha}   c_{a3}  \sqrt{2}   x   y^\nu_3  \delta_{a b} - s_{a3}   c_{\alpha}   M_{\nu_3}  \delta_{a b} \big)$\\[2mm]
${\nu l}^3_{a }$ \phantom{-} ${\nu l}^3_{b }$ \phantom{-} ${Z}_{\mu }$ \phantom{-}  &
	$\frac{1}{2}\frac{ c_{a3}^2    e}{ c_w   s_w}\gamma_{a c}^\mu \gamma_{c b}^5 $\\
\hline
\end{tabular}

\begin{tabular}{|l|l|} \hline
Fields in the vertex & Variational derivative of Lagrangian by fields \\ \hline
${\nu h}^2_{a }$ \phantom{-} ${\nu h}^2_{b }$ \phantom{-} $z$ \phantom{-}  &
	$ i   c_{a2}   s_{a2}  \sqrt{2}   y^\nu_2  \gamma_{a b}^5 $\\[2mm]
${\nu h}^2_{a }$ \phantom{-} ${\nu h}^2_{b }$ \phantom{-} ${Z'}_{\mu }$ \phantom{-}  &
	$ (1-2 s_{a2} ^2)   g^{\prime}_1\gamma_{a c}^\mu \gamma_{c b}^5 $\\[2mm]
${\nu h}^2_{a }$ \phantom{-} ${\nu h}^2_{b }$ \phantom{-} ${z'}_{}$ \phantom{-}  &
	$-\frac{ i   c_{a2}^2    M_{\nu_2}}{ x}  \gamma_{a b}^5 $\\[2mm]
${\nu h}^3_{a }$ \phantom{-} ${\nu h}^3_{b }$ \phantom{-} ${H}_{1}$ \phantom{-}  &
	$-\frac{ c_{a3}}{ x}\big( s_{a3}   c_{\alpha}  \sqrt{2}   x   y^\nu_3  \delta_{a b} - s_{\alpha}   c_{a3}   M_{\nu_3}  \delta_{a b} \big)$\\[2mm]
${\nu h}^3_{a }$ \phantom{-} ${\nu h}^3_{b }$ \phantom{-} ${H}_{2}$ \phantom{-}  &
	$-\frac{ c_{a3}}{ x}\big( s_{\alpha}   s_{a3}  \sqrt{2}   x   y^\nu_3  \delta_{a b} + c_{\alpha}   c_{a3}   M_{\nu_3}  \delta_{a b} \big)$\\[2mm]
${\nu h}^3_{a }$ \phantom{-} ${\nu h}^3_{b }$ \phantom{-} ${Z}_{\mu }$ \phantom{-}  &
	$\frac{1}{2}\frac{ e   s_{a3}^2 }{ c_w   s_w}\gamma_{a c}^\mu \gamma_{c b}^5 $\\[2mm]
${\nu h}^3_{a }$ \phantom{-} ${\nu h}^3_{b }$ \phantom{-} $z$ \phantom{-}  &
	$ i   c_{a3}   s_{a3}  \sqrt{2}   y^\nu_3  \gamma_{a b}^5 $\\
\hline
\end{tabular}

\begin{tabular}{|l|l|} \hline
Fields in the vertex & Variational derivative of Lagrangian by fields \\ \hline
${H}_{1}$ \phantom{-} ${H}_{1}$ \phantom{-} $w^+$ \phantom{-} $w^-$ \phantom{-}  &
	$-\big(2 c_{\alpha}^2  \lambda _1+ s_{\alpha}^2  \lambda _3\big)$\\[2mm]
${H}_{1}$ \phantom{-} ${H}_{1}$ \phantom{-} ${Z}_{\mu }$ \phantom{-} ${Z}_{\nu }$ \phantom{-}  &
	$\frac{1}{2}\frac{ c_{\alpha}^2    e^2 }{ c_w^2    s_w^2 }g^{\mu \nu} $\\[2mm]
${H}_{1}$ \phantom{-} ${H}_{1}$ \phantom{-} $z$ \phantom{-} $z$ \phantom{-}  &
	$-\big(2 c_{\alpha}^2  \lambda _1+ s_{\alpha}^2  \lambda _3\big)$\\[2mm]
${H}_{1}$ \phantom{-} ${H}_{1}$ \phantom{-} ${Z'}_{\mu }$ \phantom{-} ${Z'}_{\nu }$ \phantom{-}  &
	$8 s_{\alpha}^2    g^{\prime}_1{}^2   g^{\mu \nu} $\\[2mm]
${H}_{1}$ \phantom{-} ${H}_{1}$ \phantom{-} ${z'}_{}$ \phantom{-} ${z'}_{}$ \phantom{-}  &
	$-\big(2 s_{\alpha}^2  \lambda _2+ c_{\alpha}^2  \lambda _3\big)$\\[2mm]
${H}_{1}$ \phantom{-} ${H}_{2}$ \phantom{-} ${H}_{2}$ \phantom{-} ${H}_{2}$ \phantom{-}  &
	$-3 c_{\alpha}   s_{\alpha}\big(2 s_{\alpha}^2  \lambda _1-2 c_{\alpha}^2  \lambda _2+ (1-2 s_{\alpha} ^2) \lambda _3\big)$\\[2mm]
${H}_{1}$ \phantom{-} ${H}_{2}$ \phantom{-} $W^+_{\mu }$ \phantom{-} $W^-_{\nu }$ \phantom{-}  &
	$\frac{1}{2}\frac{ c_{\alpha}   e^2    s_{\alpha}}{ s_w^2 }  g^{\mu \nu} $\\
\hline
\end{tabular}

\begin{tabular}{|l|l|} \hline
Fields in the vertex & Variational derivative of Lagrangian by fields \\ \hline
$W^+_{\mu }$ \phantom{-} $W^-_{\nu }$ \phantom{-} ${Z}_{\rho }$ \phantom{-} ${Z}_{\sigma }$ \phantom{-}  &
	$-\frac{ c_w^2    e^2 }{ s_w^2 }\big(2g^{\mu \nu} g^{\rho \sigma} -g^{\mu \rho} g^{\nu \sigma} -g^{\mu \sigma} g^{\nu \rho} \big)$\\[2mm]
$W^+_{\mu }$ \phantom{-} $W^-_{\nu }$ \phantom{-} $z$ \phantom{-} $z$ \phantom{-}  &
	$\frac{1}{2}\frac{ e^2 }{ s_w^2 }  g^{\mu \nu} $\\[2mm]
$W^+_{\mu }$ \phantom{-} $w^-$ \phantom{-} ${Z}_{\nu }$ \phantom{-} $z$ \phantom{-}  &
	$\frac{1}{2}\frac{ e^2 }{ c_w}  g^{\mu \nu} $\\[2mm]
$w^+$ \phantom{-} $w^+$ \phantom{-} $w^-$ \phantom{-} $w^-$ \phantom{-}  &
	$-4 \lambda _1$\\[2mm]
$w^+$ \phantom{-} $W^-_{\mu }$ \phantom{-} ${Z}_{\nu }$ \phantom{-} $z$ \phantom{-}  &
	$\frac{1}{2}\frac{ e^2 }{ c_w}  g^{\mu \nu} $\\[2mm]
$w^+$ \phantom{-} $w^-$ \phantom{-} ${Z}_{\mu }$ \phantom{-} ${Z}_{\nu }$ \phantom{-}  &
	$\frac{1}{2}\frac{ (1-2 s_w ^2)^2    e^2 }{ c_w^2    s_w^2 }g^{\mu \nu} $\\[2mm]
$w^+$ \phantom{-} $w^-$ \phantom{-} $z$ \phantom{-} $z$ \phantom{-}  &
	$-2 \lambda _1$\\[2mm]
$w^+$ \phantom{-} $w^-$ \phantom{-} ${z'}_{}$ \phantom{-} ${z'}_{}$ \phantom{-}  &
	$- \lambda _3$\\[2mm]
${Z}_{\mu }$ \phantom{-} ${Z}_{\nu }$ \phantom{-} $z$ \phantom{-} $z$ \phantom{-}  &
	$\frac{1}{2}\frac{ e^2 }{ c_w^2    s_w^2 }g^{\mu \nu} $\\[2mm]
$z$ \phantom{-} $z$ \phantom{-} $z$ \phantom{-} $z$ \phantom{-}  &
	$-6 \lambda _1$\\[2mm]
$z$ \phantom{-} $z$ \phantom{-} ${z'}_{}$ \phantom{-} ${z'}_{}$ \phantom{-}  &
	$- \lambda _3$\\[2mm]
${Z'}_{\mu }$ \phantom{-} ${Z'}_{\nu }$ \phantom{-} ${z'}_{}$ \phantom{-} ${z'}_{}$ \phantom{-}  &
	$8 g^{\prime}_1{}^2  g^{\mu \nu} $\\[2mm]
${z'}_{}$ \phantom{-} ${z'}_{}$ \phantom{-} ${z'}_{}$ \phantom{-} ${z'}_{}$ \phantom{-}  &
	$-6 \lambda _2$\\ \hline
\end{tabular}

\end{center}

\newpage

\ 



\addtocontents{toc}{\vspace{2em}}  
\backmatter

\label{Bibliography}
\lhead{\emph{Bibliography}}  
\bibliographystyle{h-physrev}
\bibliography{Bibliography}  

\begin{thebibliography}{10}

\bibitem{Peskin:1995ev}
M.~E. Peskin and D.~V. Schroeder,
\newblock p. 842 (1995),
\newblock Reading, USA: Addison-Wesley.

\bibitem{Djouadi:2005gi}
A.~Djouadi,
\newblock Phys. Rept. {\bf 457}, 1 (2008), hep-ph/0503172.

\bibitem{Higgs:1964pj}
P.~W. Higgs,
\newblock Phys. Rev. Lett. {\bf 13}, 508 (1964).

\bibitem{Bilenky:1987ty}
S.~M. Bilenky and S.~T. Petcov,
\newblock Rev. Mod. Phys. {\bf 59}, 671 (1987).

\bibitem{Altarelli:2004za}
G.~Altarelli and F.~Feruglio,
\newblock New J. Phys. {\bf 6}, 106 (2004), hep-ph/0405048.

\bibitem{Strumia:2006db}
A.~Strumia and F.~Vissani,
\newblock (2006), hep-ph/0606054.

\bibitem{Bertone:2004pz}
G.~Bertone, D.~Hooper, and J.~Silk,
\newblock Phys. Rept. {\bf 405}, 279 (2005), hep-ph/0404175.

\bibitem{Gildener:1976ih}
E.~Gildener and S.~Weinberg,
\newblock Phys. Rev. {\bf D13}, 3333 (1976).

\bibitem{Weinberg:1978ym}
S.~Weinberg,
\newblock Phys. Lett. {\bf B82}, 387 (1979).

\bibitem{Martin:1997ns}
S.~P. Martin,
\newblock (1997), hep-ph/9709356.

\bibitem{Ellwanger:2009dp}
U.~Ellwanger, C.~Hugonie, and A.~M. Teixeira,
\newblock Phys. Rept. {\bf 496}, 1 (2010), 0910.1785.

\bibitem{Jenkins:1987ue}
E.~E. Jenkins,
\newblock Phys. Lett. {\bf B192}, 219 (1987).

\bibitem{Buchmuller:1991ce}
W.~Buchmuller, C.~Greub, and P.~Minkowski,
\newblock Phys. Lett. {\bf B267}, 395 (1991).

\bibitem{Khalil:2006yi}
S.~Khalil,
\newblock J. Phys. {\bf G35}, 055001 (2008), hep-ph/0611205.

\bibitem{Emam:2007dy}
W.~Emam and S.~Khalil,
\newblock Eur. Phys. J. {\bf C52}, 625 (2007), 0704.1395.

\bibitem{Basso:2008iv}
L.~Basso, A.~Belyaev, S.~Moretti, and C.~H. Shepherd-Themistocleous,
\newblock Phys. Rev. {\bf D80}, 055030 (2009), hep-ph/0812.4313.

\bibitem{Emam:2008zz}
W.~Emam and P.~Mine,
\newblock Erratum-ibid. {\bf G36}, 129701 (2009).

\bibitem{Huitu:2008gf}
K.~Huitu, S.~Khalil, H.~Okada, and S.~K. Rai,
\newblock Phys. Rev. Lett. {\bf 101}, 181802 (2008), 0803.2799.

\bibitem{Basso:2009hf}
L.~Basso, A.~Belyaev, S.~Moretti, and G.~M. Pruna,
\newblock JHEP {\bf 10}, 006 (2009), 0903.4777.

\bibitem{Basso:2010pe}
L.~Basso, A.~Belyaev, S.~Moretti, G.~M. Pruna, and C.~H.
  Shepherd-Themistocleous,
\newblock (2010), 1002.3586.

\bibitem{Basso:2010jt}
L.~Basso, A.~Belyaev, S.~Moretti, and G.~M. Pruna,
\newblock Phys. Rev. {\bf D81}, 095018 (2010), 1002.1939.

\bibitem{Basso:2010jm}
L.~Basso, S.~Moretti, and G.~M. Pruna,
\newblock Phys. Rev. {\bf D82}, 055018 (2010), 1004.3039.

\bibitem{Basso:2010hk}
L.~Basso, S.~Moretti, and G.~M. Pruna,
\newblock (2010), 1009.4164.

\bibitem{Basso:2010yz}
L.~Basso, S.~Moretti, and G.~M. Pruna,
\newblock Phys. Rev. {\bf D83}, 055014 (2011), 1011.2612.

\bibitem{Basso:2010si}
L.~Basso, S.~Moretti, and G.~M. Pruna,
\newblock (2010), 1012.0167.

\bibitem{Minkowski:1977sc}
P.~Minkowski,
\newblock Phys. Lett. {\bf B67}, 421 (1977).

\bibitem{VanNieuwenhuizen:1979hm}
P.~Van~Nieuwenhuizen and D.~Z. Freedman,
\newblock p. 341 (1979),
\newblock Amsterdam, Netherlands: North-Holland.

\bibitem{Yanagida:1979as}
T.~Yanagida,
\newblock (1979),
\newblock In Proceedings of the Workshop on the Baryon Number of the Universe
  and Unified Theories, Tsukuba, Japan, 13-14 Feb.

\bibitem{GellMann:1980vs}
M.~Gell-Mann, P.~Ramond, and R.~Slansky,
\newblock Print-80-0576 (CERN).

\bibitem{S.L.Glashow}
S.L. Glashow, in \emph{Quarks and Leptons}, eds. M.L\`evy et al. (Plenum, New
  York $1980$), p.~$707$.

\bibitem{Mohapatra:1979ia}
R.~N. Mohapatra and G.~Senjanovic,
\newblock Phys. Rev. Lett. {\bf 44}, 912 (1980).

\bibitem{Okada:2010wd}
N.~Okada and O.~Seto,
\newblock Phys. Rev. {\bf D82}, 023507 (2010), 1002.2525.

\bibitem{Khalil:2011tb}
S.~Khalil, H.~Okada, and T.~Toma,
\newblock (2011), 1102.4249.

\bibitem{Semenov:1996es}
A.~V. Semenov,
\newblock (1996), hep-ph/9608488.

\bibitem{Christensen:2008py}
N.~D. Christensen and C.~Duhr,
\newblock Comput. Phys. Commun. {\bf 180}, 1614 (2009), 0806.4194.

\bibitem{calchep_man}
http://www.ifh.de/$\sim$pukhov/calchep.html.

\bibitem{Pukhov:2004ca}
A.~Pukhov,
\newblock (2004), hep-ph/0412191.

\bibitem{Alwall:2007st}
J.~Alwall {\em et~al.},
\newblock JHEP {\bf 09}, 028 (2007), 0706.2334.

\bibitem{Hahn:2000jm}
T.~Hahn,
\newblock Nucl. Phys. Proc. Suppl. {\bf 89}, 231 (2000), hep-ph/0005029.

\bibitem{Chanowitz:1985hj}
M.~S. Chanowitz and M.~K. Gaillard,
\newblock Nucl. Phys. {\bf B261}, 379 (1985).

\bibitem{Fogli:2005cq}
G.~L. Fogli, E.~Lisi, A.~Marrone, and A.~Palazzo,
\newblock Prog. Part. Nucl. Phys. {\bf 57}, 742 (2006), hep-ph/0506083.

\bibitem{Fogli:2006yq}
G.~L. Fogli {\em et~al.},
\newblock Phys. Rev. {\bf D75}, 053001 (2007), hep-ph/0608060.

\bibitem{Pontecorvo:1957cp}
B.~Pontecorvo,
\newblock Sov. Phys. JETP {\bf 6}, 429 (1957).

\bibitem{Pontecorvo:1957qd}
B.~Pontecorvo,
\newblock Sov. Phys. JETP {\bf 7}, 172 (1958).

\bibitem{Maki:1962mu}
Z.~Maki, M.~Nakagawa, and S.~Sakata,
\newblock Prog. Theor. Phys. {\bf 28}, 870 (1962).

\bibitem{Pontecorvo:1967fh}
B.~Pontecorvo,
\newblock Sov. Phys. JETP {\bf 26}, 984 (1968).

\bibitem{Peskin:1990zt}
M.~E. Peskin and T.~Takeuchi,
\newblock Phys. Rev. Lett. {\bf 65}, 964 (1990).

\bibitem{Peskin:1991sw}
M.~E. Peskin and T.~Takeuchi,
\newblock Phys. Rev. {\bf D46}, 381 (1992).

\bibitem{Dawson:2009yx}
S.~Dawson and W.~Yan,
\newblock Phys. Rev. {\bf D79}, 095002 (2009), 0904.2005.

\bibitem{Buskulic:1996hz}
ALEPH, D.~Buskulic {\em et~al.},
\newblock Phys. Lett. {\bf B384}, 427 (1996).

\bibitem{Abreu:1994rp}
DELPHI, P.~Abreu {\em et~al.},
\newblock Nucl. Phys. {\bf B421}, 3 (1994).

\bibitem{Acciarri:1996um}
L3, M.~Acciarri {\em et~al.},
\newblock Phys. Lett. {\bf B385}, 454 (1996).

\bibitem{Alexander:1996ai}
OPAL, G.~Alexander {\em et~al.},
\newblock Z. Phys. {\bf C73}, 189 (1997).

\bibitem{Barate:2003sz}
LEP Working Group for Higgs boson searches, R.~Barate {\em et~al.},
\newblock Phys. Lett. {\bf B565}, 61 (2003), hep-ex/0306033.

\bibitem{Cacciapaglia:2006pk}
G.~Cacciapaglia, C.~Csaki, G.~Marandella, and A.~Strumia,
\newblock Phys. Rev. {\bf D74}, 033011 (2006), hep-ph/0604111.

\bibitem{Anthony:2003ub}
SLAC E158, P.~L. Anthony {\em et~al.},
\newblock Phys. Rev. Lett. {\bf 92}, 181602 (2004), hep-ex/0312035.

\bibitem{ew:2003ih}
LEP, {The ALEPH, DELPHI, L3, OPAL, SLD Collaborations, the LEP Electroweak
  Working Group, the SLD Electroweak and Heavy Flavour Groups},
\newblock (2003), hep-ex/0312023.

\bibitem{Azzi:2004rc}
CDF and D0 and Tevatron Electroweak Working Group, P.~Azzi {\em et~al.},
\newblock (2004), hep-ex/0404010.

\bibitem{Woods:2004zr}
SLAC E158, M.~Woods,
\newblock (2004), hep-ex/0403010.

\bibitem{Z-Pole}
{The ALEPH, DELPHI, L3, OPAL, SLD Collaborations, the LEP Electroweak Working
  Group, the SLD Electroweak and Heavy Flavour Groups},
\newblock Phys. Rept. {\bf 427}, 257 (2006), hep-ex/0509008.

\bibitem{Carena:2004xs}
M.~S. Carena, A.~Daleo, B.~A. Dobrescu, and T.~M.~P. Tait,
\newblock Phys. Rev. {\bf D70}, 093009 (2004), hep-ph/0408098.

\bibitem{Aaltonen:2008vx}
CDF, T.~Aaltonen {\em et~al.},
\newblock Phys. Rev. Lett. {\bf 102}, 031801 (2009), 0810.2059.

\bibitem{Aaltonen:2008ah}
CDF, T.~Aaltonen {\em et~al.},
\newblock Phys. Rev. Lett. {\bf 102}, 091805 (2009), 0811.0053.

\bibitem{Fogli:2008ig}
G.~L. Fogli {\em et~al.},
\newblock Phys. Rev. {\bf D78}, 033010 (2008), 0805.2517.

\bibitem{Luscher:1988gk}
M.~Luscher and P.~Weisz,
\newblock Nucl. Phys. {\bf B300}, 325 (1988).

\bibitem{Lee:1977eg}
B.~W. Lee, C.~Quigg, and H.~B. Thacker,
\newblock Phys. Rev. {\bf D16}, 1519 (1977).

\bibitem{Cynolter:2004cq}
G.~Cynolter, E.~Lendvai, and G.~Pocsik,
\newblock Acta Phys. Polon. {\bf B36}, 827 (2005), hep-ph/0410102.

\bibitem{Maalampi:1991fb}
J.~Maalampi, J.~Sirkka, and I.~Vilja,
\newblock Phys. Lett. {\bf B265}, 371 (1991).

\bibitem{Huffel:1980sk}
H.~Huffel and G.~Pocsik,
\newblock Zeit. Phys. {\bf C8}, 13 (1981).

\bibitem{Casalbuoni:1987cz}
R.~Casalbuoni, D.~Dominici, F.~Feruglio, and R.~Gatto,
\newblock Nucl. Phys. {\bf B299}, 117 (1988).

\bibitem{Aoki:2007ah}
M.~Aoki and S.~Kanemura,
\newblock Phys. Rev. {\bf D77}, 095009 (2008), 0712.4053.

\bibitem{Robinett:1986nw}
R.~W. Robinett,
\newblock Phys. Rev. {\bf D34}, 182 (1986).

\bibitem{BL_master_thesis}
L.~Basso,
\newblock {A minimal extension of the Standard Model with $B-L$ gauge
  symmetry},
\newblock Master's thesis, {Universit\`a degli Studi di Padova}, 2007,
\newblock
  {http://www.hep.phys.soton.ac.uk/$\sim$l.basso/B-L$\_$Master$\_$Thesis.pdf}.

\bibitem{Veltman:1980mj}
M.~J.~G. Veltman,
\newblock Acta Phys. Polon. {\bf B12}, 437 (1981).

\bibitem{Graudenz:1992pv}
D.~Graudenz, M.~Spira, and P.~M. Zerwas,
\newblock Phys. Rev. Lett. {\bf 70}, 1372 (1993).

\bibitem{Spira:1995rr}
M.~Spira, A.~Djouadi, D.~Graudenz, and P.~M. Zerwas,
\newblock Nucl. Phys. {\bf B453}, 17 (1995), hep-ph/9504378.

\bibitem{Gunion:1989we}
J.~F. Gunion, H.~E. Haber, G.~L. Kane, and S.~Dawson,
\newblock {\em {The Higgs hunter's guide}} (Addison Wesley, 1990).

\bibitem{CuhadarDonszelmann:2008jp}
T.~Cuhadar-Donszelmann, M.~Karagoz, V.~E. Ozcan, S.~Sultansoy, and G.~Unel,
\newblock JHEP {\bf 10}, 074 (2008), 0806.4003.

\bibitem{BahatTreidel:2006kx}
O.~Bahat-Treidel, Y.~Grossman, and Y.~Rozen,
\newblock JHEP {\bf 05}, 022 (2007), hep-ph/0611162.

\bibitem{Barger:2007im}
V.~Barger, P.~Langacker, M.~McCaskey, M.~J. Ramsey-Musolf, and G.~Shaughnessy,
\newblock Phys. Rev. {\bf D77}, 035005 (2008), 0706.4311.

\bibitem{Bhattacharyya:2007pb}
G.~Bhattacharyya, G.~C. Branco, and S.~Nandi,
\newblock Phys. Rev. {\bf D77}, 117701 (2008), 0712.2693.

\bibitem{Perez:2009mu}
P.~Fileviez~Perez, T.~Han, and T.~Li,
\newblock Phys. Rev. {\bf D80}, 073015 (2009), 0907.4186.

\bibitem{Asner:2010ve}
D.~M. Asner {\em et~al.},
\newblock (2010), 1004.0535.

\bibitem{Jadach:1988gb}
S.~Jadach and B.~F.~L. Ward,
\newblock Comput. Phys. Commun. {\bf 56}, 351 (1990).

\bibitem{:2007sg}
ILC, J.~Brau, (Ed.~) {\em et~al.},
\newblock (2007), 0712.1950.

\bibitem{Baer:1999ge}
H.~Baer, S.~Dawson, and L.~Reina,
\newblock Phys. Rev. {\bf D61}, 013002 (2000), hep-ph/9906419.

\bibitem{Castanier:2001sf}
C.~Castanier, P.~Gay, P.~Lutz, and J.~Orloff,
\newblock (2001), hep-ex/0101028.

\bibitem{Baur:2002rb}
U.~Baur, T.~Plehn, and D.~L. Rainwater,
\newblock Phys. Rev. Lett. {\bf 89}, 151801 (2002), hep-ph/0206024.

\bibitem{Baur:2009uw}
U.~Baur,
\newblock Phys. Rev. {\bf D80}, 013012 (2009), 0906.0028.

\bibitem{Baur:2003gpa}
U.~Baur, T.~Plehn, and D.~L. Rainwater,
\newblock Phys. Rev. {\bf D68}, 033001 (2003), hep-ph/0304015.

\bibitem{Plehn:2005nk}
T.~Plehn and M.~Rauch,
\newblock Phys. Rev. {\bf D72}, 053008 (2005), hep-ph/0507321.

\bibitem{Schael:2006cr}
ALEPH, S.~Schael {\em et~al.},
\newblock Eur. Phys. J. {\bf C47}, 547 (2006), hep-ex/0602042.

\bibitem{Cabibbo:1963yz}
N.~Cabibbo,
\newblock Phys. Rev. Lett. {\bf 10}, 531 (1963).

\bibitem{Kobayashi:1973fv}
M.~Kobayashi and T.~Maskawa,
\newblock Prog. Theor. Phys. {\bf 49}, 652 (1973).

\end{thebibliography}

\end{document}